\documentclass[11pt, letterpaper,openany]{article}

\usepackage[table]{xcolor}
\usepackage{amsmath}
\usepackage{amssymb}
\usepackage{amsthm}
\usepackage{graphicx}
\usepackage{hyperref}
\usepackage{url}
\usepackage{algorithm}
\usepackage{algorithmic}
\usepackage{caption}
\usepackage{subcaption}
\usepackage{booktabs}
\usepackage{authblk}
\usepackage{appendix}
\usepackage[utf8]{inputenc}
\usepackage[T1]{fontenc}
\usepackage[english]{babel}
\usepackage{geometry}
\usepackage{xcolor}
\usepackage{fancyhdr}
\usepackage{microtype}
\usepackage{parskip}
\usepackage{tabularx}
\usepackage{enumitem}
\usepackage{float}
\usepackage[numbers,sort]{natbib}
\hypersetup{
    colorlinks=true,
    linkcolor=blue,
    filecolor=magenta,      
    urlcolor=cyan,
    citecolor=red,
}

\definecolor{headercolor}{RGB}{0, 50, 100}
\title{Holistic Evaluation of GPT-4V for Biomedical Imaging}

\date{}

\newcommand*\samethanks[1][\value{footnote}]{\footnotemark[#1]}
\author[1]{Zhengliang Liu \thanks{Co-first authors.}}
\author[1,2]{Hanqi Jiang \samethanks}
\author[3]{Tianyang Zhong \samethanks}
\author[1]{Zihao Wu \samethanks}

\author[3]{Chong Ma \thanks{Co-second authors.}}
\author[1]{Yiwei Li \samethanks}
\author[4]{Xiaowei Yu \samethanks}
\author[5]{Yutong Zhang \samethanks}
\author[1,6]{Yi Pan \samethanks}
\author[1]{Peng Shu \samethanks}
\author[4]{Yanjun Lyu \samethanks}
\author[4]{Lu Zhang \samethanks}
\author[6]{Junjie Yao \samethanks}
\author[3]{Peixin Dong \samethanks}
\author[4]{Chao Cao \samethanks}
\author[6]{Zhenxiang Xiao \samethanks}
\author[6]{Huan Zhao \samethanks}
\author[7]{Jiaqi Wang \samethanks}
\author[1]{Shaochen Xu \samethanks}
\author[3]{Yaonai Wei \samethanks}

\author[8]{Jingyuan Chen}
\author[8]{Peilong Wang}
\author[1]{Haixing Dai}
\author[9]{Hao He}
\author[3]{Zewei Wang}
\author[3]{Xinyu Wang}
\author[9]{Xu Zhang}
\author[1]{Lin Zhao}
\author[9]{Yiheng Liu}
\author[10]{Kai Zhang}
\author[10]{Zhiling Yan}

\author[11]{Jun Liu}
\author[10]{Lichao Sun}
\author[9]{Ning Qiang}
\author[9]{Bao Ge}
\author[3]{Xiaoyan Cai}
\author[3]{Shijie Zhao}
\author[3]{Xintao Hu}
\author[12]{Yixuan Yuan}
\author[13]{Gang Li}
\author[7]{Shu Zhang}
\author[5]{Xin Zhang}
\author[6]{Xi Jiang}
\author[3]{Tuo Zhang}
\author[14,15,16]{Dinggang Shen}
\author[17]{Quanzheng Li}
\author[8]{Wei Liu}
\author[17]{Xiang Li}
\author[4]{Dajiang Zhu}
\author[1]{Tianming Liu\thanks{Corresponding author. E-mail: tliu@uga.edu}}

\affil[1]{School of Computing, University of Georgia, GA, USA}
\affil[2]{School of Computer and Information Technology, Beijing Jiaotong University, Beijing 100044, China}
\affil[3]{School of Automation, Northwestern Polytechnical University, Xi'an 710072, China}
\affil[4]{Department of Computer Science and Engineering, University of Texas at Arlington, TX, USA}
\affil[5]{Institute of Medical Research, Northwestern Polytechnical University, Xi'an 710072, China}
\affil[6]{School of Life Science and Technology, University of Electronic Science and Technology of China, Chengdu 611731, China}
\affil[7]{School of Computer Science, Northwestern Polytechnical University, Xi'an 710072, China}
\affil[8]{Department of Radiation Oncology, Mayo Clinic, Phoenix, Arizona, USA}
\affil[9]{School of Physics and Information Technology, Shaanxi Normal University, Xi'an 710119, China}
\affil[10]{Department of Computer Science and Engineering, Lehigh University, PA, USA}
\affil[11]{Department of Radiology, Second Xiangya Hospital, Central South University, Changsha, 410011, China}
\affil[12]{Department of Electronic Engineering, Chinese University of Hong Kong, Hong Kong 999077, China}
\affil[13]{Department of Radiology and BRIC, University of North Carolina at Chapel Hill, NC, USA}
\affil[14]{School of Biomedical Engineering, ShanghaiTech University, and Shanghai Clinical Research and Trial Center, Shanghai 201210, China}
\affil[15]{Shanghai United Imaging Intelligence Co., Ltd.}
\affil[16]{Shanghai Clinical Research and Trial Center}
\affil[17]{Department of Radiology, Massachusetts General Hospital and Harvard Medical School, MA, USA}

\begin{document}

\maketitle


\begin{abstract}
In this paper, we present a large-scale evaluation probing GPT-4V's capabilities and limitations for biomedical image analysis. GPT-4V represents a breakthrough in artificial general intelligence (AGI) for computer vision, with applications in the biomedical domain. We assess GPT-4V's performance across 16 medical imaging categories, including radiology, oncology, ophthalmology, pathology, and more. Tasks include modality recognition, anatomy localization, disease diagnosis, report generation, and lesion detection. The extensive experiments provide insights into GPT-4V's strengths and weaknesses. Results show GPT-4V's proficiency in modality and anatomy recognition but difficulty with disease diagnosis and localization. GPT-4V excels at diagnostic report generation, indicating strong image captioning skills. While promising for biomedical imaging AI, GPT-4V requires further enhancement and validation before clinical deployment. We emphasize responsible development and testing for trustworthy integration of biomedical AGI. This rigorous evaluation of GPT-4V on diverse medical images advances understanding of multimodal large language models (LLMs) and guides future work toward impactful healthcare applications.

\end{abstract}

\tableofcontents

\newpage

\listoffigures

\newpage

\section{Introduction}

Large language models (LLMs) such as ChatGPT \cite{openaiIntroducingChatGPT} and GPT-4 \cite{openai2023gpt4} have demonstrated immense progress in natural language processing~\cite{liu2023summary,zhou2023comprehensive,zhao2023brain,liu2022survey,rothman2022transformers,rahaman2023chatgpt}. However, their ability to interpret and leverage visual information remains relatively underexplored~\cite{li2023artificial,bubeck2023sparks,zhang2023segment,chen2023ma,kim2023medivista}, especially in specialized domains such as medicine and the general biomedical field \cite{li2023comprehensive,nori2023capabilities,li2023artificial,cai2023multimodal,wang2023prompt,liu2023artificial,holmes2023evaluating,ma2023impressiongpt,holmes2023benchmarking}. In this paper, we present a large-scale evaluation probing GPT-4V's capabilities and limitations for biomedical image analysis. 

\subsection{Background: The Rise of LLMs}

The advancement of LLMs has its roots in fundamental breakthroughs in deep learning for natural language processing \cite{liu2023summary,zhou2023comprehensive,liu2023radiology,liu2023radonc,liu2023transformation,liu2023evaluating,liuradiology}. Transformers \cite{vaswani2017attention}, introduced in 2017, marked a significant advancement in natural language processing by overcoming key limitations present in earlier architectures such as Recurrent Neural Networks (RNNs) \cite{grossberg2013recurrent} and Long Short-Term Memory networks (LSTMs) \cite{hochreiter1997long}. While RNNs and LSTMs were foundational in handling sequential text data, they struggled with parallelization and were constrained by the length of sequences they could effectively model. In contrast, transformer models, exemplified by BERT \cite{devlin2018bert} and GPT-2 \cite{radford2019language}, employ an attention mechanism that allows for the processing of long-range dependencies in text without the same constraints on sequence length. This capability significantly enhances the efficiency and scalability of LLMs, forming the basis for the substantial progress seen in subsequent language model developments.

Autoregressive language models such as the GPT series \cite{liu2023summary} generated text probabilistically one token at a time, allowing flexible and coherent text generation. GPT-1 \cite{radford2018improving}, developed by OpenAI in 2018, was one of the first autoregressive LLMs. More recent models such as GPT-2 \cite{radford2019language} and GPT-3 \cite{brown2020language} built on these foundations by pre-training transformer-based autoregressive LLMs on ever-increasing textual corpora scraped from the internet. For instance, GPT-3 was pre-trained on hundreds of billions of words. This enabled the models to learn nuances of natural language and world knowledge. Pre-training on massive data from diverse sources enables language models to capture a vast amount of world knowledge and linguistic patterns \cite{gu2021domain,zhao2023brain,liu2023summary}. InstructGPT \cite{ouyang2022training} built on this by incorporating instruction fine-tuning and reinforcement learning from human feedback (RLHF) into the training process. RLHF helped align the model outputs better with human preferences and values \cite{casper2023open}. Together, these innovations significantly enhanced the natural language processing capabilities of LLMs.

These advancements culminated in the release of ChatGPT in late 2022. ChatGPT leveraged the power of pre-training and RLHF to achieve impressive performance on complex language tasks like reasoning, question answering, summarization and dialogue. Its conversational abilities led ChatGPT to gain immense popularity among the public \cite{gill2023chatgpt,liu2023summary}. However, ChatGPT was limited to only text modalities.

Humans naturally process information multimodally, integrating vision, language, and auditory inputs \cite{zhao2023brain}. In addition, multimodal processing is not only a cognitive reality but also a practical necessity in many real-world applications. For instance, autonomous vehicles rely on integrating visual, spatial, and auditory data to navigate safely. In healthcare, diagnostic systems benefit from analyzing visual, textual, and sensor data to provide comprehensive patient assessments. Social media platforms use multimodal AI to understand and moderate content that includes text, images, and videos. 

This understanding prompted the extension of LLMs to encompass multimodal intelligence, capable of handling images, videos, and speech. Such advancement led to the creation of models like Kosmos-1 \cite{huang2023language}, PaLM-E \cite{driess2023palm}, and GPT-4V \cite{openaiGPT4VisionSystem}, all of which can process both images and text. For example, GPT-4V was pre-trained on vast datasets that included both text and images sourced from the internet, enabling it to develop visual capabilities alongside its linguistic proficiency. Specifically, GPT-4V enhances the visual processing strengths of the model. The advent of such multimodal LLMs has broadened the horizons and created new possibilities at the confluence of natural language processing, computer vision, and human-AI interaction.

\subsection{Motivation}
Biomedical images contain intricate visual details and domain knowledge~\cite{wang2023r2gengpt,singhal2023towards,kim2023medivista,iglesias2015multi,bertero2006inverse,qiang2023functional,bi2023community,liu2022discovering,zhang2023beam,zhang2023differentiating,ding2023deep,qiang2023deep,dai2022graph} that can be difficult even for human experts to interpret. Indeed, specialized domains like radiology~\cite{drew2013informatics,van2017visual,wu2023exploring,zhong2023chatradio,chen2023ma} and pathology~\cite{crowley2003development,jaarsma2015expertise} rely on perceiving subtle visual indicators and nuanced clinical knowledge to detect disease and guide diagnoses. For example, a radiologist must discern ambiguous masses, faint shadows, texture patterns, and other anomalies to identify cancer on a CT scan~\cite{maurer2021early,makaju2018lung,mets2011identification}. This requires extensive training to recognize disease manifestations versus normal variations in human anatomy. Similarly, a pathologist relies on detecting slight deviations in cell morphology through a microscope to diagnose illness~\cite{deture2019neuropathological,xing2016robust,uranova2001electron}. The intricacy of biomedical imagery goes far beyond generic object recognition. 

However, most benchmark datasets for evaluating AI systems consist of everyday internet images~\cite{antol2015vqa,lin2014microsoft,everingham2010pascal,deng2009imagenet,krizhevsky2009learning,lecun1998gradient,dai2023hierarchical,dai2023samaug,li2023artificial}. This motivated our work to rigorously test GPT-4V's capabilities using real-world medical imaging data. By evaluating performance on specialized tasks like localization of anatomical structures and classification of radiography modalities, our study provides empirical insights into how large language models fare on the complex visual reasoning challenges inherent in biomedical domains. Being able to automate and enhance biomedical image understanding could have profound impacts on healthcare, biological research, and medical education~\cite{yang2023dawn,cheng2023exploring,wang2023accelerating,yang2023performance,zhao2023generic,zhou2023fine,cai2023exploring}. For instance, AI models and tools that can analyze microscope images~\cite{merchant2022microscope,xing2018deep}, radiology scans~\cite{gharaibeh2022radiology,liu2023radiology}, or anatomical diagrams~\cite{cornejo2022anatomical} could help accelerate scientific discoveries, improve clinical diagnostics, and augment medical training. Similarly, in the realm of medical robotics~\cite{rodrigues2022surgical}, the integration of sophisticated AI systems capable of interpreting complex medical imagery in real-time can significantly advance the precision and adaptability of robotic surgeries and interventions.

\begin{figure}
\centering

  \includegraphics[width=16cm]{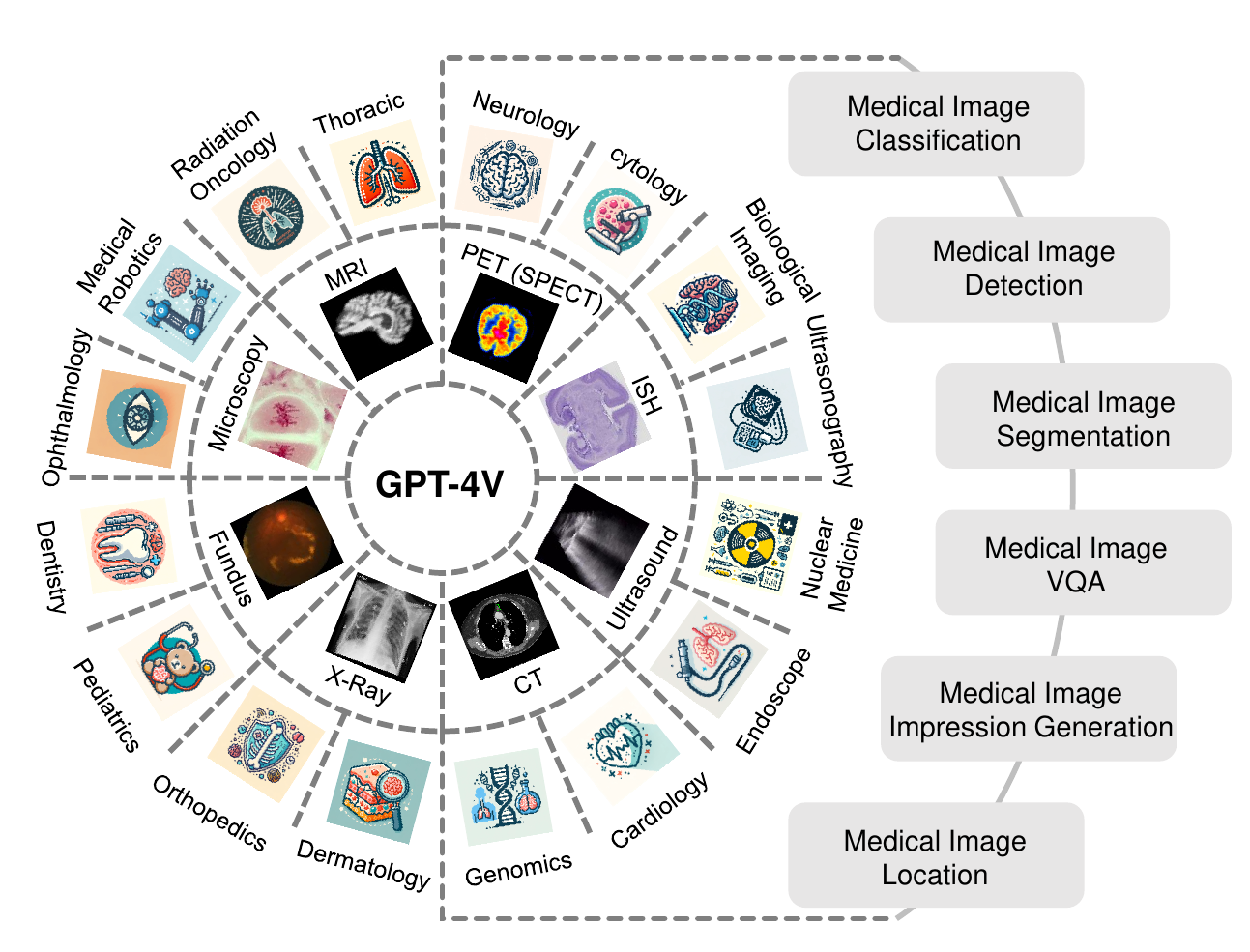}

\caption[Schematic Overview of the Evaluation Methodology]{\textbf{Schematic Overview of the Evaluation Methodology.} The periphery of the diagram delineates the array of medical departments subjected to our analysis, whereas the core layer graphically represents the medical imaging modalities under scrutiny. The efficacy of GPT-4V was methodically gauged across a spectrum of six distinct medical tasks.}
\label{fig:main}      
\end{figure}

\subsection{Study Objectives}
In our study, we aimed to evaluate GPT-4V across a diverse set of capabilities required for real-world medical imaging tasks. A comprehensive visualization of the evaluation is depicted in Figure \ref{fig:main}, where the structural hierarchy of the testing scope is illustrated. For various modalities including X-ray, MRI, CT, and microscopy images, we assessed GPT-4V on its ability to recognize the imaging modality itself. We tested the localization of anatomical structures within medical images, which requires an understanding of human anatomy. Additionally, we benchmarked image classification skills through diagnosis and disease identification from medical images, a crucial application for clinical use. Finally, we explored GPT-4V's capacity for generating free-text reports summarizing image findings, which could provide tremendous workflow efficiencies if automated successfully. By probing performance on modality recognition, region localization, image classification, and report generation, our experiments provide a comprehensive portrait of GPT-4V's strengths and weaknesses at the core skills necessary for biomedical imaging applications. Section 3 details our rigorous assessment across each of these capabilities using diverse benchmark medical imaging data.

Our study encompasses an expansive range of biomedical imaging domains, spanning various medical specialties, biological research, and healthcare applications. We evaluate GPT-4V across chest radiography data, including X-rays from the MIMIC dataset \cite{johnson2019mimic}, to probe interpretive skills for pulmonary conditions. Additionally, we test capabilities in neuroimaging by assessing performance on MRI data for detecting neurological pathologies. Oncological imaging analysis is examined through GPT-4V's interpretation of MRI and CT scans used in cancer radiotherapy planning and monitoring. Beyond radiology, we explore GPT-4V's skills in cytopathology for cancer diagnosis, using public data of microscopic cell imaging. Ophthalmological imaging, medical robotics, and other specialty areas are incorporated to provide a comprehensive perspective.

Importantly, our work also assesses capabilities on data modalities highly relevant in biological research, such as optical microscopy for cellular imaging~\cite{bagheri2022new}, in situ hybridization for gene expression~\cite{leung2022current}, and functional MRI~\cite{loh2022probing} in neuroscience studies. This expands the evaluation of GPT-4V's interpretive skills beyond clinical radiology into wider biomedical imaging applications. We believe our heterogeneous datasets better represent the true diversity of visual reasoning tasks required in real-world medical and biological settings compared to benchmarks constrained to a single specialty. By holistically evaluating GPT-4V across this expansive range of biomedical imaging data, our findings provide multifaceted insights into the current capabilities and limitations of large language models for supporting various medical specialties, scientific research, and potential healthcare applications.

\subsection{Potential Applications and Implications}
As powerful AGI systems such as GPT-4V evolve to tackle complex tasks like medical imaging, upholding rigorous validation and ethical principles becomes increasingly crucial~\cite{liu2023deid,rayhan2023ethical,zhao2023brain,zhang2023one}. Real-world deployment of biomedical AGI demands extensive testing across diverse patient populations and clinical settings to characterize strengths and limitations~\cite{mishra2022external}. We must proactively assess models for potential biases along lines of race, gender, age, and other factors that may impact equitable performance~\cite{thapa2023chatgpt}. Maintaining transparency about AI capabilities and thought processes can help uphold accountability if issues emerge. Ultimately, biomedical AGI should aim to augment clinician expertise and efficiency without compromising personalized care. Human discretion, oversight, and responsibility must remain central to any high-stakes medical application of AI. By emphasizing comprehensive validation, earnest bias evaluation, and clinician-centered design, our research aims to responsibly advance biomedical AI in a manner that earns public trust and optimism about these transformative technologies.

By evaluating GPT-4V on a wide range of biomedical images, we identify both promising capabilities and potential limitations of LLMs for this vital application domain. Our findings will help guide the responsible development and deployment of AI for healthcare and biological research. The remainder of this paper presents our experimental setup, results, and discussion of GPT-4V's proficiencies and areas for improvement in biomedical visual intelligence.

\section{Scope of the Study and Used Public Datasets}
\label{sec:scope}

In our study, we aim to explore and evaluate GPT-4V's capabilities and limitations across various domains of medical imaging. Below is a comprehensive list of the domains we have included in our research.

\subsection{Chest Radiography}
\label{subsec:chest-radiography}
Assessing GPT-4V's ability to interpret chest X-rays, including data from MIMIC-CXR~\cite{johnson2019mimic}, CheXpert~\cite{irvin2019chexpert}, ChestXray2017~\cite{kermany2018labeled}, COVID-Qu-Ex~\cite{tahir3122958covid}, OpenI~\cite{demner2016preparing}, SIIM-ACR~\cite{SIIM-ACR}, and NIH Chest X-rays~\cite{wang2017chestx} datasets. The MIMIC-CXR~\cite{johnson2019mimic} dataset contains 377,110 chest X-ray images and 227,835 diagnostic reports, with each study having a corresponding multi-class label (14 common chest disease labels extracted by CheXpert Laberler~\cite{irvin2019chexpert}). We tested the ability of GPT-4V to directly output multi-class labels based on the images. Additionally, we evaluated GPT-4V's capability to generate structured diagnostic reports (including findings and impression sections) based on the images. From the CheXpert~\cite{irvin2019chexpert} dataset, we selected a subset of frequently encountered categories (Atelectasis, Cardiomegaly, Edema, Consolidation, Pleural Effusion) in zero-shot classification tasks and attempted direct classification using GPT-4V on the corresponding images. The ChestXray2017~\cite{kermany2018labeled} dataset consists of 5,863 chest X-ray images, including two classes: pneumonia and normal. We conducted binary classification tasks based on images using GPT-4V. Similarly, for the COVID-Qu-Ex~\cite{tahir3122958covid} dataset, which provides binary classification labels and mask regions of COVID-19 lesions, we tested GPT-4V's ability to diagnose and locate COVID-related abnormalities in chest X-ray images. The OpenI~\cite{demner2016preparing} dataset consists of 3,955 chest X-ray images with corresponding diagnostic reports. We evaluated GPT-4V's capability to generate diagnostic reports on this dataset. The SIIM-ACR~\cite{SIIM-ACR} dataset provides 2,379 chest X-ray images for pneumothorax segmentation. We tested GPT-4V's ability to directly locate pneumothorax lesions based on the images. The NIH Chest X-rays~\cite{wang2017chestx} dataset contains 33,920 chest X-ray images, each with corresponding multi-class labels (14 classes). A small subset of images in this dataset also includes bounding box information for diseases. Similar to the COVID-Qu-Ex dataset, we tested GPT-4V's multi-class classification ability and its capability to locate lesion regions based on the images.

\subsection{Neuroimaging}
\label{subsec:neuroimaging}
In order to verify whether GPT-4V can understand neuronal image datasets and brain images, we tested on SEU-ALLEN's R1741~\cite{peng2021morphological} dataset and the Allen Institute's mouse brain whole-brain atlas. The R1741 dataset is a whole-brain neuron reconstruction dataset completely manually annotated and reconstructed on whole-brain-level fMOST images of the mouse brain. It has several major advantages. First, the resolution of the fMOST~\cite{gong2016high} image is particularly high. In addition, the neurons in the R1741 dataset are all labeled using sparse labeling technology. The quality of the neuron images is particularly easy to identify. The image quality is compared with similar images. Second, R1741 neurons are all manually labeled and repeatedly checked, and their reliability is also very high. Third, because the size of whole-brain images can often reach dozens of terabytes, special visualization tools (such as Vaa3D~\cite{peng2014extensible}) are required to visualize them. This also means that during the training process of GPT-4V, there are almost no opportunities to encounter similar problems. For images, verification on such datasets is a great test of GPT-4V’s emergent abilities. The Allen Brain Institute's whole-brain atlas~\cite{yao2023high} is a template map of mouse brain regions that is highly recognized in the field. Mouse brain regions generally have specific name abbreviations. We tested whether GPT-4V can obtain the brain regions from the image. The relationship between accurate information and adjacent brain areas is very meaningful for testing whether GPT-4V has knowledge transfer and synthesis capabilities. 

Since whole-brain images are generally displayed in 3D, in this test, we selected parts with clearer imaging, sparser neuron fibers, and fewer interference signals to generate 2D screenshots and give them to GPT-4V for interpretation. In the task, we provide GPT-4V with additional information about neuron imaging and neuron skeleton reconstruction to help the model understand the task. GPT-4V will be required to first understand the original neuron image, then superimpose the image of the neuron reconstruction results, and finally evaluate the quality of the neuron reconstruction results before and after optimization. The image understanding ability of GPT-4V was tested by gradually increasing the difficulty of understanding the neuron reconstruction results.

The experimental results show that it is difficult for GPT-4V to directly recognize neuron images without additional information. GPT-4V will mistake neurons for cell fiber diagrams, and after prompts, it can provide knowledge of neuron images. On the one hand, this shows that GPT-4V rarely comes into contact with neuron images during the training process. On the other hand, it also shows that GPT-4V can have a good understanding of neuron images after being given certain background knowledge. 

After superimposing the neuron reconstruction results on the image, GPT-4V can also understand and evaluate the quality of the neuron results from a very professional perspective, which is very surprising. GPT-4V can detect very detailed problems in neuron reconstruction results, such as intersections in neuron reconstruction results, incomplete reconstruction of terminals, incorrect connections in neuron reconstruction results, etc. In addition, for the brain template map, GPT-4V can understand the information marked on the brain template map and can also provide relevant background information, which shows that GPT-4V can master basic neurobiology knowledge.

\subsection{Oncological Imaging for Radiotherapy}
\label{subsec:oncological-imaging-radiotherapy}
We presented an in-depth evaluation of GPT-4V using the Burdenko Glioblastoma Progression Dataset (Burdenko-GBM-Progression)\cite{burdenko_gbm} and the Glioma Image Segmentation for Radiotherapy Dataset (GLIS-RT)\cite{gooya2012glistr} on The Cancer Imaging Archive (TCIA) website\cite{cancerimagingarchive2013}, a comprehensive resource of biomedical imaging. These datasets stand out for its voluminous collection of imaging data, encompassing an array of high-resolution Magnetic Resonance Imaging (MRI) and Computed Tomography (CT) scans. Each case in the dataset has included MR sequences and CT scans with the radiotherapy targets, Gross Tumor Volume (GTV), Clinical Target Volume (CTV) as well as organs at risk annotated, making it a treasure trove for testing multimodal language models. 

Moreover, we provided an assessment of GPT-4V utilizing the Large-Scale CT and PET/CT Dataset for Lung Cancer Diagnosis dataset (Lung-PET-CT-Dx)\cite{lungpetctdx2020}. This dataset contribute to the test by providing its voluminous collection of imaging data, encompassing an array of high-resolution computed tomography (CT) and positron emission tomography-computed tomography (PET/CT) scans. The Lung-PET-CT-Dx dataset’s extensive annotations and diverse imaging modalities make it a pivotal dataset for evaluating AI models like GPT-4V in the domain of lung cancer imaging.

Finally, we conducted a brief test of GPT-4V using breast X-ray images sourced from the Digital Database for Screening Mammography dataset (DDSM)\cite{rose2006web}. The results of this evaluation are included in Appendix ~\ref{appendix:radonc} for reference.

\subsection{Cytopathology in Cancer Diagnosis}
\label{subsec:cytopathology-cancer}
In the field of oncological cytopathology, our study utilized the LC25000 dataset~\cite{borkowski2019lung}, which encompasses 25,000 color images across five categories, namely colorectal adenocarcinoma, benign colorectal tissue, lung adenocarcinoma, lung squamous cell carcinoma, and benign lung tissue. In addition, we employed the Acute Lymphoblastic Leukemia (ALL) image dataset~\cite{aria2021acute}, containing 3,256 images of peripheral blood smears from patients suspected of having ALL.

The aim of our experiment was to challenge the GPT-4V model to diagnose cancer within cytopathological images by evaluating cellular morphology, nuclear features, cell arrangement, and cellular heterogeneity, among other aspects. The results demonstrated GPT-4V's exceptional proficiency and accuracy in the recognition of cytopathological imagery. The model was adept at identifying cellular structures and, by integrating changes in these structures with pathological knowledge, it could analyze and diagnose pathological alterations within the images. This capability holds significant potential application in distinguishing between normal and diseased cancer cells. However, GPT-4V's performance in determining the stage of cancer progression requires further enhancement. Future research will focus on improving the model's accuracy in the diagnostic staging of pathology.

Our findings indicate that GPT-4V possesses substantial potential to aid in clinical pathology diagnostics. Nascent multimodal foundation models such as GPT-4V can assist physicians in the preliminary selection of cell samples requiring further examination, thus enhancing diagnostic efficiency and alleviating the workload on medical professionals. In the realm of medical education and training, GPT-4V can support medical students and clinicians in learning to identify key features within cytopathological imagery. With ongoing technological advancements, this tool is poised to play a pivotal role in precision medicine, tele-diagnosis, and the formulation of personalized treatment plans.

\subsection{Ophthalmological Imaging}
\label{subsec:ophthalmological-imaging}
Ophthalmological imaging can encompass various modalities to aid in the diagnosis and detection of eye diseases, such as Fundus Fluorescein Angiography (FFA) and Optical Coherence Tomography (OCT). In this study, we use Color Fundus Photography (CFP) images, which is the most common imaging modality in ophthalmology, to explore the potential of GPT-4V in the field of ophthalmological imaging. Through color fundus photography, ophthalmologists are able to directly visualize structures such as blood vessels, the retina, the optic disc, and the macula. This aids in the timely detection and diagnosis of various retinal diseases, including diabetic retinopathy (DR) and glaucoma. We conducted tests to evaluate GPT-4V's performance in diagnosing four distinct ophthalmological diseases, including age-related macular degeneration (AMD) evaluated using the iChallenge-AMD dataset \cite{dt4f-rt59-20}, diabetic retinopathy evaluated using the iDRiD dataset \cite{h25w98-18}, pathological myopia (PM) evaluated with the iChallenge-PM dataset \cite{55pk-8z03-19}, and glaucoma evaluated using the iChallenge-GON dataset \cite{ORLANDO2020101570, li2020development}. Furthermore, we explored GPT-4V's capabilities in Optic Disc localization and segmentation.

We employed the GPT-4V model to analyze ophthalmological images, comparing its analysis descriptions and identifying specific pathological features with ground truth to evaluate its ability to identify ophthalmological diseases. Our experimental results indicate that GPT-4V possesses the capacity to identify structures like the optic disc, blood vessels, and lesions such as microaneurysms while generating textual reports. However, as it cannot provide precise location coordinates or segmentation masks, we cannot precisely judge its diagnostic accuracy. This ability can serve as a preliminary screening tool in clinical settings, guiding healthcare professionals in making informed diagnostic decisions. Although GPT-4V's utility as a direct diagnostic tool may be limited at present, it shows promise in enhancing diagnostic efficiency.

\subsection{Medical Robotics Imaging}
\label{subsec:medical-robotics-imaging}
To rigorously examine the capability of GPT-4V in terms of robotic imagery, we use the endoscopic video dataset CholecT50~\cite{nwoye2023cholectriplet2021} of laparoscopic cholecystectomy surgery, which consists of 50 videos, 45 of which are from the Cholec80 dataset, and 5 of which are from an internal dataset of the same surgical procedure. The dataset is labeled with ternary information in the format <instrument, verb, target>, detailing the instrument visible in the video, the action being performed by the instrument, and the surgical site involved in the instrument, respectively. Using the GPT-4V model, the researchers sought to test its ability to decode these triplets and attempted to evaluate its zero-shot performance in deciphering robotic images.

Throughout the experimental phase, the GPT-4V model devised cues to elicit descriptions of movements (expressed through verbs) and guesses about possible anatomical sites, based on the images provided. The model was then instructed to summarize its analysis in three terms: instrument name, surgical action, and anatomical site~\cite{nwoye2022rendezvous}.

The experimental results show that GPT-4V can grasp some common sense knowledge such as the name of the medical device and the action in progress. Still, it is limited by the consistency problem of language comprehension and cannot accurately use the unified terminology to give the answer. Meanwhile, in the refined medical field, it is difficult for GPT-4V to give accurate answers, for example, when inquiring about the specific parts of the anatomy, GPT-4V can only give possible answers based on the color and texture of the anatomical parts, but the specific conclusions need to be provided with more hints of information, which need to be guided by the relevant experts.

\subsection{Neurological Disease Imaging}
\label{subsec:neurological-disease-imaging}
To explore GPT-4V's potential in interpreting neurological disease imaging, we leverage the extensive Alzheimer's Disease Neuroimaging Initiative (ADNI) dataset, a vital resource in Alzheimer's disease research \cite{Petersen2010ADNI, Jack2008ADNIMRI}. GPT-4V, known for its multimodal capabilities, holds significant potential for extracting valuable insights from the wealth of neuroimaging data within the ADNI dataset. These MRI images are instrumental in advancing our understanding of Alzheimer's disease by allowing researchers to track changes in the brain as the disease progresses.

One promising avenue of exploration entails using GPT-4V to analyze MRI images for structural changes indicative of Alzheimer's disease. This involves utilizing the enormous knowledge of the model to recognize specific patterns, such as the presence of amyloid plaques or regions of brain atrophy commonly associated with the disease. It is of great interest to assess the model's capability to provide automated evaluations of these patterns in new MRI scans, which can assist in the early detection and monitoring of Alzheimer's disease progression.

Additionally, GPT-4V could be employed in generating textual reports based on MRI images. By analyzing the images and extracting relevant information, the model could produce detailed reports summarizing the observed findings, potentially reducing the burden on radiologists and clinicians in the diagnostic process. Furthermore, GPT-4V's capabilities extend beyond mere pattern recognition. With its natural language understanding, the model can assist in analyzing MRI findings and extracting relevant information to produce detailed reports summarizing observed findings.

The integration of GPT-4V in the analysis and interpretation of ADNI MRI images could open new horizons in the field of neuroimaging research and contribute to a more comprehensive understanding of Alzheimer's disease. While the potential is promising, it is important to emphasize that any application in the clinical setting would require rigorous validation, adherence to ethical guidelines, and collaboration with domain experts to ensure the highest quality and accuracy in the analysis and interpretation of these critical medical images.

\subsection{Biological Imaging}
\label{subsec:biological-imaging}
Aiming at assessing GPT-4V's understanding of complex biological imaging data, specifically the optical cell imaging data, we mainly conduct experiments on The Cell Image Library~\cite{Wong} and Cell Tracking Challenge~\cite{mavska2023cell}. The Cell Image Library is a repository for optical images of cells from a variety of organisms. It hosts a vast array of cell images from multiple species, offering a broad spectrum for comparative studies. These images are curated and standardized, ensuring that GPT-4V has access to images that are clear, detailed, and consistent in quality, which is critical for accurate analysis and interpretation. High-resolution images enable the identification of subtle cellular structures and processes that are pivotal for breakthroughs in understanding cell biology. Additionally, the images are often annotated by experts, which provides reliable metadata and context for testing GPT-4V's ability to comprehend optical cell imaging data. This curation process enhances the reliability and validity of the data, making it a robust resource for advanced biomedical image analysis. Here in this experiment, we select optical images of a diploid lily during the cell division stage, which is the basis of all life activities. With ground truth descriptions provided, we compare these descriptions with the contents generated by GPT-4V to assess whether they match up to each other.

For Cell Tracking Challenge, it comprises a comprehensive collection of time-lapse sequences of varying imaging modalities, cell types, and cell shapes designed to benchmark cell tracking algorithms, understanding the mechanobiology of cell migration and its multiple implications in both normal tissue development and many diseases. Utilizing this dataset in this optical cell imaging research offers multifaceted benefits, including access to a standardized, diverse set of time-lapse images that ensure comparability across our GPT-4V study. Furthermore, the dataset contains either 2D+Time datasets or 3D+Time datasets. Modalities in both spatial and time dimensions can comprehensively evaluate GPT-4V's ability to extract kinetic and morphic features of objects in the visual input. In this experiment, we select images from HeLa cells on a flat glass and human hepatocarcinoma-derived cells expressing the fusion protein YFP-TIA-1. GPT-4V is asked to provide the location information of cells in each image in the form of bounding boxes. As GPT-4V cannot directly generate images from commands, these bounding boxes are visualized manually to check the cell segmentation performance.

Our experimental results indicate that the GPT-4V model demonstrates a rudimentary comprehension of visual inputs, successfully identifying objects within images and leveraging prompts to establish connections with domain-specific knowledge in optical cell imaging, articulating findings in the appropriate scientific nomenclature. In the task of cell division stage recognition, GPT-4V was able to discern specific cellular structural components within images, such as chromosomes, plates, and poles, and conducted a preliminary spatial analysis of these components, assessing whether they were aligned or exhibited crossover. This suggests that GPT-4V can not only recognize individual elements within an image but also make basic judgments regarding their spatial orientation, integrating domain-specific terminology from optical cell imaging prompts in its descriptions.

However, the model displays inaccuracies, such as omitting or misjudging critical positional information. For instance, a stage of cell division that is correctly identified in the ground truth as "Telophase I" is erroneously recognized by GPT-4V as "Prophase I" due to a misidentification of decondensation as condensation. The limitations of GPT-4V are further manifested in cell segmentation tasks, where its deficiencies in the identification of minute scales and spatial comprehension are apparent. Grounding boxes derived from GPT-4V, once visualized and encapsulated in red outlines, reveal a coarse distribution characteristic, and upon closer inspection, the segmentation outcomes are largely ineffective. This reveals that while GPT-4V possesses foundational capabilities in the interpretation and processing of visual information, further improvements are still required to enhance its performance in professional domains.

\subsection{Cardiac Imaging}
\label{subsec:cardiac-imaging}
In order to evaluate GPT-4V's capability for interpreting cardiac images, we conducted the experiment based on the ADCD (Automated Cardiac Diagnosis Challenge) dataset \cite{bernard2018deep}. The overall ACDC dataset was created from real clinical exams acquired at the University Hospital of Dijon. Acquired data were fully anonymized and handled within the regulations set by the local ethical committee of the Hospital of Dijon (France). The dataset is composed of 150 exams (all from different patients) divided into 5 evenly distributed subgroups (4 pathological plus 1 healthy subject group). Furthermore, each patient comes with the following additional information: weight, height, as well as diastolic and systolic phase instants. The training database is composed of a total of 100 patients, divided equally into five groups as follows: healthy patients; patients with previous myocardial infarction; patients with dilated cardiomyopathy; patients with hypertrophic cardiomyopathy; and patients with abnormal right ventricle. The testing database is within 50 patients and 10 patients for each group.

Since cardiac cine-MRIs consist of a series of sequential frames and the maximum number of images GPT-4V accepts for each prompt is four, we combine the slices of diastolic and systolic phases into two individual images. During the task, these images are provided to GPT-4V with additional information. In detail, our prompts include the modality of the images as well as the phase of these two images. The goal is to explore whether GPT-4V has the capability to analyze cardiac cine-MRIs and is able to find cardiac abnormality.

Experiment results of our test suggest that if not including the modality of the image, GPT-4V will easily recognize them to be abdominal CT images. With modality in the prompts, GPT-4V can provide abundant information for a general approach to how a specialist might analyze such images. However, its capability for analyzing cardiac cine-MRIs seems not to be stable because there exist cases where it refuses to show observations of the images. As for observations it generates, a general correct conclusion is given for normal cases. It can also figure out potential cardiac problems such as myocardial infarction. But accuracy of abnormality recognition is low. It lacks specific analysis based on the aspects it mentions on how a specialist will focus on. The limited number of input images is the main barrier to cine-MRI analysis in our experiment.

\subsection{Ultrasound Imaging}
\label{subsec:ultrasound-imaging}
When it comes to ultrasound images, our experiment employs two distinct datasets: COVIDx-US \cite{COVIDxUS2021}, a pivotal contribution to COVID-19 research, providing the largest known open-access collection of lung ultrasound images and videos for AI-enhanced diagnostic initiatives. It encompasses 242 videos and 29,651 images from various patient scenarios, including confirmed COVID-19 cases, other lung conditions like pneumonia, and normal findings. Each entry is paired with a lung ultrasound score, facilitating detailed analyses. Secondly, the Breast Ultrasound Dataset \cite{AlDhabyani2020} is an essential resource for advancing breast cancer research through machine learning. It comprises 780 ultrasound images in PNG format, collected from 600 female patients aged 25 to 75 in 2018. Categorized into normal, benign, and malignant classes, these images are paired with ground truth data for aiding in classification, detection, and segmentation tasks.

The GPT-4V model is tasked with analyzing and interpreting ultrasound images from both datasets to identify the condition depicted in each image. This approach tests the model's ability to discern nuanced medical imaging features that differentiate various health states. By presenting each image to GPT-4V and asking it to provide relevant information about the observed condition, we aim to assess the potential of advanced artificial intelligence in supporting medical diagnosis, with a focus on its accuracy in classifying ultrasound imagery according to disease state.

The findings of our experiment indicate that while GPT-4V does not possess the capability to diagnose medical conditions from ultrasound images explicitly, it has demonstrated proficiency in identifying features within the clinically relevant images. In the case of breast ultrasound analysis, GPT-4V successfully highlighted anomalies that warrant further investigation, suggesting an innate ability to detect irregular patterns or deviations from normality. This capability could be exceptionally useful in clinical practice, serving as a preliminary screening tool that assists medical professionals in pinpointing areas of concern. The model's guidance on key indicators within ultrasound imagery could potentially streamline the diagnostic process, enabling clinicians to focus on specific image regions that may exhibit signs of pathological changes. Moreover, this assists in educational settings, providing medical trainees with an AI-based framework for learning to discern critical features in ultrasound diagnostics. While GPT-4V's current utility as a direct diagnostic tool may be limited, its value in enhancing diagnostic efficiency and education in medical imaging is promising.

\subsection{Nuclear Medicine Imaging}
\label{subsec:nuclear-medicine-imaging}
In this section, we aim to assess the capabilities of GPT-4V, a state-of-the-art language model, in interpreting nuclear medicine imaging of the brain. To accomplish this task, the Harvard Medical Image Fusion Dataset \cite{hricak2021medical} is utilized, which comprises a comprehensive collection of 269 positron emission computed tomography (PET) images and 357 single-photon emission computed tomography (SPECT) images. Our evaluation mainly focuses on GPT-4V's ability to generate well-structured diagnostic reports based on the provided images. To further investigate its performance, we specifically select a subset of frequently encountered brain conditions for zero-shot diagnostic report generation tasks. Through direct analysis using GPT-4V, insights into its effectiveness in this context would be obtained.

In our experimental findings, we observe that GPT-4V demonstrates promising performance in generating diagnostic reports based on PET images. It exhibits the ability to recognize structures and modal information, enabling the generation of textual diagnostic reports. However, it is important to note that some errors are also observed. These results suggest a substantial degree of semantic alignment between the findings generated by GPT-4V and real diagnostic reports, indicating its potential in this domain. 

When it comes to generating diagnostic reports based on SPECT images, GPT-4V's performance in zero-shot testing is relatively good. During the evaluation of the diagnosis, GPT-4V can analyze the surface information of the image, but it fails to provide deep clinical information in this context. We attribute these experimental findings to the training regimen of GPT-4V. The model is primarily trained on general images and free text, allowing it to establish extensive relationships between images and text. However, for the task of unconventional images such as SPECT, it performs poorly in the field of nuclear medicine images. Therefore, it can be inspired that GPT-4V would benefit from further enhancement through advanced reasoning, multi-modal fusion, and artificial general intelligence
techniques.

\subsection{Endoscopic Imaging}
\label{subsec:endoscopic-imaging}
Assessing GPT-4V’s ability to interpret endoscopic images, including data from Kvasir-SEG \cite{jha2020kvasir}, m2caiSeg \cite{maqbool2020m2caiseg}, CholecSeg8k \cite{twinanda2016endonet}, and CVC-ClinicDB \cite{bernal2015wm} datasets. The Kvasir-SEG \cite{jha2020kvasir} dataset contains 1000 polyp images and their corresponding ground truth from the Kvasir Dataset v2. We tested the ability of GPT-4V to directly output the location of polyps based on the images. Additionally, we evaluated GPT-4V’s capability to generate structured diagnostic reports based on the images. From the m2caiSeg \cite{maqbool2020m2caiseg} dataset, we selected a series of endoscopic images during surgery in zero-shot classification tasks and attempted direct classification using GPT-4V on the corresponding images. The CholecSeg8k \cite{twinanda2016endonet} dataset contains 8080 images of laparoscopic cholecystectomy, including thirteen classes. We conducted multi-classification tasks based on images using GPT-4V. Similarly, for the CVC-ClinicDB \cite{bernal2015wm} dataset, which provides 612 static images extracted from colonoscopy videos, we conducted an assessment of the diagnostic and localization capabilities of GPT-4V for identifying intestinal polyp-associated abnormalities in colonoscopy images. 

Based on our experimental findings, we have observed promising performance of GPT-4V in generating diagnostic reports based on endoscopic images. As shown in Figure \ref{fig:endoscopic_1}, it showcases a remarkable proficiency in identifying intricate organizational structures, thereby facilitating the generation of highly informative textual diagnostic reports. To our astonishment, GPT-4V not only proficiently identified the lesion's precise location within the image, but also astutely detected and inferred the presence of a light source reflection, thereby exceeding our expectations, indicated in Figure \ref{fig:endoscopic_4}. These findings suggest that GPT-4V exhibits a high level of accuracy in understanding images, indicating its promising potential in this field.

However, in the context of multi-classification tasks involving endoscopic images, GPT-4V's performance exhibits room for improvement, as it falls short of perfection. Overall, GPT-4V demonstrates commendable performance; but it does not exhibit flawless accuracy in identifying all categories though successfully recognizing the majority of them, as illustrated in Figure \ref{fig:endoscopic_4}. There could be several contributing factors to this phenomenon, including but not limited to insufficient training of GPT-4V on relevant specialized datasets, low image resolution, and inappropriate contextual cues. Leveraging the formidable image comprehension capabilities of GPT-4V, it holds immense potential in aiding physicians with diagnostic tasks, thereby paving the way for significant advancements in the field. 

GPT-4V has been proven to be highly effective in accurately detecting abnormal areas in endoscopic images through a single segmentation task. Additionally, it is capable of generating diagnostic reports and providing relevant recommendations based on its findings. While GPT-4V may not have attained perfect scores in multi-class segmentation tasks, its exceptional potential in this domain cannot be overlooked. In conclusion, supported by experimental evidence, we have compelling grounds to assert that GPT-4V holds promise as a valuable tool for medical assistance in the realm of endoscopy.

\subsection{Dermatological Imaging}
\label{subsec:dermatological-imaging}
\begin{figure}[H]
    \centering
    \includegraphics[width = \textwidth]{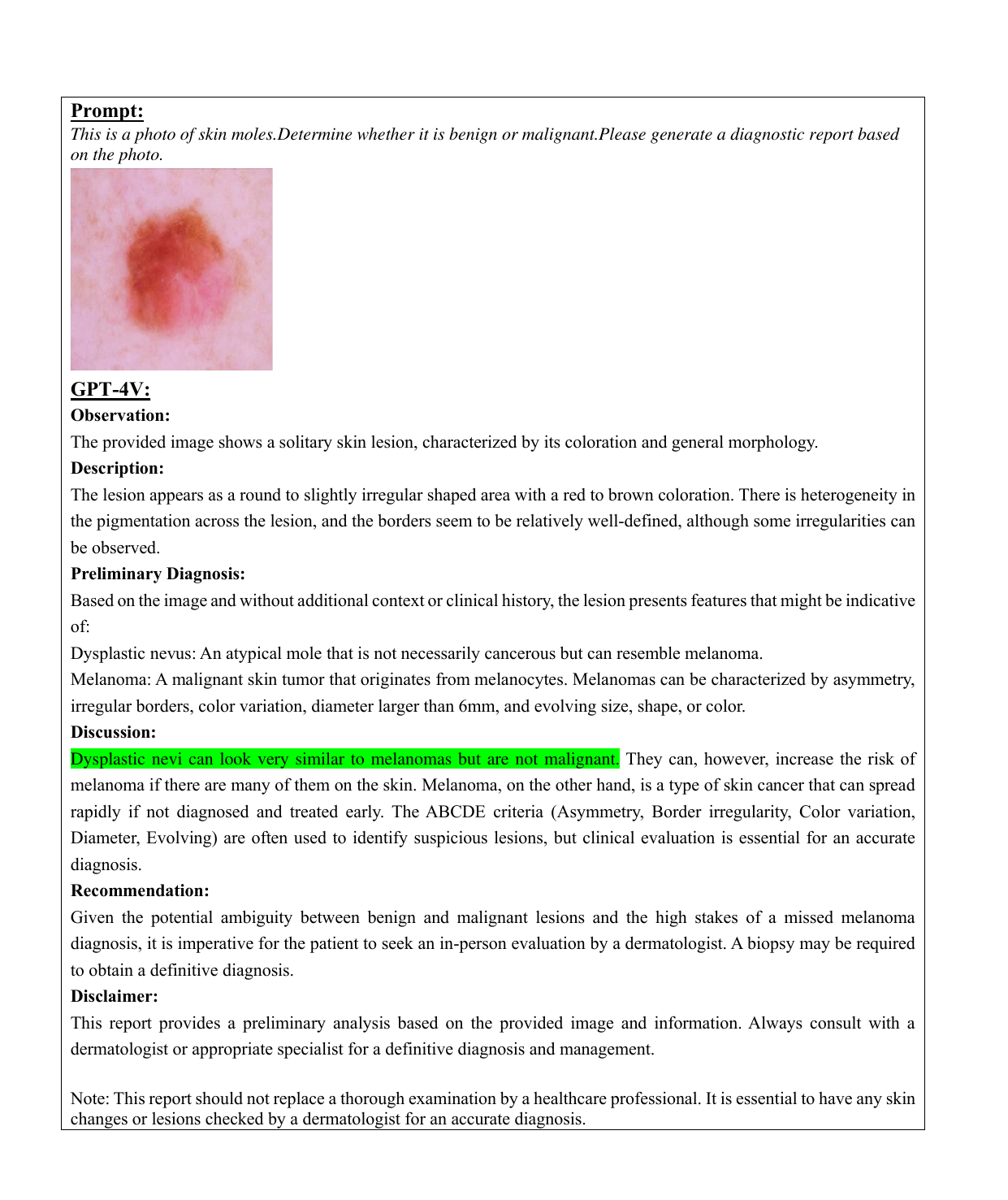}
    \caption[Dermatological Imaging: Case 2]{\textbf{Skin: Case 3.} One of the specific examples of benign skin moles. GPT-4V also has a high accuracy in judging benign and malignant skin diseases.}
    \label{fig:skin_6}
\end{figure}

In this section, our goal is to assess GPT-4V's ability to diagnose skin diseases, including two aspects:

\begin{itemize}
    \item The ability to identify different types of skin diseases. Due to the wide variety of skin diseases with diverse symptoms and appearances, some diseases may even manifest differently in different individuals and may change over time and under different environmental conditions. Additionally, the quality and resolution of images can impact accuracy. Low-resolution, blurry, or unclear images can lead to inaccurate diagnoses. Therefore, accurately identifying the type of skin disease is a significant challenge for GPT-4V.
    \item The ability to determine whether a skin disease is malignant (cancer) or benign (non-cancerous)(as shown in Figure \ref{fig:skin_6}). Some benign and malignant skin lesions may appear very similar visually, making it difficult even for professional doctors to differentiate. This requires GPT-4V to possess highly precise image analysis and pattern recognition capabilities to make accurate diagnoses. Some benign skin lesions may undergo secondary changes, such as ulcers, infections, or inflammation, making them appear more like malignant lesions in images. This can be misleading for GPT-4V. Furthermore, the overlap of skin diseases and clinical variations can lead to misjudgments by GPT-4V.
\end{itemize}

\begin{figure}[H]
    \centering
    \includegraphics[width = \textwidth]{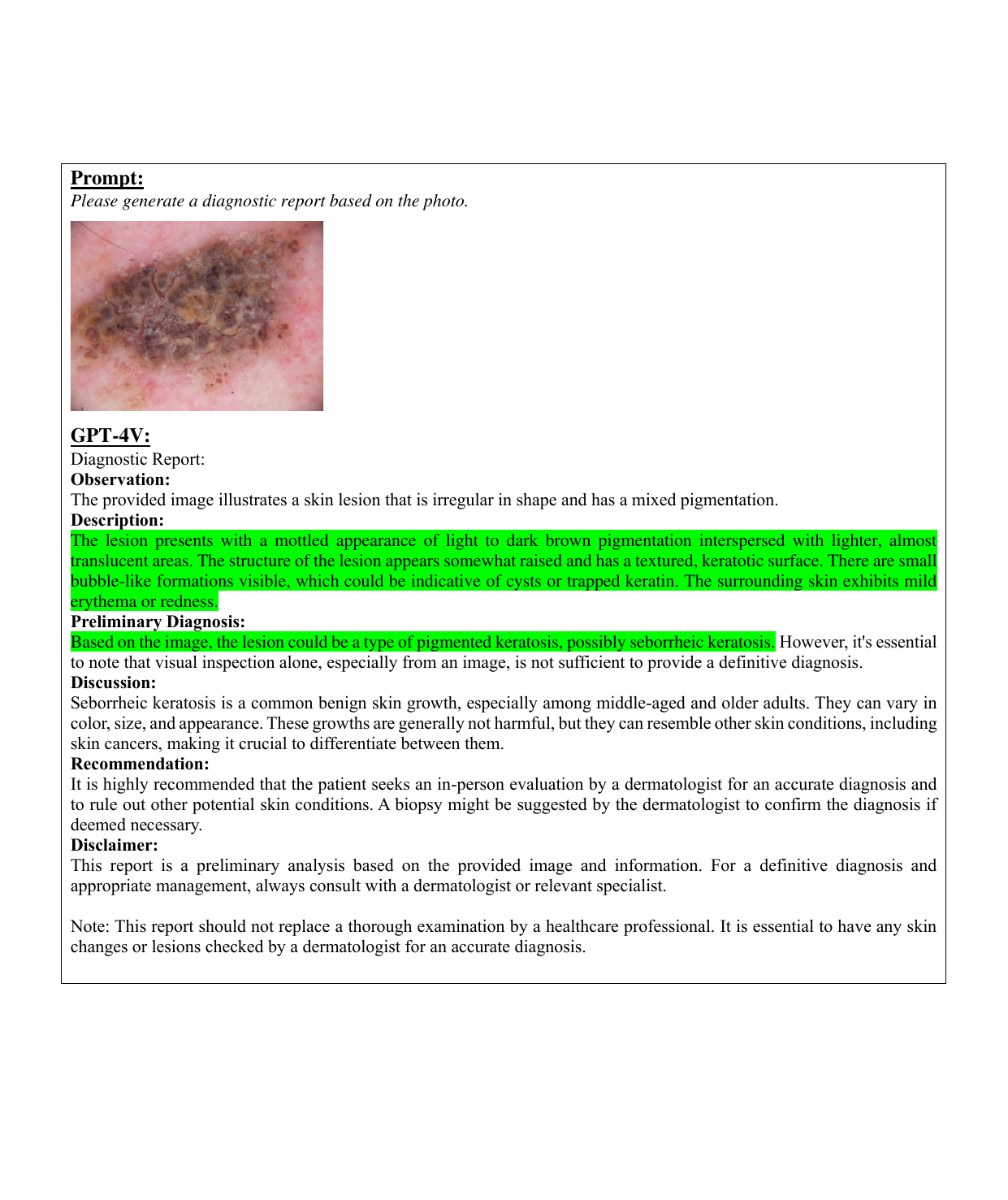}
    \caption[Dermatological Imaging: Case 1]{\textbf{Skin: Case 1.} One of the specific examples of benign pigmented keratoses. Without notifying the disease type in advance, GPT-4V still gave an accurate and detailed diagnosis report.}
    \label{fig:skin_2}
\end{figure}

In this work, we used the ISIC dataset \cite{Rotemberg2021} to examine the diagnostic capabilities of GPT-4V. The ISIC dataset is a public medical image dataset mainly used for dermatology diagnosis and research. This dataset contains a large number of dermatology images, covering various skin diseases and lesions. From the test results, it can be observed that, for the sake of thoroughness, GPT-4V often provides multiple possible diagnoses that include the correct disease type (as shown in Figure \ref{fig:skin_4}). For skin diseases with clear features, GPT-4V can directly and accurately provide a diagnosis (as shown in Figure \ref{fig:skin_5}). Moreover, GPT-4V exhibits relatively high accuracy in assessing disease severity. Overall, GPT-4V's diagnostic reports for skin diseases have several advantages:

\begin{itemize}
    \item Accuracy: GPT-4V's descriptions accurately reflect the visual characteristics of skin diseases, including the color, size, shape, texture, and distribution of lesions, which are crucial for accurate diagnosis (as shown in Figure \ref{fig:skin_3}).
    \item Completeness: GPT-4V's descriptions are comprehensive and encompass all visible symptoms and features while minimizing subjectivity. This aids doctors in obtaining comprehensive and objective information to better understand the condition.
    \item Standardization: GPT-4V uses standardized terminology and expression to ensure that doctors can understand and compare descriptions of different cases (as shown in Figure \ref{fig:skin_2}).
\end{itemize}

Based on these advantages, GPT-4V can serve as a powerful auxiliary tool in the field of dermatology for the diagnosis of skin diseases.

\subsection{Genetic Imaging}
\label{subsec:genetic-imaging}
The ability of  GPT-4V's ability to interpret complex genetic research data could be tested from two aspects: gene expression research data and GWAS research data.

As for gene expression research, the in situ hybridization (ISH) image technology \cite{doi:10.1177/030098589803500301} is widely used as a direct way to detect gene expression levels. Specifically, RNA in situ hybridization (RNA-ISH) spatial transcriptomics technology enables the detection of gene expression patterns within individual cells while maintaining the spatial context of the tissue \cite{chen2015spatially, burgess2019spatial}.

In this work, the GPT-4V model has been tested to analyze and interpret the ISH images to yield key information. The whole approach could be viewed in two parts. In the first part, the experiments tested the model's ability to discern medical imaging features and further identify medical image types and species based on such features. In the second part, more complicated tasks have been applied to meticulously investigate and observe the ability of the model to extract tissue atlas information and perform alignment analysis between the tissue’s structural template image and ISH gene expression images. By applying the proper prompt to GPT-4V following the test images and providing detailed guidance to maximize the comprehension abilities of the model, our objective is to evaluate the capabilities of advanced artificial intelligence in extracting information from image data generated in genetic research and conducting subsequent analyses on this data. The open-access image dataset of Marmoset Gene Atlas\footnote{\href{https://gene-atlas.brainminds.jp}{https://gene-atlas.brainminds.jp}} \cite{shimogori2018digital,kita2021cellular} was used in this work. The whole image data consists of Nissl-dye Marmoset brain atlas data and ISH-rich expression data of 1897 genes.

 The findings of the first part of the experiments indicate that GPT-4V does possess the capability to recognize and extract structural information from Nissl-staining images. In the structural images analysis of Gene case 1 (Figure \ref{fig:gene_1}) and Gene case 2 (Figure \ref{fig:gene_2}), GPT-4V successfully highlighted the species and organ of the image, as well as recognized the cortical area of the sample marmoset’s brain slide. It is noted that besides yielding correct information through the input image, the GPT-4V also explained the criteria for interpretation, e.g., it pointed out that the Nissl-stain could reveal the cell bodies, the convoluted structure of brain folding and the RNA-rich region of neuron cells. These results suggest that the model’s innate ability to correctly recognize structural features of cells and tissue.

 The findings of the second part of the experiments indicate that GPT-4V does possess the capability to recognize the annotated atlas image of certain issues. Most noteworthy, under proper prompt, the model could perform alignment analysis to extract and merge information from a pair of contrastive images. In the structural images analysis of Gene case 3 (Figure \ref{fig:gene_3_2}) and Gene case 4 (Figure \ref{fig:gene_4}), after correctly extracting the IDs of cortical regions from the annotated atlas image, we designed an experiment to guide the model into an alignment task. By concatenating the pictures together, GPT-4V gets an input of both the atlas information and ISH-rich expression intensity information at the same time. With the detailed prompt which provided an explanation of the setting of the sub-image and the relation between the two parts of the image, the model highlighted the rich expression region based on intensity in the form of the region ID provided by the structure atlas. Although the generated results are not as comprehensive as those identified by human experts, this experiment suggests the huge potential of GPT-4V to merge and compare information from different types of images, serving as a preliminary tool to automatically analyze the image results from genetic research.

Genome-wide association studies (GWAS) seek to discover links between genotypes and phenotypes by examining variations in the allele frequencies of genetic variants among individuals from inheritance-related populations that differ phenotypically. In the experimental workflow of a GWAS, conducting the statistical test for association is directly connected to the final results of the study. The genetic association theory relies on the biometrical model and employs either linear or logistic regression models to assess associations, depending on whether the phenotype is continuous \cite{Hirschhorn2005,balding2006tutorial,uffelmann2021genome,VISSCHER20175}.
In the past 10 years, the Manhattan plot has been widely used in works of literature as a conventional way to present the statistical association relationship between the phenotype and the variant across the whole genome (or part of the genome). Here, we aim to assess the potential of advanced artificial intelligence in extracting information from Manhattan plot results of GWAS results and thus generate text descriptions of the association relationship. The image data used in our experiment is acquired from open access dataset ENIGMA 3\footnote{\href{https://enigma.ini.usc.edu/research/gwasma-of-cortical-measures}{https://enigma.ini.usc.edu/research/gwasma-of-cortical-measures}}, which is a genome-wide association meta-analysis of brain magnetic resonance imaging data from 51,665 individuals, focus on brain surface measurements of the whole cortex and 34 regions with functional specializations \cite{grasby2020genetic}. The Manhattan plot of two phenotypes, the surface area of the precentral region and the surface area of the full surface of the human brain, are chosen to perform the test.

The findings of our experiment indicate that GPT-4V does not possess the capability to explicitly characterize the information in the Manhattan plots (Figure \ref{fig:gene_5} and \ref{fig:gene_6}). In both tests Gene case 5 and Gene case 6, The model identifies part of the chromosomes in which the significant variants are most located. However, the most variant-rich distributed chromosome 15 and chromosome 12 have been missed according to the generated text summary from GPT-4V. Although the generated results differ significantly from the results identified by experts, the GPT-4V does show its reasoning process and criteria, e.g., the meaning of the Manhattan plot and P-value. This suggests that the concealed ability of GPT-4V is waiting for future exploration, which could serve as a tool to automatically analyze the image results from genetics research. However, due to the image size limitation of GPT-4V (the image result data for genetic research is rich in information and requires HD image with large size) and shortage of open-accessed GWAS results imaging data, only limited exploration could have proceeded with implementation in this work.

\subsection{Orthopedic and Pediatric Imaging}
\label{subsec:orthopedic-pediatric-imaging}
In this section, we conduct research on the use of GPT-4V in human bone X-ray films. The research direction in this field is very broad. Firstly, we use the MURA dataset, which is an open dataset for deep learning research in musculoskeletal radiology. It consists of over 40000 X-ray images of individual body parts, covering eight main body parts, including the hand, wrist, elbow, shoulder, chest, knee, ankle, and buttocks. On this dataset, GPT-4V needs to identify the corresponding bone tissue in the X-ray film, and then analyze any abnormal areas in the image, such as whether the bone is intact, whether there are fracture issues, and the presence of non bone tissue in the X-ray.Next, we hope to explore the deeper capabilities of GPT-4V in the study of bone X-ray films. We know that the bone structure at joints is very complex, including gaps between joints, bone density, mass, and the presence of dislocations or implants between joints. Therefore, we found the Asptic Loose Hip Implant X-ray dataset and the Digital Knee X-ray dataset to test the ability of GPT-4V.

Through testing, we found that without prompt, GPT-4V can correctly identify this X-ray from a certain part of the human body, and then determine whether it is normal based on bone integrity, bone density, etc., with more accurate results. At the same time, GPT-4V is also capable of identifying complex joints, locating joint positions, and determining whether the surrounding tissues are normal. However, for problems between joints, such as fluid accumulation or foreign objects, GPT-4V cannot make a judgment solely from X-ray images. Overall, GPT-4V can complete simple diagnosis, but there is still room for improvement in complex situations. It can serve as an auxiliary diagnostic tool to help doctors complete the diagnosis.

\subsection{Dental Imaging}
\label{subsec:dental-imaging}
Panoramic X-ray is widely used in the diagnosis of dental diseases. It can provide a panoramic view of the patient's mouth to assist dentists in their treatment. In this study, we aim to explore the recognition ability of GPT-4V for panoramic X-ray images.

The experiment is divided into two parts. Firstly, the GPT-4V is tested without any additional prompt information. This section uses the MICCAI 2023 Challenge: STS - Tooth Segmentation Task Based on 2D Panoramic Images dataset \cite{zhang2023children}, which contains 3000 labeled Panoramic images of teeth. Secondly, we conducted tests to evaluate the performance of GPT-4V in tooth counting tasks and diagnosing four different types of abnormal dental diseases, including caries, deep caries, periapical lesions, and impacted teeth. This section uses the DENTEX2023 dataset \cite{hamamci2023diffusion}. The DENTEX dataset companies pharmaceutical dental X-rays observed from three different institutions using standard clinical conditions but varying equipment and imaging protocols, resulting in diverse image quality reflecting heterogeneous clinical practice The dataset includes X-rays from patients aged 12 and above, randomly selected from the hospital's database to ensure patient privacy and confidentiality The train dataset contains 1005 X-rays fully labeled for abnormal tooth detection with quadrant, tooth enumeration, and diagnosis classes.

Through experiments, we found that GPT-4V can recognize teeth, implants, and surrounding bones, and simply determine whether they are normal. For the specified four dental diseases and tooth counting tasks, GPT-4V cannot provide accurate answers. However, it provides a relevant judgment basis and potential areas of concern, indicating its ability to identify abnormal tooth areas, but this ability still needs further improvement. GPT-4V currently has limited capabilities in dental image recognition and diagnosis, but it still has great development prospects in assisting dentists in diagnostic tasks.

\subsection{Testing Procedure}
In the test phase, model reasoning and evaluation, clinging to the power of GPT-4V, we exerted it to generate causal inferences on the prompts and compared them with the test labels to evaluate GPT-4V's performance. The GPT-4V was subjected to a comprehensive testing procedure to evaluate its performance and capabilities. The testing process followed rigorous protocols and adhered to the high standards of scientific evaluation. The objective is to assess the GPT-4V's proficiency in various medical field tasks and its ability to generate coherent and contextually appropriate responses.

To begin, a diverse set of benchmark datasets was carefully curated, encompassing a wide range of medical image tasks such as classification, detection, segmentation, visual question answering (VQA), impression generation, and medical image location. These datasets were selected to cover different medical domains and levels of complexity, ensuring a comprehensive evaluation of the GPT-4V's capabilities. This allowed for a contextual understanding of the LLM's relative strengths and weaknesses, providing valuable insights into its advancements and potential applications.

For each test case, the GPT-4V model was concurrently presented with medical imagery alongside specific prompts. Within the responses generated by GPT-4V, erroneous information was denoted by red markings, correct information by green, and areas of uncertainty were indicated in yellow. Color-coded sections within the "Reference Answer" were employed to provide a basis for judging the accuracy of GPT-4V's output. Moreover, an exhaustive explanation and reasoning process accompanied each case, facilitating a deeper understanding among evaluators of how GPT-4V approached the core issue and formulated its inferences.

For each task, an assessment was conducted to encapsulate the capabilities exhibited by GPT-4V within the domain, alongside current limitations and deficiencies, to ensure the reliability of the test outcomes. These evaluations not only highlighted the precision of GPT-4V in executing specific medical tasks but also unveiled its constraints when dealing with intricate medical scenarios. Such profound analysis has yielded a lucid comprehension of the practical applicability and prospective enhancement trajectories for GPT-4V.



\section{Related Work}

\subsection{Foundation Models}
The Transformer architecture, known for its flexibility and cross-modal capabilities, has become a widely adopted backbone for language and vision models. Initially applied in the field of natural language processing~\cite{vaswani2017attention}, the Transformer addressed the issue of sequential text processing and introduced a global attention mechanism, achieving state-of-the-art performance in diverse tasks \cite{liu2023radonc,liu2023context,rezayi2022clinicalradiobert,liao2023mask,cai2022coarse,zhao2022embedding,ding2022accurate,rezayi2022agribert,zhao2023generic,wu2023exploring,holmes2023evaluating,liu2023transformation,tang2023policygpt,dai2023auggpt}. Building upon the Transformer architecture, models like  Transformer-based BERT achieved remarkable performance across multiple language processing tasks. With the increase in training data and model parameters, large-scale Transformer-based language models emerged. Models such as GPT-1~\cite{radford2018improving}, GPT-2~\cite{radford2019language}, GPT-3~\cite{brown2020language}, GPT-4~\cite{openai2023gpt}, Llama~\cite{touvron2023llama}, Llama2~\cite{touvron2023llama}, PaLM~\cite{chowdhery2022palm}, and others have progressively improved performance across various natural language processing tasks. Notably, ChatGPT~\cite{openaiIntroducingChatGPT} and GPT-4 have demonstrated impressive zero-shot reasoning abilities, producing satisfactory results without fine-tuning on downstream tasks, thanks to large-scale training data and techniques like Reinforcement Learning from Human Feedback (RLHF). Concurrently, research has focused on fine-tuning large language models for specific domains, such as medical text-related tasks. Examples include early attempts like ChestXrayBERT~\cite{cai2021chestxraybert} and Clinical-BERT~\cite{huang2019clinicalbert}, which fine-tuned BERT models using medical text data, as well as Radiology-Llama2~\cite{liu2023radiologyllama2} and DoctorGLM~\cite{xiong2023doctorglm}, which fine-tuned Llama2 and ChatGLM~\cite{du2022glm} models with a greater emphasis on medical text data. These models have shown exceptional performance in medical question-answering and diagnostic report-generation tasks. Additionally, research~\cite{ma2023impressiongpt} has explored the use of dynamic and interactive prompt designs, instead of fine-tuning, to achieve optimal performance of large language models like GPT in specific domains. 

In the field of computer vision, Vision Transformer (ViT)~\cite{dosovitskiy2020image} utilized the Transformer architecture for image processing. It demonstrated that the architecture can be directly applied to images without additional modifications, providing strong theoretical support for subsequent multimodal frameworks. Building upon ViT, models like DeiT~\cite{touvron2021training}, Swin Transformer~\cite{liu2021swin}, MAE~\cite{he2022masked}, MoCo-v3~\cite{chen2021empirical}, and others have further improved performance across various visual tasks. MaskViT~\cite{gupta2022maskvit} introduced a masking mechanism to enable the model to learn semantically relevant features with limited samples. EG-ViT~\cite{ma2023eye} utilized eye-tracking information from radiologists to assist ViT in disease diagnosis and address shortcut learning issues. The same team proposed SGT~\cite{ma2023rectify}, which employed more general saliency information to guide ViT's classification tasks. CP-ViT~\cite{yu2023core} leveraged the Core-Periphery (CP) organization in human brain networks to guide the information communication mechanism in ViT's self-attention for vision tasks. Instruction-ViT~\cite{xiao2023instruction} utilized multi-modal prompts to guide the fine-tuning of ViT models. Similar to the field of natural language processing, as the availability of image data increased, researchers began training ViT models with larger-scale image data and more parameters, such as ViT-G, ViT-E, and ViT-22B. These models further improved performance on image-related tasks. In subsequent research, ViT is often used as the foundational image encoder embedded within other model architectures. Models like SAM~\cite{kirillov2023segment}, CLIP~\cite{radford2021learning}, BLIP~\cite{li2022blip}, ViTL~\cite{kim2021vilt}, CoCa~\cite{yu2022coca}, and BEiT-v3~\cite{wang2023image}, among others, utilize ViT as their image encoder. In the medical field, models like MedCLIP~\cite{wang2022medclip} and GLoRIA~\cite{huang2021gloria} fine-tune ViT encoders and text encoders using extensive medical image-text data, achieving remarkable performance in various medical tasks such as visual question answering, image diagnosis, and image-text retrieval.

\subsection{Multi-modal Models}
Language models like ChatGPT and GPT-4~\cite{openai2023gpt4} have significantly impacted the field by demonstrating exceptional skills in understanding and producing natural language text. This advancement has motivated researchers to extend beyond mere text-based applications and explore multimodal data~\cite{liu2022survey}, including images, videos, and audio. This broadening scope enables a more comprehensive analysis and amalgamation of diverse types of information.

Since its introduction in 2022, ChatGPT, building upon InstructGPT, has demonstrated remarkable proficiency in a range of tasks, including reasoning and text generation \cite{wei2023chat2brain,liu2023pharmacygpt,dou2023towards,tang2023policygpt,liu3surviving,guan2023cohortgpt,zhong2023chatabl,zhao2023meta,dai2023ad,shi2023mededit,rezayi2023exploring}. Its effectiveness is largely due to extensive training data and an advanced reinforcement learning approach.This enables ChatGPT to produce high-quality text \cite{dai2023auggpt}, redact sensitive information \cite{liu2023deid,liu3surviving,liao2023differentiate}, perform advanced reasoning \cite{zhong2023chatabl}, engage in nuanced dialogue \cite{zhou2023comprehensive}, generate creative content \cite{liu2023transformation} and assist in medical diagnosis \cite{liu2023pharmacygpt,liu3surviving}. These capabilities have positioned it among the first models to achieve near-human performance in a wide array of tasks \cite{liu2023summary,liu2022survey,liao2023differentiate,zhao2023brain}. 

The natural human inclination towards processing multimodal inputs \cite{zhao2023brain} has guided the evolution of AI models from text-centric systems like ChatGPT to multimodal frameworks. Recognizing the vast potential of integrating multiple forms of data processing, researchers have introduced models that not only understand text but also interpret and respond to visual cues. Recent multimodal models such as MiniGPT-4~\cite{zhu2023minigpt} and mPLUG-OWL~\cite{ye2023mplugowl} demonstrated their capabilities in tasks like image captioning and handwriting recognition. By integrating visual encoders with language models, these initial versions provided a foundation for further multimodal research. Subsequent models, including Visual ChatGPT~\cite{wu2023visual} and MMREACT~\cite{yang2023mmreact}, incorporated visual foundational models into their design, enhancing language models' ability to understand visuals during interactions and improving their response to visual inputs.

Subsequent advancements led researchers to concentrate on developing models that could be trained end-to-end, which improved both their efficiency and effectiveness. Flamingo~\cite{flamingo2022} and BLIP-2~\cite{li2023blip2} showed that it's possible to use a unified architecture for processing both text and visual information. Models like PaLM-E~\cite{driess2023palme} represent frameworks that are trainable end-to-end, leveraging extensive multimodal datasets. They can handle various tasks without the need for task-specific fine-tuning due to their ability to combine and learn from multimodal data.

To optimize user interaction, models such as MM-GPT~\cite{gong2023multimodalgpt} and Otter~\cite{li2023otter} have improved interface design and user experience. Models like BLIP-2~\cite{li2023blip2} are advancing in aligning visual features with text, enhancing the efficiency and accuracy of information retrieval. An emerging trend is streamlined connectivity, as exemplified by LLaVA~\cite{liu2023llava}, which introduces visual features into LLMs through trainable fully connected layers, showing that even simpler structures can effectively process visual information and offer flexibility in design.

GPT-4V, launched by OpenAI in 2023, marks a notable advancement in the development of foundational multimodal models, demonstrating its capability to interpret and create visual content. Its effectiveness in areas such as medical imaging recognition~\cite{wu2023gpt4vision} indicates a wide range of potential applications.

As technology advances, the ability of large language models (LLMs) to process multimodal data is expanding the horizons of artificial intelligence~\cite{zhao2023brain}. This progress is paving the way for the development of intelligent interactive systems capable of understanding complex information. It is expected that as these models are increasingly applied across various domains, their importance in areas such as task adaptability, multimodal integration, and performance evaluation will continue to grow.

\subsection{Instruction Tuning in Large Multi-modal Models}
Instruction tuning (IT) in large multi-modal models is a burgeoning field of study focusing on enabling the capabilities of these models to understand and execute specific tasks. This technique is particularly vital for enhancing the performance of multi-modal models in terms of several modalities of data, such as text and images. At its core, instruction tuning involves the use of a dataset consisting of instruction and output pairs. These pairs are used in a supervised learning context to further train pre-existing models. The instructions act as prompts, and the outputs represent the desired responses or actions that the model should learn to produce. Through this process, the model is tuned to better align with human expectations and interpretive frameworks, increasing its utility across a range of applications \cite{zhang2023instruction}.

Recent advancements in visual instruction tuning~\cite{wang2023review} are exemplified by open-source projects like LLaVA~\cite{liu2023llava} and MiniGPT-4~\cite{zhu2023minigpt}, which have made notable progress despite the scale of parameters they apply with 13 billion parameters or fewer. The multi-modal tasks these models undertake are significant because they involve the procedure that interprets images through cross-modality feature spaces. The technique of instruction tuning also enables the integration of several downstream tasks into a unified, generic model \cite{wei2021finetuned}. By training these multi-modal large language models with the user's own well-defined collection of datasets, generally ranging from 10,000 text-image pairs as the minimum data size to 1 million pairs or larger, these multi-modal models can outperform in particular downstream tasks. Moreover, instruction tuning is also being explored as a means to leverage feedback from more powerful LLMs, such as InstructGPT \cite{ouyang2022training}, to align model behaviors with human preferences in a cost-effective manner. 
InstructBLIP accomplishes a variety of downstream visual tasks by constructing different language instructions to the vision-language model \cite{li2023blip2}. Previous research has demonstrated that the OFA model can enhance their zero-shot capabilities in multiple multimodal tasks with the guidance of five expert-crafted instructions \cite{wang2022ofa,xu2022multiinstruct}. In addition, Instruction-ViT with multi-modal prompts accomplishes a variety of visual downstream tasks based on ViT rather than the LLM \cite{xiao2023instruction}.

There is also a focus on creating large and diverse datasets such as the Multi-Modal In-Context Instruction Tuning (MIMIC-IT) dataset, which consists of millions of multimodal instruction-response pairs. Such datasets are pivotal in refining instruction tuning, with an emphasis on the diversity and creativity of instructions allowing multi-modal models to learn desired features from the images.

\subsection{Emergent Abilities and Zero-shot Learning} 
Large language models (LLMs) sometimes produce emergent abilities when processing tasks~\cite{wei2022emergent}. These abilities suddenly appear after the model reaches a certain complexity threshold, rather than improving smoothly and gradually. For example, larger models are capable of decoding movies based on emojis~\cite{ozturkler2022thinksum}, performing complex mathematical operations, and even generating executable computer code. Researchers are trying to understand how and why these emergent capabilities arise, which could reveal deep questions about artificial intelligence and machine learning, such as whether complex models are actually doing something entirely new, or whether they are simply becoming statistically unusual.

These large models, such as GPT-3~\cite{brown2020language} and Google's PaLM~\cite{chowdhery2022palm}, exhibit unexpected behavior due to the large number of parameters. However, these models also come with unpredictable risks, including biases and inaccuracies that crop up~\cite{dai2023ad}. Some models can even enable logical thinking. Researchers now have to not only identify more emergent capabilities but also understand how and why they occur. Similarly, for more advanced models with more parameters such as GPT-4~\cite{bubeck2023sparks}, LLAMA2~\cite{touvron2023llama}, etc., the models show significantly stronger zero-shot capabilities~\cite{liu2023evaluating}. And after specific fine-tuning~\cite{liu2023tailoring}, these large language models can still have strong independent understanding capabilities in highly professional industries~\cite{liu2023radiology,liu2023radonc}. Similar phenomena have also appeared in the fields of medicine, humanities, art, linguistics, history, etc~\cite{liu2023transformation} allowing people to truly see the dawn of AGI. 

With the emergence of more and more multi-modal large languages~\cite{zhu2023minigpt,ye2023mplugowl}, questions have been raised about whether LLMs can produce higher-level understanding and reasoning capabilities for more diverse data. As we all know, pictures contain more potential information than text. Verifying whether LLMs can also produce zero-shot images is an important criterion for judging whether LLMs have real emergent abilities. Especially in more professional fields, because in some fields such as medical and biological fields, there will be a large amount of undisclosed data. These images have not been seen during the training process of GPT-4V. Testing whether GPT-4V can understand these images is a good example of testing whether GPT-4V has zero-shot analysis capabilities for multi-modal information.

\subsection{LLM for Medical Robotics}
In the vanguard of modern medicine, the advent of robotics and computer-aided healthcare systems heralds a transformative era~\cite{bhatt2018trends}. These sophisticated tools have been pivotal in automating detection and identification processes, significantly enhancing efficiency and mitigating costs. A prime example of this innovation is the nuanced identification of surgical instruments and operative sites during complex procedures, marking each surgical stage with unprecedented precision~\cite{zhao2019real,ward2021computer,van2021gesture,garrow2021machine}, such as da Vinci Surgical System (Intuitive Surgical, Inc., Sunnyvale, CA), which designed to facilitate complex surgery using a minimally invasive approach. Yet, the integration of medical robotics is not without its hurdles. One significant challenge is the nuanced differences in robotic tasks compared to the broader natural language tasks~\cite{anil2023palm}. The specificities of terminology consistency and understanding across diverse datasets pose a unique set of difficulties that must be carefully navigated~\cite{rodrigues2022surgical}.

LLM have proven their mettle across various domains, and their robust natural language understanding, extensive common-sense knowledge, and intricate theoretical reasoning skills are now finding their place in the robotics field~\cite{wu2023tidybot}. Recent research underscores the impressive zero-shot capabilities of LLMs and their capacity for executing intricate long-term tasks with minimal prompting~\cite{zhang2023building,gadre2023cows,vemprala2023chatgpt}. These capabilities are instrumental in bridging the gap between natural language and executable robotic commands. Specifically, the integration of Large Language Models (LLMs) into robotics enhances their capabilities in various tasks such as reasoning, planning, manipulation, and navigation \cite{vemprala2023chatgpt,huang2023instruct2act,szot2023large,wang2023prompt,lin2023text2motion,xie2023text2reward,yang2023octopus}. Moreover, the zero-shot learning abilities of LLMs combined with large-scale cross-embodiment data, evolve robots from being specialized to generalists. The recent release of Google's RT-X \cite{padalkar2023open} is an advanced general-purpose robotics model based on the foundations of Palm-E \cite{driess2023palm}, RT-1 \cite{brohan2022rt}, and RT-2 \cite{brohan2023rt}. This model demonstrates the ability to perform novel skills with unseen objects, thereby advancing Artificial General Intelligence (AGI) into more practical real-world applications in both industrial and domestic environments. Furthermore, intelligence modeled after LLMs can discern the current state of affairs and engage in autonomous planning through dynamic human-computer interactions~\cite{yu2023language}.

The GPT-4V, with its enriched common-sense knowledge base, complex reasoning faculties, and acute visual comprehension, emerges as a frontrunner for the reasoning and communication modules within robotic systems~\cite{mai2023llm}. Its sophisticated human-computer interaction capabilities not only facilitate seamless communication but also enable efficient task completion through feedback-driven task planning. GPT-4V's capabilities to execute zero-shot tasks and to communicate and collaborate effectively elevate its potential as a significant asset in the ever-evolving landscape of medical robotics. These strengths of the GPT-4V highlight its potential as a revolutionary tool in the medical robotics field, promising a future where human expertise and Artificial General Intelligence (AGI) \cite{zhao2023brain} collaboratively advances the development of general-purpose medical robotic systems, thereby elevating the standards of healthcare excellence.

\subsection{Multi-modality Large Language Models Reasoning}
In the past few years, multi-modality large language models have developed astonishing contextual understanding, reasoning, and image comprehension abilities. However, constrained by the limited training corpus of large language models, this leads to a potential lack of in-depth knowledge in specific domains during the inference process. The challenges in abstract reasoning stem from the intricate nature of integrating knowledge across multiple domains or disciplines, the nuances of logical thinking, and the finite incorporation of world knowledge. When encountering complex problems (especially images), even state-of-the-art models still regularly produce logical mistakes. For instance, the large-scale language model faces challenges in accurately solving high dimension mathematics problems~\cite{lightman2023let} and understanding images. Hence, enhancing the logical inference capability of multi-modality large language models is of paramount importance. 

Based on the training mechanism and cognitive framework of the multi-modality large language model, its deficiencies are evident. Specifically, addressing intricate problems necessitates deliberate and meticulous contemplation. Nevertheless, owing to the limited guidance provided to large language models during the training phase, they continue to be restricted by token-level, left-to-right decision-making processes throughout the inference phase~\cite{yao2023tree}. When confronted with intricate reasoning challenges, humans commonly employ diverse cognitive abilities and rely on interactions with tools, knowledge, and external environmental information to accomplish complex tasks. However, achieving this level of capability in large language models remains challenging~\cite{xie2023olagpt}.

In light of these constraints, numerous approaches have been put forth to emulate human cognitive processes. These include the cumulative reasoning (CR)~\cite{zhang2023cumulative} that divides tasks into smaller components for reasoning, and the OlaGPT~\cite{xie2023olagpt} that simulates the framework of human cognitive architecture. In addition, there is the Chain-of-Thought (CoT)~\cite{wei2022chain} that provides step-by-step solutions and the Tree-of-Thought (ToT)~\cite{yao2023tree} that models the solving process as a thought search. Similarly, process supervision models are employed to identify and mitigate the influence of hallucinations on the reasoning capacity of large language models~\cite{lightman2023let}.

While LLMs possess a certain degree of graphic comprehension capability, current iterations of LLMs exhibit inherent limitations in executing precise mathematical computations, multi-step logical reasoning, perceiving spatial and topological factors, as well as processing time-intensive operations~\cite{zhang2023graph}. Consequently, the performance of large language models in tasks involving graphic comprehension is unsatisfactory.

To address the deficiencies observed in large language models regarding image comprehension, approaches such as Graph-ToolFormer~\cite{zhang2023graph} and I-JEPA~\cite{assran2023self} have been embraced as solutions to rectify the limitations associated with weak reasoning capabilities and excessive dependency on manual data when dealing with intricate graph data in multi-modality large language models.

\subsection{Evaluation in High-Risk Domains}
GPT-4V has demonstrated impressive capabilities in integrating visual functionalities into large language models (LLMs). Similar to GPT-4~\cite{openai2023gpt4}, GPT-4V underwent pretraining on a large dataset of text and image data from the Internet as well as licensed sources of data, and subsequently fine-tuned using Reinforcement Learning from Human Feedback (RLHF)~\cite{ouyang2022training,christiano2017deep} to better align with human preferences.

Despite its noteworthy performance in general domains, GPT-4V exhibits limitations within certain high-risk domains, particularly in medical tasks. OpenAI's GPT-4V report\footnote{\href{https://openai.com/research/gpt-4v-system-card}{https://openai.com/research/gpt-4v-system-card}} reveals key findings after evaluating GPT-4V on medical tasks, including inconsistencies when interpreting medical imaging, vulnerabilities due to lack of contextual information, factual errors, and hallucinations. Specifically, GPT-4V may fail to identify key information within images, such as text, characters, symbols, or other pivotal visual details that are crucial for medical interpretation. It can also be unreliable in identifying medications, chemical compounds, or toxic substances when analyzing images. For real-world medical applications, the current GPT-4V is not yet sufficiently reliable for tasks such as medical diagnosis, treatment recommendation, or other clinical functions due to these inconsistencies and the potential risks to patient safety. This underscores the necessity for additional research and development to enable the model to reliably interpret medical images and provide sound medical advice across diverse contexts.

Our research conducts a thorough evaluation of GPT-4V's proficiency across a range of biomedical scenarios. We provide detailed findings for each imaging modality in Section \ref{Exp}, offering a more fine-grained analysis of its capabilities.

\newpage
\section{Experiments and Observation}
\label{Exp}

\subsection{Chest Radiography}

\begin{figure}[H]
    \centering
    \includegraphics[width = \textwidth]{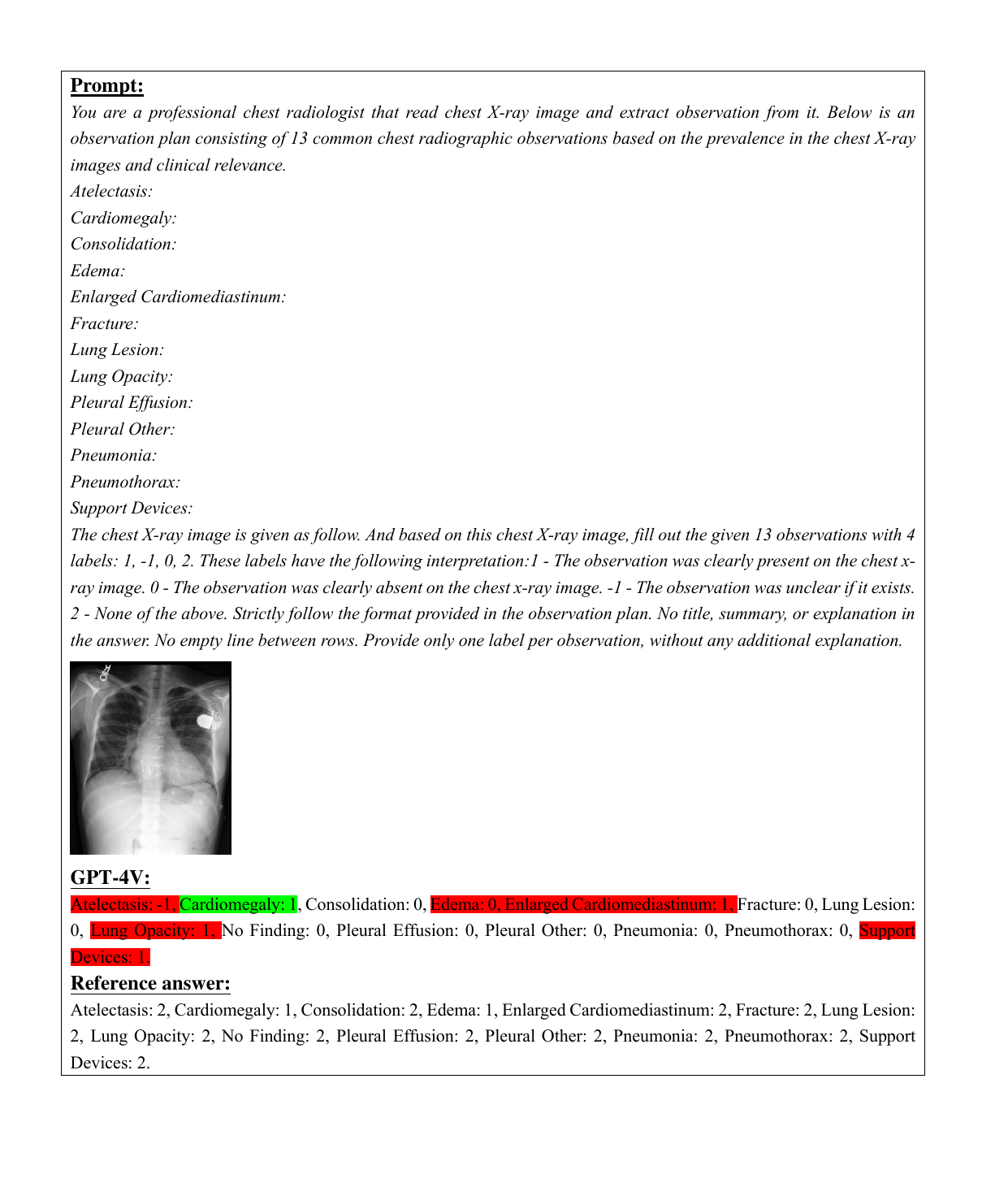}
    \caption[Chest Radiography: Case 1]{\textbf{Chest: Case 1.} A demonstration case of Classification task on MIMIC-CXR dataset. Green denotes the correct classification. Red in the figure denotes the incorrect classification.}
    \label{fig:chest_1}
\end{figure}
\newpage

\begin{figure}[H]
    \centering
    \includegraphics[width = \textwidth]{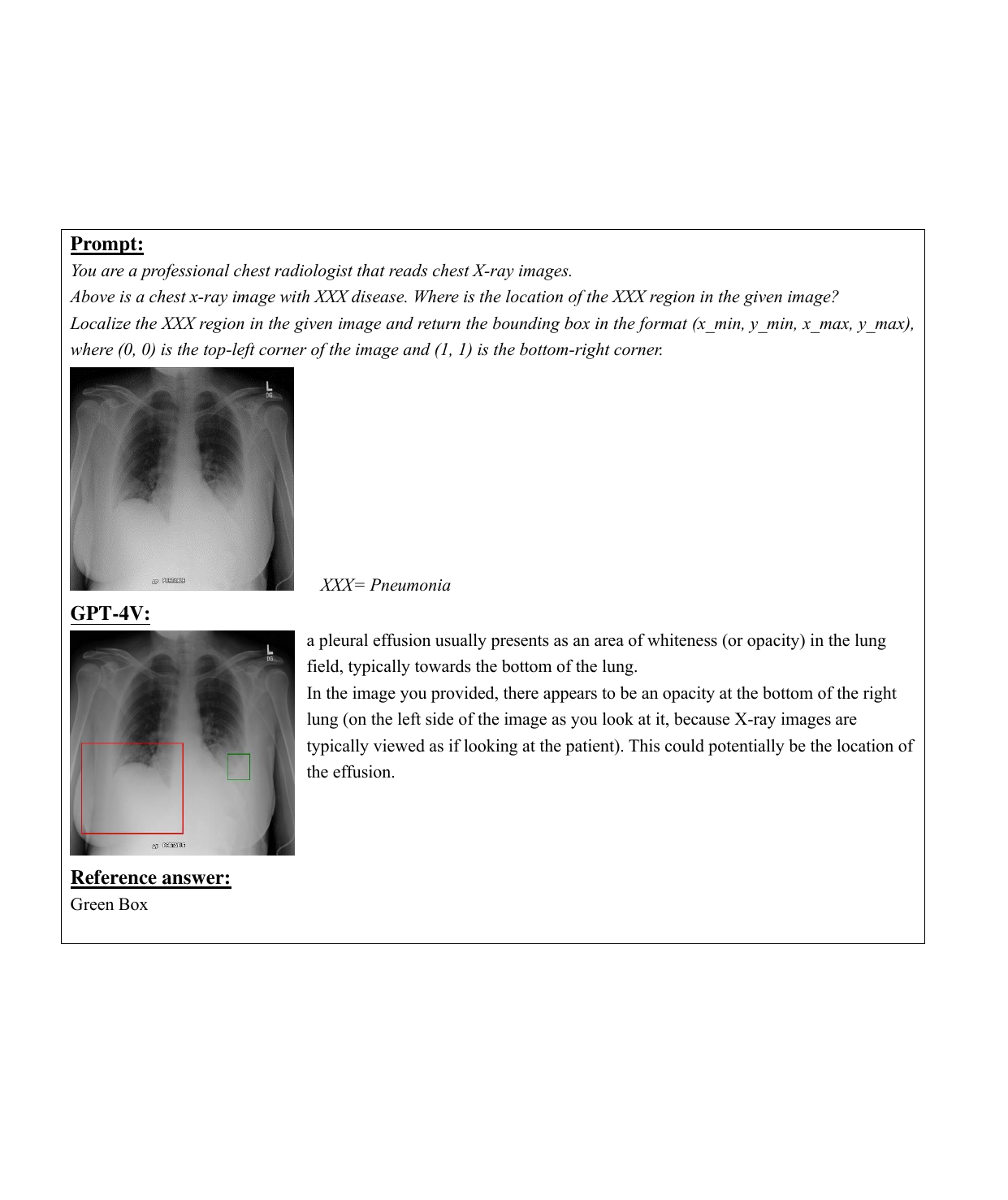}
    \caption[Chest Radiography: Case 2]{\textbf{Chest: Case 2.} A demonstration case of Localization task on NIH Chest X-rays dataset. The red box is the corresponding diagnosis region. The GPT-4V’s explanation about the image is located on the right side of the image. In addition, we provide a ground-truth box (Green Box) of the Pneumonia area for comparison.}
    \label{fig:chest_28}
\end{figure}
\newpage

The 13 categories in the prompt of Figure~\ref{fig:chest_1} represent common diseases found in chest X-rays. It can be seen that providing direct diagnoses for these 14 categories (13 diseases + No Finding) solely based on images is challenging for GPT-4V. Except correctly identifying Cardiomegaly in Figure~\ref{fig:chest_1}, the model struggled to accurately recognize the remaining diseases. In Figure~\ref{fig:chest_28}, we present an example of localization task on NIH Chest X-rays dataset. The NIH Chest X-rays dataset provides multiple disease classification labels along with location information of the lesions. It is noted that Pneumonia is not diagnosed correctly. Although it is difficult for GPT-4V to directly locate the lesions in the chest X-ray images, it can be seen from the explanations that GPT-4V has a certain understanding of the structure of the chest radiograph. We present more examples in the appendix, including more cases of 14-class classification test on MIMIC-CXR dataset, diagnostic reports generation task on MIMIC-CXR and OpenI datasets, binary and 5-class classification task on COVID-Qu-Ex, ChestXray2017 and CheXpert datasets, disease localization task on COVID-Qu-Ex, SIIM-ACR and NIH Chest X-rays datasets. 

Based on the Figure \ref{fig:chest_4}, \ref{fig:chest_5} and \ref{fig:chest_6}, it can be observed that GPT-4V performs well in the task of generating diagnostic reports, particularly in the generation of the Findings section. Most of the descriptions generated by GPT-4V based on the images align with the corresponding information in the ground truth. This demonstrates the strong image comprehension and image captioning capabilities of GPT-4V. The five categories provided by the CheXpert dataset (Figure~\ref{fig:chest_7}, \ref{fig:chest_8}, \ref{fig:chest_9}, \ref{fig:chest_10} and \ref{fig:chest_11}) are determined as the ground truth by multiple radiology experts. This dataset is also frequently used for zero-shot classification testing of diagnostic models. As seen from the examples in the appendix, GPT-4V is not proficient in the task of generating diagnostic results based on input images. The ChestXray2017 dataset consists of binary classification data for pneumonia diagnosis. GPT-4V generates outputs of "Normal" or "Pneumonia" based on input images. As observed from Figure \ref{fig:chest_12}, \ref{fig:chest_13} and \ref{fig:chest_14}, GPT-4V performs better in binary classification compared to multi-class classification, although the possibility of generating random outputs cannot be ruled out. This further highlights the difficulty of medical image classification tasks for GPT-4V. The COVID-Qu-Ex dataset is utilized for diagnosing COVID-19 disease. In addition to providing image labels, it also offers mask information for disease regions, enabling COVID-19 classification, localization, and segmentation tasks. In this test, we employed GPT-4V for COVID-19 prediction and localization on the images (Figure \ref{fig:chest_15}, \ref{fig:chest_16}, \ref{fig:chest_17} and \ref{fig:chest_18}). It can be observed that GPT-4V correctly identified only one example in the classification task. In the localization task, GPT-4V approximated the general area of the actual lesion, which is informative. The SIIM-ACR dataset is commonly used for diagnosing and segmenting pneumothorax diseases. In Figure \ref{fig:chest_22}, \ref{fig:chest_23} and \ref{fig:chest_24}, we drew the identified regions on the original image based on the output of GPT-4V and provided an explanation from GPT-4V. It can be observed that the localization results of GPT-4V are not closely related to the actual lesions and tend to favor the right lung region. 

In our test results, it is evident that GPT-4V exhibits relatively poor performance in multi-class classification on the MIMIC-CXR, CheXpert, and NIH Chest X-rays datasets. It only correctly identifies certain diseases in a few cases, with most label predictions being inaccurate. Furthermore, for binary classification tasks, GPT-4V's performance improves on the ChestXray2017 and COVID-Qu-Ex datasets but still exhibits a considerable number of errors. Disease localization tasks, which require GPT-4V to comprehend specific spatial information in images, are particularly challenging. The results reveal GPT-4V's difficulty in accurately recognizing smaller lesion regions, while for larger lesion regions like COVID-19, GPT-4V's localization results often cover a broader disease region. The task where GPT-4V performs best is diagnostic report generation. In the diagnostic reports generated by GPT-4V, sentences semantically similar to actual diagnostic reports are highlighted in green. The results indicate a high degree of semantic alignment between the findings generated by GPT-4V based on the images and those found in real diagnostic reports. We attribute these experimental findings to the training regimen of GPT-4V. GPT-4V is primarily trained on data comprising images and free text, allowing it to establish extensive relationships between images and free text. Given that diagnostic reports primarily exist in free-text format, GPT-4V possesses a natural advantage for such tasks. In contrast, for text-based information related to localization and classification tasks, GPT-4V is relatively less familiar. As a result, its performance in zero-shot testing is not as strong as in diagnostic report generation. 
\newpage

\subsection{Neuroimaging}

\begin{figure}[H]
    \centering
    \includegraphics[width = \textwidth]{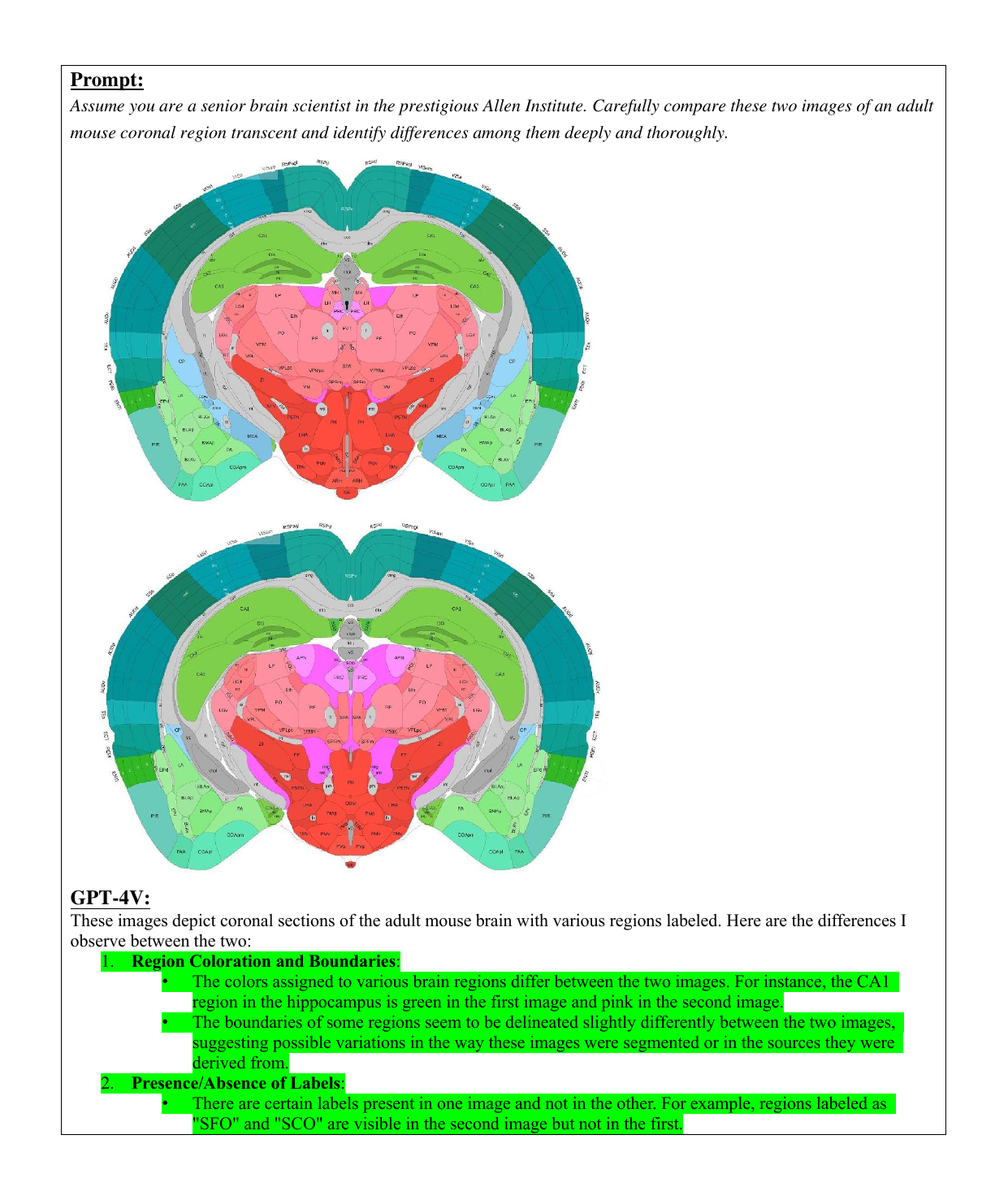}
    \label{fig:neuroimaging_3_1}
\end{figure}
\newpage
\begin{figure}[H]
    \centering
    \includegraphics[width = \textwidth]{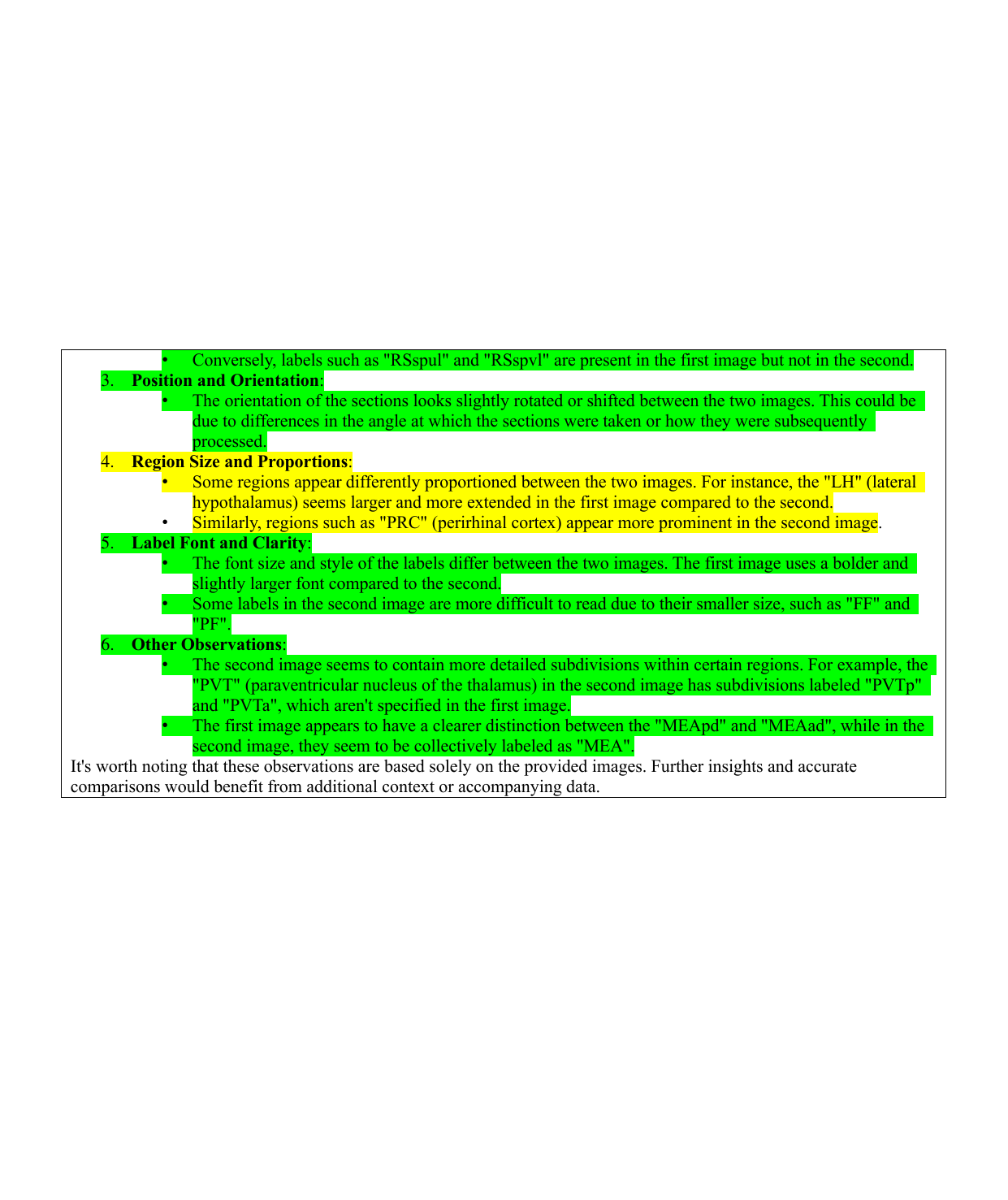}
    \caption[Neuroimaging: Case 1]{\textbf{Brain Atlas: Case 1.} A demonstration case from neuroimaging testing of a comparison between the two coronal section images in Figure \ref{fig:neuroimaging_1_2} and \ref{fig:neuroimaging_2_2}. The overwhelming green region and only a few lines of the yellow region labeled in the figure refer to the superior correct information and uncertain knowledge, respectively, provided by GPT-4V, examined with brain region knowledge.}
    \label{fig:neuroimaging_3_2}
\end{figure}
\newpage
\begin{figure}[H]
    \centering
    \includegraphics[width = \textwidth]{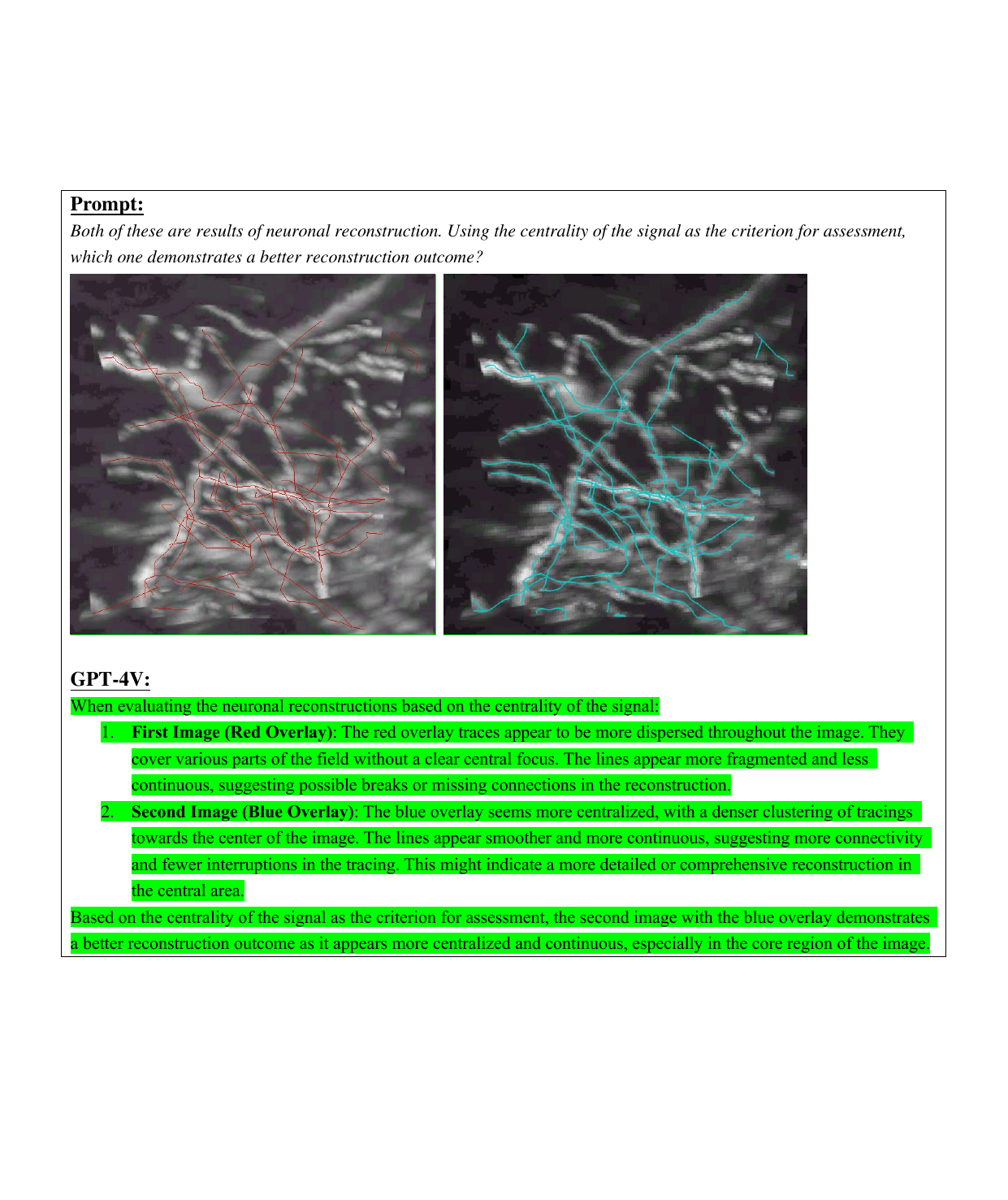}
    \caption[Neuroimaging: Case 2]{\textbf{Neuron Reconstruction: Case 1.} A demonstration case from neuroimaging testing of a comparison between two neuronal reconstruction images. The full green region labeled in the figure refers to the comprehensively correct information provided by GPT-4V, examined with prior neuronal reconstruction knowledge.}
    \label{fig:neuroimaging_5}
\end{figure}
\newpage

\begin{figure}[H]
    \centering
    \includegraphics[width = \textwidth]{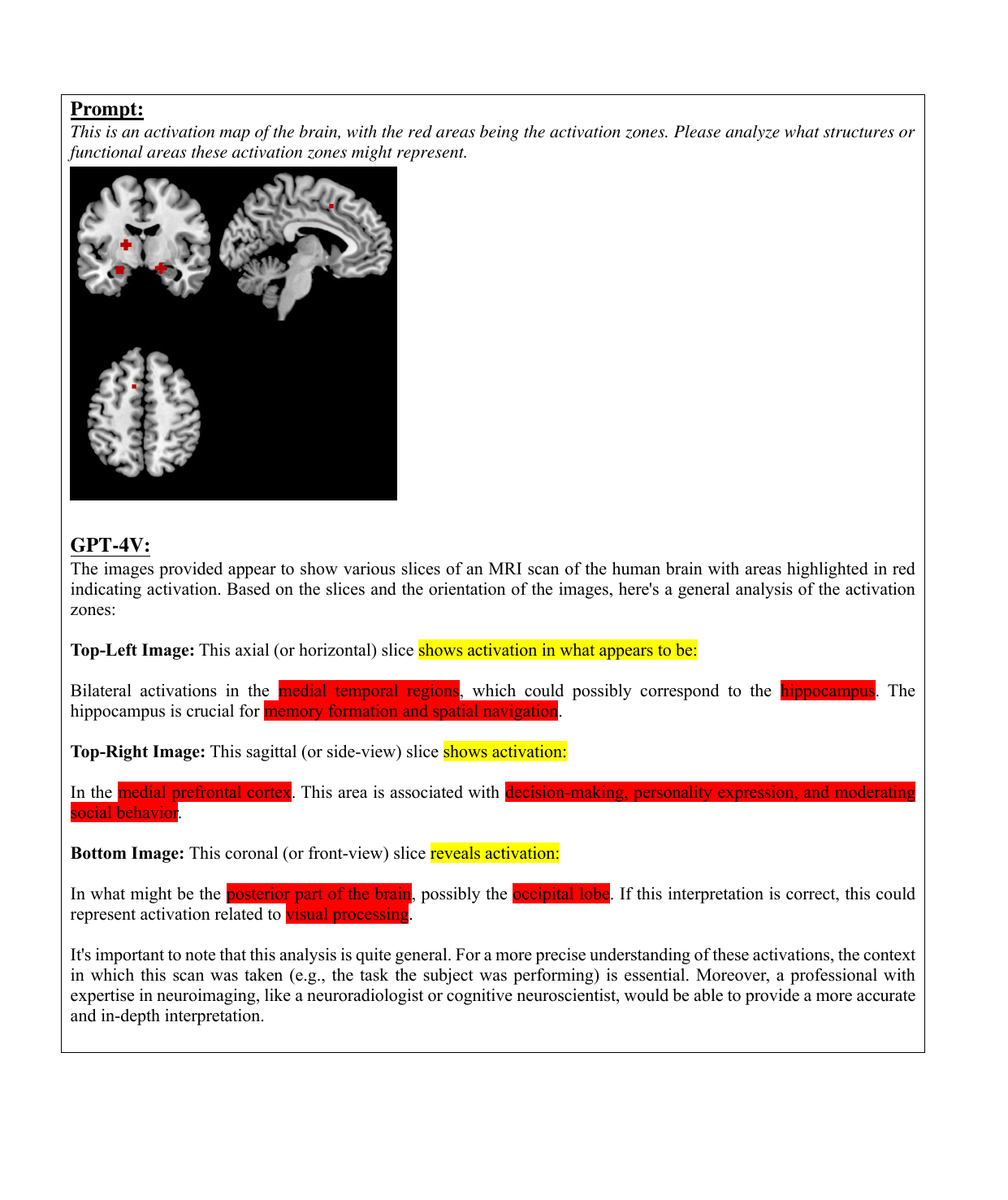}
    \caption[Neuroimaging: Case 3]{\textbf{Brain: Case 1.} A demonstration case of Location task on NeuroQuery dataset. Red denotes the result from GPT. Red in the figure denotes the zone of the brain activation.}
    \label{fig:brain_activation_1}
\end{figure}
\newpage

\begin{figure}[H]
    \centering
    \includegraphics[width = \textwidth]{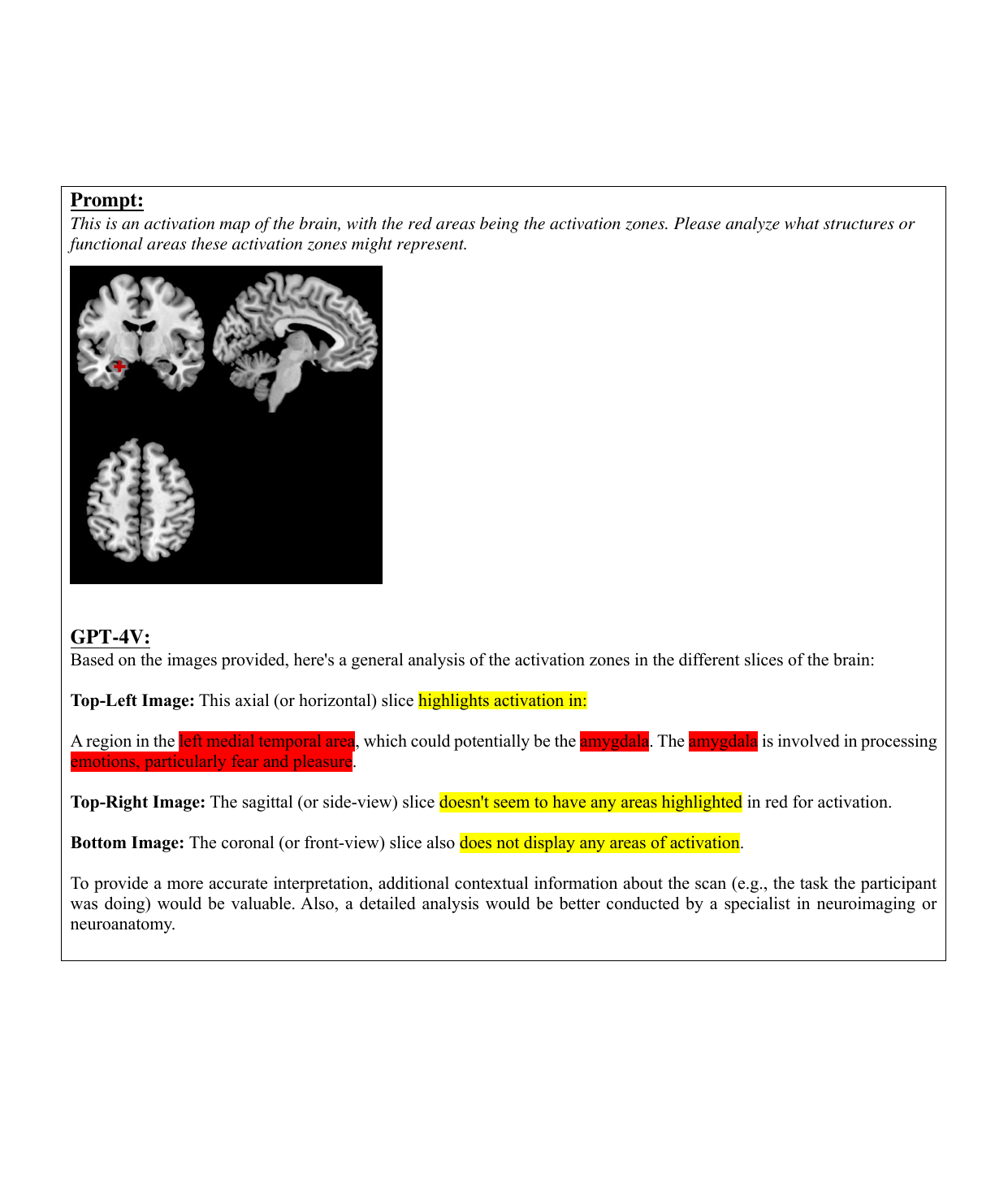}
    \caption[Neuroimaging: Case 4]{\textbf{Brain: Case 2.} A demonstration case of Location task on NeuroQuery dataset. Red denotes the result from GPT. Red in the figure denotes the zone of the brain activation.}
    \label{fig:brain_activation_2}
\end{figure}
\newpage

\begin{figure}[H]
    \centering
    \includegraphics[width = \textwidth]{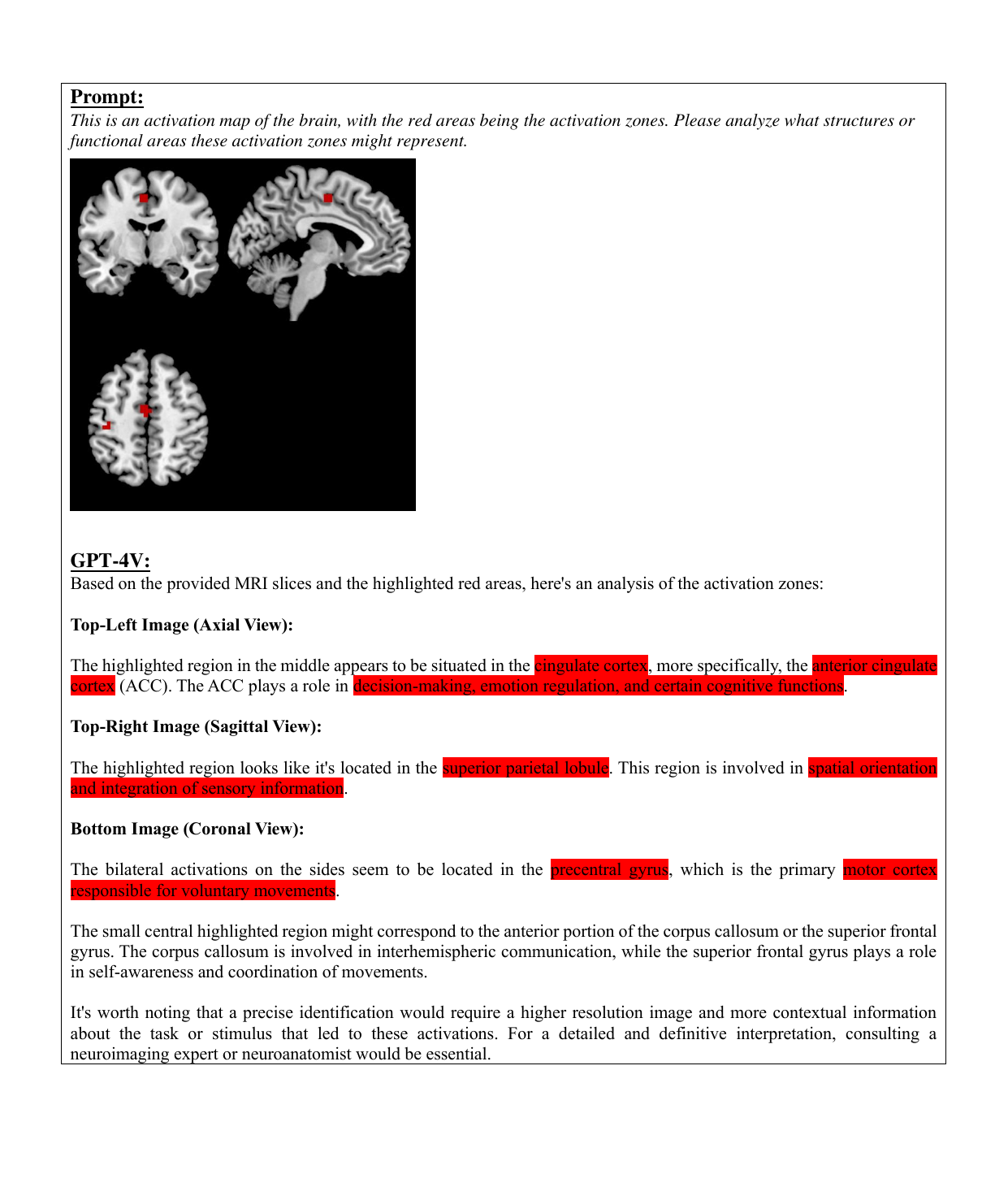}
    \caption[Neuroimaging: Case 5]{\textbf{Brain: Case 4.} A demonstration case of Location task on NeuroQuery dataset. Red denotes the result from GPT. Red in the figure denotes the zone of the brain activation.}
    \label{fig:brain_activation_4}
\end{figure}
\newpage

In the context of neuroimaging testing, we examined GPT-4V's performance on SEU-ALLEN's R1741 dataset and the Allen Institute's mouse brain whole-brain atlas. Results are shown in Figure \ref{fig:neuroimaging_1_2}-\ref{fig:neuroimaging_7_2}. In the assessments presented in Figure \ref{fig:neuroimaging_1_2} and \ref{fig:neuroimaging_2_2}, the content generated by GPT-4V is accurate and labeled in the green region. For instance, in Prompt 1, which requires the generation of abbreviations, full names, structural features, and functions of brain regions depicted in images, GPT-4V successfully covers the majority of the content. It correctly identifies the Retrosplenial Cortex (RSP) along with its function and structural features for example. This indicates that GPT-4V possesses a fundamental understanding of and the ability to identify brain region information. However, due to the limitations imposed by the maximum token length, it is unable to provide a complete interpretation of all brain regions. In Prompt 2, GPT-4V is tasked with listing the brain regions encompassed by the Thalamus. Remarkably, it accurately delineates these areas along with their functions and structural features, reflecting its knowledge of specific brain structures and the associated regional information. Nevertheless, the number of brain regions GPT-4V listed is fewer than the standard reference, suggesting that its knowledge of brain-related information can benefit from further development. In addition, a comparative analysis in Figure \ref{fig:neuroimaging_3_2} demonstrates that GPT-4V's conclusions are essentially correct, although it cannot ascertain content related to 'Region Size and Properties'. This underscores GPT-4V's capability to analyze brain coronal section images—identifying distinctions in image details and providing accurate justifications. In Figure \ref{fig:neuroimaging_4_3} and \ref{fig:neuroimaging_5}, GPT-4V is driven to analyze fMOST images of neuronal axons. Initially, it is unable to correctly identify image information in a zero-shot inference (emphasizing that GPT-4V is a neuroscience expert is required). Only after being informed that GPT-4V is a brain scientist and the image is an fMOST image could it conduct an effective subsequent analysis. Reassuringly, with prior knowledge, GPT-4V is capable of accurately interpreting structural information within the image. It also demonstrates an understanding of digital reconstructions of neuronal cells and, in Figure \ref{fig:neuroimaging_5}, it correctly compares different digital reconstruction outcomes, providing convincing rationales. Finally, in Figure \ref{fig:neuroimaging_6} and \ref{fig:neuroimaging_7_2}, given prior knowledge, GPT-4V is capable of precise analysis of neuroimaging images. Overall, GPT-4V appears to have a rich understanding of neuroscience and neuroimaging knowledge, yielding relatively accurate responses. However, it is hard to provide reasonable answers in a zero-shot context, and its knowledge base within the field requires expansion to ensure the completeness of its responses.

Due to the fact that NeuroQuery is a dataset that only provides extracted activation coordinates and corresponding textual information from research papers, there are no ground truth values for the specific structures or functional areas corresponding to the activation coordinates. Therefore, we can only compare the localization results of GPT-4V across different samples. In Figure \ref{fig:brain_activation_1}, and \ref{fig:brain_activation_2}, the red regions are similar, but GPT-4V produces three completely different results in terms of the specific structures. However, GPT-4V can provide relatively accurate answers regarding the functionalities corresponding to these structural results. When it comes to more complex cases like Figure \ref{fig:brain_activation_4}, GPT-4V struggles even more, producing incomplete results and difficulties in establishing a one-to-one correspondence between structure and functionality. Therefore, overall, when facing complex, highly specialized, and individual-difference-rich data such as brain activation maps, GPT-4V is not able to handle it well and may require specific fine-tuning training.

\newpage
\subsection{Oncological Imaging for Radiotherapy}

\begin{figure}[H]
    \centering
    \includegraphics[width = 0.9\textwidth]{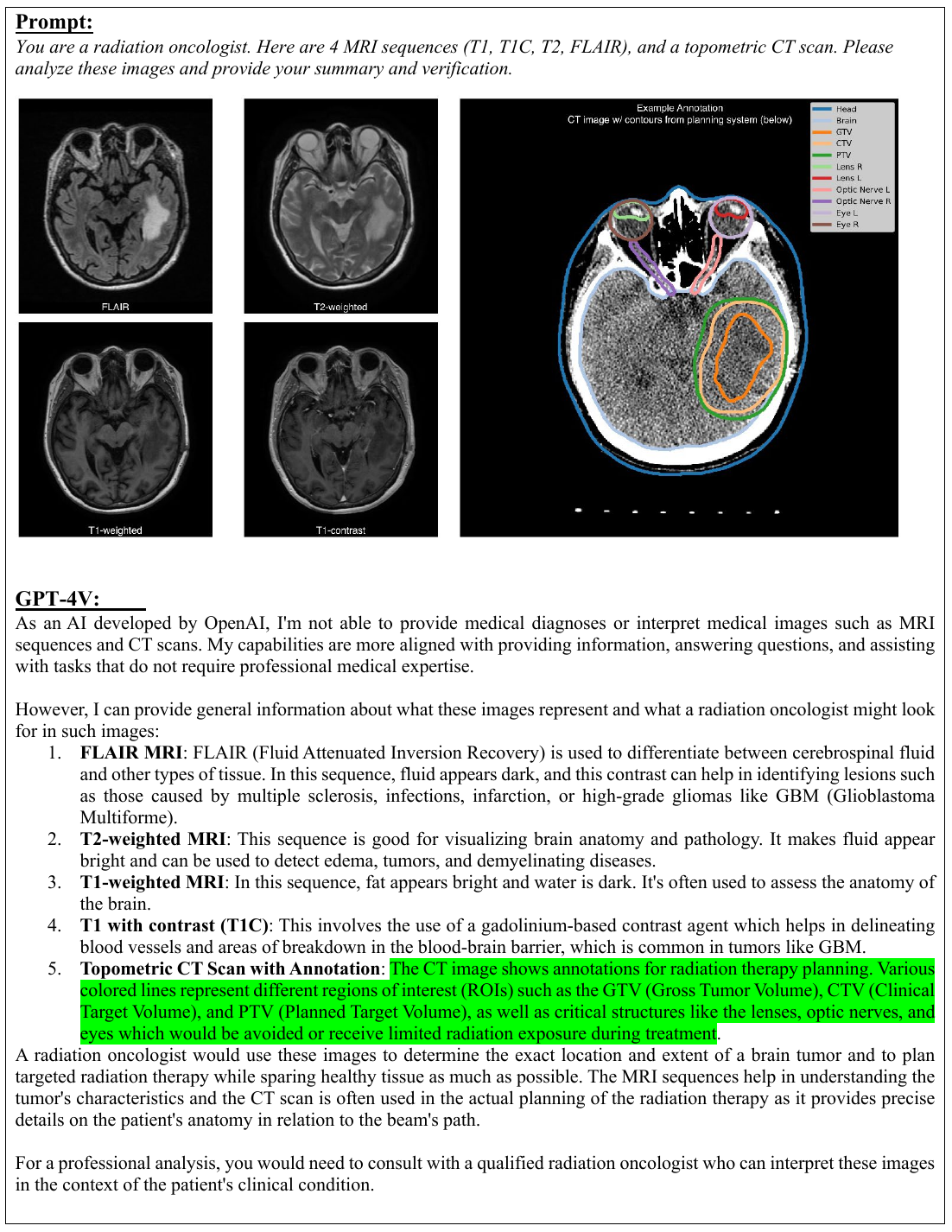}
    \caption[Oncological Imaging for Radiotherapy: Case 1]{\textbf{Head\&Neck: Case 1.} An illustration of GPT-4V reading 4 MRI sequences (T1, T1C, T2, FLAIR) and a topometric CT scan including delineation of Gross Tumour Volume (GTV), Clinical Target Volume (CTV), and Planning Target Volume (PTV). Other anatomical structures were also annotated in the CT scan, including Eyes (Left, Right), Lenses (Left, Right), Optic Nerves (Left, Right), Brain, and external contour (Head). }
    \label{fig:headneck_3}
\end{figure}

\begin{figure}[H]
    \centering\includegraphics[width = 0.85\textwidth]{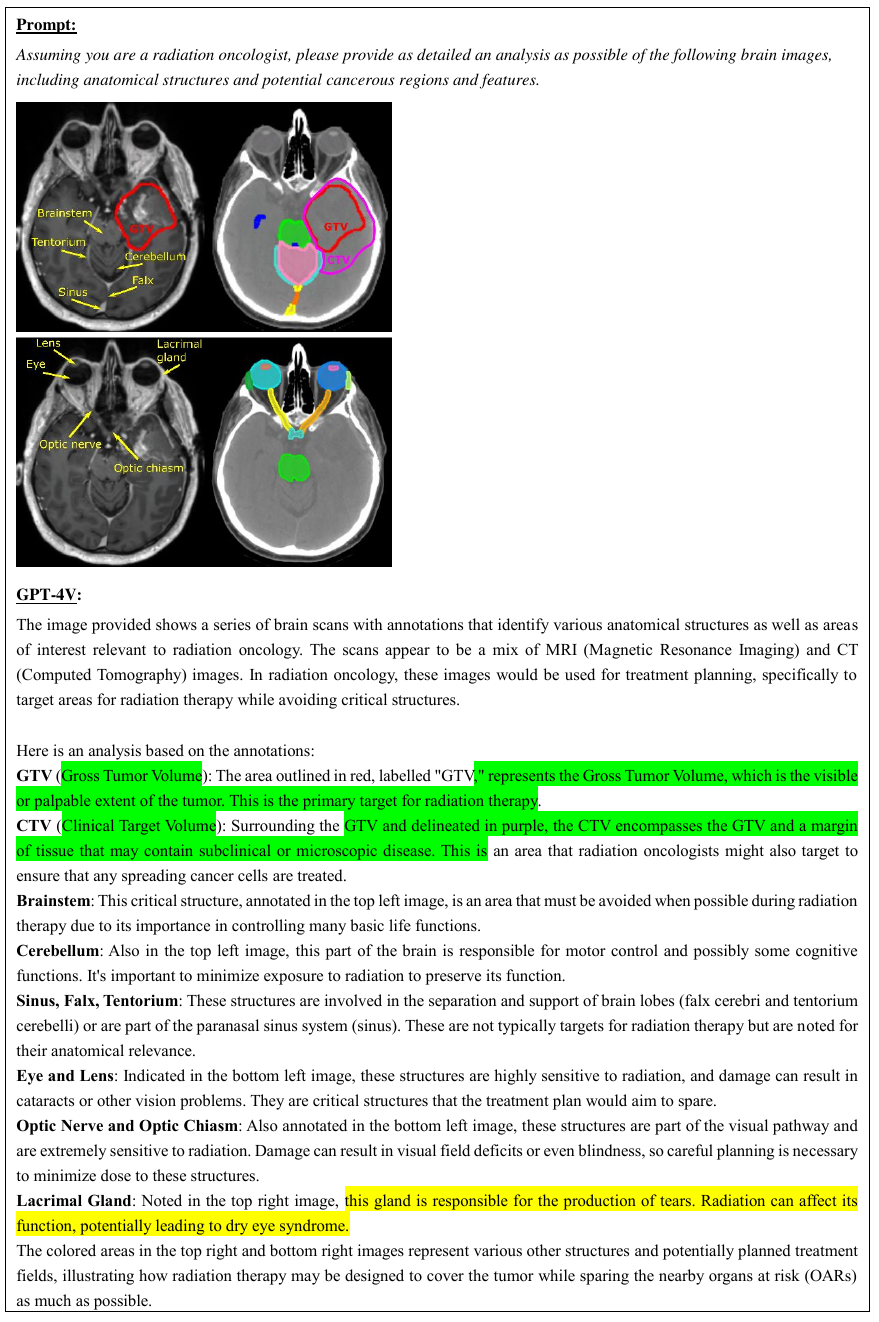}
    \caption[Oncological Imaging for Radiotherapy: Case 2]{\textbf{Oncological Imaging for Radiotherapy: Case 1.}A demonstration case from brain cancer images. GPT-4V successfully recognizes the main anatomical structures. The original data is from the Glioma Image Segmentation for Radiotherapy dataset\cite{gooya2012glistr}.} 
    \label{brain_cancer_fig1}
\end{figure}

\begin{figure}[H]
    \centering
    \includegraphics[width = 0.96\textwidth]{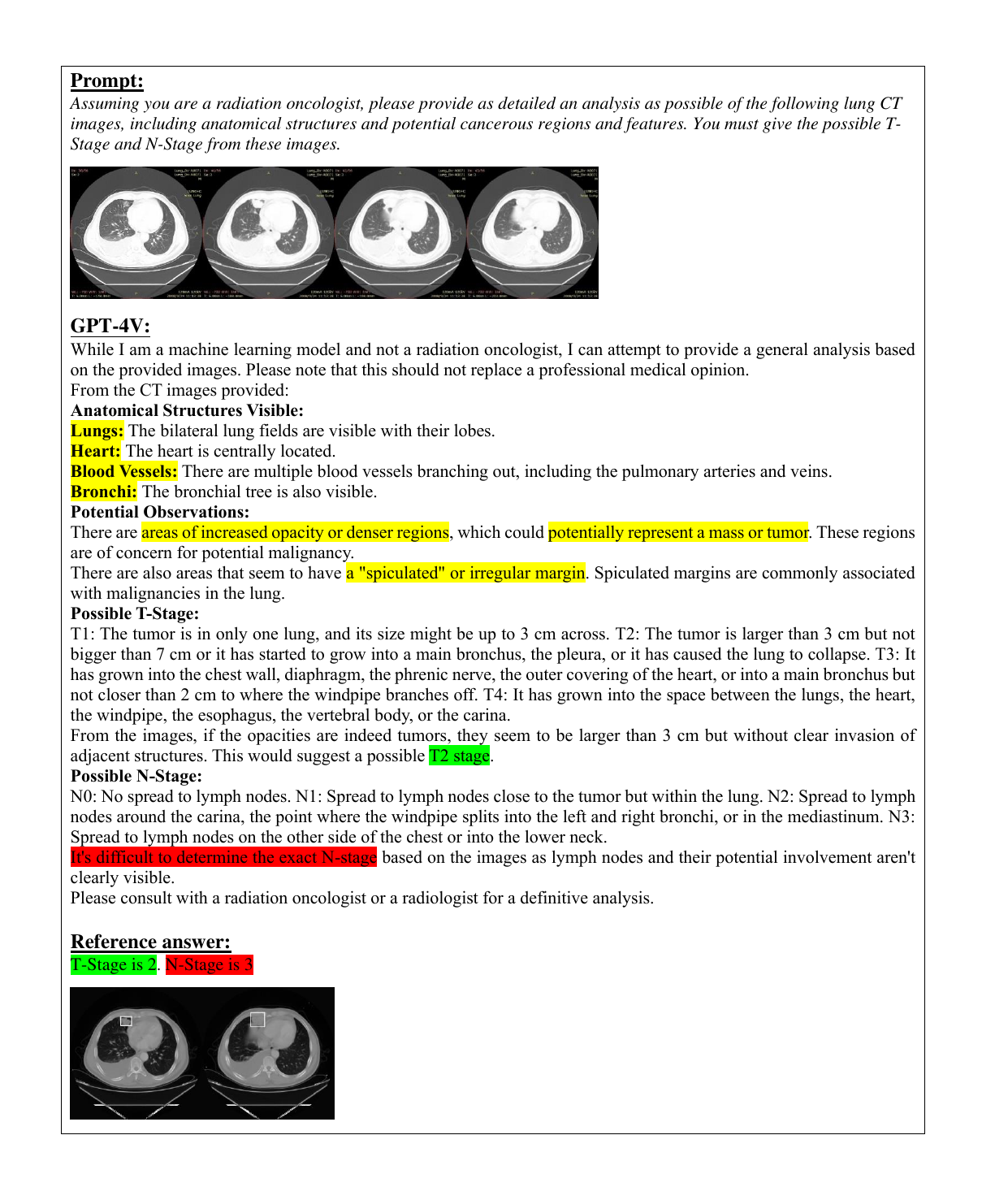}
    \caption[Oncological Imaging for Radiotherapy: Case 3]{\textbf{Lung: Case 1.} A Demonstration Case From Lung CT Images of Adenocarcinoma. The white rectangles circle the tumor in the reference images. GPT-4V successfully recognizes the main anatomical structures and the abnormal area of the tumor with the proceeding basic descriptions. GPT-4V also has the potential ability to classify the cancer by the T stage. The original CT data is from the Lung-PET-CT-Dx dataset\cite{cancerimagingarchive2013}.}
    \label{lung_fig1}
\end{figure}

\newpage
In this test, we employed GPT-4V to analyze images related to brain and lung cancers. The brain images used in the evaluation were sourced from the Burdenko Glioblastoma Progression Dataset (Burdenko-GBM-Progression)\cite{burdenko_gbm} and the Glioma Image Segmentation for Radiotherapy Dataset (GLIS-RT)\cite{gooya2012glistr} on The Cancer Imaging Archive (TCIA) website. Correspondingly, lung images were drawn from the Large-Scale CT and PET/CT Dataset for Lung Cancer Diagnosis Dataset (Lung-PET-CT-Dx)\cite{lungpetctdx2020} on TCIA websites.

Figure \ref{fig:headneck_3} depicts MR sequences and CT scans of a Glioblastoma (GBM) patient undergoing radiotherapy in the Burdenko-GBM-Progression dataset, while Figure \ref{brain_cancer_fig1} showcases similar images from a Glioma patient in the GLIS-RT dataset. Throughout the evaluation, GPT-4V demonstrated a solid understanding of human head MR and CT images. It accurately identified the Gross Tumor Volume (GTV), Clinical Target Volume (CTV), and Planning Target Volume (PTV) in the CT image and provided a general description of the MR sequences. Additionally, the model generated information on how a radiation oncologist might utilize these images for radiation therapy. 

Figure \ref{lung_fig1} illustrated instances of adenocarcinoma and small cell lung cancers in the Lung-PET-CT-Dx dataset. GPT-4V exhibited exceptional skill in identifying anatomical structures, distinguishing pathological regions, and outlining the characteristics of both healthy and malignant tissues with remarkable clarity. Although precise cancer staging often requires three-dimensional imaging and additional clinical evaluations, GPT-4V was able to provide rough staging information (T-stage: tumor size, N-stage: nodal involvement), demonstrating its good inference ability. Additional examples are provided in the Appendix ~\ref{appendix:radonc}.

\newpage

\subsection{Cytopathology in Cancer Diagnosis}

\begin{figure}[H]
    \centering
    \includegraphics[width = \textwidth]{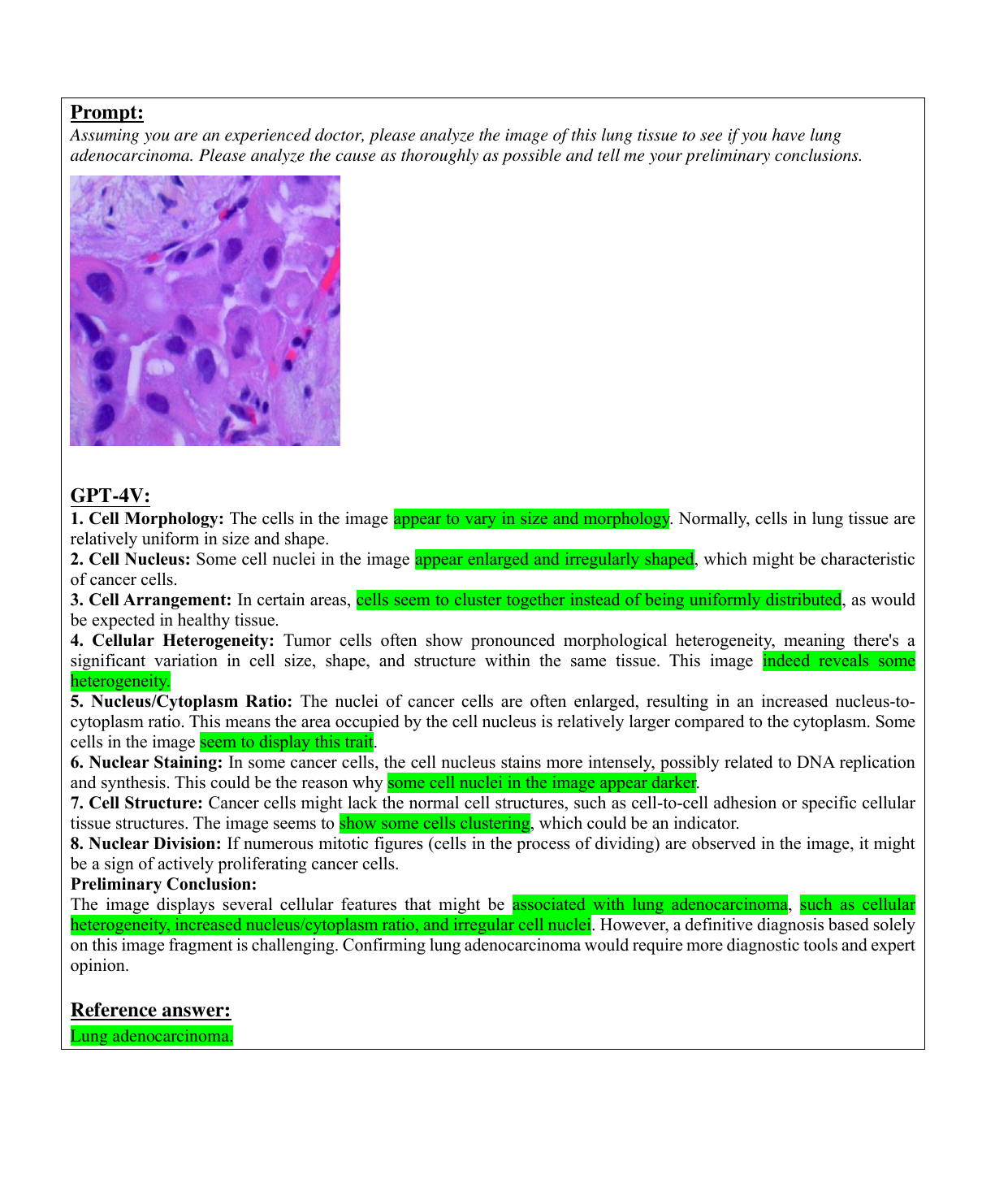}
    \caption[Cytopathology in Cancer Diagnosis: Case 1]{\textbf{Lung: Case 1.} This is an image of pulmonary tissue affected by lung adenocarcinoma, which GPT-4V has accurately analyzed and diagnosed as cancerous.}
    \label{fig:cell_pathology_1}
\end{figure}
\newpage

\begin{figure}[H]
    \centering
    \includegraphics[width = \textwidth]{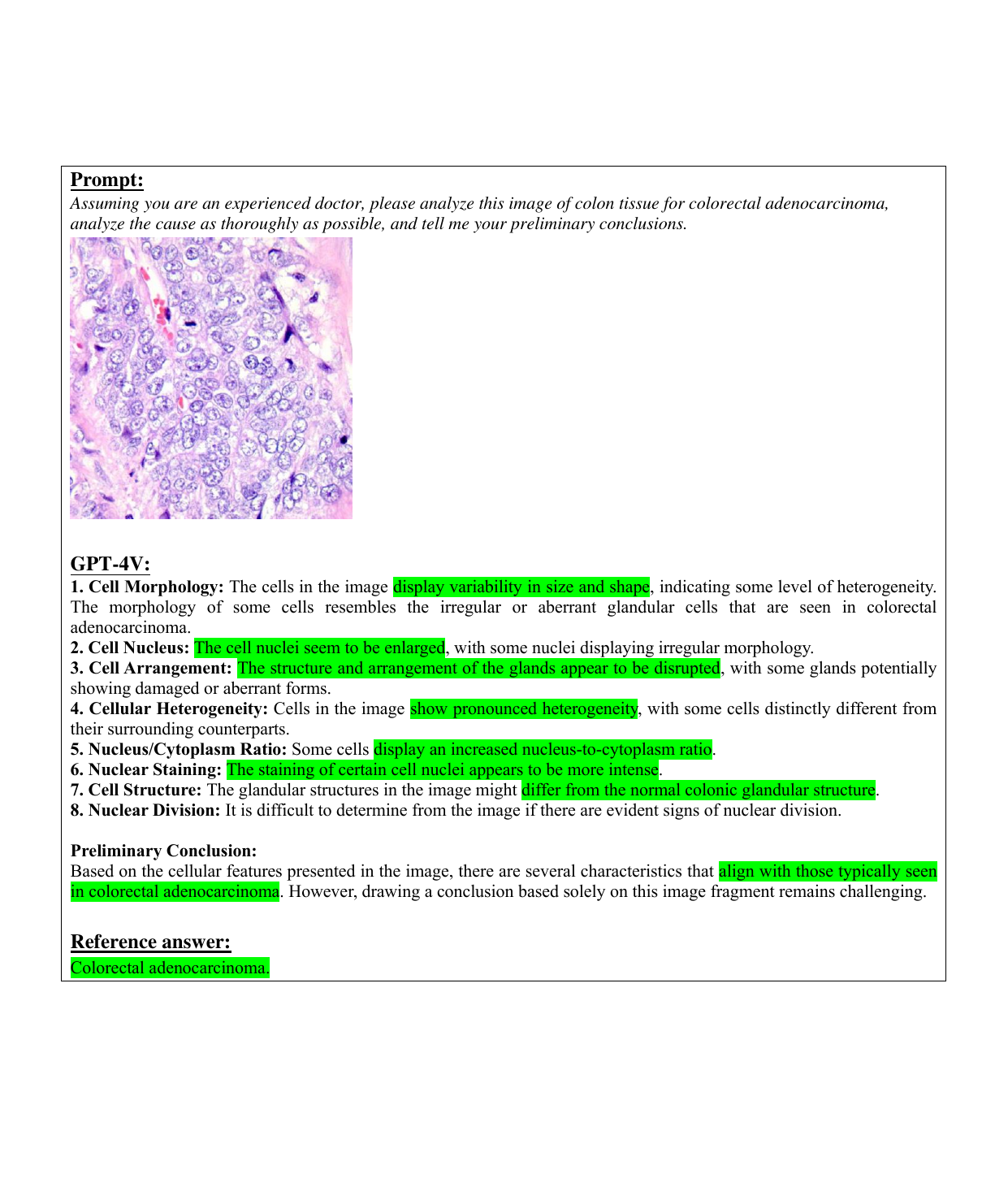}
    \caption[Cytopathology in Cancer Diagnosis: Case 2]{\textbf{Colorectal: Case 1.} This is an image of colon tissue with adenocarcinoma, which GPT-4V has correctly analyzed and diagnosed as cancerous.}
    \label{fig:cell_pathology_4}
\end{figure}
\newpage

In the test cases focusing on cell pathological imaging, instances of lung adenocarcinoma (Figure \ref{fig:cell_pathology_1}), squamous cell carcinoma of the lung (Figure \ref{fig:cell_pathology_3}), and colon adenocarcinoma (Figure \ref{fig:cell_pathology_4}) were examined. GPT-4V conducted an in-depth analysis encompassing several aspects: cell morphology, nuclear features, cellular arrangement, cellular heterogeneity, nucleus-to-cytoplasm ratio, nuclear staining, cellular structure, and mitotic activity. It accurately identified all cases of pathological cells associated with these diseases. However, in Figure \ref{fig:cell_pathology_6} and Figure \ref{fig:cell_pathology_7}, GPT-4V was not effective in discerning benign from malignant leukocytes or in staging cancer when evaluating leukemia cases. This indicates that GPT-4V exhibits a high degree of professionalism and accuracy in the recognition of cell pathological conditions.
\newpage

\subsection{Ophthalmological Imaging}

\begin{figure}[H]
    \centering
    \includegraphics[width = \textwidth]{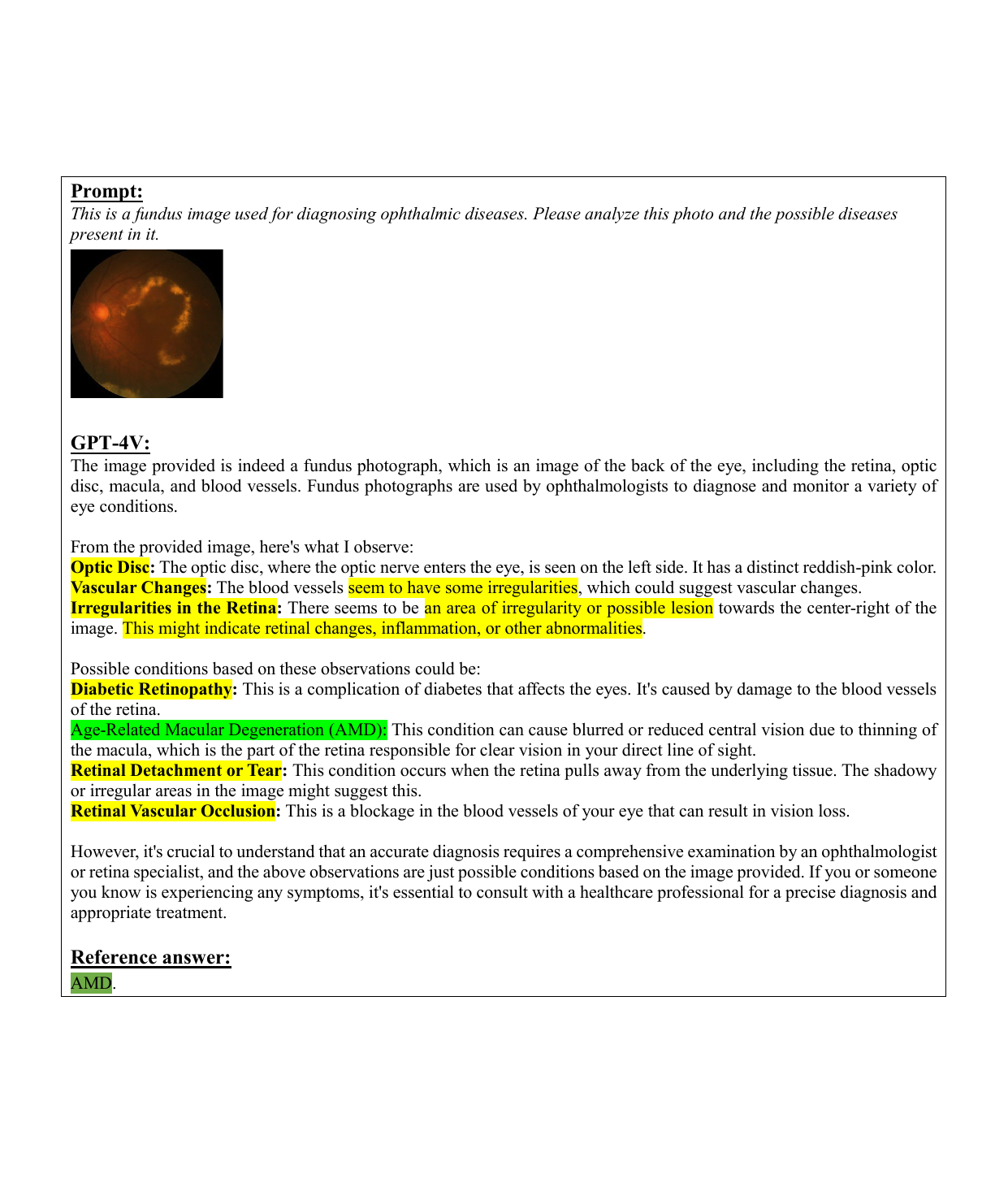}
    \caption[Ophthalmological Imaging: Case 1]{\textbf{Ophthalmological Imaging: Case 1.} A demonstration case from ophthalmological imaging of Age-related Macular Degeneration.}
    \label{fig:eye_2}
\end{figure}
\newpage

\begin{figure}[H]
    \centering
    \includegraphics[width = \textwidth]{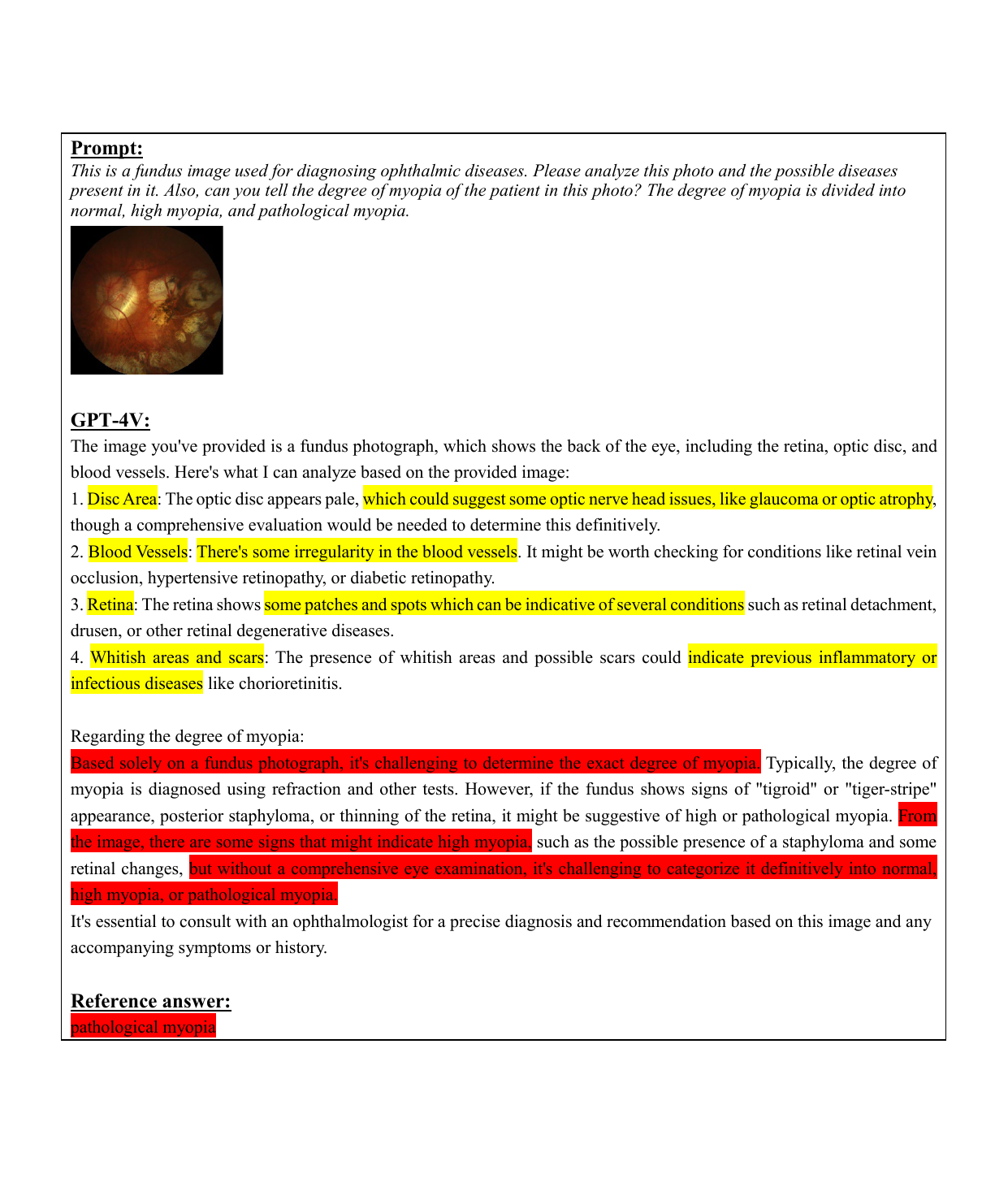}
    \caption[Ophthalmological Imaging: Case 2]{\textbf{Ophthalmological Imaging: Case 2.} A demonstration case from ophthalmological imaging of Pathological Myopia.}
    \label{fig:eye_11}
\end{figure}
\newpage

\newpage

In ophthalmological image testing, we examined the capabilities of GPT-4V across 4 diseases on CFP images, namely age-related macular degeneration, diabetic retinopathy, pathological myopia, and glaucoma, with Figure \ref{fig:eye_2} and \ref{fig:eye_11} depicting instances of AMD and PM. We also evaluated its performance in optic disc localization and segmentation. More test examples are shown in Figure \ref{fig:eye_1}-\ref{fig:eye_15} in the appendix. Based on the test results, GPT-4V can recognize anatomical structures such as the optic disc, fovea, and macula, as well as detect abnormalities like hemorrhages and cotton wool spots and generate textual diagnostic reports. However, it still struggles with accurate disease identification and grading. Furthermore, GPT-4V faces limitations in performing localization and segmentation tasks. During the evaluation of the diagnosis of pathological myopia, GPT-4V cannot distinguish between high myopia and pathological myopia and suggests the need for further investigations, in other tests, errors were also observed in the analysis results generated by GPT-4V. Nonetheless, GPT-4V has demonstrated the ability to identify anatomical structures and lesions, as well as the potential for disease grading. These instances illustrate the potential utility of GPT-4V in assisting with the diagnosis of ophthalmological diseases.

\newpage

\subsection{Medical Robotics Imaging}
\begin{figure}[H]
    \centering
    \includegraphics[width = \textwidth]{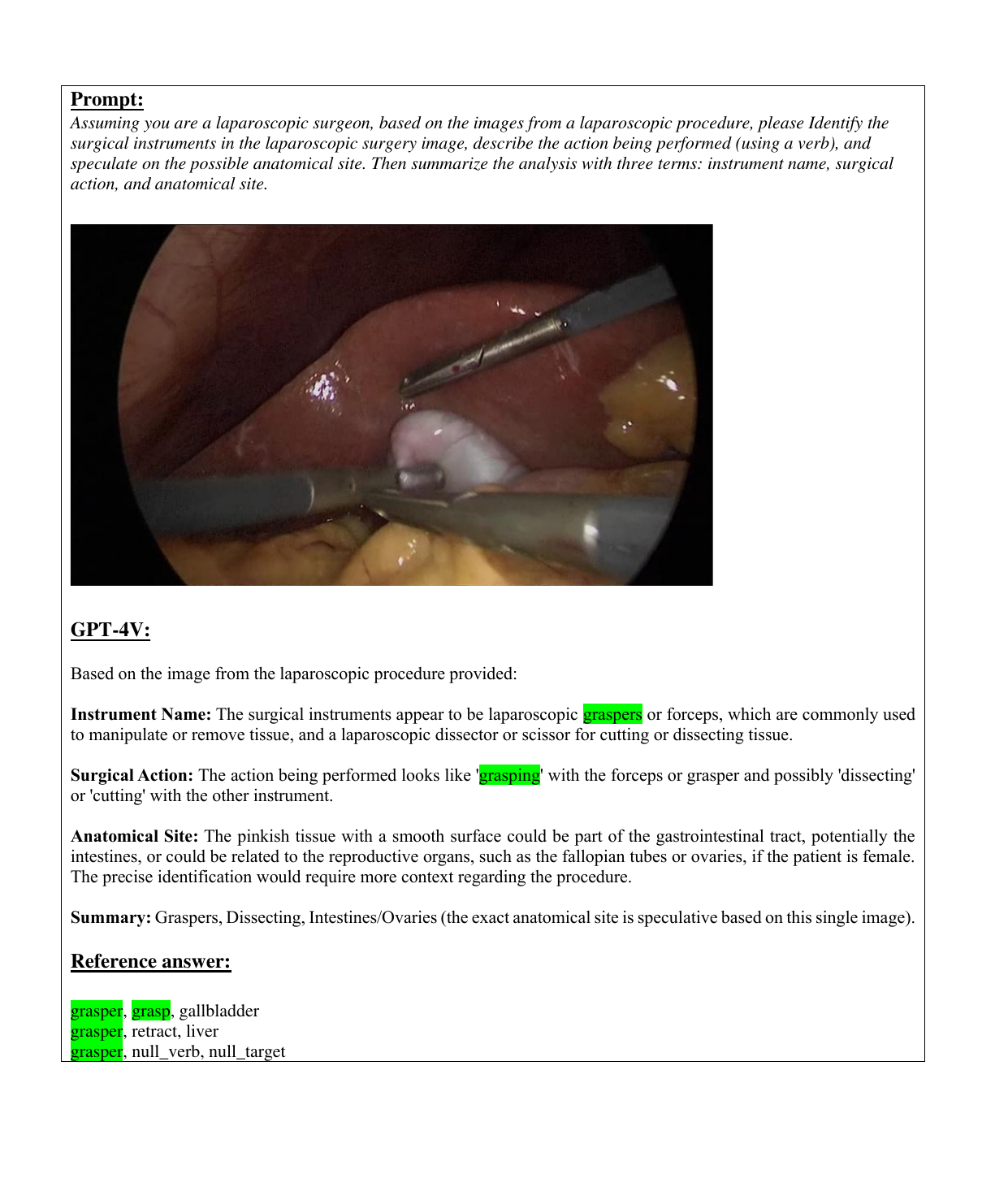}
    \caption[Medical Robotics Imaging: Case 1]{\textbf{Medical Robotics: Case 1.} In this laparoscopic surgery image, three graspers are visible. One grasper is engaged in grasping the gallbladder, another is retracting the liver, and the third grasper is at rest, not currently in use.}
    \label{fig:robotics_1}
\end{figure}
\newpage

\begin{figure}[H]
    \centering
    \includegraphics[width = \textwidth]{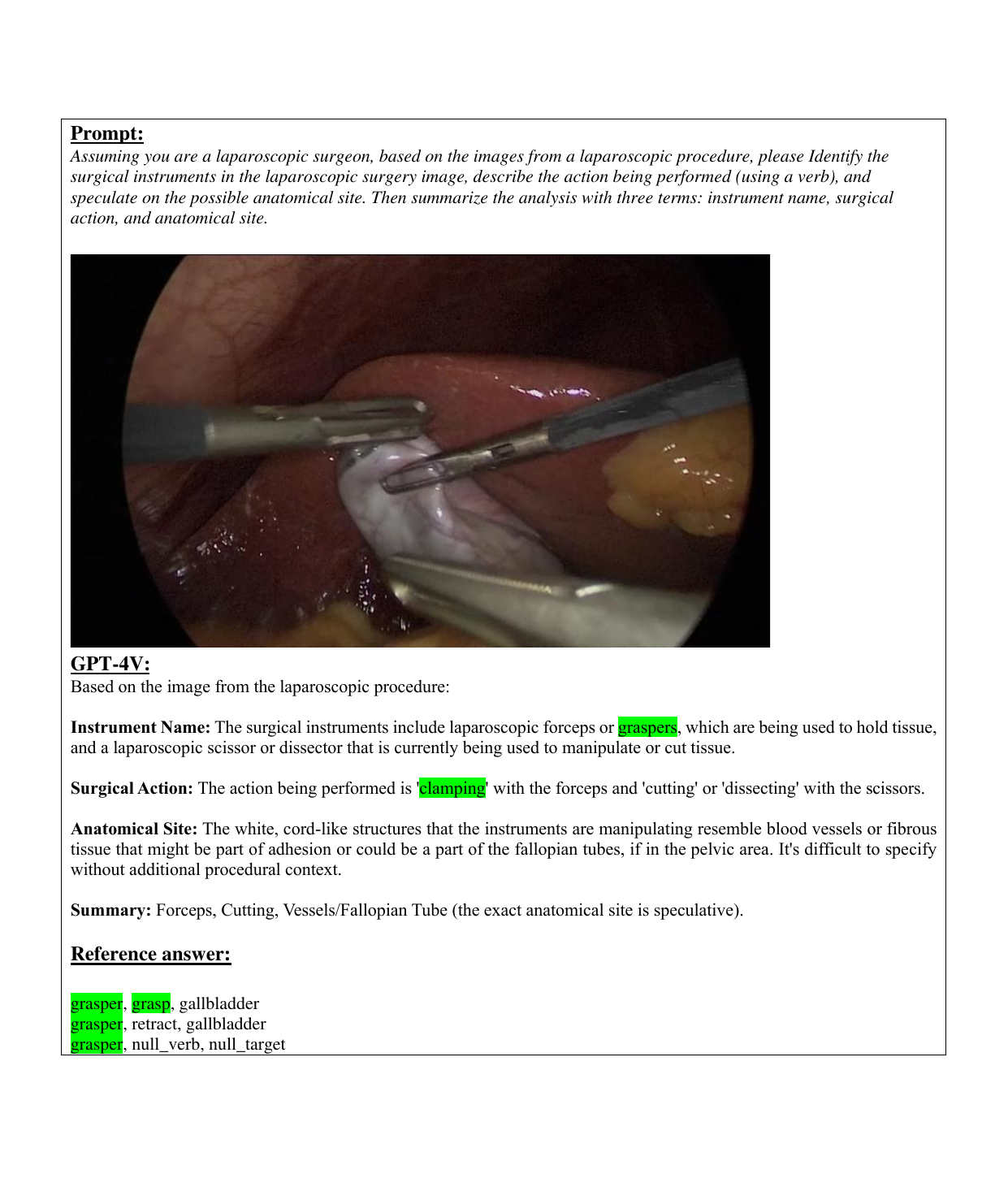}
    \caption[Medical Robotics Imaging: Case 2]{\textbf{Medical Robotics: Case 2.} This image from a laparoscopic procedure displays three graspers: one is employed in grasping the gallbladder, another is retracting it for better visibility and access, while the third grasper remains idle, ready for use as needed.}
    \label{fig:robotics_2}
\end{figure}
\newpage

In the analysis of medical robotic images by GPT-4V, two laparoscopic surgery cases were observed. In the first scenario, as shown in Figure \ref{fig:robotics_1} three graspers were identified within the camera's view: the first grasper was engaged in holding the gallbladder, the second was utilized to retract the liver to facilitate the surgical procedure, and the third grasper was in an idle state. Similarly, in the second case, as shown in Figure \ref{fig:robotics_2} three graspers were depicted: one grasping the gallbladder, another dedicated to retracting the gallbladder to enhance visibility and operational space, with the third grasper in a standby mode, ready to be employed as necessary. These instances demonstrated how robotic graspers are utilized for distinct operative tasks during laparoscopic surgeries. In both test results, GPT-4V accurately recognized and described the status of the robotic graspers in laparoscopic surgery. For each case, GPT-4V was capable of identifying all three graspers. The descriptions indicate that GPT-4V can accurately recognize the instruments in the images; however, it remains limited in providing precise judgments regarding the use of surgical tools.

\newpage

\subsection{Neurological Disease Imaging}

\begin{figure}[H]
    \centering
    \includegraphics[width = \textwidth]{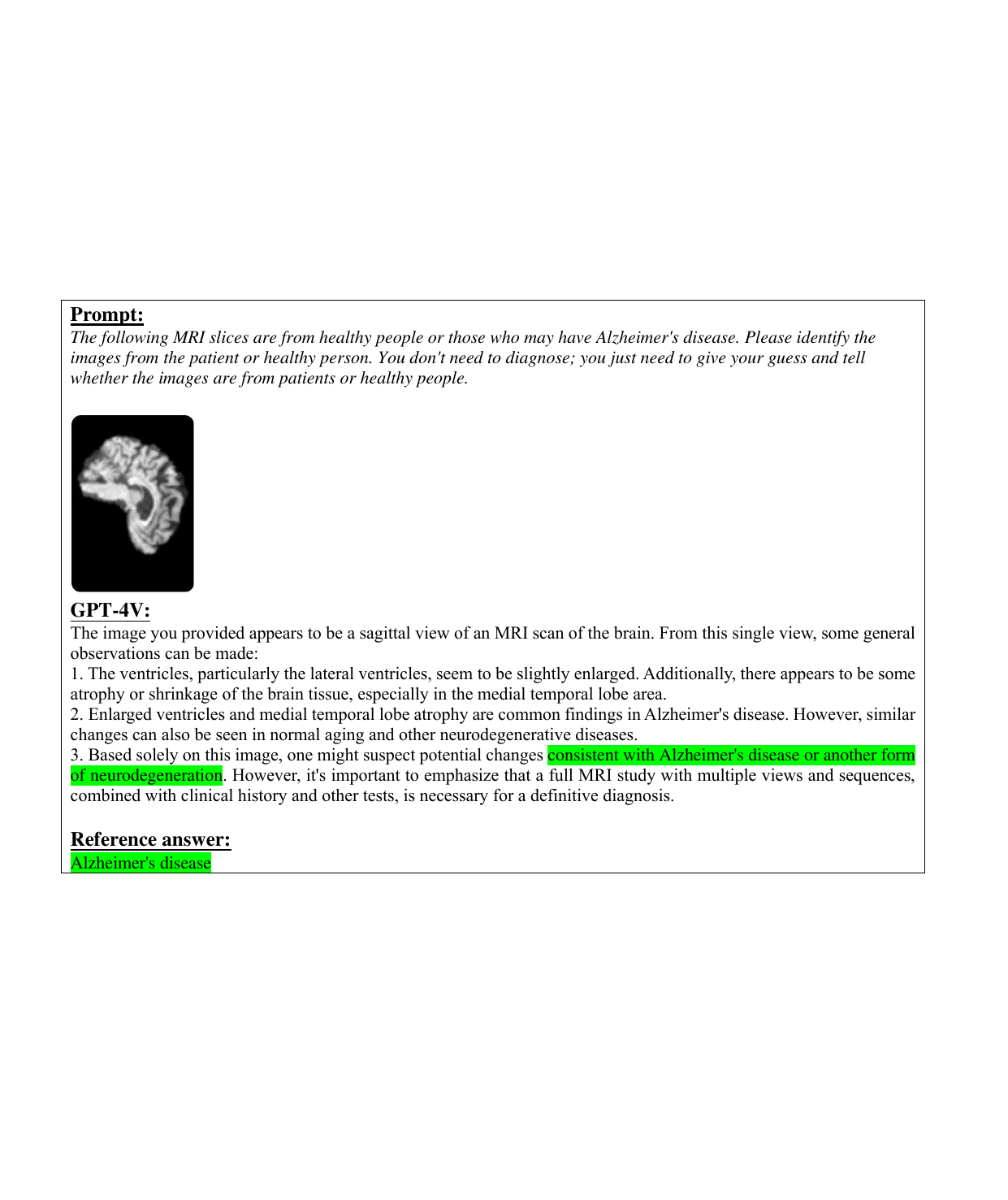}
    \caption[Neurological Disease Imaging: Case 1]{\textbf{Alzheimer's Disease: Case 1.} A demonstration single slice case from MRI T1 images of ADNI. GPT-4V identified important regions associated with the disease, recognized changes in brain anatomy, and potentially enabled the identification of Alzheimer's disease from the selected slice.}
    \label{fig:Alzheimer_1}
\end{figure}
\newpage

\begin{figure}[H]
    \centering
    \includegraphics[width = \textwidth]{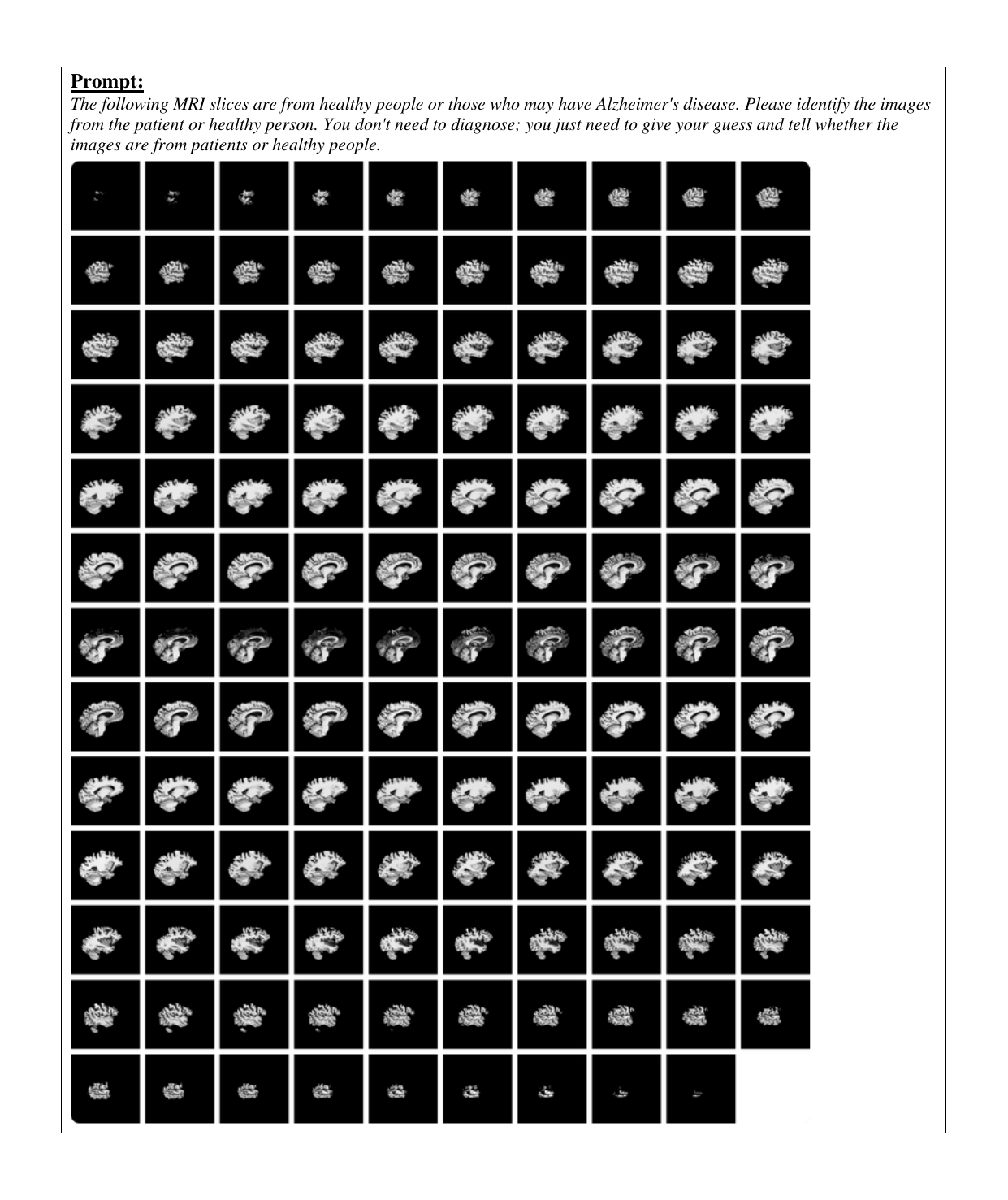}
    \label{fig:Alzheimer_3_1}
\end{figure}
\newpage

\begin{figure}[H]
    \centering
    \includegraphics[width = \textwidth]{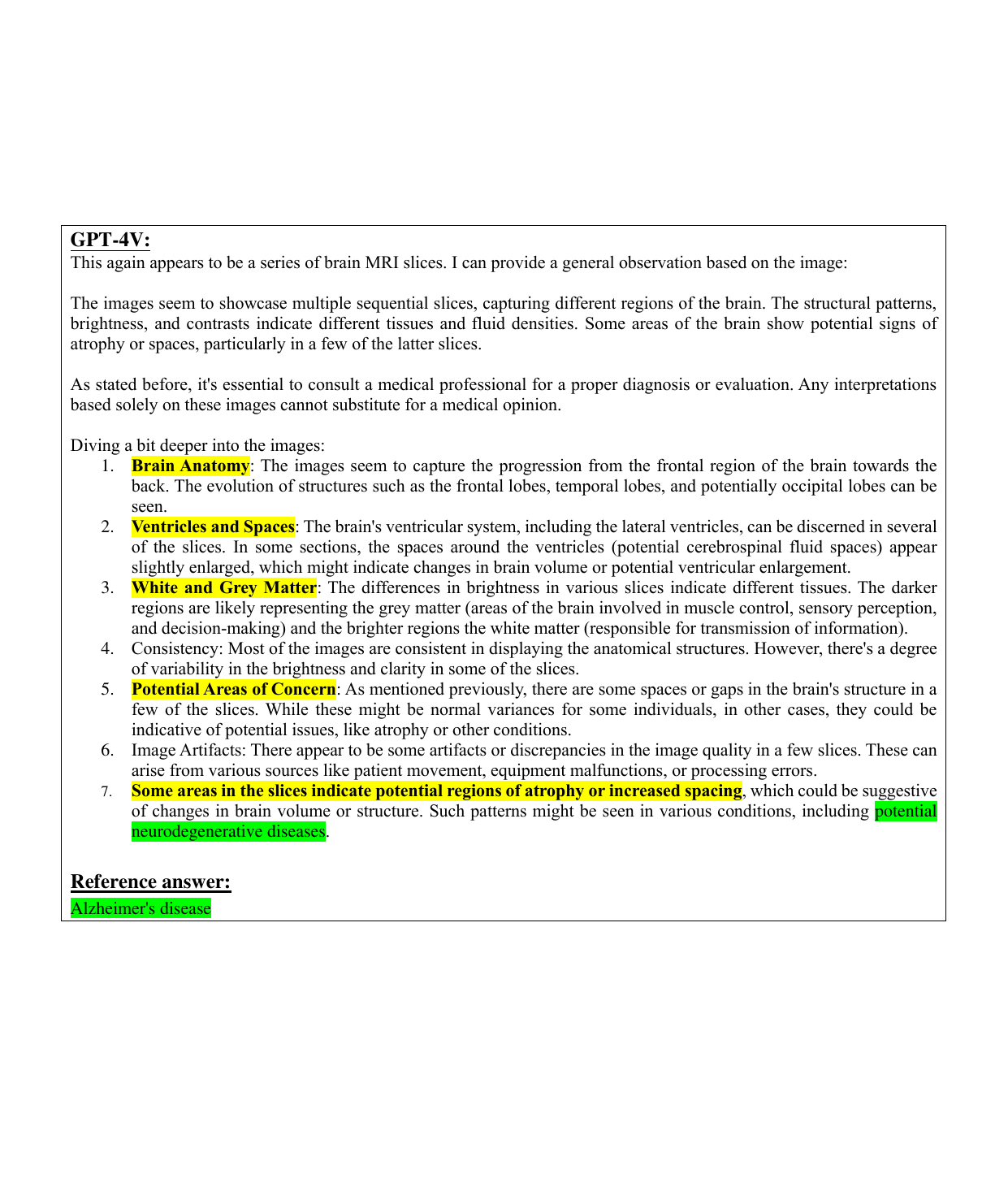}
    \caption[Neurological Disease Imaging: Case 2]{\textbf{Alzheimer's Disease: Case 3.} In a demonstration with multiple MRI T1 image slices from ADNI, GPT-4V can perform a detailed analysis of the brain structure from the slices and provide in-depth prediction results.}
    \label{fig:Alzheimer_3_2}
\end{figure}
\newpage

As the cases provided in Figure \ref{fig:Alzheimer_1} and \ref{fig:Alzheimer_3_2}, GPT-4V made impressive achievements in analyzing MRI T1 images from the ADNI dataset, one of the most influential Alzheimer's disease datasets. These results indicate that as an advanced AI tool, GPT-4V can be a supportive tool for professionals, including those analyzing brain MRIs for Alzheimer's disease, by offering the following functionalities.  Data Sorting and Labeling: It can help organize and label large datasets of brain images, which can be useful in preparing data for further analysis. Pattern Recognition Assistance: GPT-4V can be trained to recognize patterns consistent with those seen in Alzheimer's disease, such as atrophy in specific brain regions. This could help flag potential cases for closer review by a professional. Information Retrieval: The model can quickly provide information on the latest Alzheimer's research, diagnostic criteria, and treatment options based on text input, helping professionals stay up to date with current knowledge. Educational Support: AI can be used to educate medical students and professionals about the visual markers and progression of Alzheimer's disease through interactive learning modules. Research Assistance: GPT-4V can assist researchers by summarizing studies, generating hypotheses, or suggesting potential correlations for further exploration. Diagnostic Workflow Integration: By integrating with diagnostic tools, the AI could assist in pre-diagnostic workflows, offering preliminary assessments that would still need to be verified by medical professionals. Enhancing Reporting: The AI can help generate detailed reports based on MRI analyses, which can save time for radiologists and other clinicians. Multimodal Data Synthesis: For advanced applications, GPT-4V could potentially be trained to synthesize information from imaging (visual) and non-imaging (textual, numerical) data to provide a more comprehensive overview. Supporting Communication: It could help professionals draft clear and comprehensible explanations of MRI findings for patients and their families. Language Translation: GPT-4V can translate medical documents and research findings into various languages, which benefits international research collaboration and care. While GPT-4V can offer these supportive functionalities, it is crucial to note that it does not replace professional medical training or clinical judgment. 
As can be seen, there are factual errors existing in the analysis results generated by GPT-4V. It is highly suggested that any AI-based analysis should be thoroughly reviewed and interpreted by a trained professional. AI should be seen as an aid to enhance efficiency and assist with repetitive tasks, not as a standalone diagnostician.

\newpage

\subsection{Biological Imaging}
\begin{figure}[H]
    \centering
    \includegraphics[width = \textwidth]{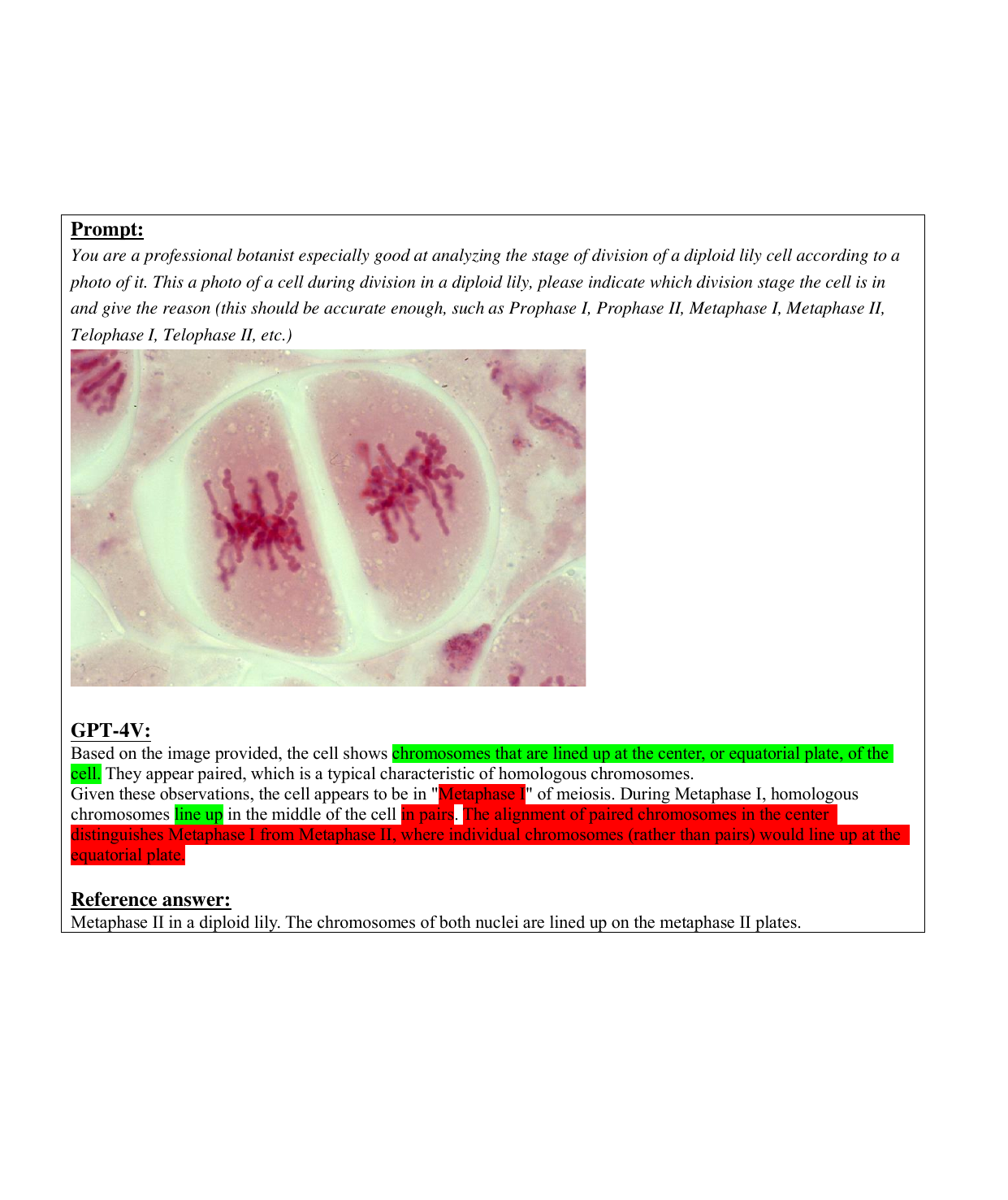}
    \caption[Biological Imaging: Case 1]{\textbf{OCI: Case 1.} A demonstration case from optical cell imaging (OCI) testing of an optical diploid lily cell image. The green region labeled in the figure refers to the correct information provided by GPT-4V, while the red region indicates the wrong answer generated.}
    \label{fig:oci_1}
\end{figure}
\newpage
\begin{figure}[H]
    \centering
    \includegraphics[width = \textwidth]{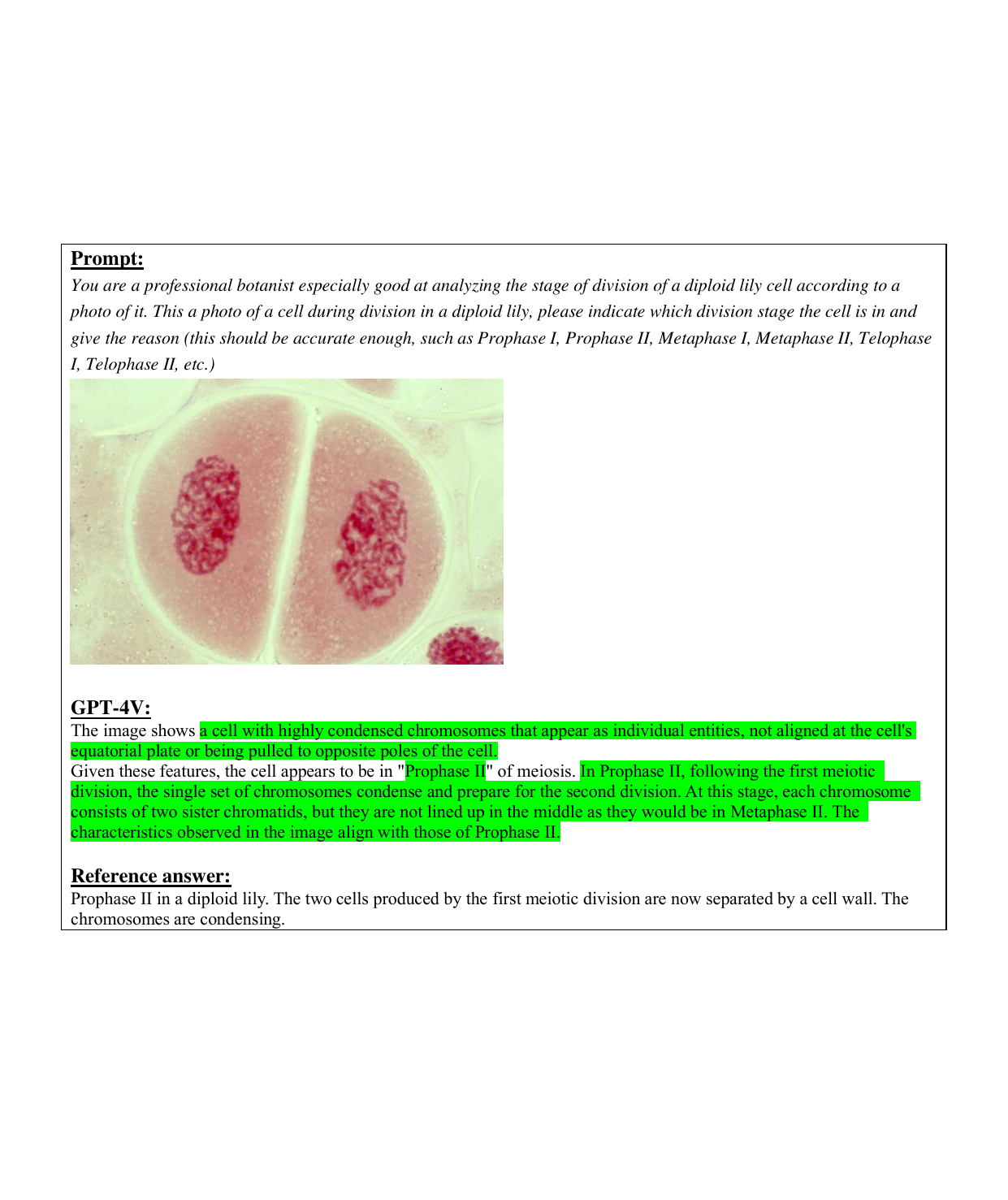}
    \caption[Biological Imaging: Case 2]{\textbf{OCI: Case 2.} A demonstration case from optical cell imaging (OCI) testing of an optical diploid lily cell image. The full green region labeled in the figure refers to the comprehensively correct information provided by GPT-4V.}
    \label{fig:ocig_2}
\end{figure}
\newpage
Figure \ref{fig:oci_1}-\ref{fig:ocig_2} (above), and \ref{fig:ocig_3}-\ref{fig:ocig_8} (in Appendix) showcase the ability of GPT-4V to distinguish the specific cell division stage given the corresponding optical diploid lily cell image. In Figure \ref{fig:oci_1}, GPT-4V's answer does not match the reference answer. In terms of stage discrepancy (SDP), the ground truth (GT) states that the image represents "Metaphase II" while GPT-4V interprets it as "Metaphase I". These are distinct stages of meiosis with different chromosomal alignments. For chromosome pairing (CMP), GT mentions that the chromosomes of both nuclei are lined up, implying individual chromosomes are aligned at the metaphase plates. However, GPT-4V emphasizes the chromosomes appear paired, indicating tetrads or homologous chromosomes, which is characteristic of Metaphase I. With regards to cell division context (CDC), while both descriptions discuss chromosomes lining up at the metaphase plate, the context is different. GT speaks of Metaphase II with individual chromosomes aligned, while GPT-4V describes the alignment of paired homologous chromosomes, a feature of Metaphase I. Similarly, in Figure \ref{fig:ocig_3}, GPT-4V's answer does not match the content of GT. GT specifies that the cell is in "Telophase I", while GPT-4V suggests the cell is in "Prophase I". Referring to CMP, GT mentions that the chromosomes have reached the poles and are beginning to de-condense. In contrast, GPT-4V states that the chromosomes are condensed and are typical of the early stages of cell division. Considering CDC, GPT-4V provides detailed characteristics of "Prophase I" such as synapsis, crossing over, and the beginning of the breakdown of the nuclear envelope. These characteristics do not align with the features of "Telophase I" as described in GT. However, in Figure \ref{fig:ocig_2} and \ref{fig:ocig_4}, GPT-4V's results match the contents in their corresponding GTs. In Figure \ref{fig:ocig_2}, GPT-4V's description correctly matches GT in identifying the cell division stage as Prophase II in a diploid lily, with both descriptions noting the presence of highly condensed chromosomes characteristic of this phase and the separation of cells by a cell wall following the first meiotic division. GPT-4V elaborates that the chromosomes are not yet aligned at the equatorial plate, distinguishing this stage from Metaphase II, which is consistent with GT observation of the chromosomes preparing for the second division, thus confirming the accuracy of the generated content. In Figure \ref{fig:ocig_4}, the generated content in GPT-4V does not fully match the ground truth content. GT accurately describes the cell as being in "Metaphase I of meiosis," a stage characterized by the arrangement of chromosome pairs (bivalents) along the metaphase plate. However, it also specifies that due to the "polar view," the metaphase plate is not visible, but the bivalents and crossover regions are clear. GPT-4V's description recognizes the pairing of chromosomes as tetrads, which is consistent with GT, but incorrectly suggests that these tetrads are roughly aligned in the center, indicating the metaphase plate is visible, which contradicts GT's mention of the metaphase plate not being apparent due to the viewing angle. Therefore, while GPT-4V correctly identifies the stage of meiosis, it does not acknowledge the specific observational limitations noted in GT, namely the polar viewpoint and the consequent non-visibility of the metaphase plate. In Figure \ref{fig:ocig_5}-\ref{fig:ocig_8}, we conduct experiments to explore the capability of GPT-4V to identify specific locations within an optical cell image (using bounding boxes) to facilitate cell segmentation. The results consistently reveal that GPT-4V possesses limited recognition ability for small cellular structures and a constrained understanding of their spatial information. The clusters of bounding boxes provide only a restricted representation of the spatial distribution of the cells. The delineated areas within each bounding box are found to be exceedingly approximate, to the extent that they are ineffective for extracting useful information. Notably, these outcomes are consistent across images of varying resolutions, specifically 512x512 and 1024x1024 pixels, indicating an inability to achieve effective cell segmentation for individual cells.

\newpage

\subsection{Cardiac Imaging}
\begin{figure}[H]
    \centering
    \includegraphics[width = \textwidth]{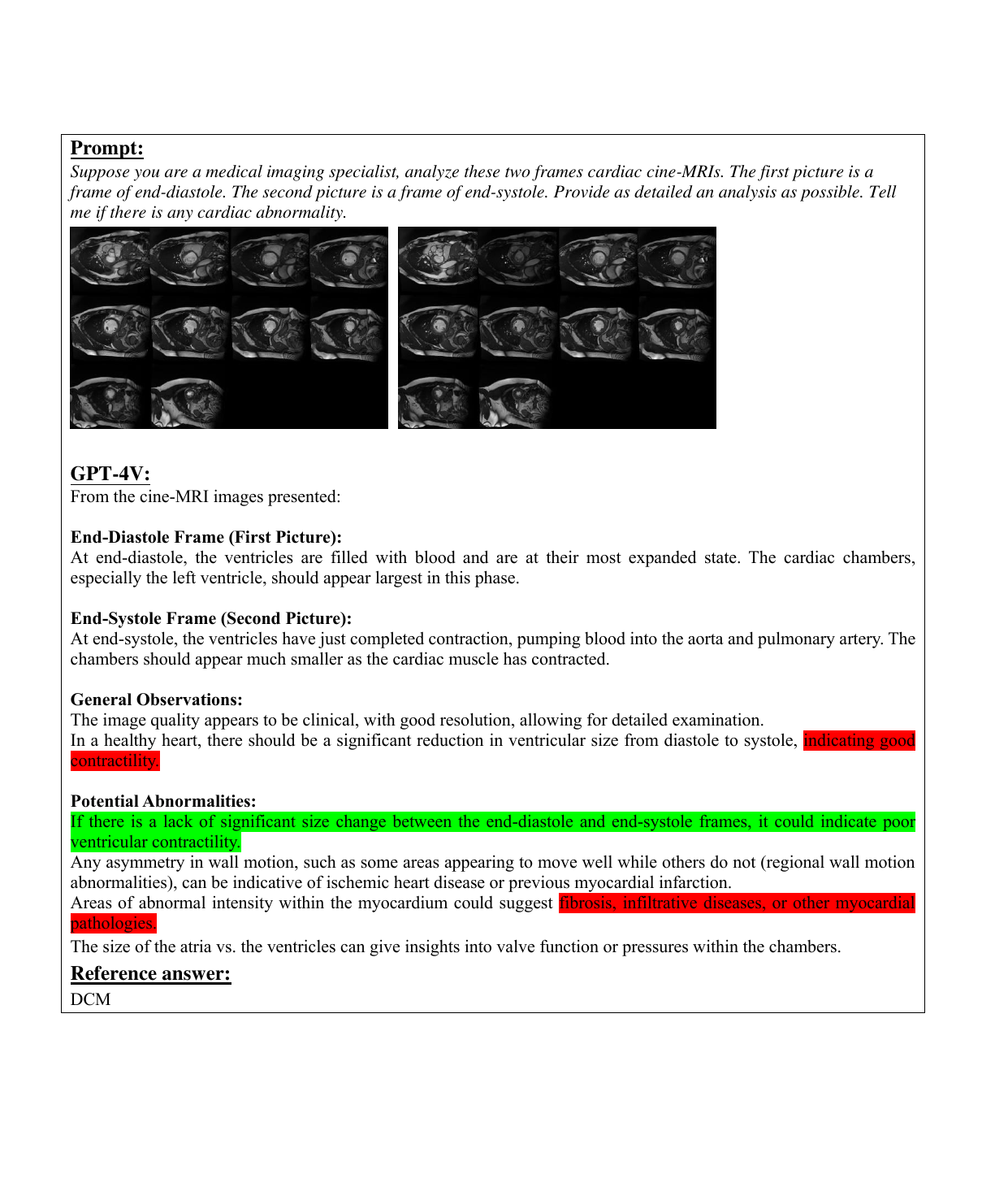}
    \caption[Cardiac Imaging: Case 1]{\textbf{Heart: Case 1.} In this case, GPT-4V has the wrong observation for good contractility. However, it shows contradictive results in potential abnormalities of poor ventricular contractility. Besides, there is no fibrosis, infiltrative diseases, or other myocardial pathologies in these two frames. The only label of cine-MRI images is DCM.}
    \label{fig:heart_CMR_1}
\end{figure}
\newpage
\begin{figure}[H]
    \centering
    \includegraphics[width = \textwidth]{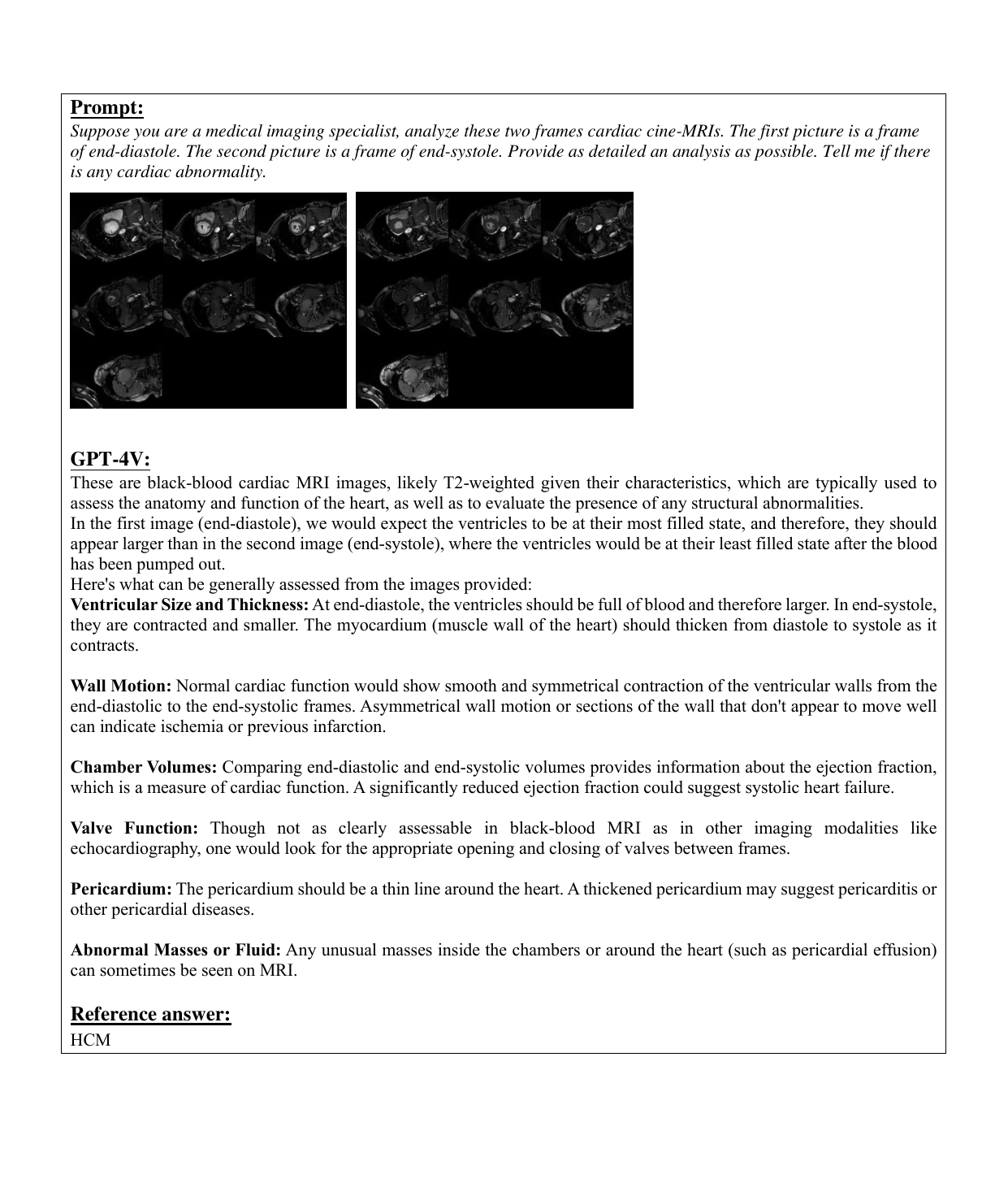}
    \caption[Cardiac Imaging: Case 2]{\textbf{Heart: Case 2.} In this case, with the same prompt, GPT-4V refused to analyze the images. However, it only gives the method and aspects of how to analyze cardiac cine-MRI.}
    \label{fig:heart_CMR_2}
\end{figure}

\newpage
GPT-4V has the ability to analyze cardiac cine-MRIs. It has basic knowledge of how to analyze this kind of image. As shown in Figure \ref{fig:heart_CMR_1}, it can capture some correct observations for both normal hearts as well as hearts with potential problems such as DCM. However, the analysis is quite brief and general with low accuracy. In order to get these results, a specific prompt with modality, frame details, and body parts is necessary. Otherwise, GPT-4V will recognize these images as other CTs. Sometimes GPT-4V refuses to answer the question but only gives some aspects that help medical imaging specialists to analyze. One example is shown in Figure \ref{fig:heart_CMR_2}. In its observations, it also includes content on how to analyze such images. More cases can be viewed in Appendix A.8 Cardiac Imaging. Besides, the maximum number of images GPT-4V accepts is four. Therefore, GPT-4V struggles to analyze time sequence data such as cardiac cine-MRIs which need the whole slices of end-diastole and end-systole (approximately 300 images as a complete circle). If inserting all images into one combing image, the resolution is too small and hard for GPT-4V to analyze. This is one of the limitations that GPT-4V has in analyzing continuous procedures.

\newpage

\subsection{Ultrasound Imaging}

\begin{figure}[H]
    \centering
    \includegraphics[width = \textwidth]{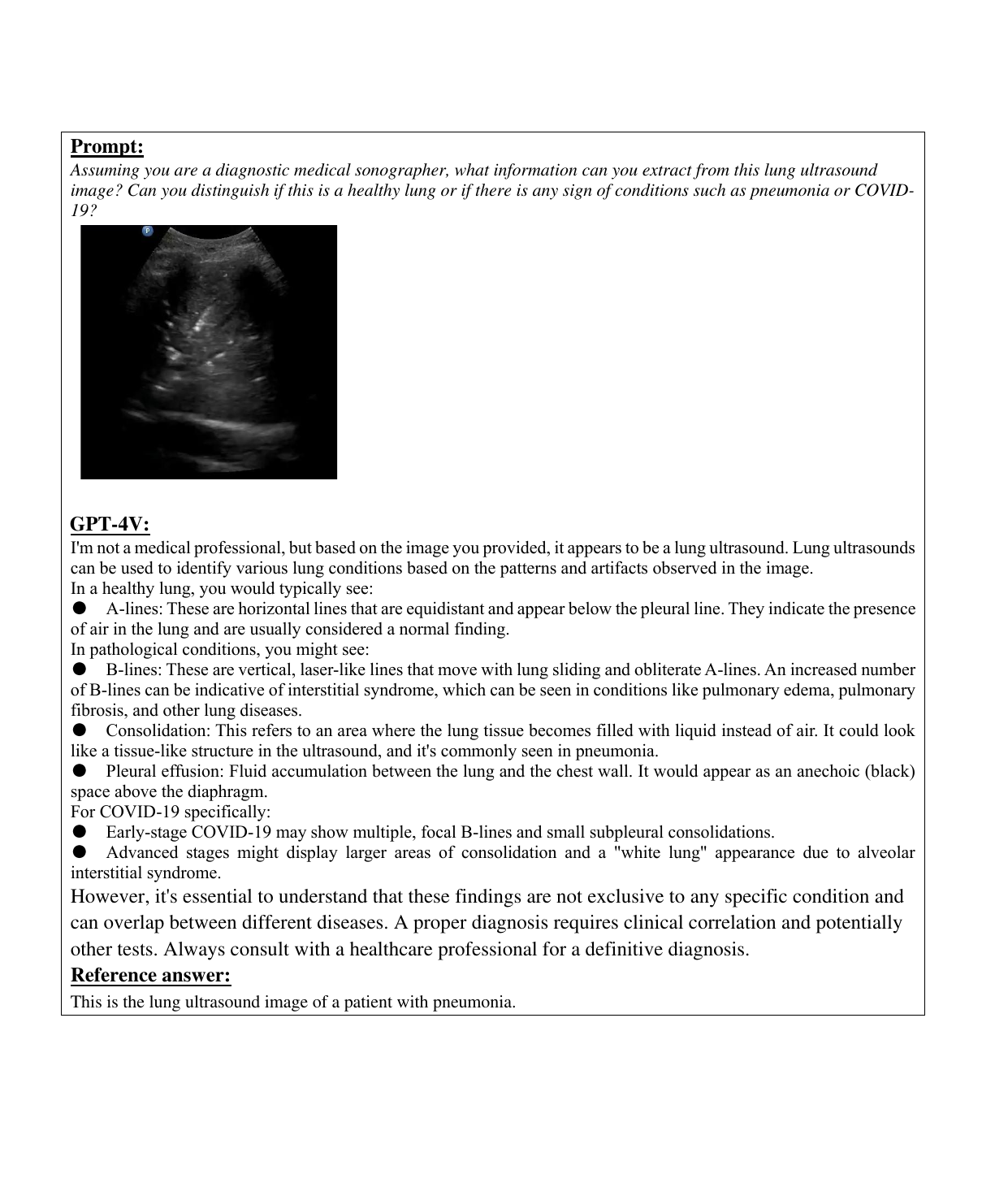}
    \caption[Ultrasound Imaging: Case 1]{\textbf{Ultrasound: Case 1.} A demonstration case on a lung ultrasound image from the COVIDx-US dataset. The image depicts a lung under pneumonia, but GPT-4V was unable to provide any information based on this particular image.}
    \label{fig:ultrasound_1}
\end{figure}
\newpage

\begin{figure}[H]
    \centering
    \includegraphics[width = \textwidth]{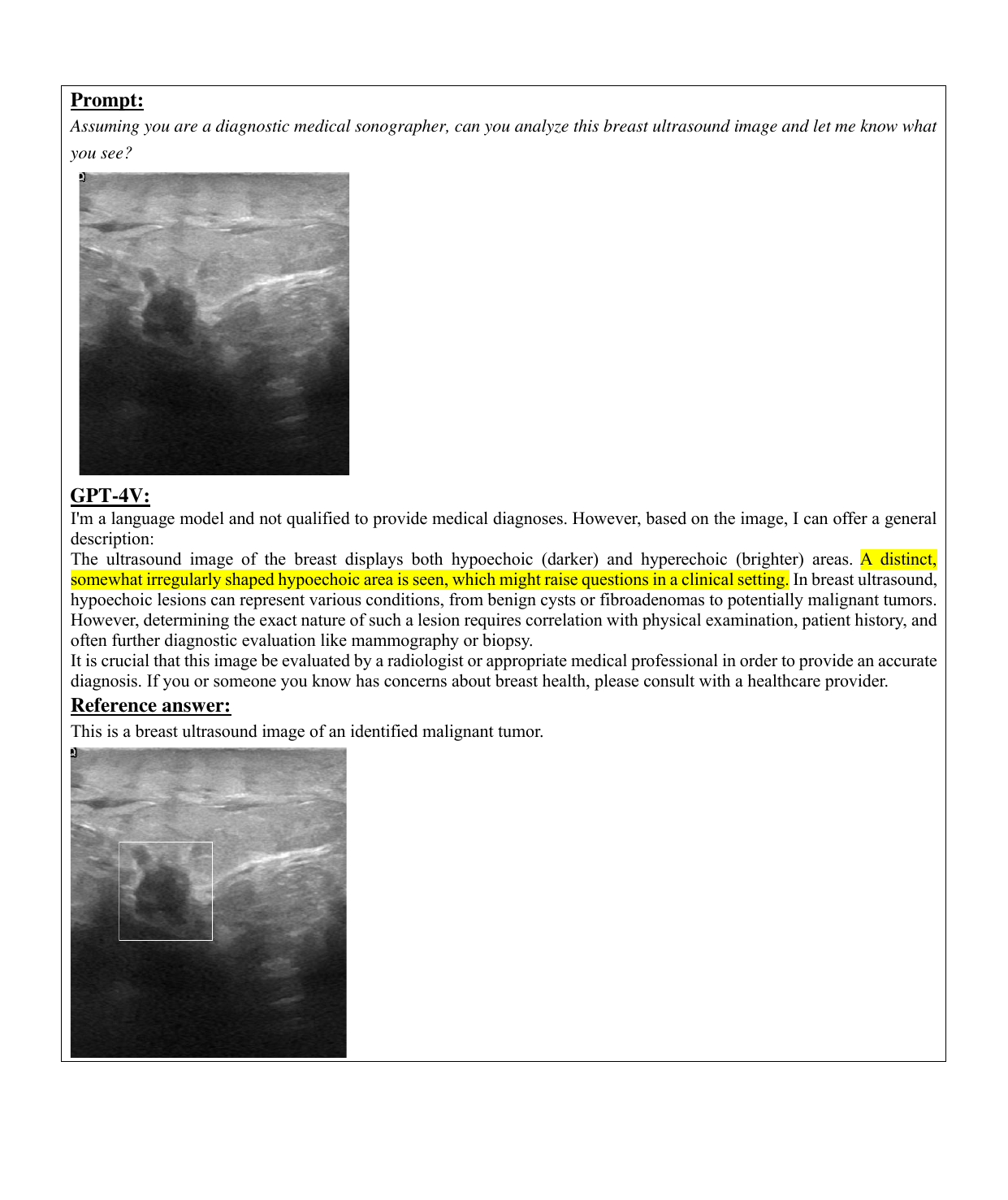}
    \caption[Ultrasound Imaging: Case 2]{\textbf{Ultrasound: Case 2.} A demonstration case on a breast ultrasound image. The image depicts a malignant tumor found in the breast region. GPT-4V was able to identify the anomaly on the image but seemed uncertain about the actual diagnosis (highlighted in yellow).}
    \label{fig:ultrasound_2}
\end{figure}
\newpage

Our findings reveal that GPT-4V's capacity for medical diagnosis through direct image classification is limited. The model did not reliably identify the specific conditions—normal, pneumonia, or COVID-19 in lung ultrasounds (Figure \ref{fig:ultrasound_1}), nor could it consistently distinguish between normal, benign, and malignant in breast ultrasounds. However, GPT-4V demonstrated a noteworthy ability to detect and describe salient features within the ultrasound images that are pertinent to clinical diagnosis and analysis.

In the context of breast ultrasound imaging, GPT-4V showed a particular aptitude for recognizing abnormalities (Figure \ref{fig:ultrasound_2}), which it flagged as points of interest for further clinical evaluation. This aspect of the AI's performance suggests that it has the potential to serve as an adjunct tool for radiologists and clinicians, helping to focus attention on abnormal image features that may require more detailed assessment. The model's ability to guide users toward these irregularities could be instrumental in the preliminary stages of image review, especially in settings with high volumes of diagnostic imaging where prioritization is critical.

Additionally, the insights provided by GPT-4V on characteristic ultrasound features can be invaluable in educational environments. By outlining what to look for in an ultrasound, GPT-4V can augment the training of medical students and residents, equipping them with enhanced pattern recognition skills before they engage in hands-on diagnostic work.

In conclusion, while GPT-4V does not currently meet the threshold for independent medical imaging diagnosis, its utility lies in its role as a supportive tool to highlight potential areas of interest in ultrasound images. Its use could potentially lead to increased diagnostic accuracy and efficiency in clinical workflows and contribute significantly to medical education. As AI technology continues to evolve, it is anticipated that its integration into medical diagnostics will expand, refining the capabilities of AI-assisted tools like GPT-4V in healthcare settings.

\newpage

\subsection{Nuclear Medicine Imaging}

\begin{figure}[H]
    \centering
    \includegraphics[width = \textwidth]{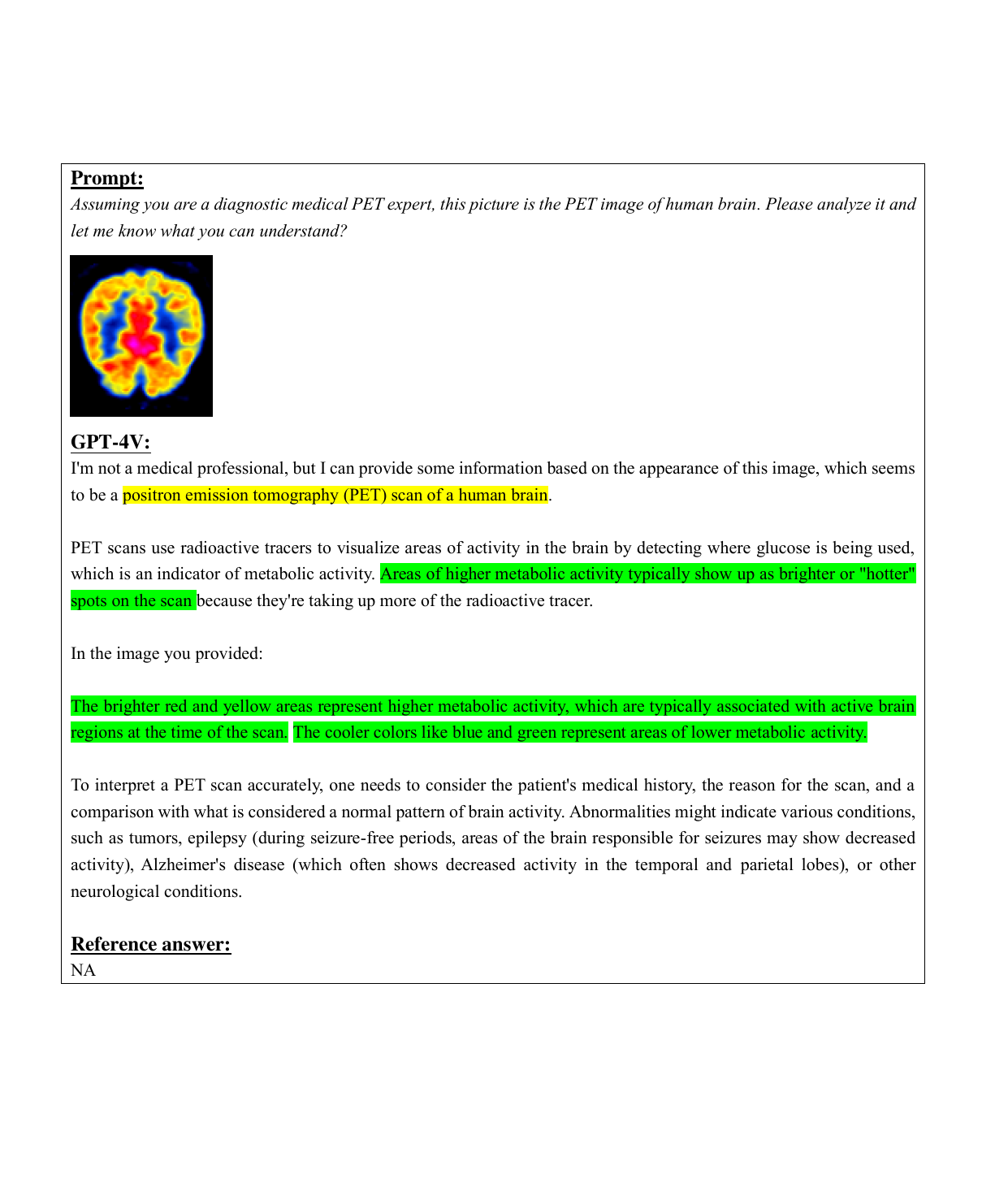}
    \caption[Nuclear Medicine Imaging: Case 1]{\textbf{Nuclear Medicine: Case 1.} A demonstration case of Report Generation Task on Harvard-Medical-Image-Fusion-Datasets with PET image.}
    \label{fig:nuclear_1}
\end{figure}
\newpage

\begin{figure}[H]
    \centering
    \includegraphics[width = \textwidth]{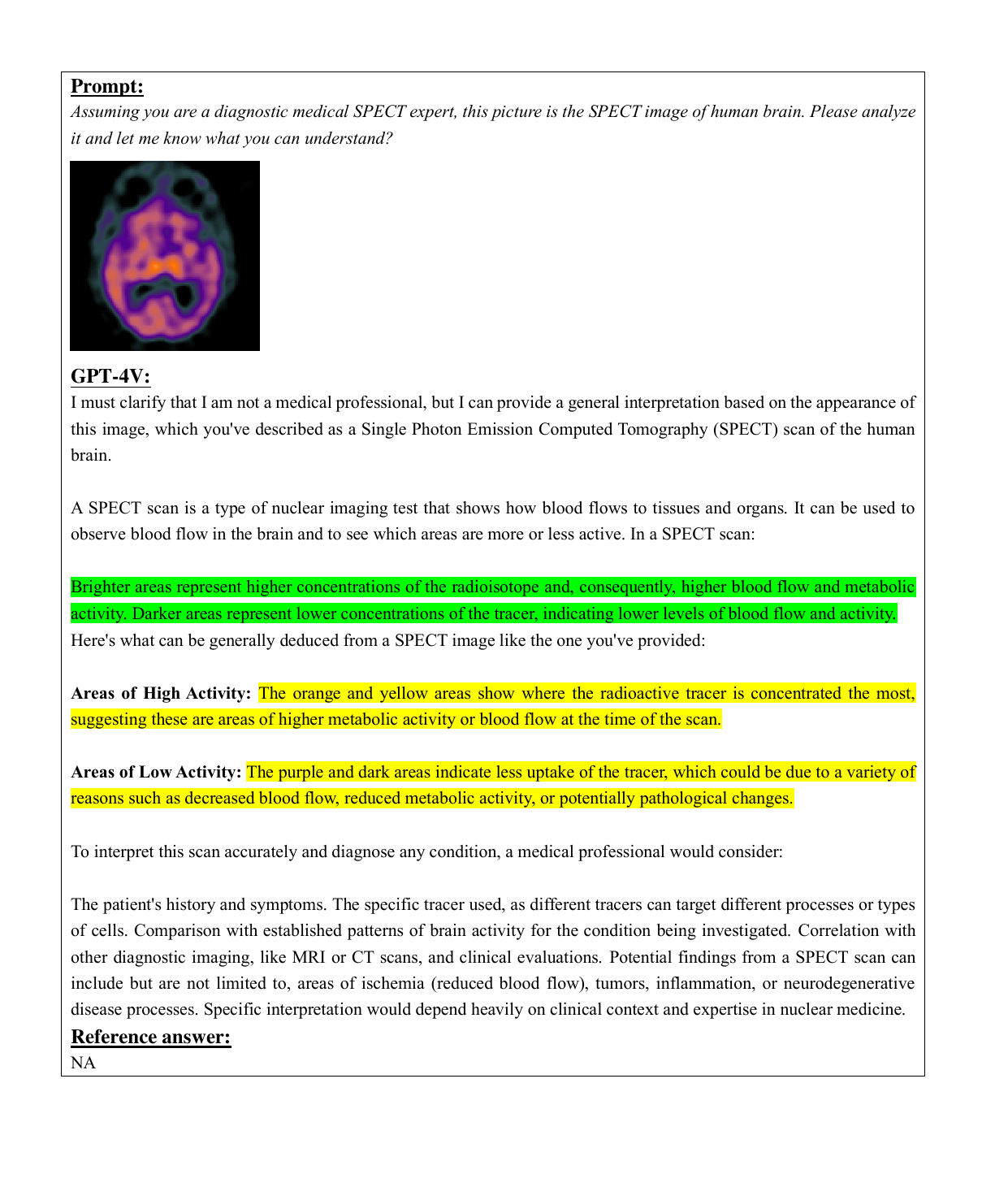}
    \caption[Nuclear Medicine Imaging: Case 2]{\textbf{Nuclear Medicine: Case 2.} A demonstration case of Report Generation Task on Harvard-Medical-Image-Fusion-Datasets with SPECT image.}
    \label{fig:nuclear_3}
\end{figure}
\newpage

In nuclear medicine imaging testing, we examined the capabilities of GPT-4V across PET (Figure \ref{fig:nuclear_1} and \ref{fig:nuclear_2}) and SPECT (Figure \ref{fig:nuclear_3} and \ref{fig:nuclear_4}). We also evaluated its performance in report generation task. Based on the test results, GPT-4V can recognize structures and modal information, and generate textual diagnostic reports. However, it still struggles with accurate disease identification and grading. During the evaluation of the diagnosis, GPT-4V can analyze the surface information of the image, but the deep clinical information cannot be provided and suggesting the need for further investigations, in other tests, errors were also observed in the analysis results generated by GPT-4V. Nonetheless, GPT-4V has demonstrated the ability to identify anatomical structures and lesions, as well as the potential for disease grading. These instances illustrate the potential utility of GPT-4V in assisting with the diagnosis of nuclear medicine imaging diseases.

\newpage

\subsection{Endoscopic Imaging}

\begin{figure}[H]
    \centering
    \includegraphics[width = \textwidth]{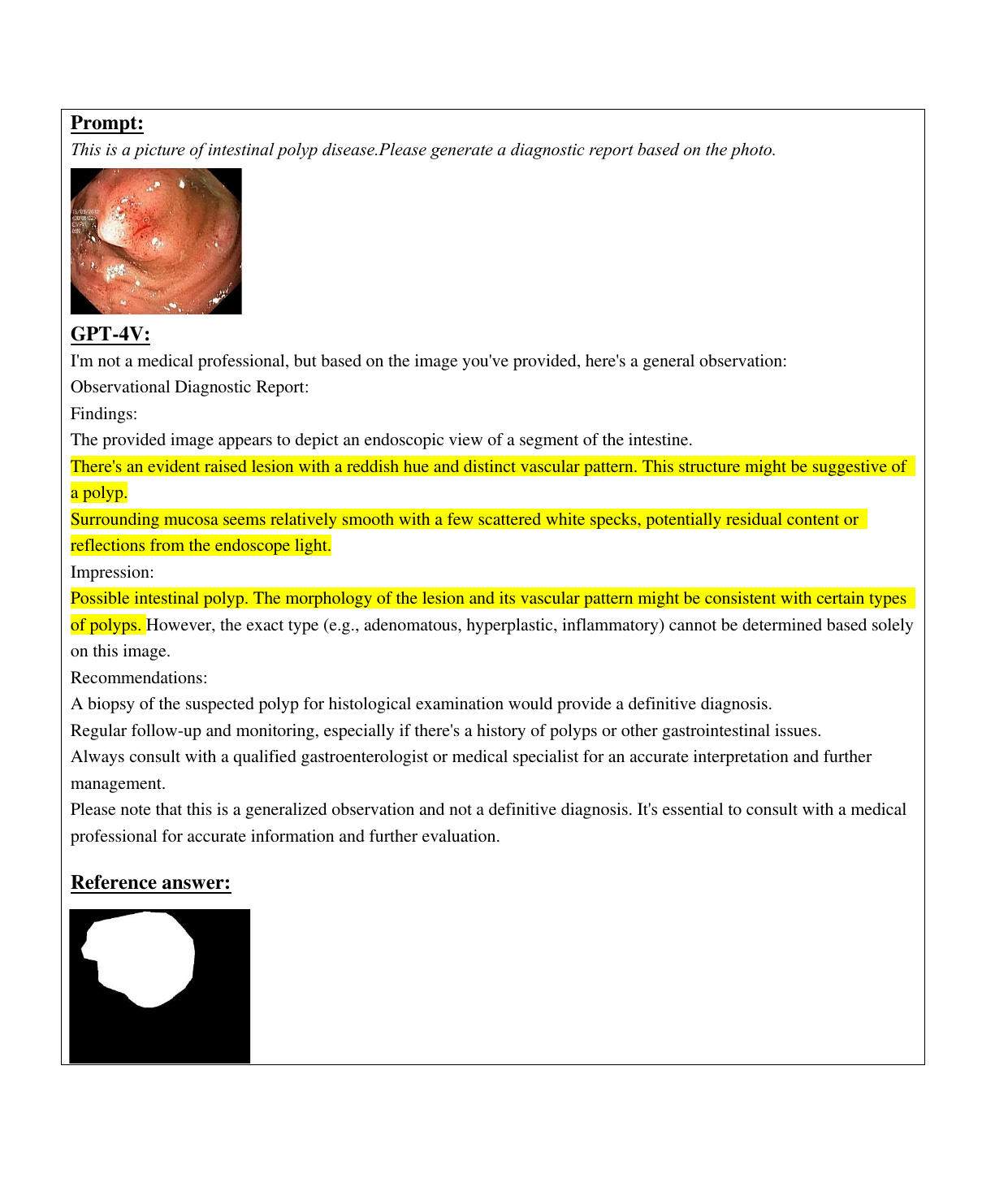}
    \caption[Endoscopic Imaging: Case 1]{\textbf{Endoscopic: Case 1.} A demonstration case of endoscopic imaging for intestinal polyp segmentation. Yellow represents the identified relevant information.}
    \label{fig:endoscopic_1}
\end{figure}
\newpage




\begin{figure}[H]
    \centering
    \includegraphics[width = \textwidth]{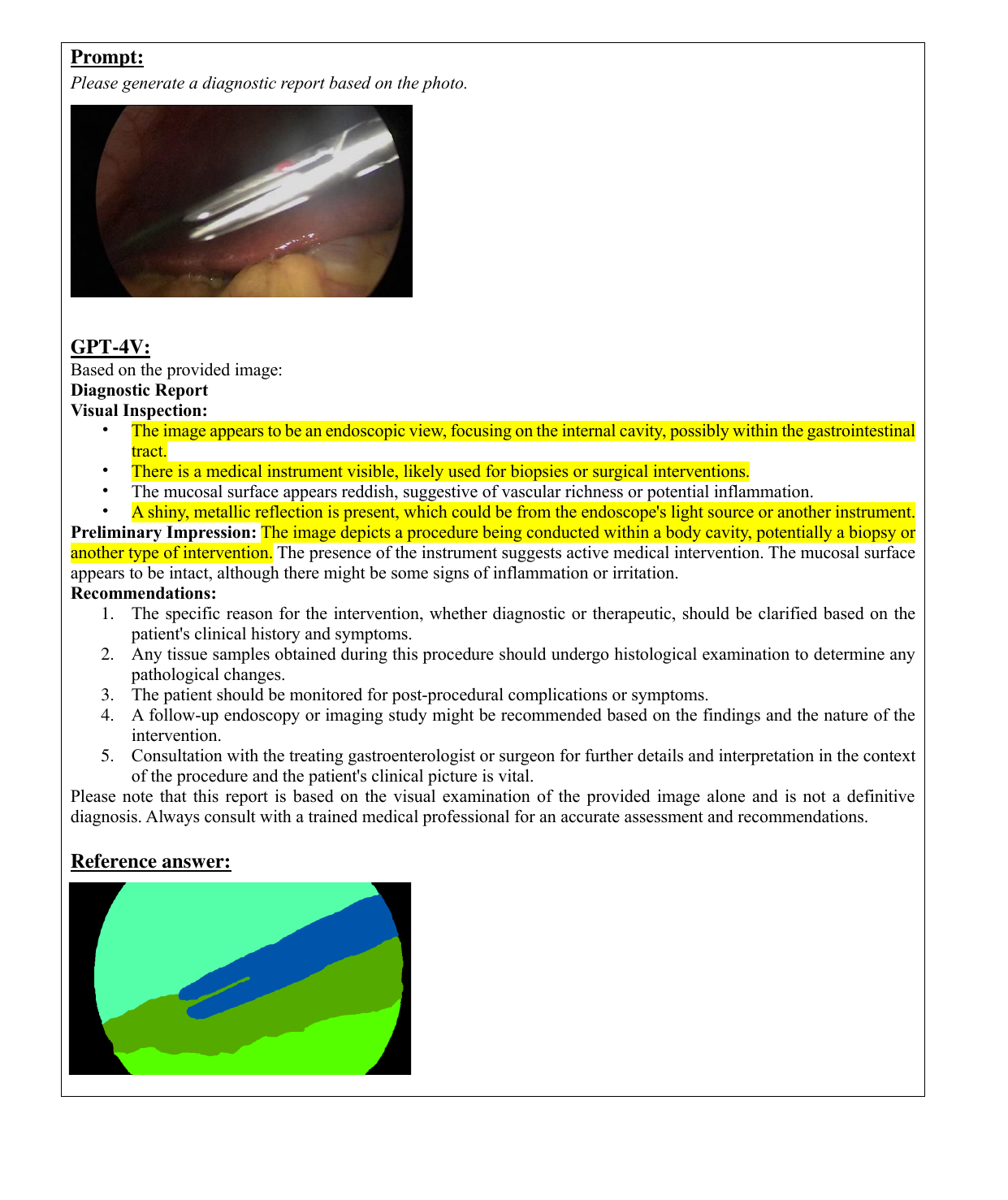}
    \caption[Endoscopic Imaging: Case 2]{\textbf{Endoscopic: Case 2.} Demonstration case of endoscopic image segmentation during surgery. Yellow represents the identified relevant information.}
    \label{fig:endoscopic_4}
\end{figure}
\newpage

In endoscopic imaging testing, we examined the capabilities of GPT-4V across intestinal polyp disease and related diseases observed by endoscope. We also evaluated its performance in clinical diagnostic report generation task. Based on the test results, GPT-4V can recognize possible disease focus based on observation results in the given polyp photos. Surprisingly, it is sensible to subtle differences such as "an evident raised lesion" in Figure \ref{fig:endoscopic_1}. However, it still struggles with accurate disease identification and grading. Most of the contents are uncertain labeled by yellow region. Thus in the "Impression" part, "Possible intestinal polyp" is also recognized as an uncertain answer. In Figure \ref{fig:endoscopic_4}, \ref{fig:endoscopic_5}, \ref{fig:endoscopic_6}, \ref{fig:endoscopic_7}, and \ref{fig:endoscopic_8}, GPT-4V can to some extent achieve normal zero-shot performance through the given prompt "Please generate a diagnostic report based on the photo. However, when taking further explorations on the answers, uncertainties occur as well. In Figure \ref{fig:endoscopic_5}, even though the polyp features are clearly explained in text, these features cannot certainly assure the pathological findings, such as "vascular prominence", are certain. In Figure \ref{fig:endoscopic_6}, biopsy forceps, mucosal surface are not certainly displayed, while the yellowish areas cannot decisively determine fatty deposits or other changes. "No overt lesions, ulcers, or growths" in Figure \ref{fig:endoscopic_7}, and "Some minor irregularities or raised areas" in Figure \ref{fig:endoscopic_8} both contribute to the uncertainties showcased in GPT-4V's answers. Definitely, GPT-4V has demonstrated the capability to identify polyp features, such as lesions, irregularities, reddish, etc. Nevertheless, the accuracy of this identification has great potential to evolve to relieve the practicality pressure due to uncertainties. Another notable finding can be derived from the comparison among Figure \ref{fig:endoscopic_1} and the other figures (Figure \ref{fig:endoscopic_4}, \ref{fig:endoscopic_5}-\ref{fig:endoscopic_8}) is that prior background information ("This is a picture of intestinal polyp disease" in Figure \ref{fig:endoscopic_1}) does not significant affect the result as other experiments do. It can be observed from Figure \ref{fig:endoscopic_1} that many of the contents still remain uncertain (marked by yellow region) as the remaining figures do.

\newpage

\subsection{Dermatological Imaging}


\newpage




\newpage

Summarize:
\begin{itemize}
    \item OpenAI's privacy policy may result in the test of GPT-4V not being comprehensive enough. In this test, if there are no privacy policy restrictions, GPT-4V has a strong zero-shot ability to process medical images(As shown in Figure \ref{fig:skin_1}).
    \item GPT-4V has high accuracy and comprehensiveness in diagnosing skin diseases with obvious characteristics. It is not accurate enough to diagnose some skin diseases with similar characteristics, but the diagnostic results still have a high reference value.
    \item The high accuracy of GPT-4V in describing the appearance of skin diseases reflects its exceptionally strong image processing capabilities. This foundation paves the way for applications in clinical practice, remote medical consultations, medical education, and training, among others.
\end{itemize}
\newpage

\subsection{Genetic Imaging}

\begin{figure}[H]
    \centering
    \includegraphics[width = \textwidth]{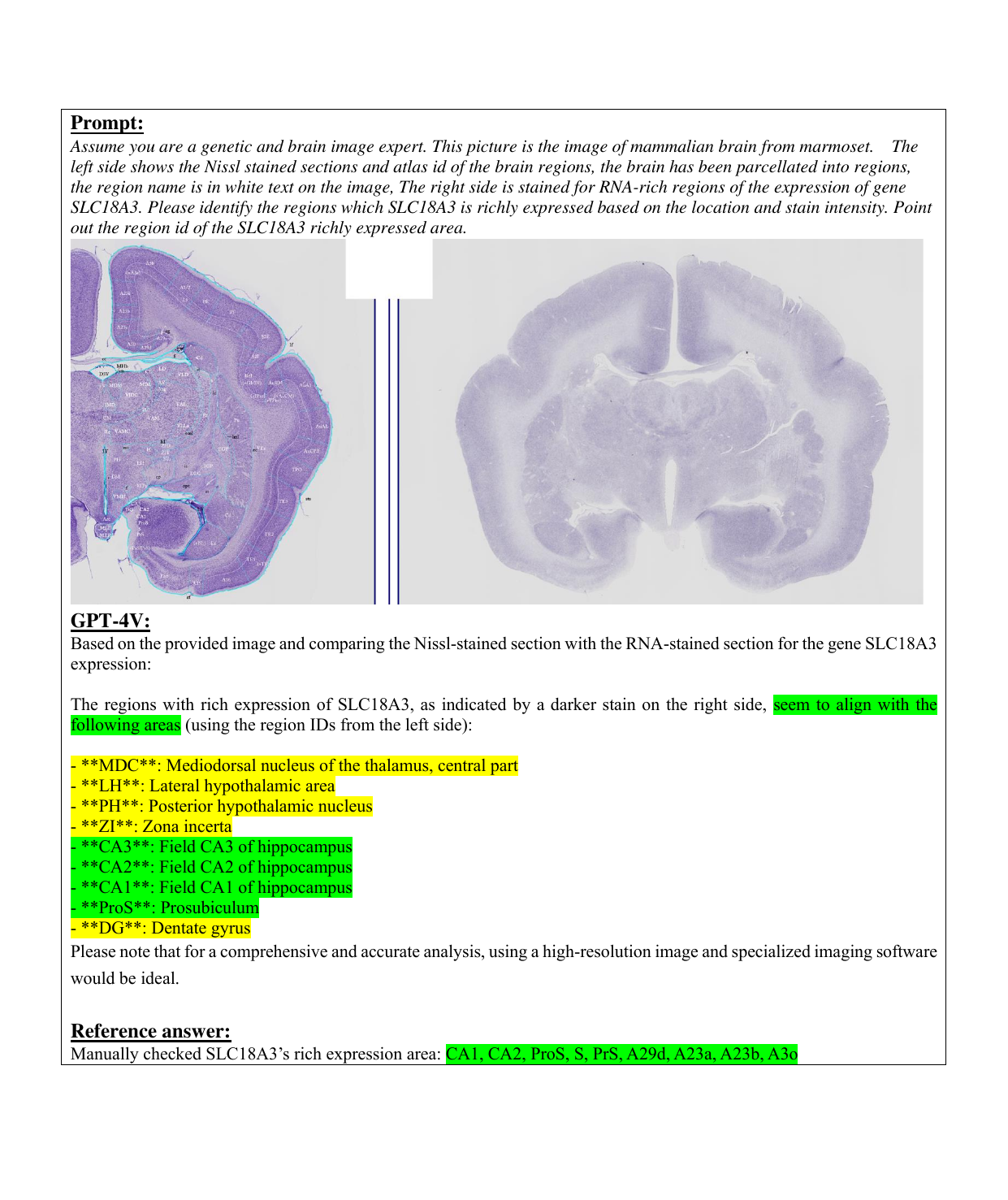}
    \caption[Genetic Imaging: Case 1]{\textbf{Gene: Case 4.} A demonstration case of recognizing the rich expression regions of certain genes from ISH images. Yellow denotes that correct information has been recognized by performing image alignment tasks between the Nissl-stain atlas image and gene expression ISH images.}
    \label{fig:gene_4}
\end{figure}
\newpage

\begin{figure}[H]
    \centering
    \includegraphics[width = \textwidth]{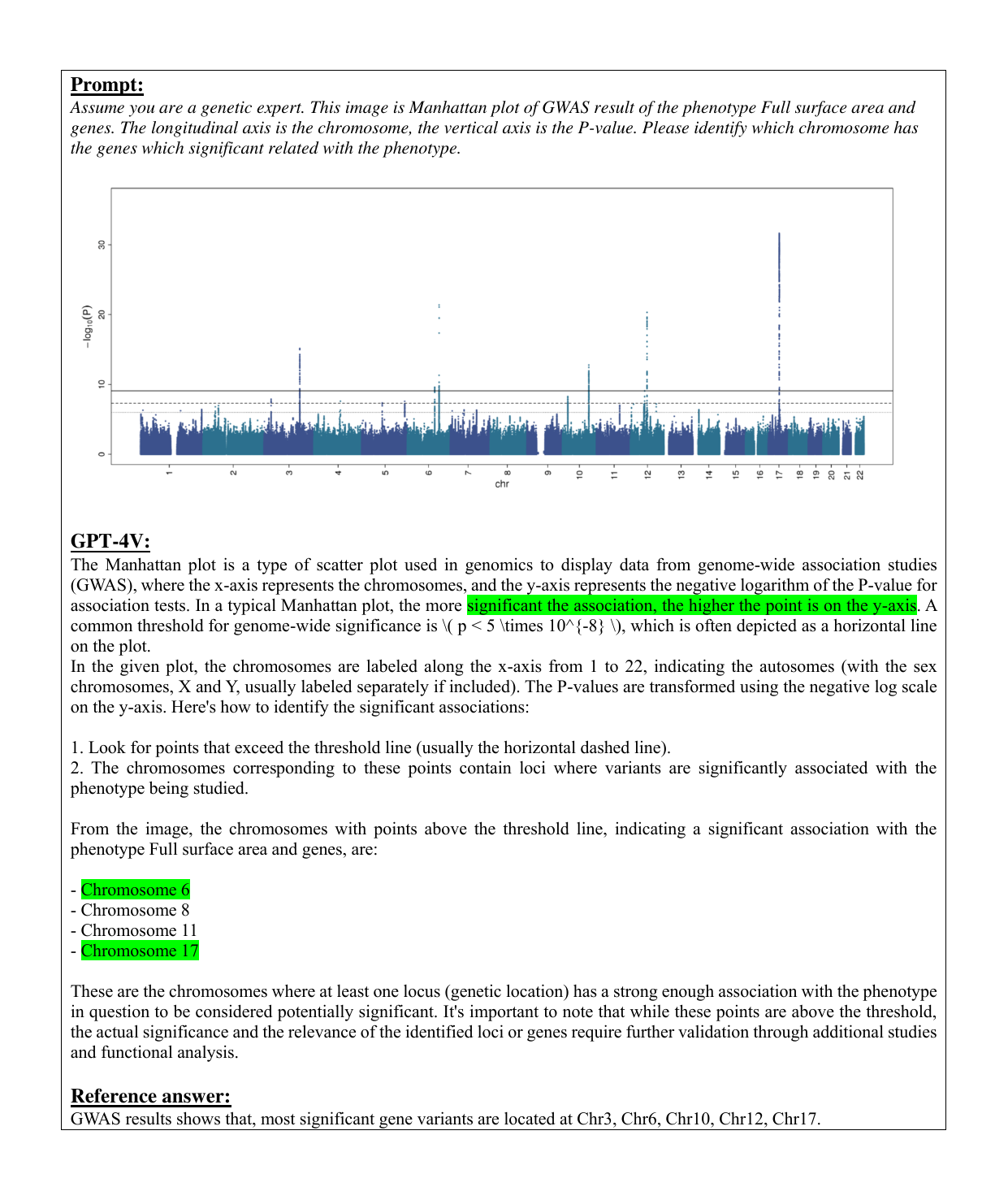}
    \caption[Genetic Imaging: Case 2]{\textbf{Gene: Case 6.} A demonstration case of recognizing the Manhattan plot of GWAS results of phenotype: the surface area of the full surface in the human brain. Yellow denotes that correct information has been recognized.}
    \label{fig:gene_6}
\end{figure}
\newpage

The findings of GPT-4V’s capability in understanding nissl-stain images(Figure \ref{fig:gene_1}, \ref{fig:gene_2}, \ref{fig:gene_3_2}), ISH images (Figure \ref{fig:gene_4}) and GWAS results images(Figure \ref{fig:gene_5} and \ref{fig:gene_6}) shows that, under proper prompt guidance, GPT-4V have: 1) strong proficiency in interpret the structural information in nissl-stain and ISH image, e.g. distinguishing the species, recognizing the atlas and annotations, making comparison between nissl-stain image and ISH image according to their structural correspondence. 2) fair capability in parsing the GWAS result plots from literature articles. It could be observed that the model are merging the extracted information from text prompts and the extracted information from the input genetic-research images together to optimize its answers regarding to particular instruction of task.

\newpage

\subsection{Orthopedic and Pediatric Imaging}

\begin{figure}[H]
    \centering
    \includegraphics[width = \textwidth]{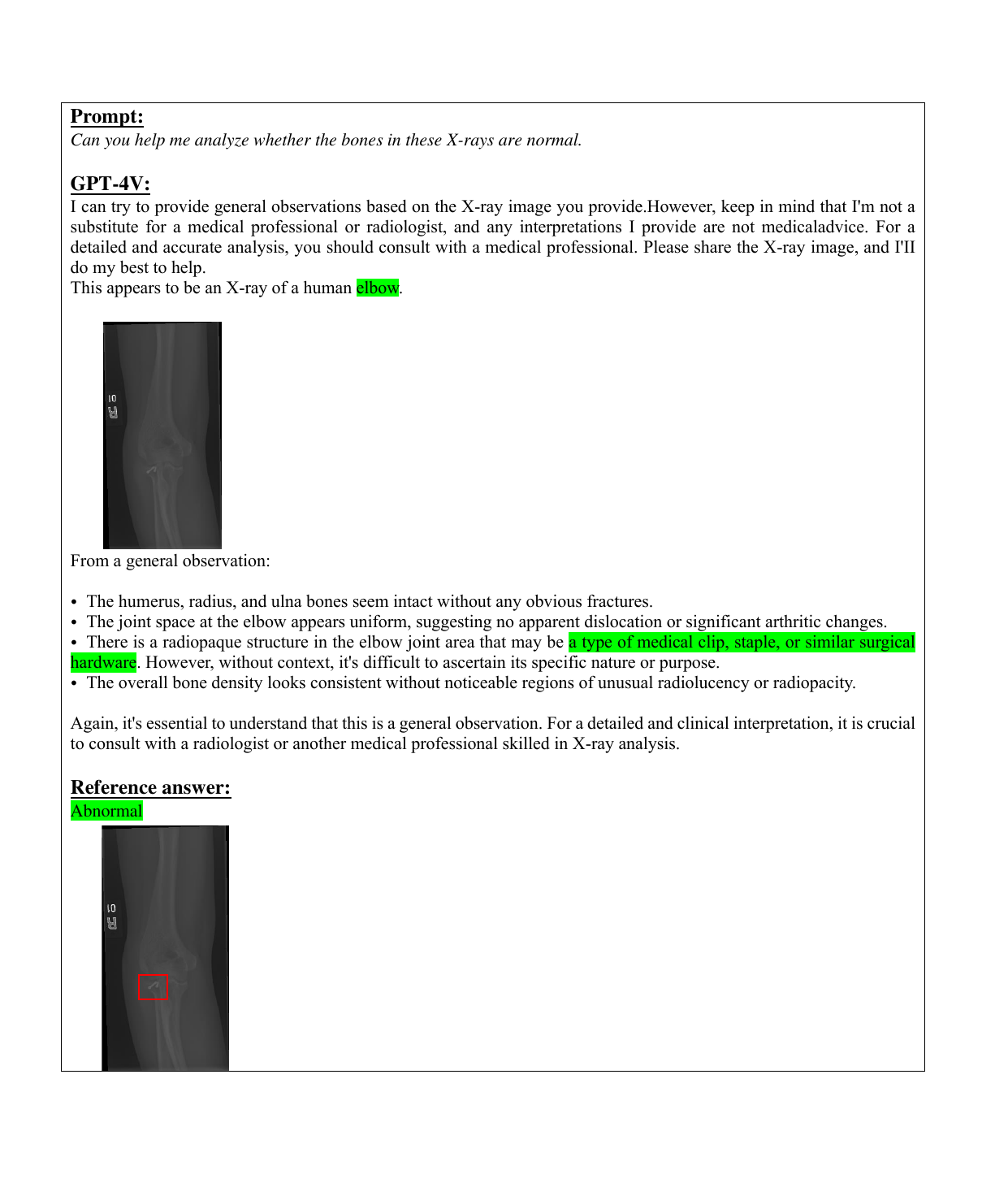}
    \caption[Orthopedic and Pediatric Imaging: Case 1]{\textbf{Orthopedic and Pediatric Imaging: Case 3.} Elbow X-ray images from the MURA dataset. The green color in the analysis section indicates correct information recognition, while the red box indicates abnormal areas.}
    \label{fig:bone_3}
\end{figure}
\newpage

\begin{figure}[H]
    \centering
    \includegraphics[width = \textwidth]{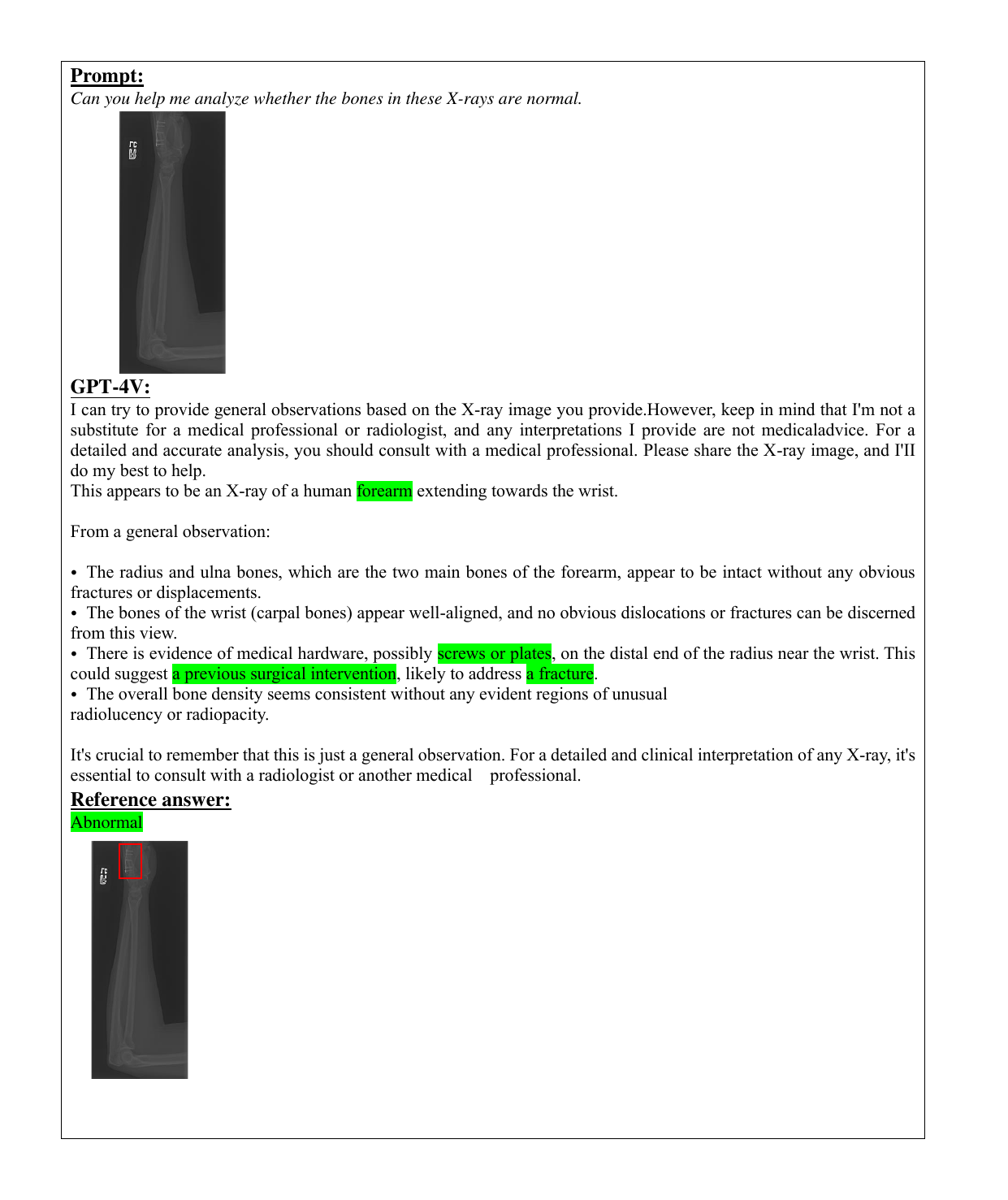}
    \caption[Orthopedic and Pediatric Imaging: Case 2]{\textbf{Orthopedic and Pediatric Imaging: Case 4.} Forearm X-ray images from the MURA dataset. The green color in the analysis section indicates correct information recognition, while the red box indicates abnormal areas.}
    \label{fig:bone_4}
\end{figure}
\newpage

\begin{figure}[H]
    \centering
    \setlength{\abovecaptionskip}{-3cm}
    \includegraphics[width = \textwidth]{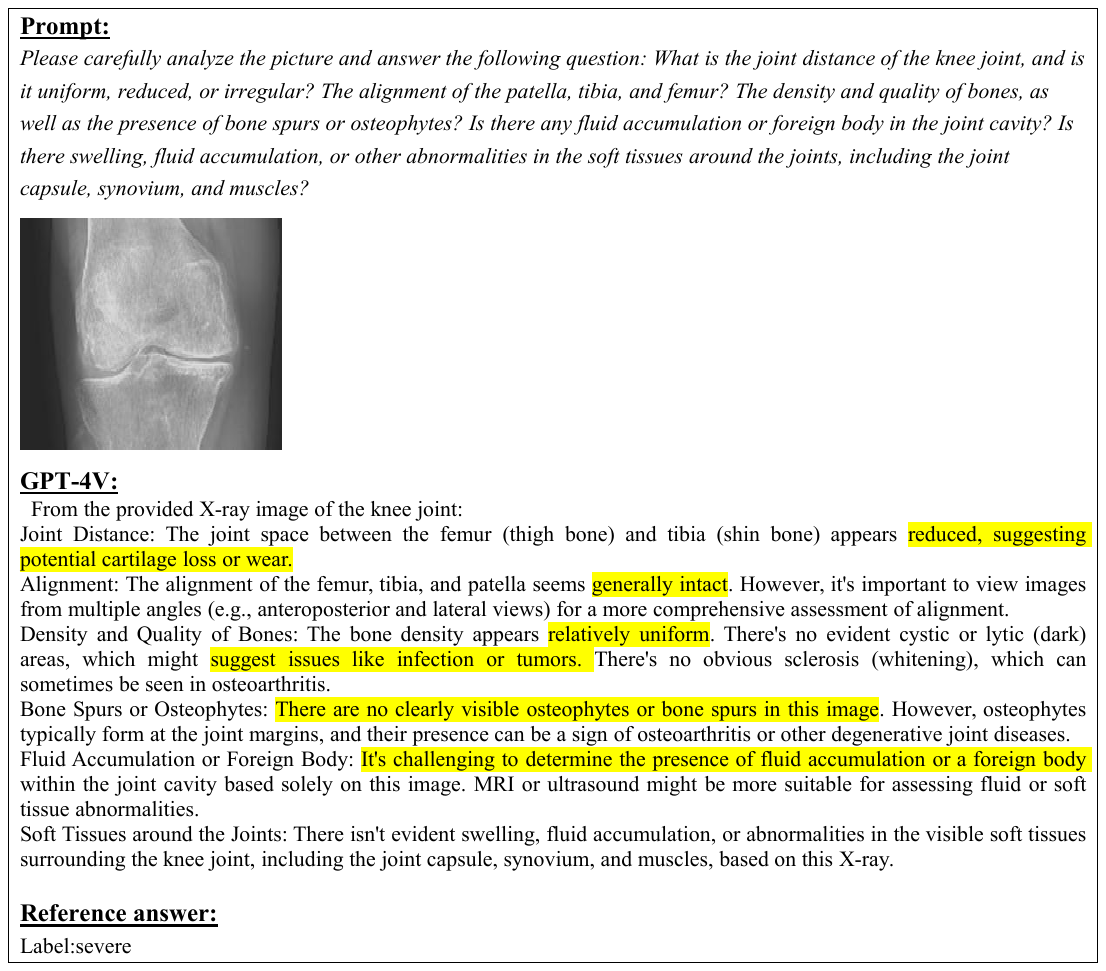}
    \caption[Orthopedic and Pediatric Imaging: Case 3]{\textbf{Orthopedic and Pediatric Imaging: Case 5.} For patients who had undergone total hip arthroplasty surgery, an anteroposterior (AP) view of the X-ray images for the patients with fixed (control group)  hip implants were shown above, while the X-ray images having a wire or plate attached to the implant were excluded.}
    \label{fig:bone_5}
\end{figure}
\newpage

\begin{figure}[H]
    \centering
    \setlength{\abovecaptionskip}{-3cm}
    \includegraphics[width = \textwidth]{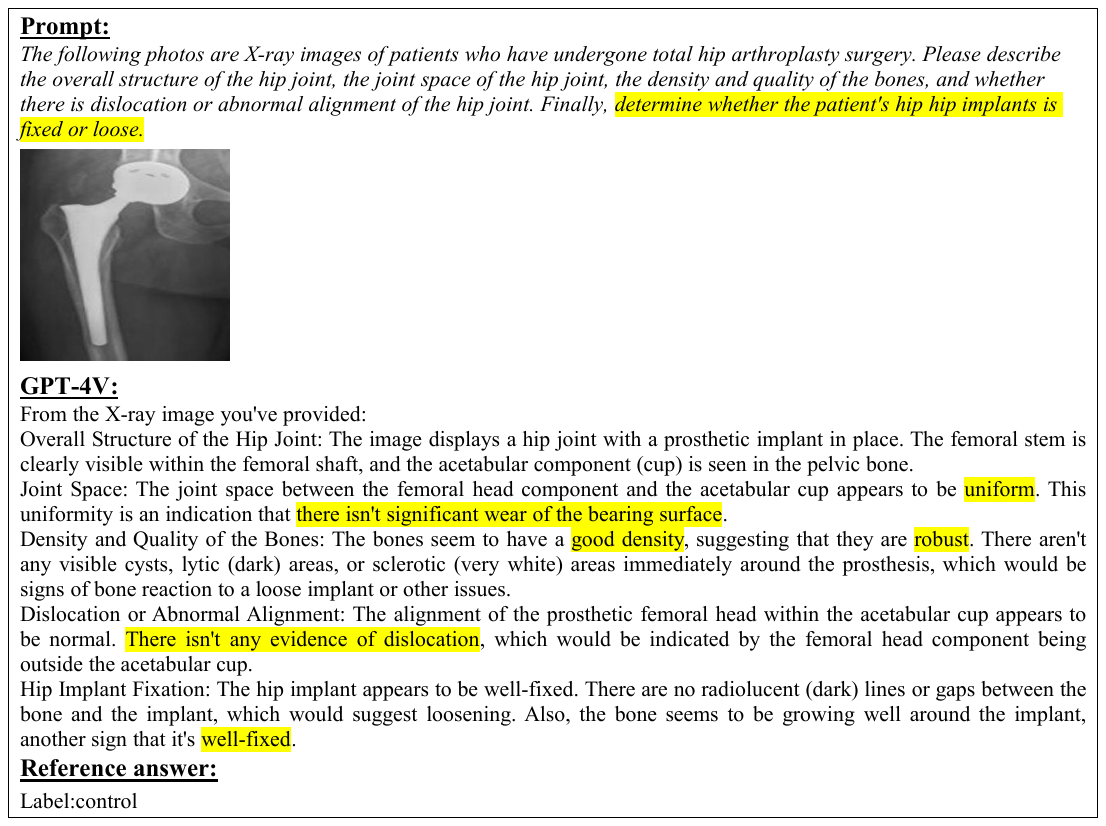}
    \caption[Orthopedic and Pediatric Imaging: Case 4]{\textbf{Orthopedic and Pediatric Imaging: Case 6.} The radiographic knee X-ray image is manually annotated labeled as per Kellgren and Lawrence grades by 2 medical experts, which is labeled as severe.}
    \label{fig:bone_6}
\end{figure}
\newpage

We collected orthopedic and pediatric X-ray image data on different parts of human body such as elbow (Figure \ref{fig:bone_3}),forearm (Figure~\ref{fig:bone_4}),hip implants (Figure~\ref{fig:bone_5}) and knee (Figure~\ref{fig:bone_6}) and used the promote engineering to get GPT-4v to recognize the X-ray images. Finally, we came to the following conclusions: GPT-4V can clearly identify Joint Space, Density and Quality of the Bones, Dislocation or Abnormal Alignment in different orthopedic X-ray images. But it’s difficult to classify Bone Spur or Osteophytes, Fluid Accumulation or Foreign Body, and Soft Tissues around the Joints for GPT-4V in some cases.

\newpage

\subsection{Dental Imaging}
\begin{figure}[H]
    \centering
    \includegraphics[width = \textwidth]{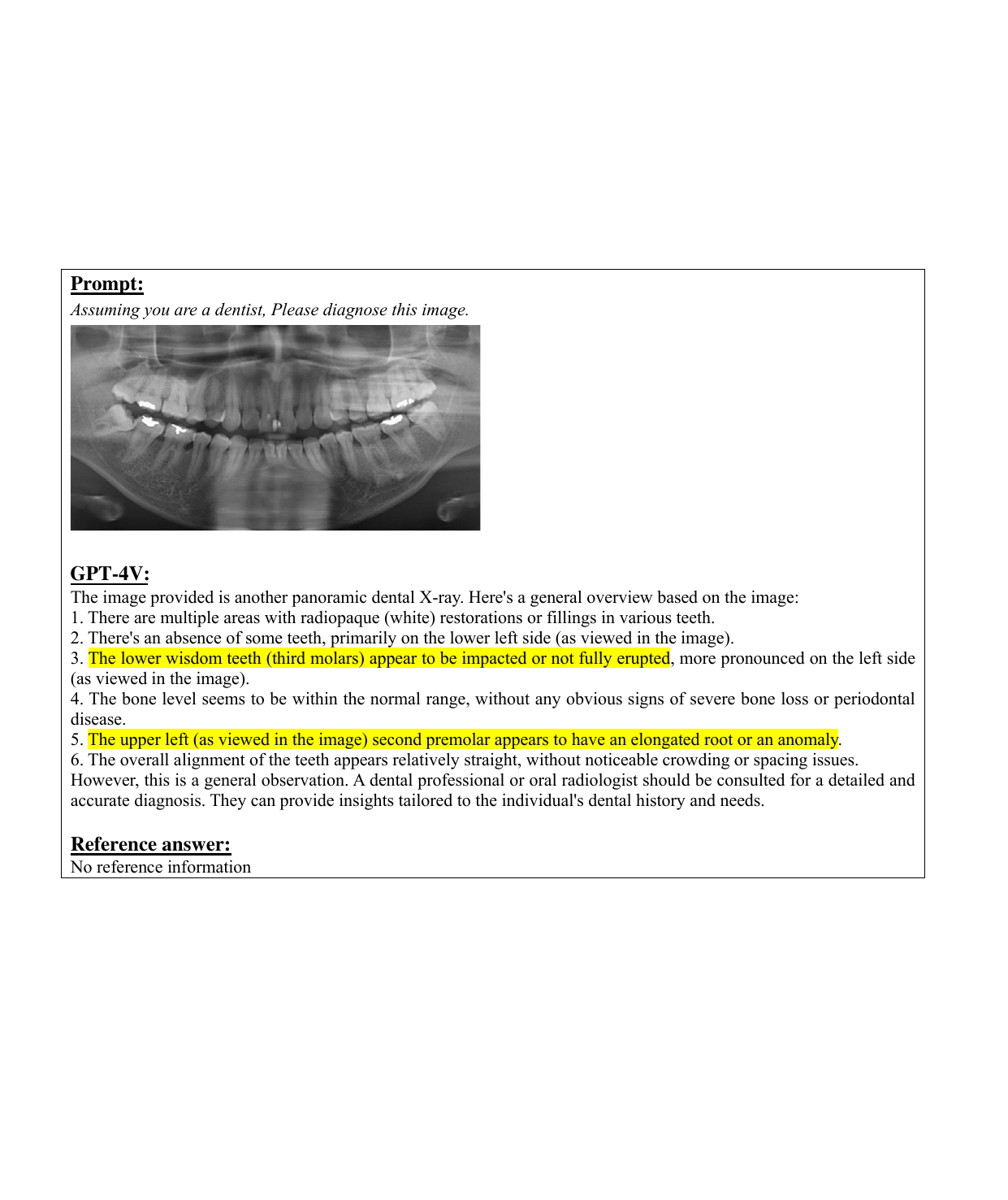}
    \caption[Dental Imaging: Case 1]{\textbf{Dental Image: Case 1.} A demonstration case from lung X-rays dental images. Yellow denotes that abnormal areas are recognized. }
    \label{fig:dental_3}
\end{figure}
\newpage

\begin{figure}[H]
    \centering
    \includegraphics[width = \textwidth]{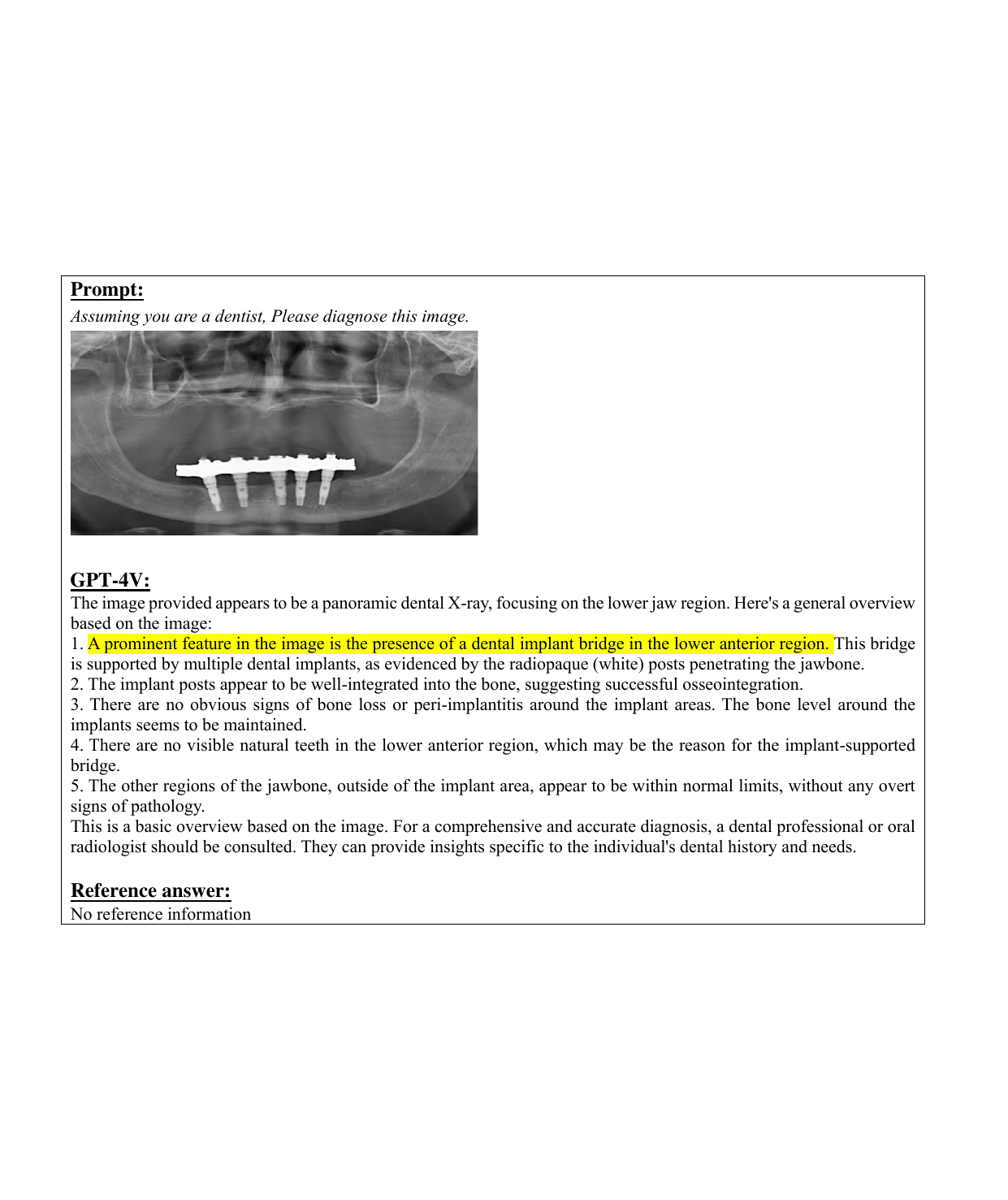}
    \caption[Dental Imaging: Case 2]{\textbf{Dental Image: Case 2.} A demonstration case from lung X-rays dental images. Yellow denotes that abnormal areas are recognized. }
    \label{fig:dental_4}
\end{figure}
\newpage

\begin{figure}[H]
    \centering
    \includegraphics[width = \textwidth]{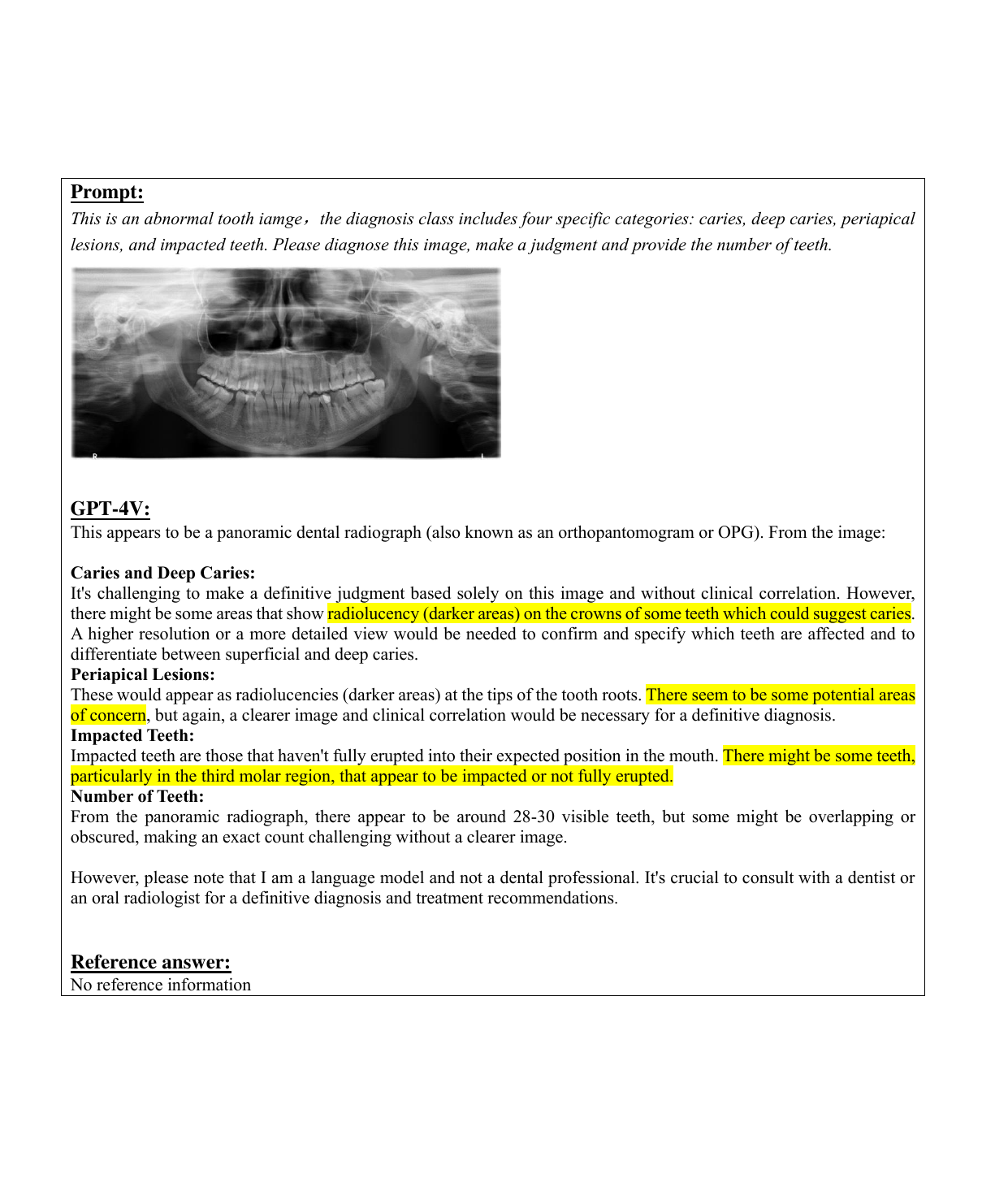}
    \caption[Dental Imaging: Case 3]{\textbf{Dental Image: Case 3.} A demonstration case from lung X-rays dental images. Yellow indicates the identification of these four abnormal teeth in caries, deep caries, periapical lesions, and impacted teeth. }
    \label{fig:dental_6}
\end{figure}
\newpage

Without any additional prompt information for testing, we found that GPT-4V can recognize these images as panoramic dental X-rays and diagnose them(Figure \ref{fig:dental_3} and \ref{fig:dental_4}). In this case, GPT-4V mainly focuses on the integrity of teeth and surrounding bones, as well as the presence of implants, while also determining the presence of wisdom teeth, significant bone loss, and periodontal disease. When four disease categories are set to require GPT-4V for diagnosis(Figure \ref{fig:dental_6}), for more obvious abnormalities, GPT-4V can provide diagnostic information and some potential areas of concern, but due to the lack of standards in the dataset, we cannot determine whether the diagnosis is correct. As for the tooth counting task, GPT-4V can only give a general judgment due to the presence of factors such as overlap or impacted teeth. It is worth mentioning that we tried to make GPT-4V perform image segmentation and annotation tasks, but it does not have this ability at present. Overall, GPT-4V can provide some auxiliary diagnostic information during the diagnostic process, but its current level of development is still difficult to replace the diagnosis of professional dentists.

\newpage

\section{Conclusion}
In this study, we evaluated the performance of the latest multimodal language model, GPT-4V, specifically in the context of biomedical imaging. We conducted this assessment using 16 different types of medical data, which include Chest Radiography, Neuroimaging, Oncological Imaging for Radiotherapy, Cytopathology in Cancer Diagnosis, Ophthalmological Imaging, Medical Robotics Imaging, Neurological Disease Imaging, Biological Imaging, Cardiac Imaging, Ultrasound Imaging, Nuclear Medicine Imaging, Endoscopic Imaging, Dermatological Imaging, Genetic Imaging, Orthopedic and Pediatric Imaging, and Dental Imaging. These data sources cover a wide range of medical modalities commonly used in clinical practice and biomedical research, such as X-ray, Computed Tomography (CT), Magnetic Resonance Imaging (MRI), Positron Emission Tomography (PET), In situ Hybridization (ISH), Genome-wide Association Study (GWAS), Single-photon emission computed tomography (SPECT), and Ultrasound.

We systematically evaluated the capabilities of GPT-4V across a range of clinical tasks using diverse multimodal biomedical imaging datasets. These tasks encompassed imaging analysis, anatomy recognition, disease diagnosis, report generation, and disease localization. Our observations revealed that GPT-4V excels in distinguishing between different medical image modalities and anatomical structures, demonstrating strong proficiency in these areas. Furthermore, in some tests, the model displayed a remarkable capacity for analyzing biomedical research results, providing valuable insights and criteria consistent with expert knowledge. However, GPT-4V faces challenges in accurately diagnosing diseases and generating comprehensive medical reports. In the worst-case scenarios, it was observed to "hallucinate" facts and make errors in reasoning, leading to generated responses that overlooked critical information from the image input. These findings underscore the potential and limitations of large multimodal models in biomedical imaging applications, emphasizing the need for further exploration and refinement of GPT-4V's capabilities through prompt engineering to enhance its effectiveness in real-world clinical decision support. It's important to note that GPT-4V is designed to avoid direct answers related to disease stage diagnosis and prediction, which may contribute to some of the observed limitations.

This work sheds light on potential future works using large language models in multimodality biomedical applications. When considering the avenues for future research, there are several key areas deserving exploration. For example: \\
\begin{itemize}
  \item \textbf{Multimodal Integration}: A critical research direction involves enhancing techniques to integrate and contextualize data from various biomedical modalities. This facilitates a better understanding of the relationships between different types of biomedical data. \\
 \item \textbf{Data Augmentation}: Investigating methods to enrich training data with additional clinical context, patient histories, and domain-specific knowledge to improve the model's diagnostic capabilities and report generation. \\
 \item \textbf{Ethical Considerations}: Delving into the ethical concerns associated with AGI deployment in healthcare, including issues such as patient privacy, informed consent, and transparency in decision-making \cite{liu3surviving,liu2023deid}. Ensuring compliance with medical ethics and regulatory guidelines is of paramount importance. \\
 \item \textbf{Clinical Validation}: Conducting extensive clinical validation studies to assess the real-world impact of models like GPT-4V on patient outcomes, healthcare costs, and clinical workflows. \\
 \item \textbf{Data Generation}: In specific domains, such as genetics research, GPT-4V's ability to interpret chart images enables the batch generation of text data from literature sources. \\
 \item \textbf{Imaging-Genetics Analysis}: Leveraging GPT-4V's capability for multi-modal data processing in the domain of imaging-genetics analysis, which heavily relies on technology for handling diverse data types and advancing multiscale integration. \\
 \item \textbf{Multi-Modal Transformation in Bioinformatics}: The potential for AGI models to merge sequence data (text) and 3D structural data (graphics and images) in bioinformatics analysis.
\end{itemize}

In terms of practical applications for GPT-4V in clinical settings, several use cases come to the fore: \\
\begin{itemize}
  \item \textbf{Assisting Radiologists}: GPT-4V can be a valuable tool for radiologists, aiding them in the initial screening of medical images and enhancing the speed and accuracy of diagnoses. \\
 \item \textbf{Automated Report Generation}: The model can be employed to generate preliminary medical reports based on imaging data, leading to time savings for healthcare professionals and streamlining the reporting process. \\
 \item \textbf{Teaching Tool}: GPT-4V can serve as an educational resource for medical students and residents, helping them understand and interpret various medical imaging modalities. \\
 \item \textbf{Second Opinion}: Offering a second opinion or support for healthcare professionals, particularly in cases involving rare conditions or complex diagnoses.
\end{itemize}

The GPT series represents significant progress toward artificial general intelligence (AGI), with ongoing research pushing the boundaries of this field. In conclusion, while models like GPT-4V signify substantial advancements in AI and AGI, there remain noteworthy challenges and opportunities for further research and application, especially in the intricate realm of medical diagnosis, treatment, and prognosis. The future holds promise for AGI in healthcare, but careful consideration of ethical, regulatory, and clinical aspects is essential as we move forward.

Finally, it should be pointed out that GPT-4V is an OpenAI's commercial product that is not open-sourced, expensive to access, and not transparent to AGI researchers, biomedical scholars, and clinical physicians. Also, it is extremely challenging, although not completely impossible, to upload clinical medical images to OpenAI's APIs for interpretation and report generation. Therefore, our vision is that the academia should explore open-source solutions of developing and deploying GPT-4V type of AGI systems for biomedical imaging research in the near future.

\newpage
\appendix
\section{Appendix}
\subsection{Chest Radiography}

\begin{figure}[H]
    \centering
    \includegraphics[width = \textwidth]{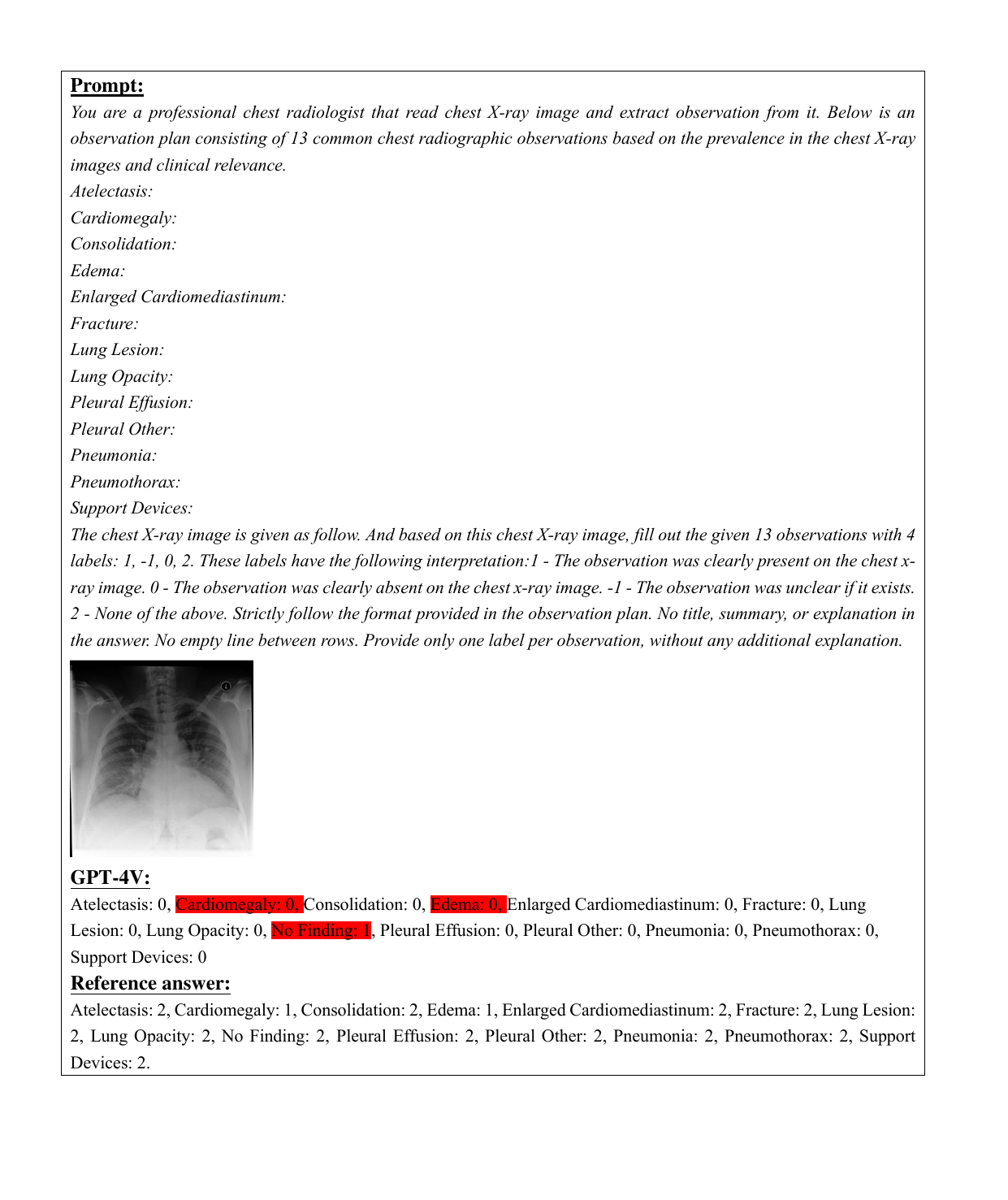}
    \caption[Chest Radiography: Case 3]{\textbf{Chest: Case 3.} A demonstration case of Classification task on MIMIC-CXR dataset. Green denotes the correct classification. Red in the figure denotes the incorrect classification.}
    \label{fig:chest_2}
\end{figure}
\newpage

\begin{figure}[H]
    \centering
    \includegraphics[width = \textwidth]{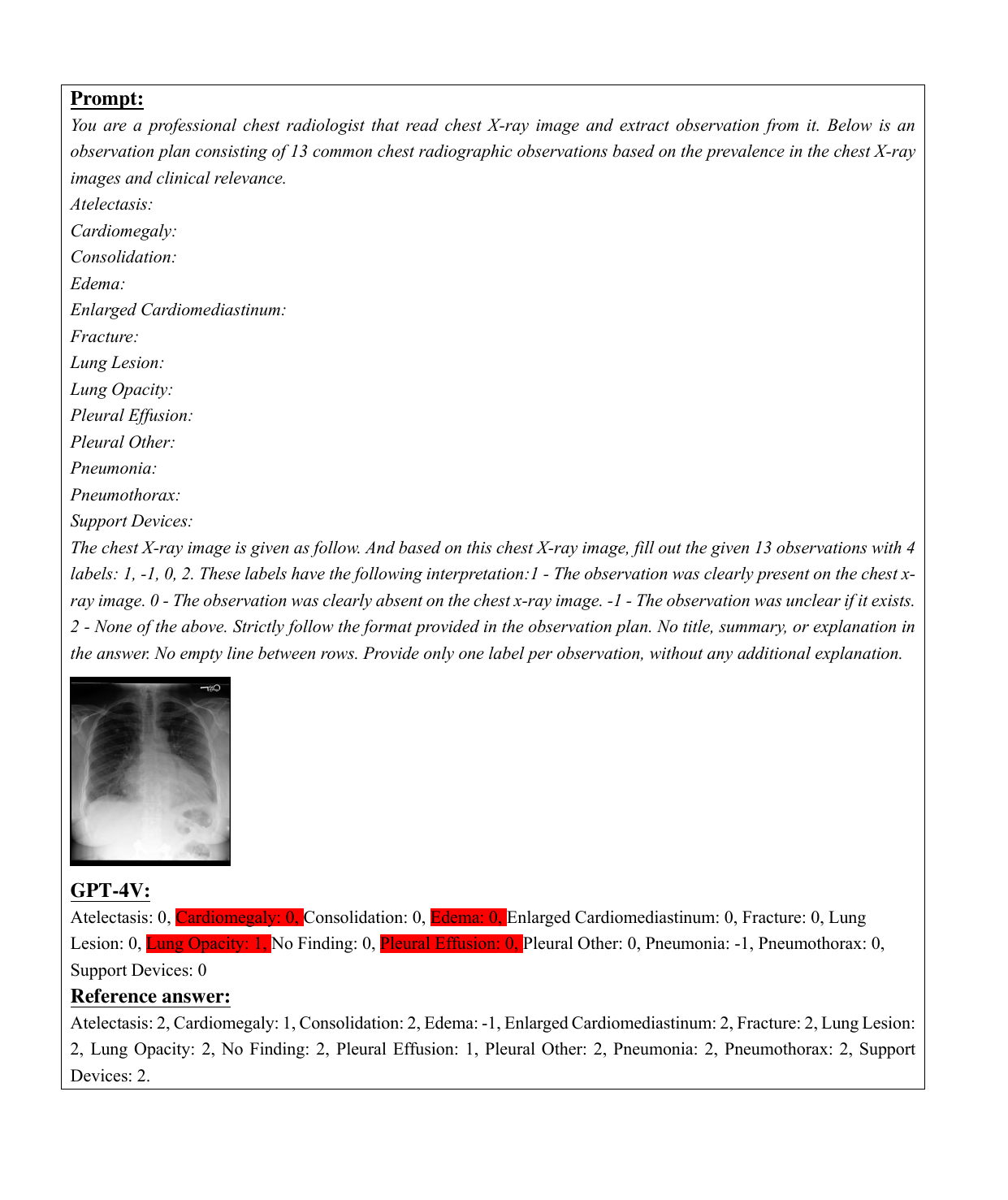}
    \caption[Chest Radiography: Case 4]{\textbf{Chest: Case 4.} A demonstration case of Classification task on MIMIC-CXR dataset. Green denotes the correct classification. Red in the figure denotes the incorrect classification.}
    \label{fig:chest_3}
\end{figure}
\newpage

\begin{figure}[H]
    \centering
    \includegraphics[width = \textwidth]{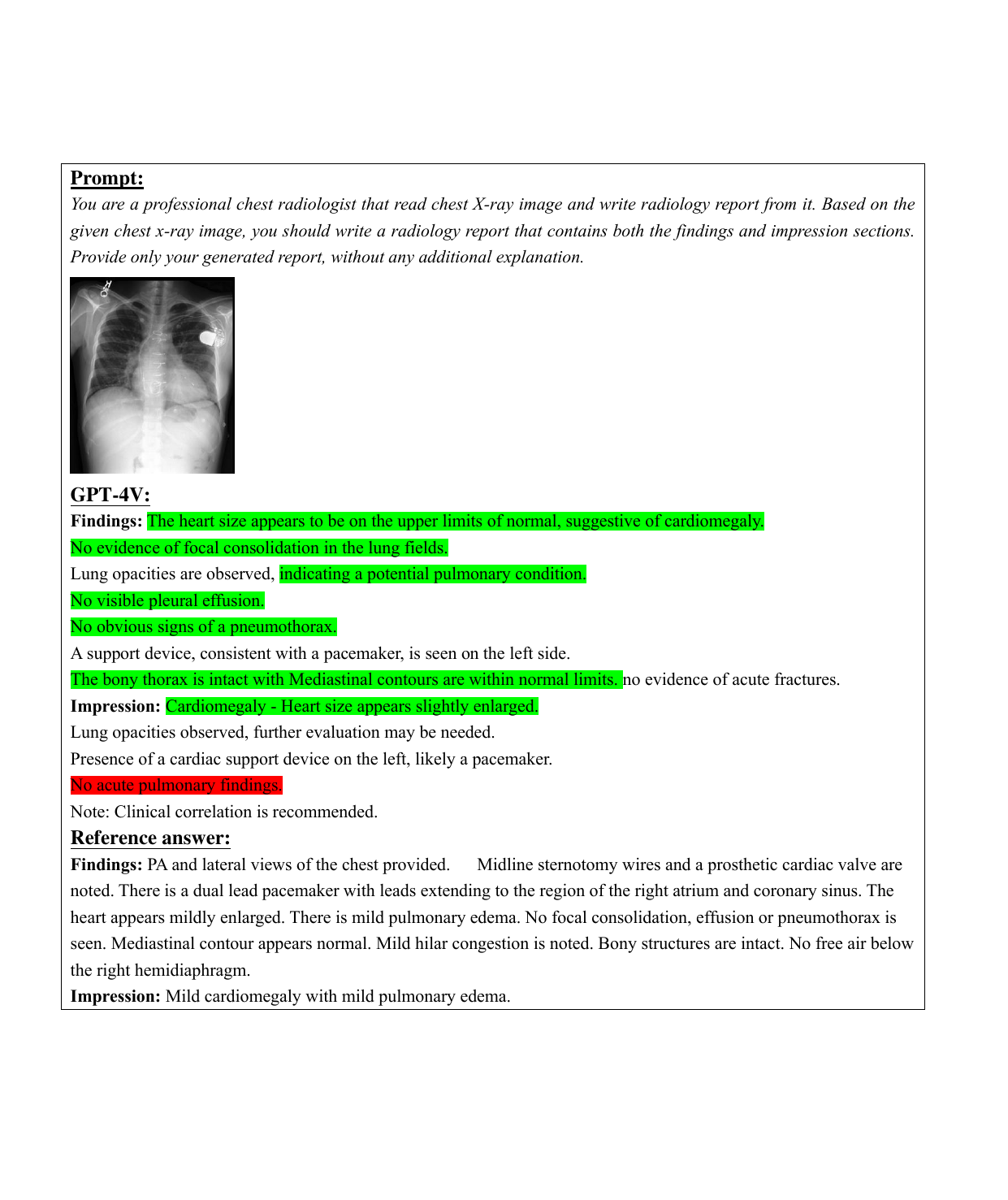}
    \caption[Chest Radiography: Case 5]{\textbf{Chest: Case 5.} A demonstration case of Report Generation Task on MIMIC-CXR dataset. Green, yellow, and red represent correct, uncertain, and incorrect information identified respectively.}
    \label{fig:chest_4}
\end{figure}
\newpage

\begin{figure}[H]
    \centering
    \includegraphics[width = \textwidth]{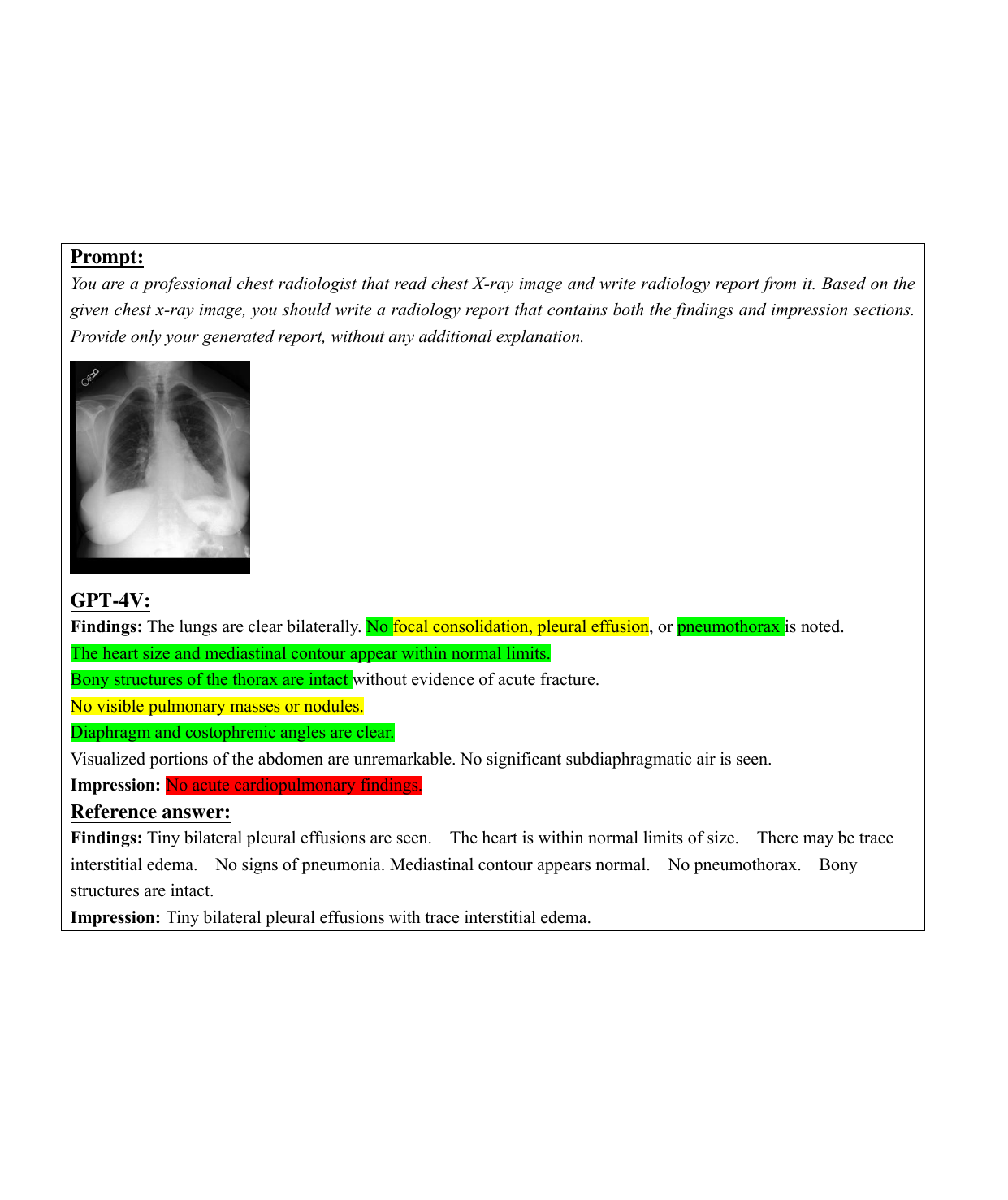}
    \caption[Chest Radiography: Case 6]{\textbf{Chest: Case 6.} A demonstration case of Report Generation Task on MIMIC-CXR dataset. Green, yellow, and red represent correct, uncertain, and incorrect information identified respectively.}
    \label{fig:chest_5}
\end{figure}
\newpage

\begin{figure}[H]
    \centering
    \includegraphics[width = \textwidth]{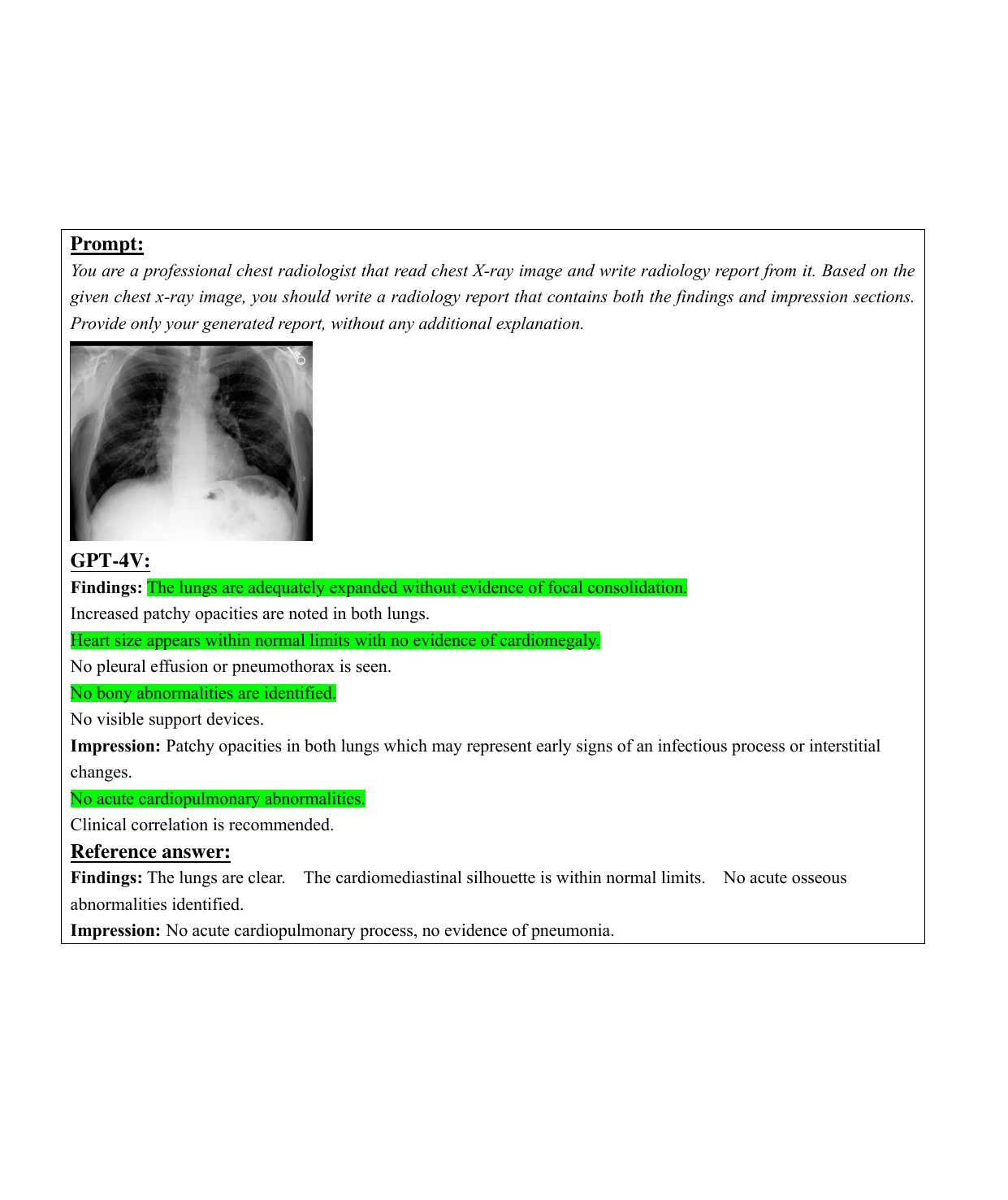}
    \caption[Chest Radiography: Case 7]{\textbf{Chest: Case 7.} A demonstration case of Report Generation Task on MIMIC-CXR dataset. Green, yellow, and red represent correct, uncertain, and incorrect information identified respectively.}
    \label{fig:chest_6}
\end{figure}
\newpage

\begin{figure}[H]
    \centering
    \includegraphics[width = \textwidth]{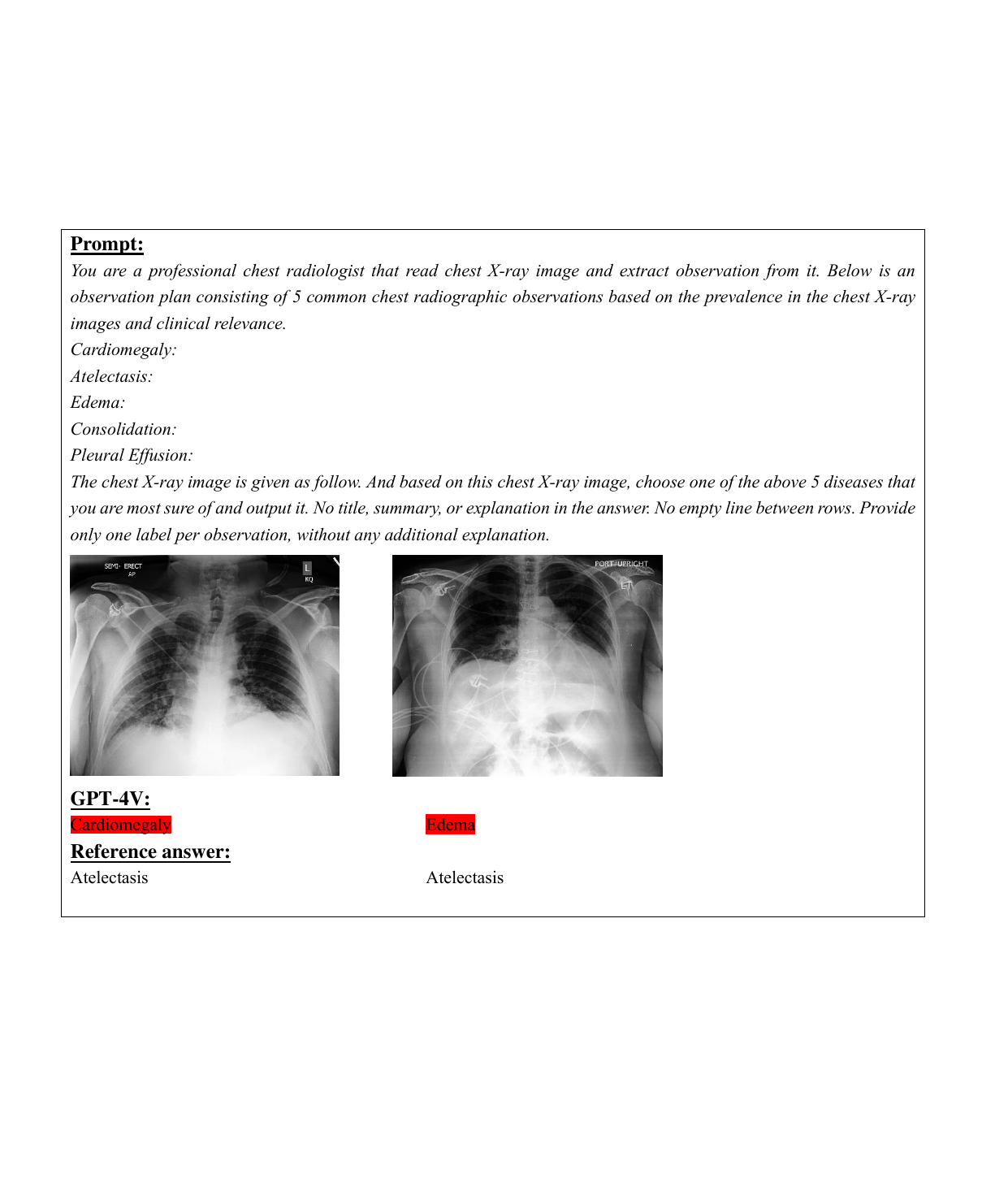}
    \caption[Chest Radiography: Case 8]{\textbf{Chest: Case 8.} Two demonstration cases of Classification task on CheXpert dataset. Green denotes the correct classification. Red in the figure denotes the incorrect classification.}
    \label{fig:chest_7}
\end{figure}
\newpage

\begin{figure}[H]
    \centering
    \includegraphics[width = \textwidth]{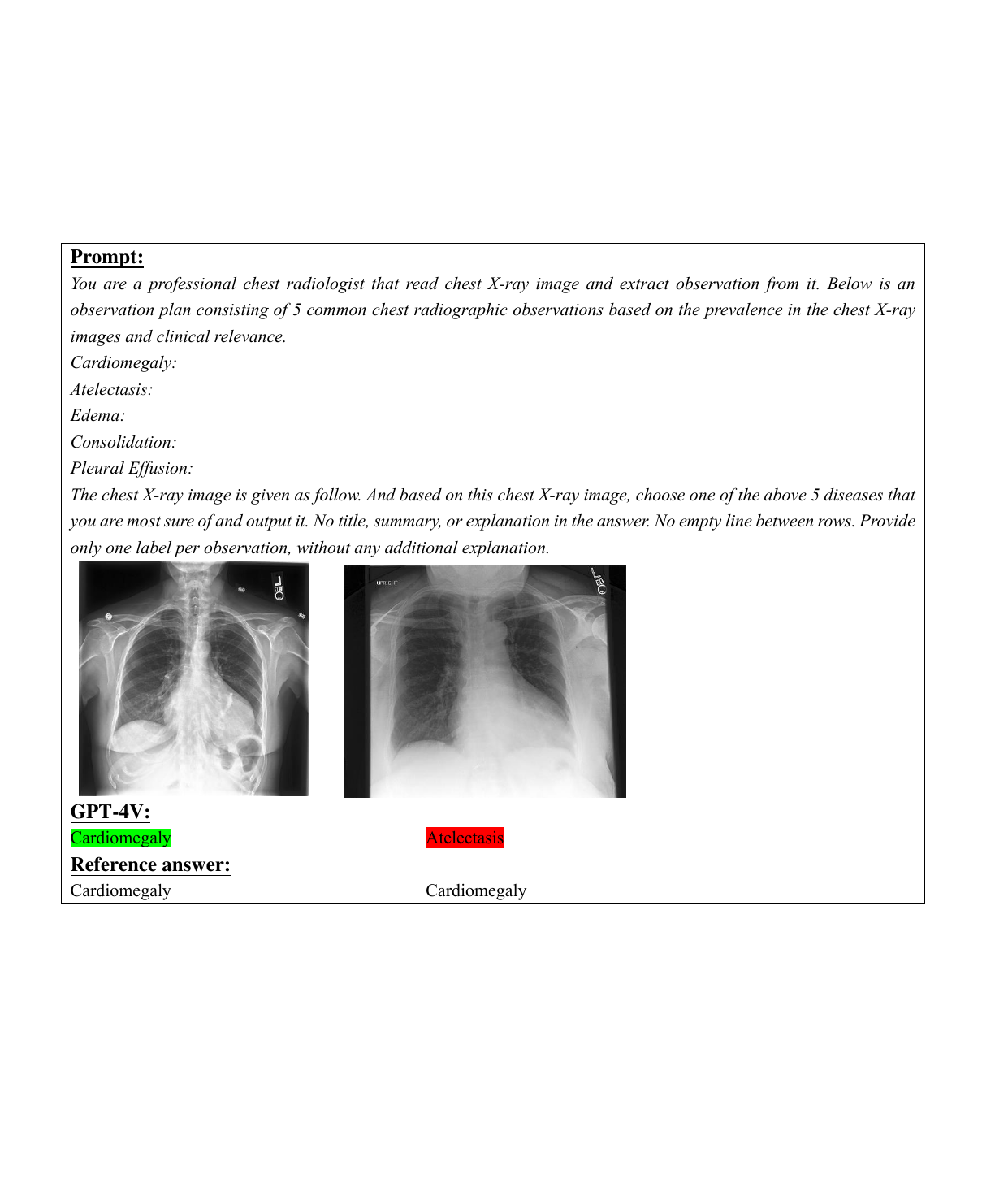}
    \caption[Chest Radiography: Case 9]{\textbf{Chest: Case 9.} Two demonstration cases of Classification task on CheXpert dataset. Green denotes the correct classification. Red in the figure denotes the incorrect classification.}
    \label{fig:chest_8}
\end{figure}
\newpage

\begin{figure}[H]
    \centering
    \includegraphics[width = \textwidth]{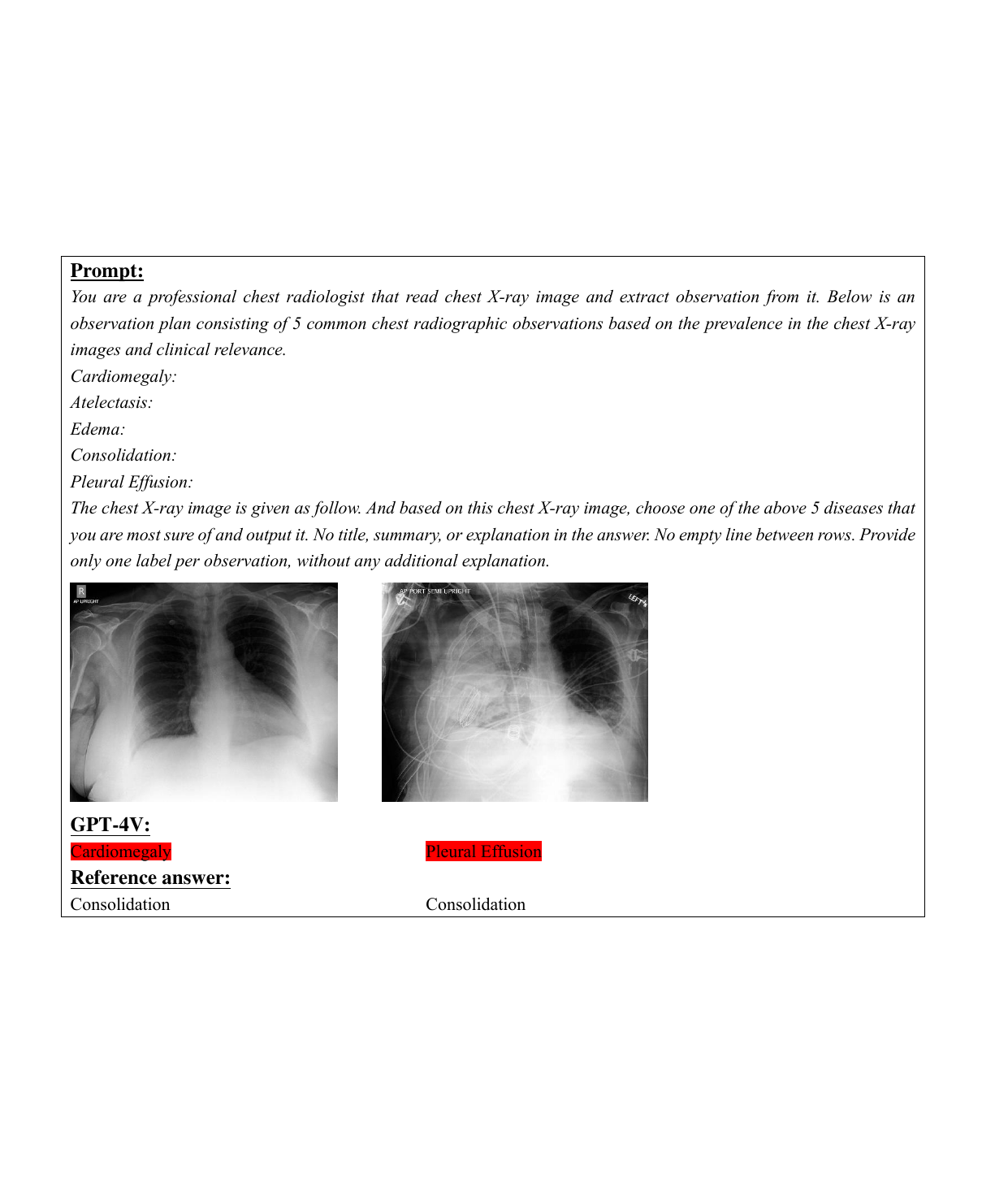}
    \caption[Chest Radiography: Case 10]{\textbf{Chest: Case 10.} Two demonstration cases of Classification task on CheXpert dataset. Green denotes the correct classification. Red in the figure denotes the incorrect classification.}
    \label{fig:chest_9}
\end{figure}
\newpage

\begin{figure}[H]
    \centering
    \includegraphics[width = \textwidth]{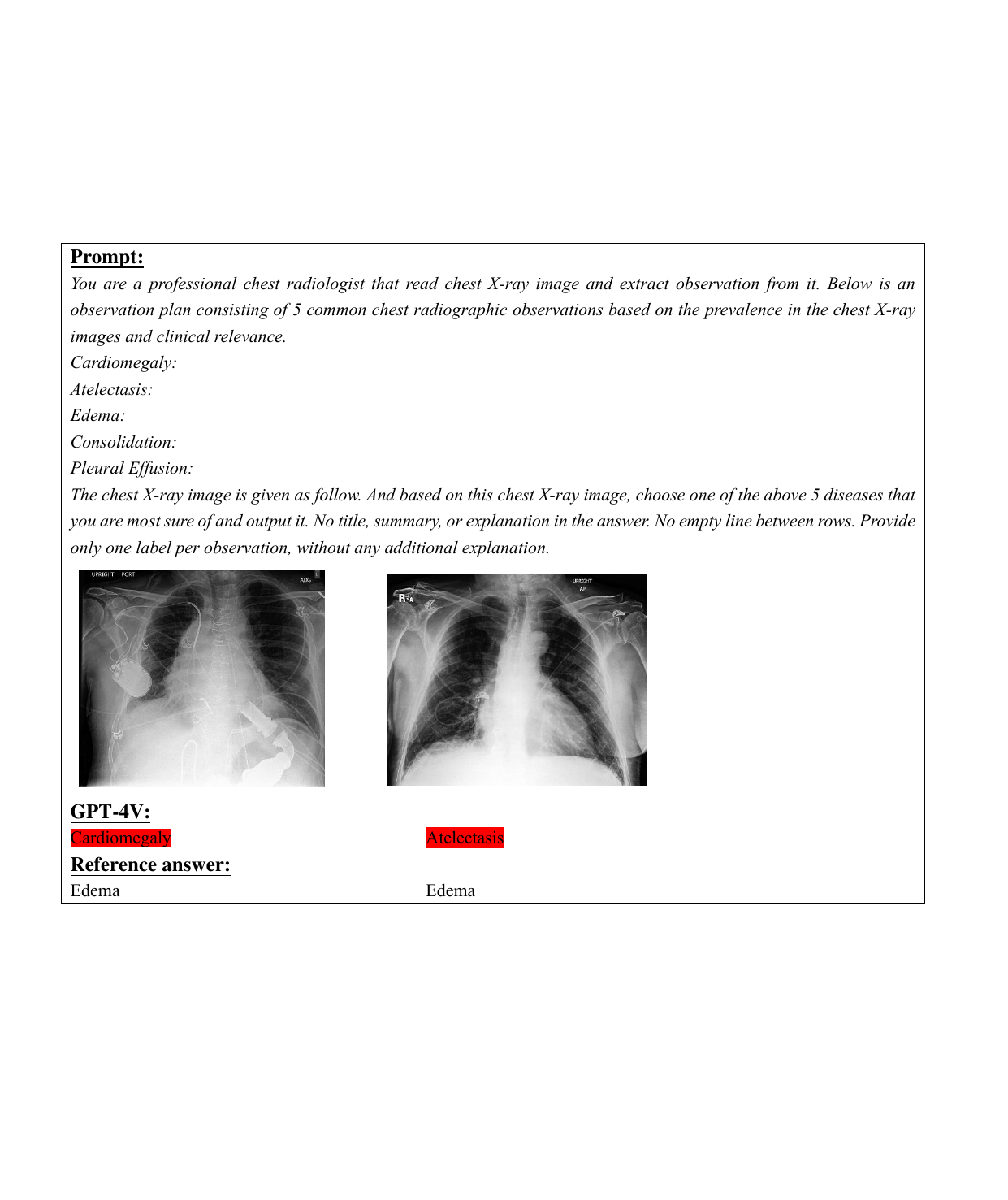}
    \caption[Chest Radiography: Case 11]{\textbf{Chest: Case 11.} Two demonstration cases of Classification task on CheXpert dataset. Green denotes the correct classification. Red in the figure denotes the incorrect classification.}
    \label{fig:chest_10}
\end{figure}
\newpage

\begin{figure}[H]
    \centering
    \includegraphics[width = \textwidth]{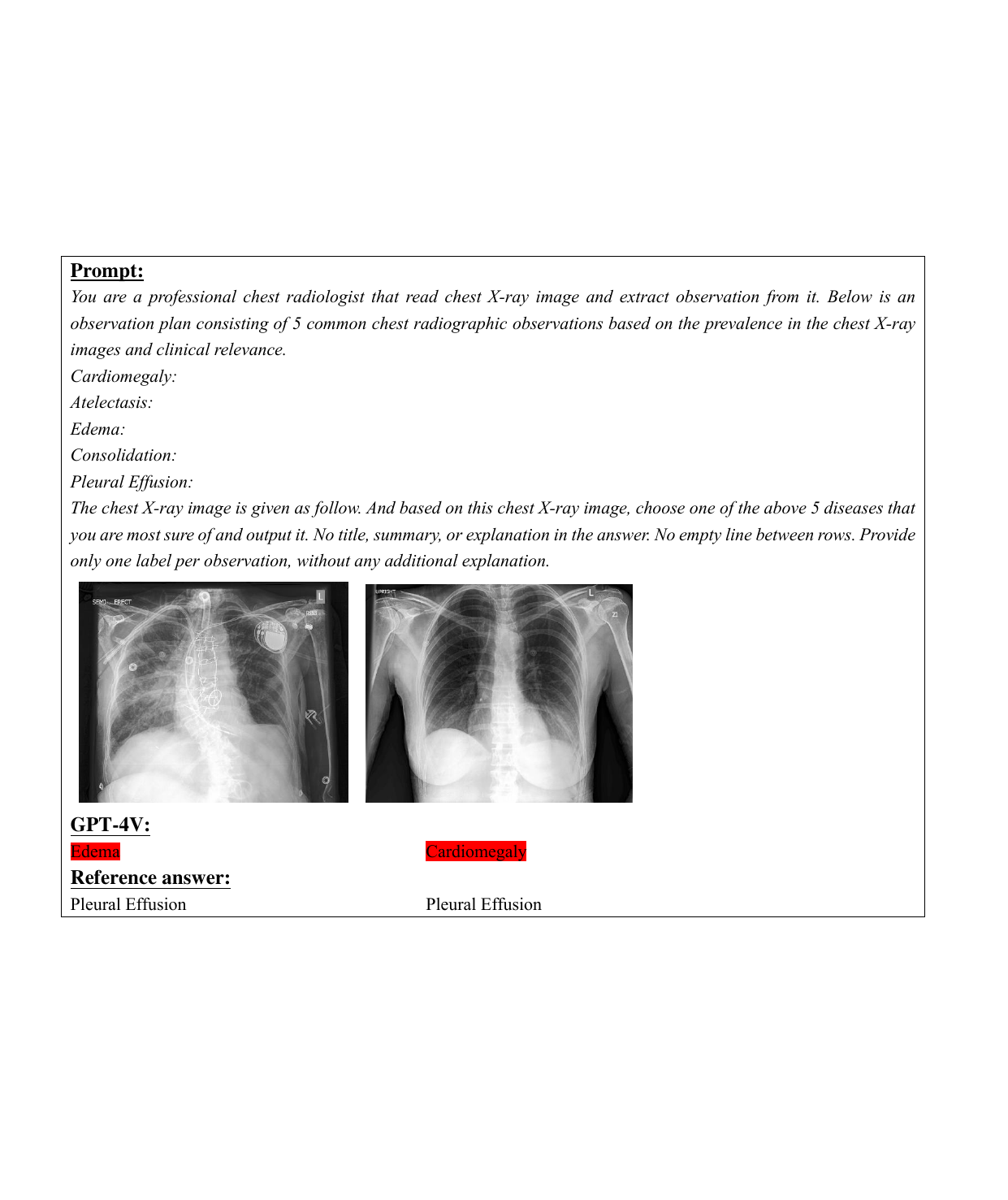}
    \caption[Chest Radiography: Case 12]{\textbf{Chest: Case 12.} Two demonstration cases of Classification task on CheXpert dataset. Green denotes the correct classification. Red in the figure denotes the incorrect classification.}
    \label{fig:chest_11}
\end{figure}
\newpage

\begin{figure}[H]
    \centering
    \includegraphics[width = \textwidth]{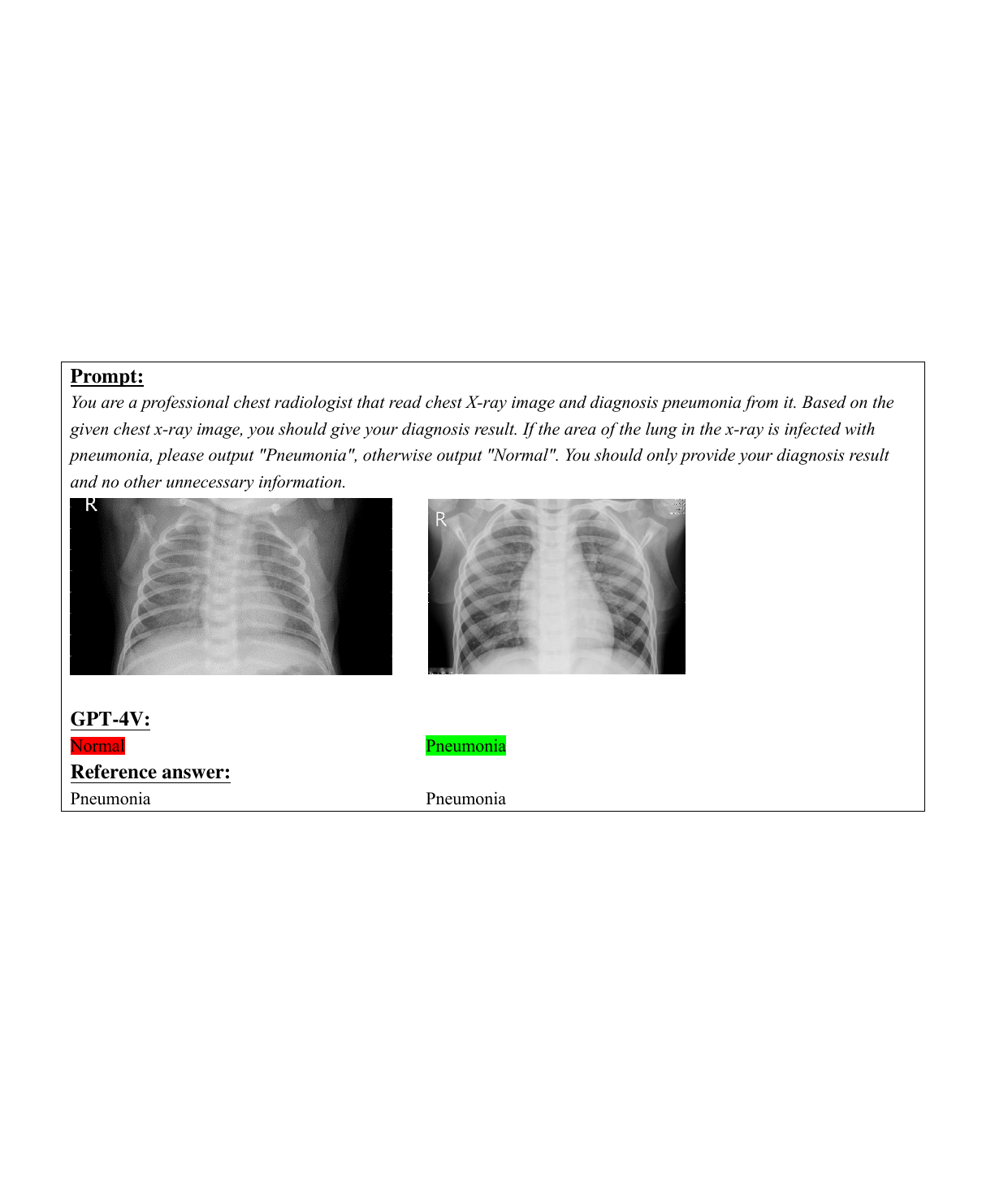}
    \caption[Chest Radiography: Case 13]{\textbf{Chest: Case 13.} Two demonstration cases of Classification task on ChestXray2017 dataset. Green denotes the correct classification. Red in the figure denotes the incorrect classification.}
    \label{fig:chest_12}
\end{figure}
\newpage

\begin{figure}[H]
    \centering
    \includegraphics[width = \textwidth]{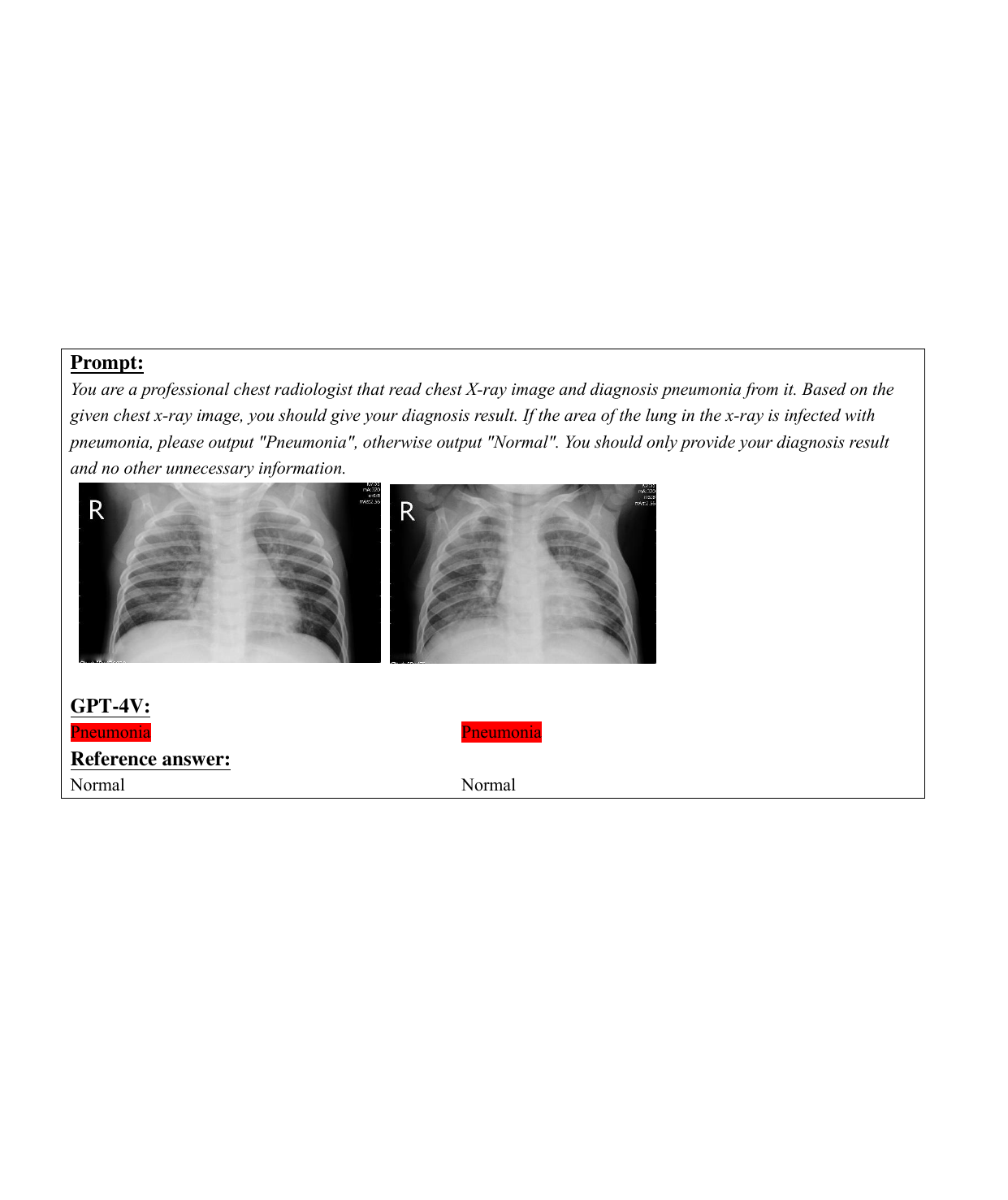}
    \caption[Chest Radiography: Case 14]{\textbf{Chest: Case 14.} Two demonstration cases of Classification task on ChestXray2017 dataset. Green denotes the correct classification. Red in the figure denotes the incorrect classification.}
    \label{fig:chest_13}
\end{figure}
\newpage

\begin{figure}[H]
    \centering
    \includegraphics[width = \textwidth]{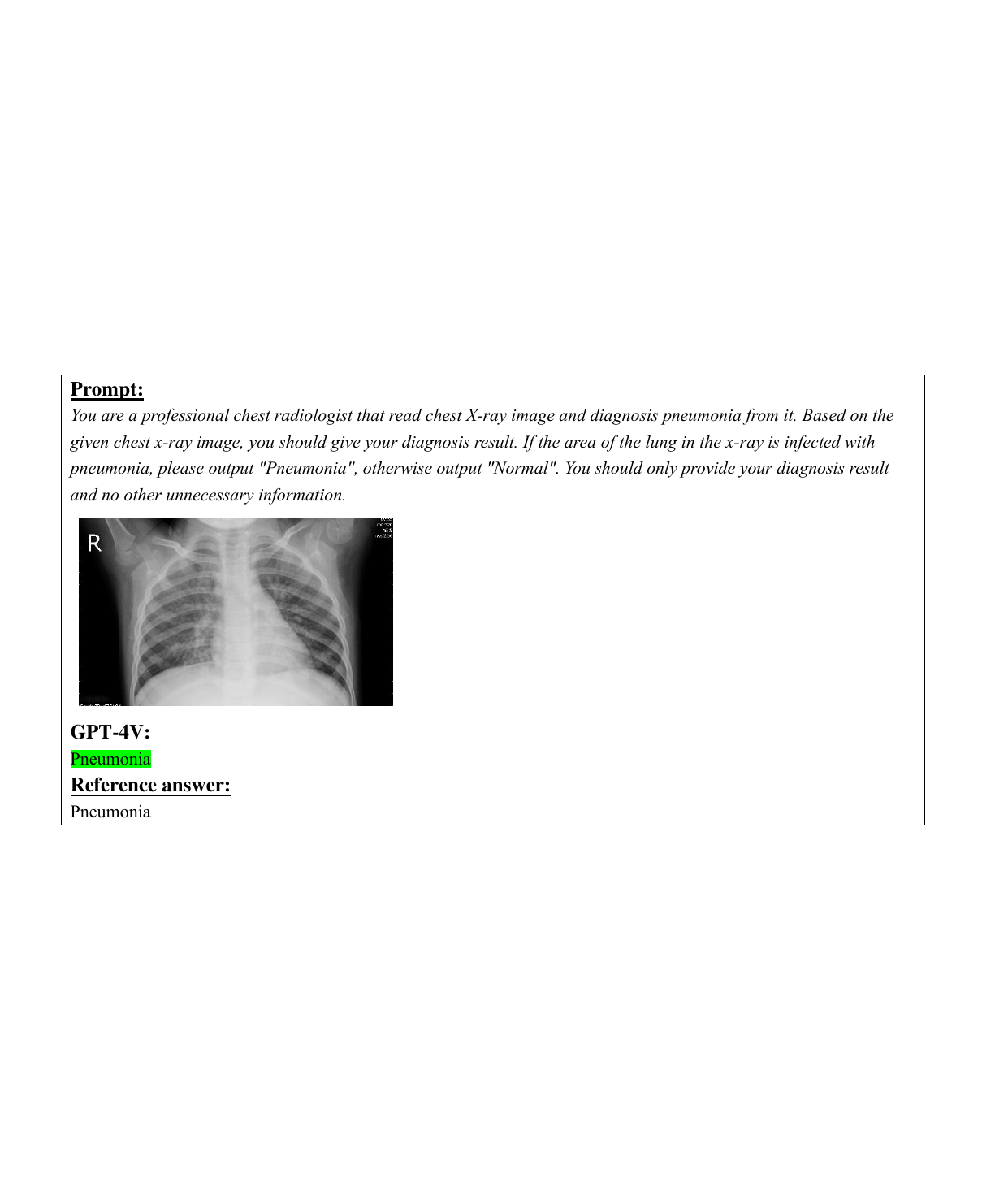}
    \caption[Chest Radiography: Case 15]{\textbf{Chest: Case 15.} A demonstration case of Classification task on ChestXray2017 dataset. Green denotes the correct classification. Red in the figure denotes the incorrect classification.}
    \label{fig:chest_14}
\end{figure}
\newpage

\begin{figure}[H]
    \centering
    \includegraphics[width = \textwidth]{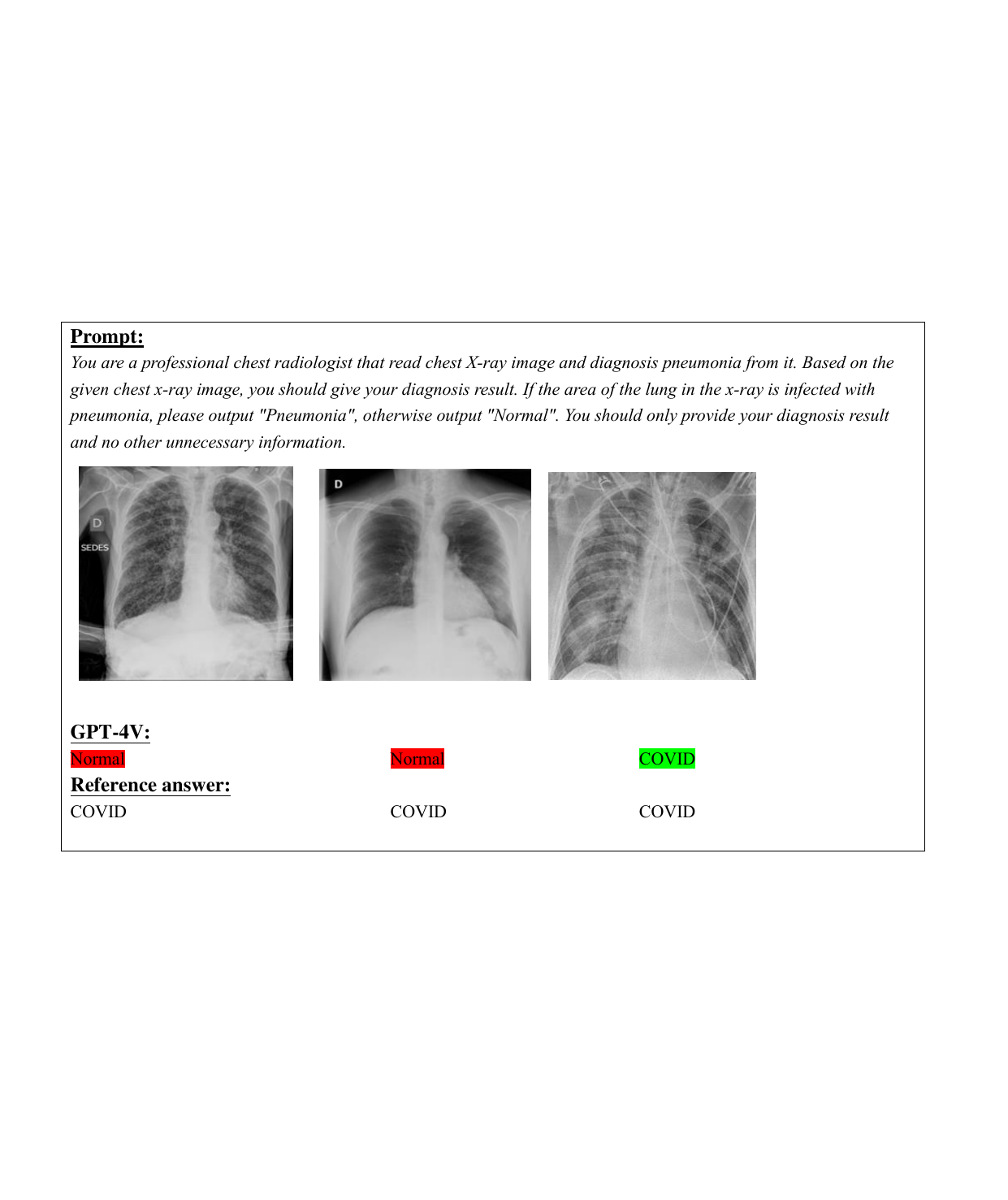}
    \caption[Chest Radiography: Case 16]{\textbf{Chest: Case 16.} Demonstration cases of Classification task on COVID-Qu-Ex dataset. Green denotes the correct classification. Red in the figure denotes the incorrect classification.}
    \label{fig:chest_15}
\end{figure}
\newpage

\begin{figure}[H]
    \centering
    \includegraphics[width = \textwidth]{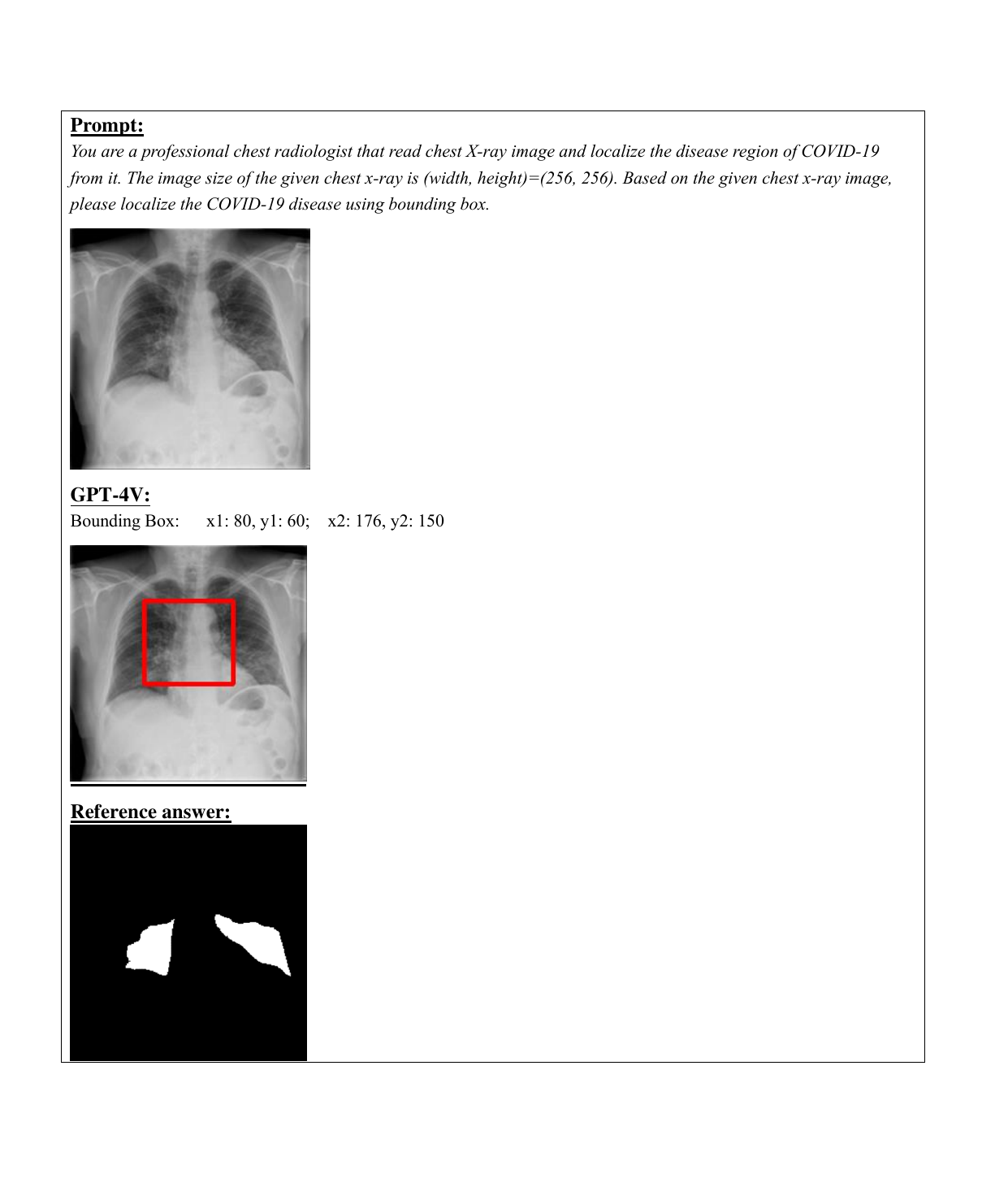}
    \caption[Chest Radiography: Case 17]{\textbf{Chest: Case 17.} A Demonstration case of Detection task on COVID-Qu-Ex dataset. The output of GPT-4V is the coordinate values of the bounding box. The red box is the corresponding image region. In addition, we provide a ground-truth mask image of the lesion area for comparison.}
    \label{fig:chest_16}
\end{figure}
\newpage

\begin{figure}[H]
    \centering
    \includegraphics[width = \textwidth]{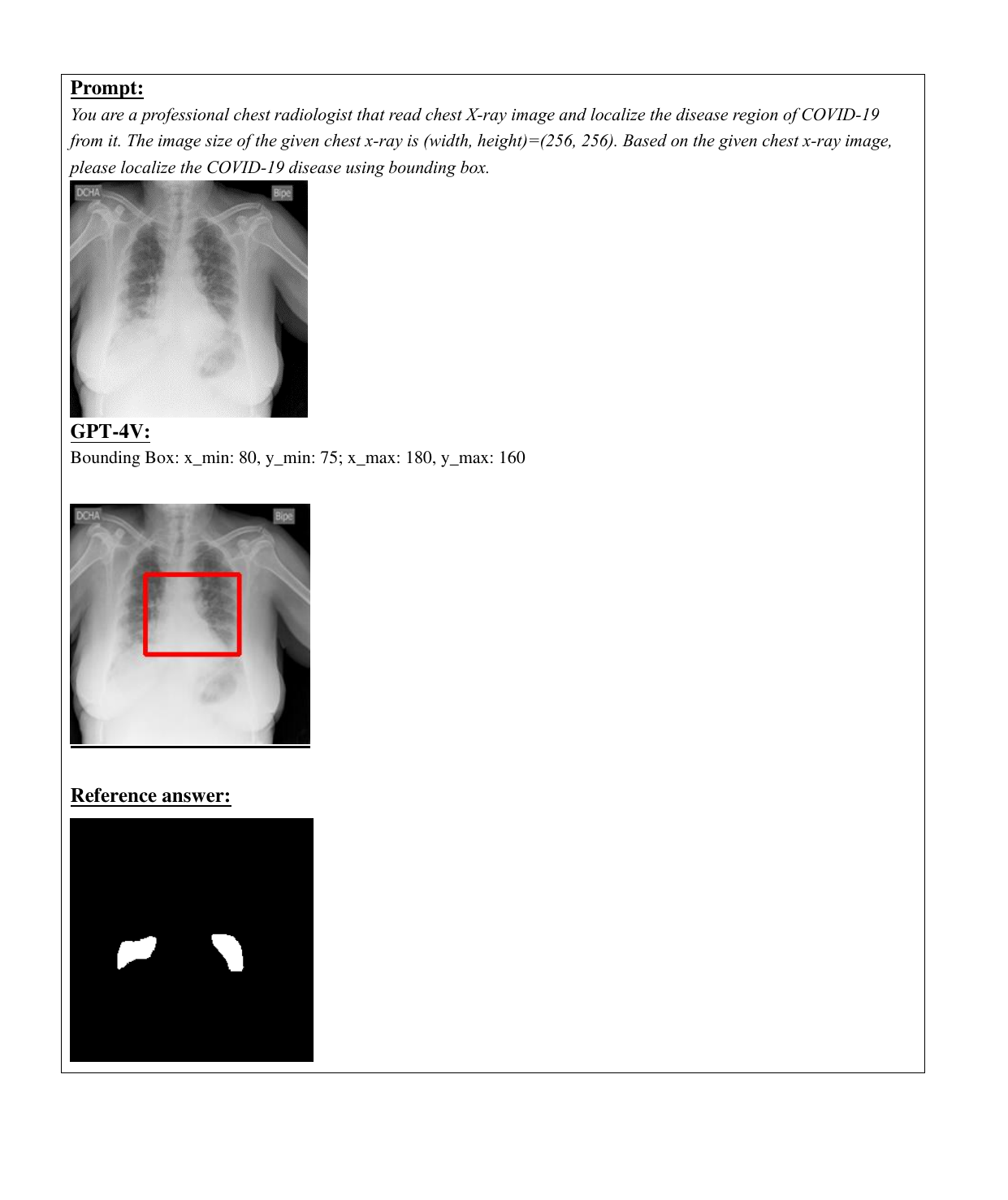}
    \caption[Chest Radiography: Case 18]{\textbf{Chest: Case 18.} A Demonstration case of Detection task on COVID-Qu-Ex dataset. The output of GPT-4V is the coordinate values of the bounding box. The red box is the corresponding image region. In addition, we provide a ground-truth mask image of the lesion area for comparison.}
    \label{fig:chest_17}
\end{figure}
\newpage

\begin{figure}[H]
    \centering
    \includegraphics[width = \textwidth]{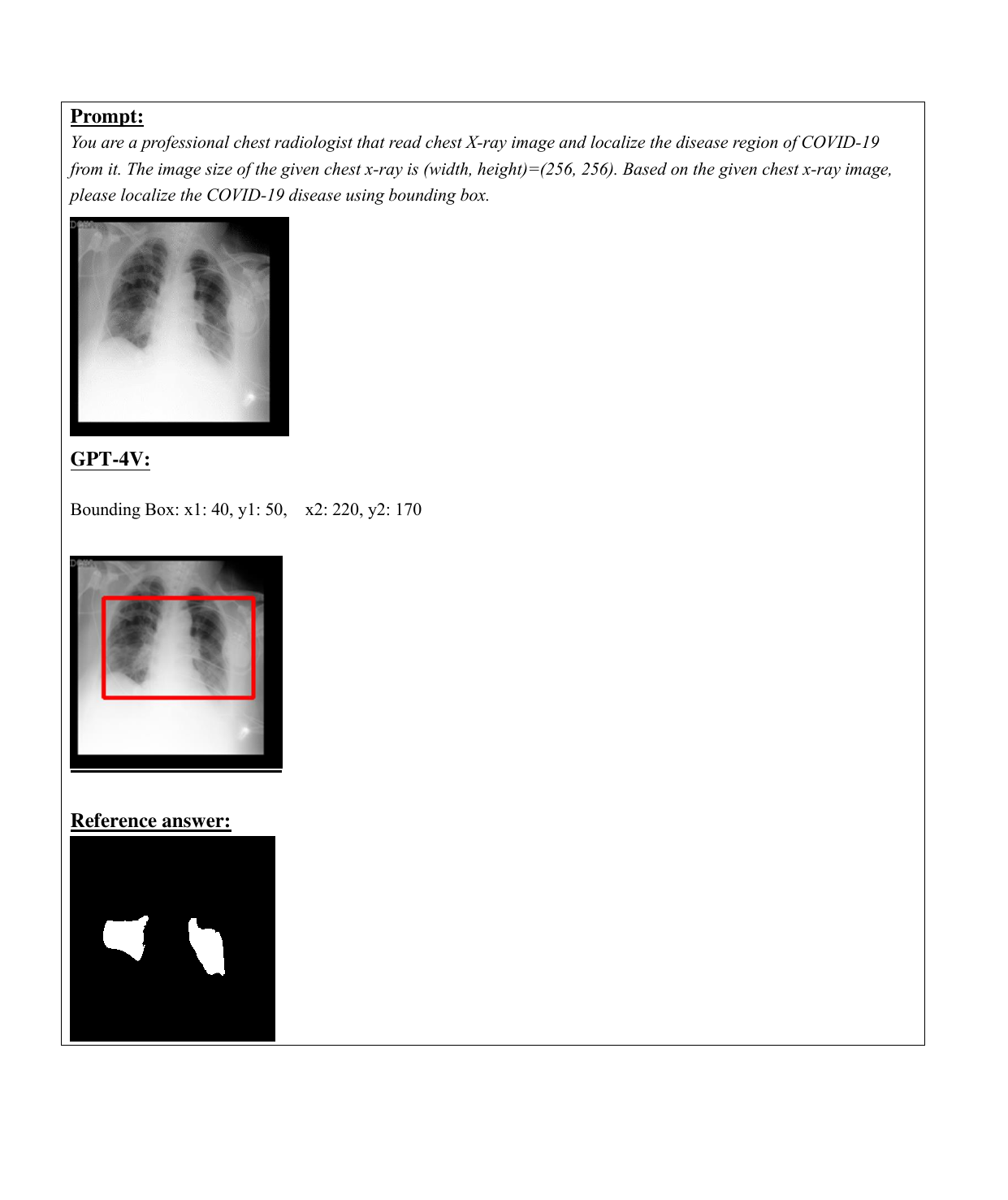}
    \caption[Chest Radiography: Case 19]{\textbf{Chest: Case 19.} A Demonstration case of Detection task on COVID-Qu-Ex dataset. The output of GPT-4V is the coordinate values of the bounding box. The red box is the corresponding image region. In addition, we provide a ground-truth mask image of the lesion area for comparison.}
    \label{fig:chest_18}
\end{figure}
\newpage

\begin{figure}[H]
    \centering
    \includegraphics[width = \textwidth]{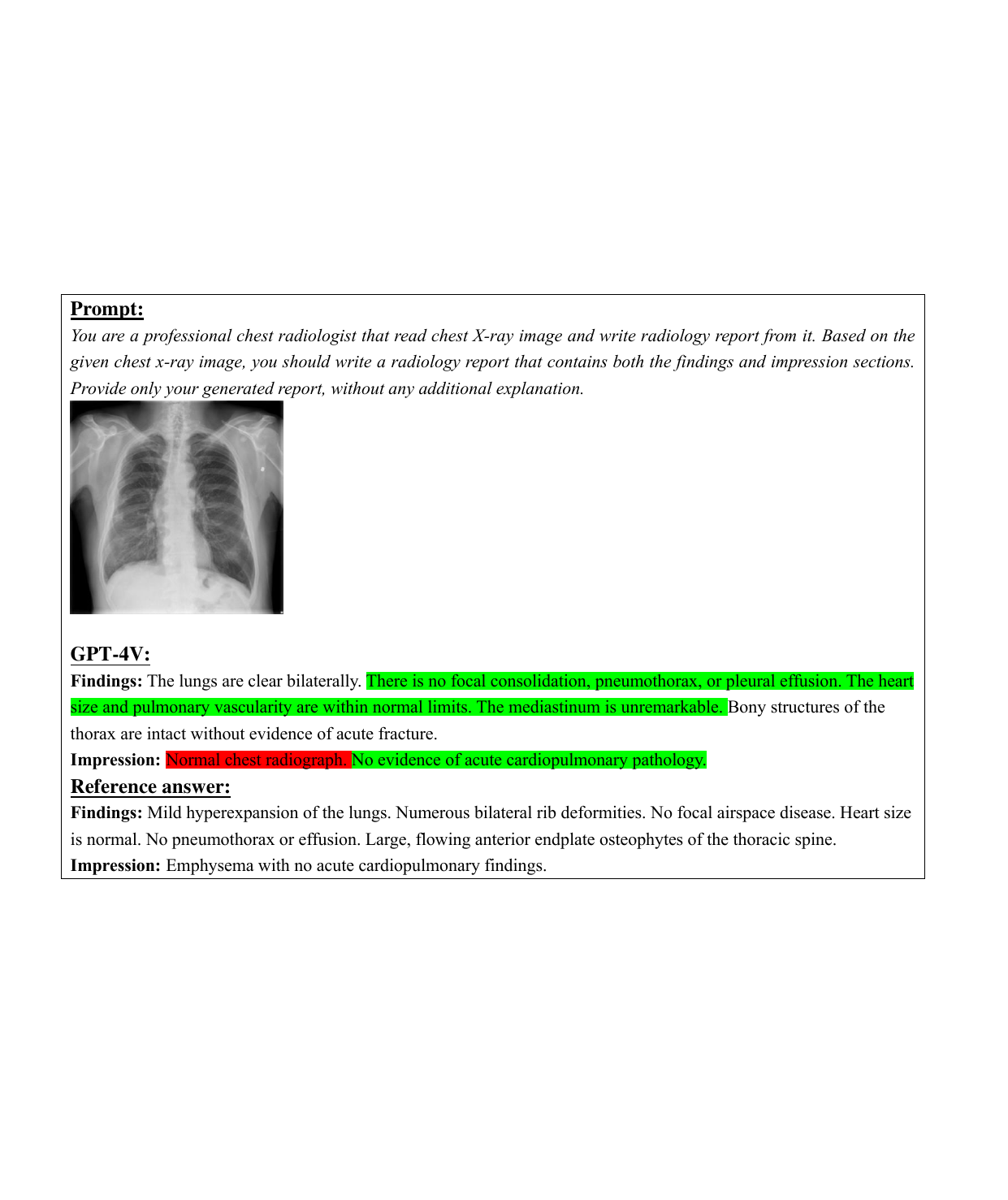}
    \caption[Chest Radiography: Case 20]{\textbf{Chest: Case 20.} A demonstration case of Report Generation Task on OpenI dataset. Green, yellow, and red represent correct, uncertain, and incorrect information identified respectively.}
    \label{fig:chest_19}
\end{figure}
\newpage

\begin{figure}[H]
    \centering
    \includegraphics[width = \textwidth]{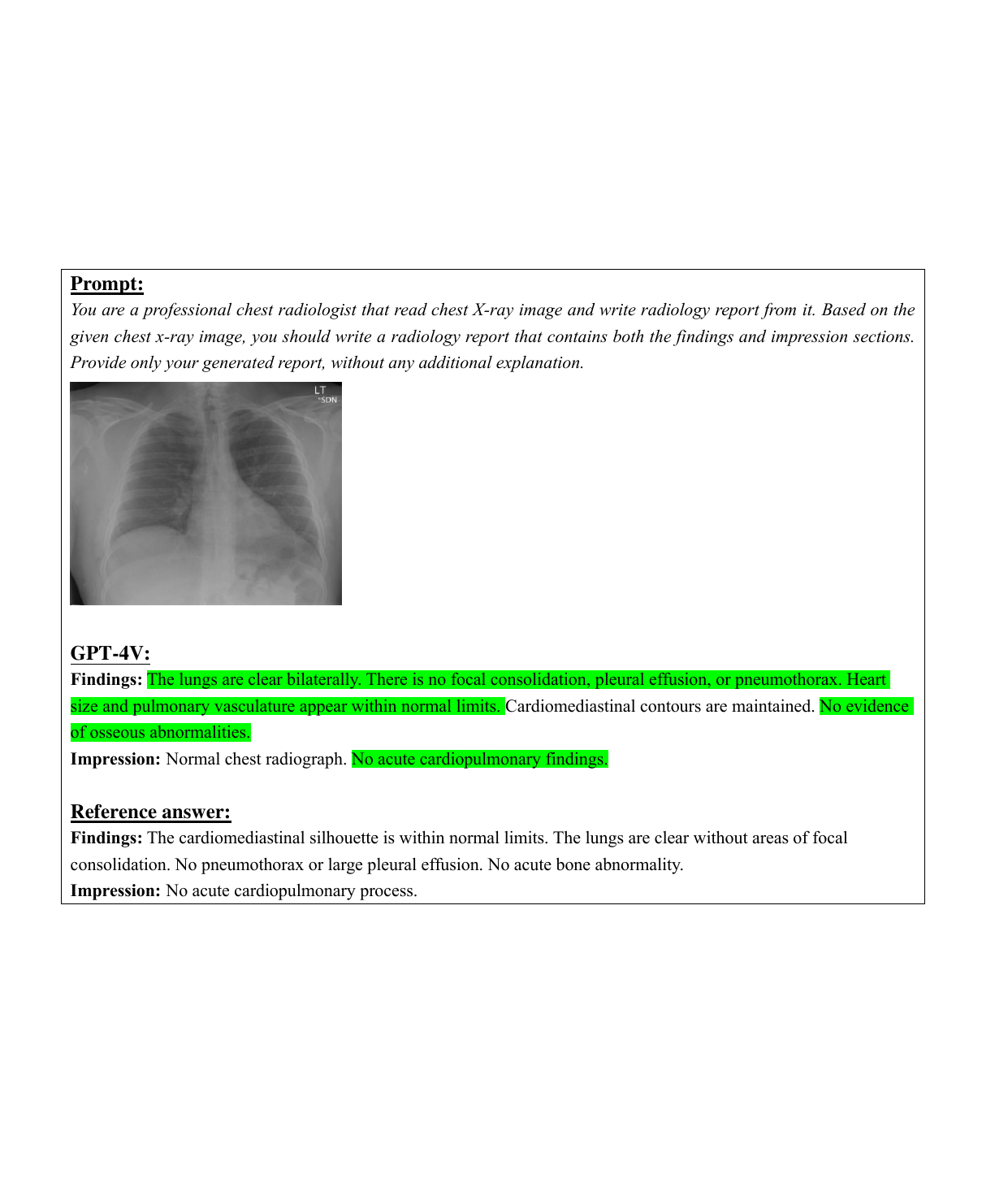}
    \caption[Chest Radiography: Case 21]{\textbf{Chest: Case 21.} A demonstration case of Report Generation Task on OpenI dataset. Green, yellow, and red represent correct, uncertain, and incorrect information identified respectively.}
    \label{fig:chest_20}
\end{figure}
\newpage

\begin{figure}[H]
    \centering
    \includegraphics[width = \textwidth]{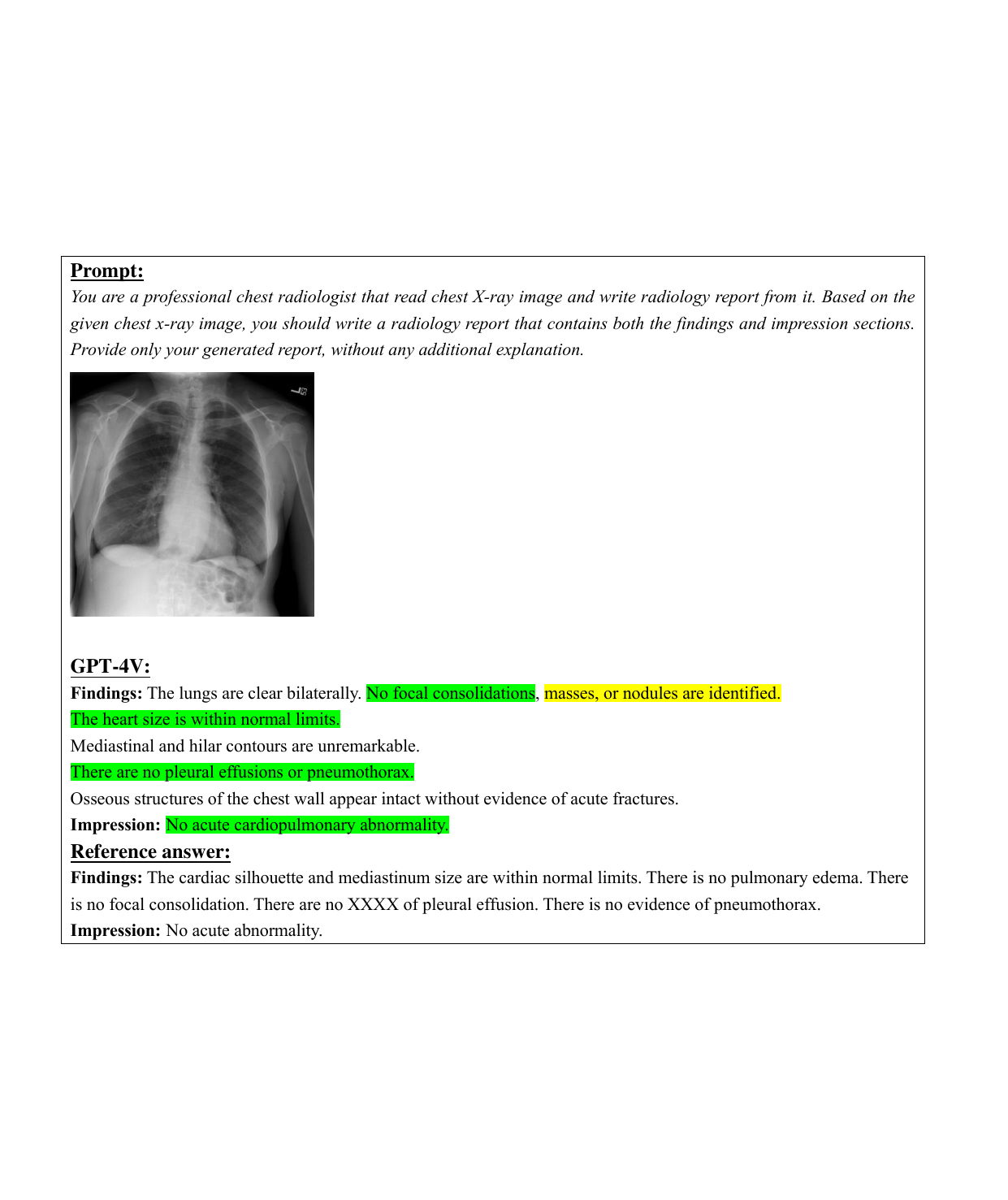}
    \caption[Chest Radiography: Case 22]{\textbf{Chest: Case 22.} A demonstration case of Report Generation Task on OpenI dataset. Green, yellow, and red represent correct, uncertain, and incorrect information identified respectively.}
    \label{fig:chest_21}
\end{figure}
\newpage

\begin{figure}[H]
    \centering
    \includegraphics[width = \textwidth]{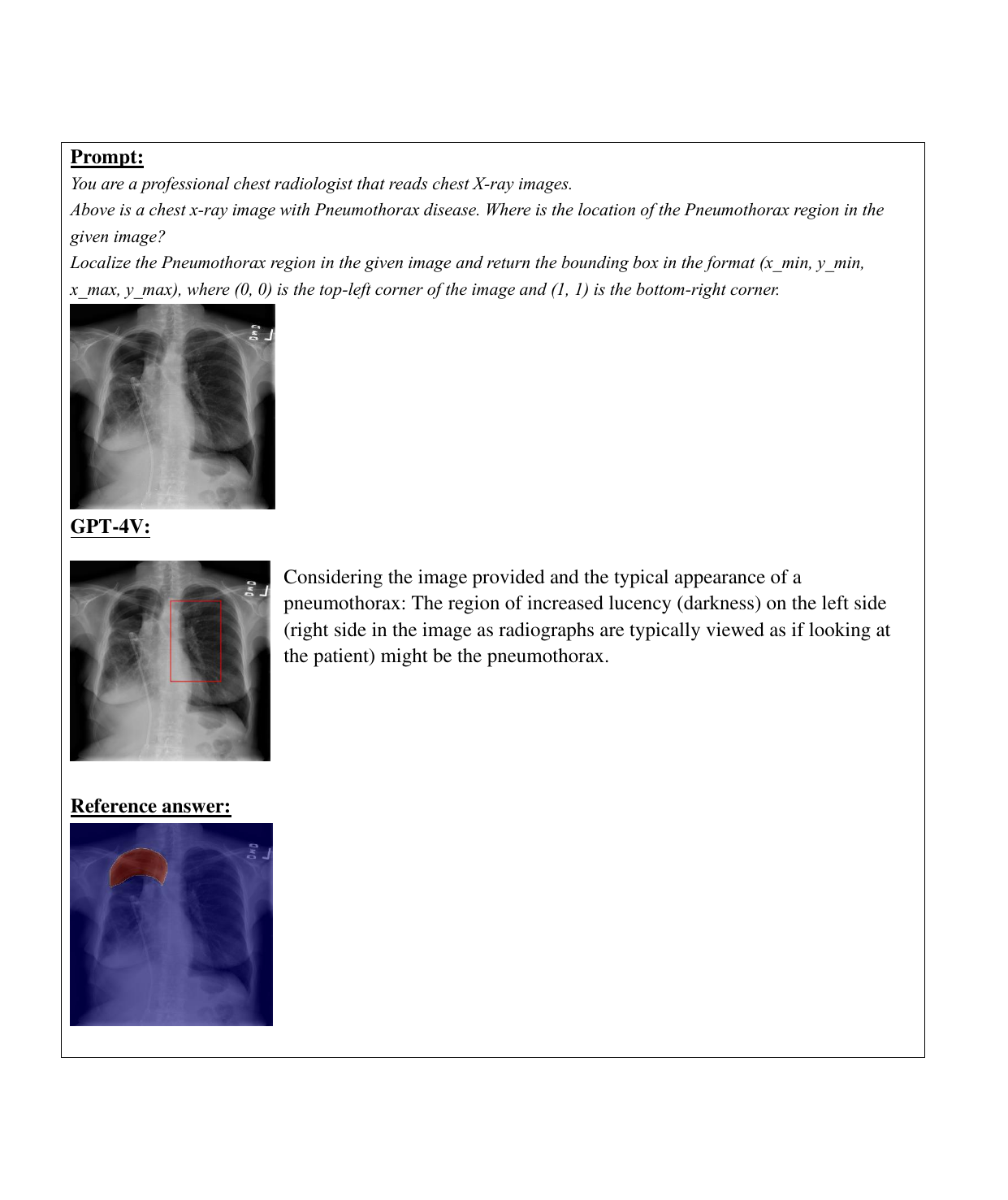}
    \caption[Chest Radiography: Case 23]{\textbf{Chest: Case 23.} A Demonstration case of Detection task on SIIM-ACR dataset. The red box is the corresponding image region. The GPT-4V’s explanation about the image is located on the right side of the image. In addition, we provide a ground-truth mask image of the lesion area for comparison.}
    \label{fig:chest_22}
\end{figure}
\newpage

\begin{figure}[H]
    \centering
    \includegraphics[width = \textwidth]{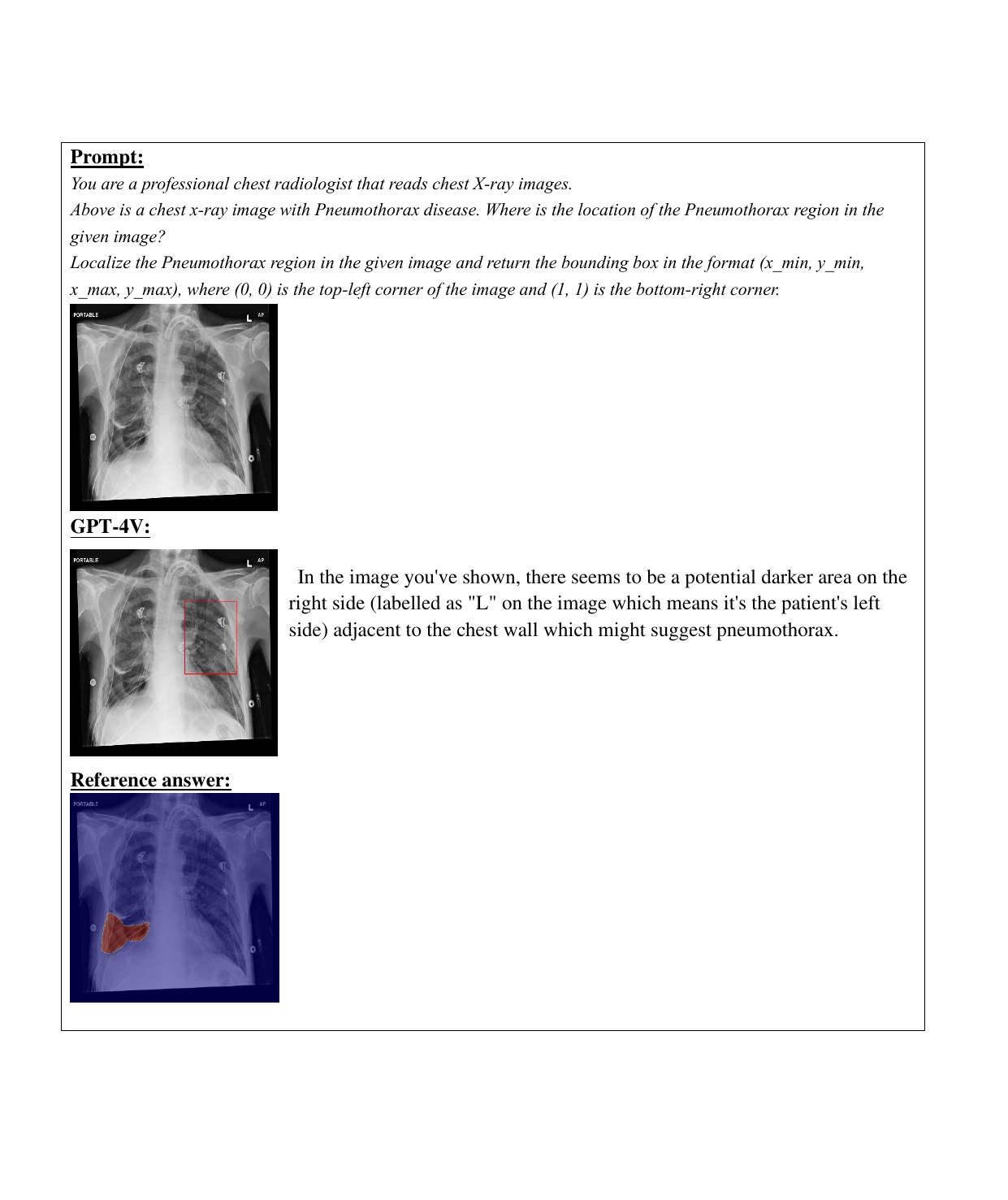}
    \caption[Chest Radiography: Case 24]{\textbf{Chest: Case 24.} A Demonstration case of Detection task on SIIM-ACR dataset. The red box is the corresponding image region. The GPT-4V’s explanation about the image is located on the right side of the image. In addition, we provide a ground-truth mask image of the lesion area for comparison.}
    \label{fig:chest_23}
\end{figure}
\newpage

\begin{figure}[H]
    \centering
    \includegraphics[width = \textwidth]{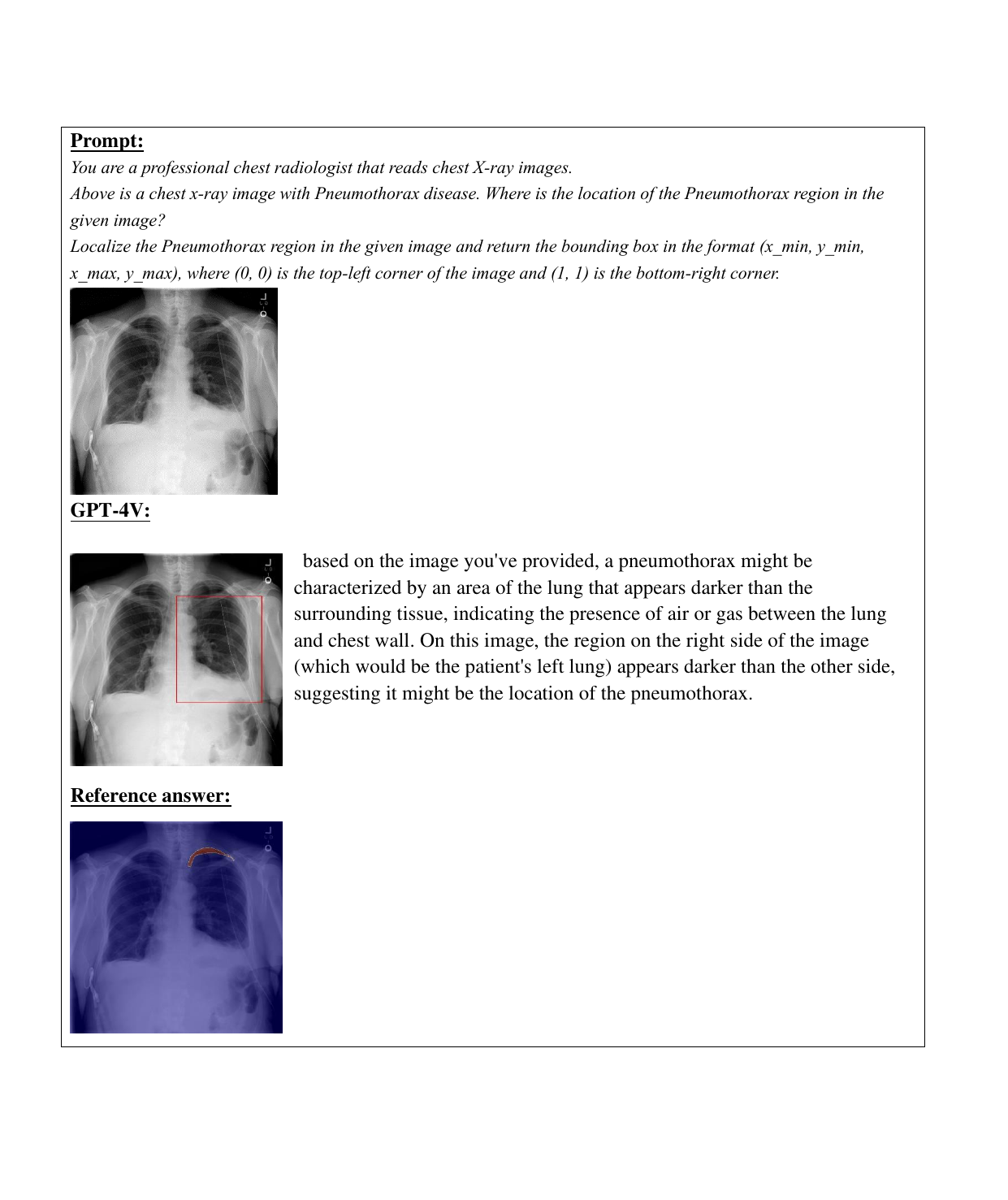}
    \caption[Chest Radiography: Case 25]{\textbf{Chest: Case 25.} A Demonstration case of Detection task on SIIM-ACR dataset. The red box is the corresponding image region. The GPT-4V’s explanation about the image is located on the right side of the image. In addition, we provide a ground-truth mask image of the lesion area for comparison.}
    \label{fig:chest_24}
\end{figure}
\newpage

\begin{figure}[H]
    \centering
    \includegraphics[width = \textwidth]{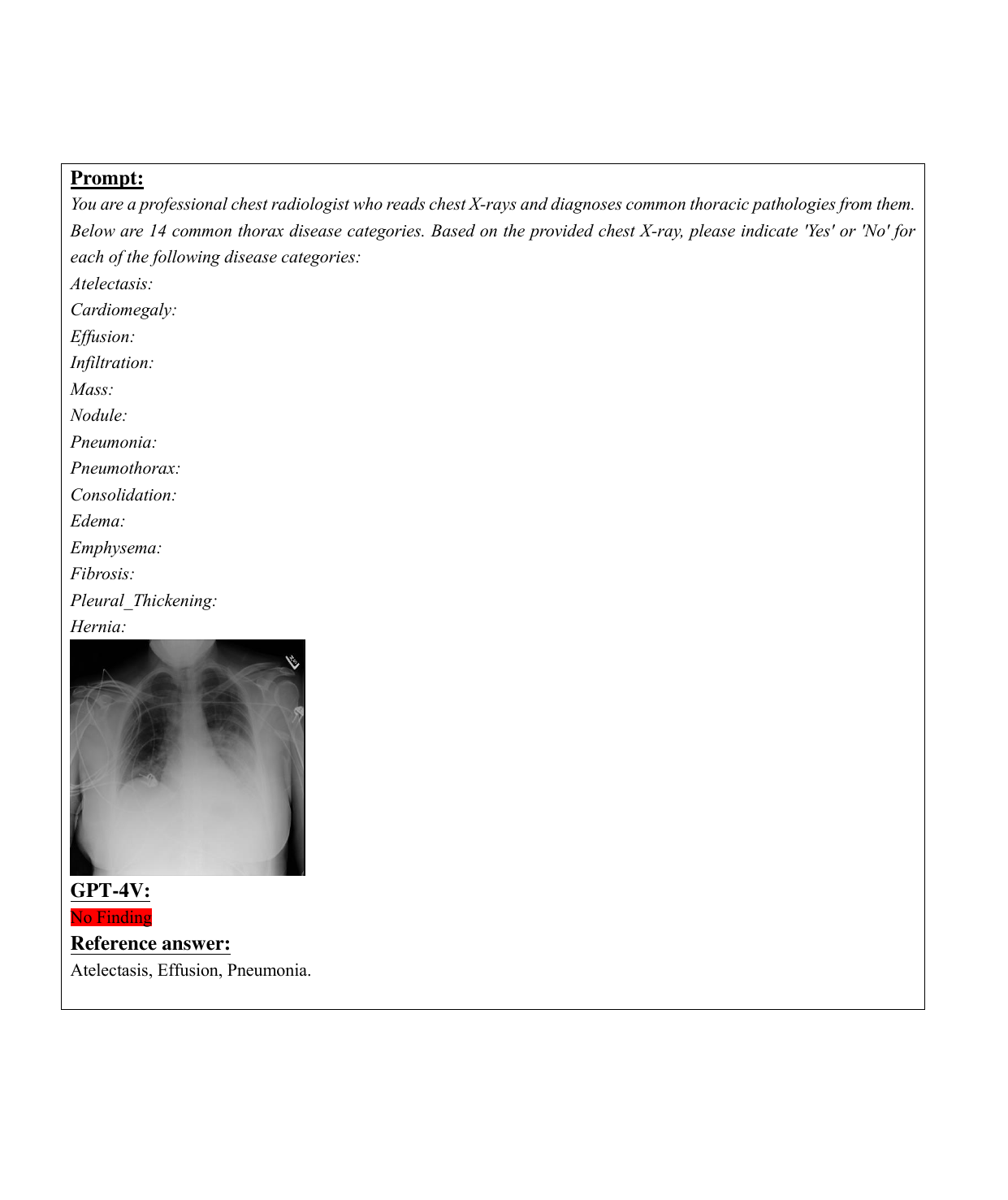}
    \caption[Chest Radiography: Case 26]{\textbf{Chest: Case 26.} Demonstration cases of Classification task on NIH dataset. Green denotes the correct classification. Red in the figure denotes the incorrect classification.}
    \label{fig:chest_25}
\end{figure}
\newpage

\begin{figure}[H]
    \centering
    \includegraphics[width = \textwidth]{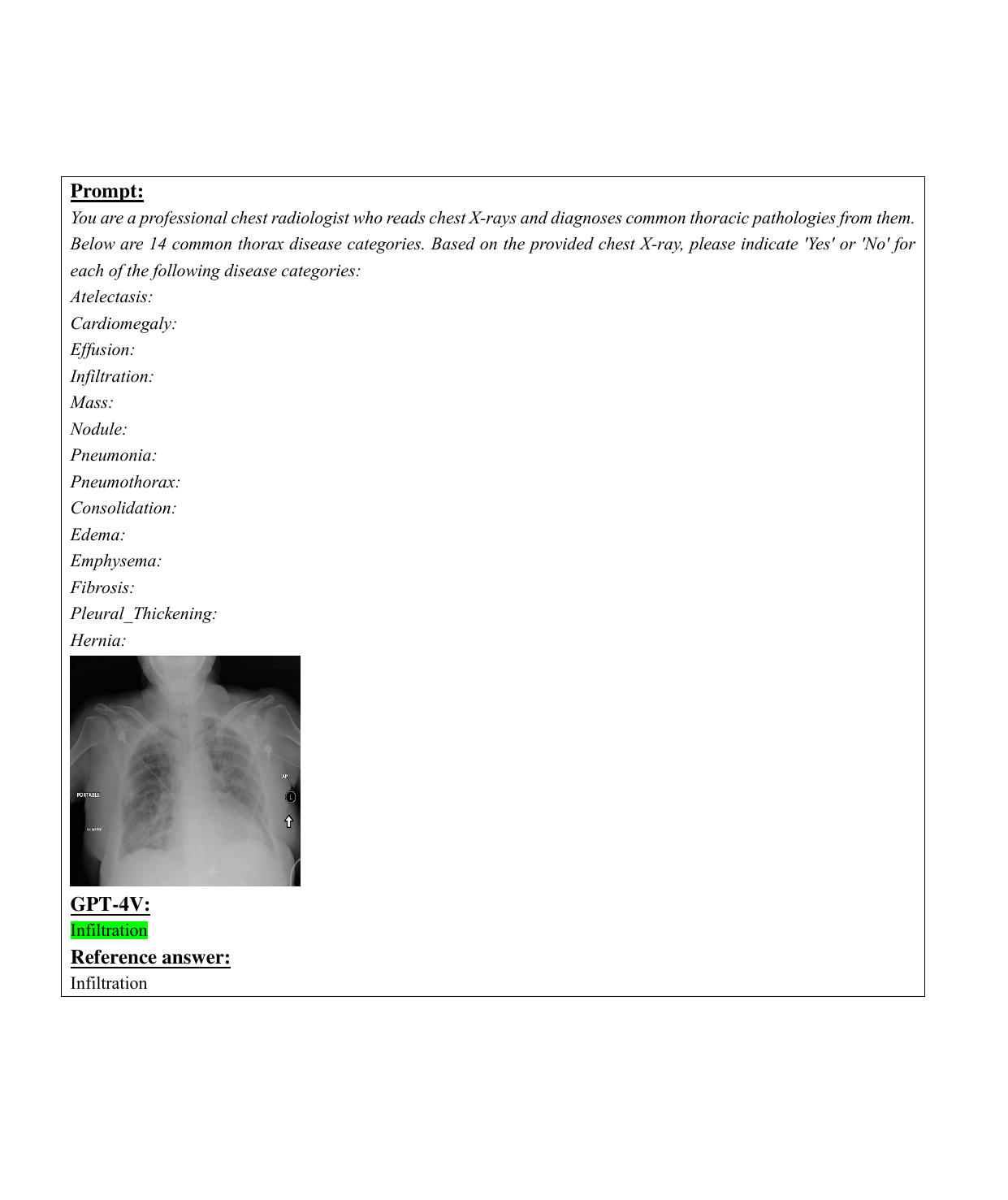}
    \caption[Chest Radiography: Case 27]{\textbf{Chest: Case 27.} Demonstration cases of Classification task on NIH dataset. Green denotes the correct classification. Red in the figure denotes the incorrect classification.}
    \label{fig:chest_26}
\end{figure}
\newpage

\begin{figure}[H]
    \centering
    \includegraphics[width = \textwidth]{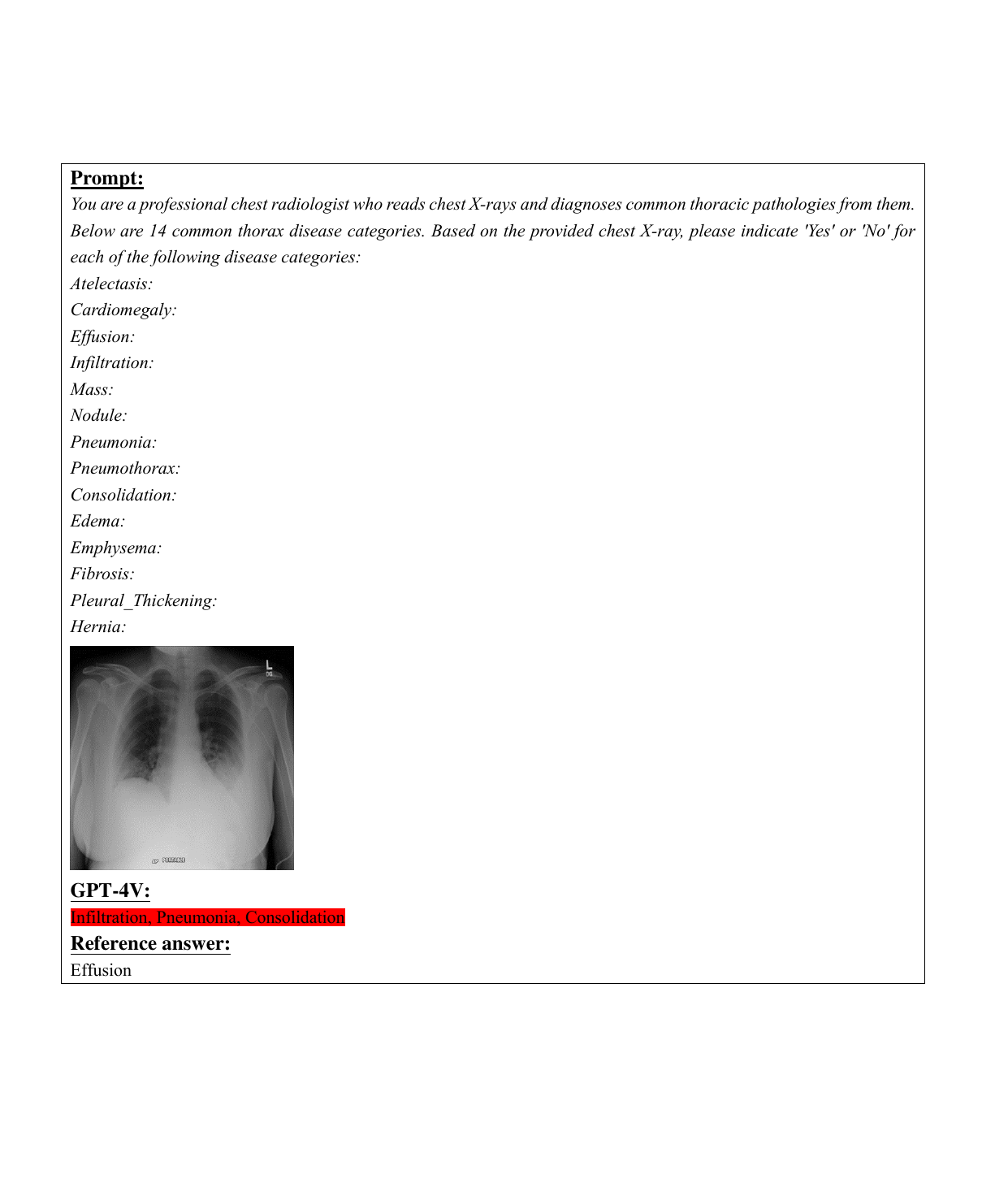}
    \caption[Chest Radiography: Case 28]{\textbf{Chest: Case 28.} Demonstration cases of Classification task on NIH dataset. Green denotes the correct classification. Red in the figure denotes the incorrect classification.}
    \label{fig:chest_27}
\end{figure}
\newpage

\begin{figure}[H]
    \centering
    \includegraphics[width = \textwidth]{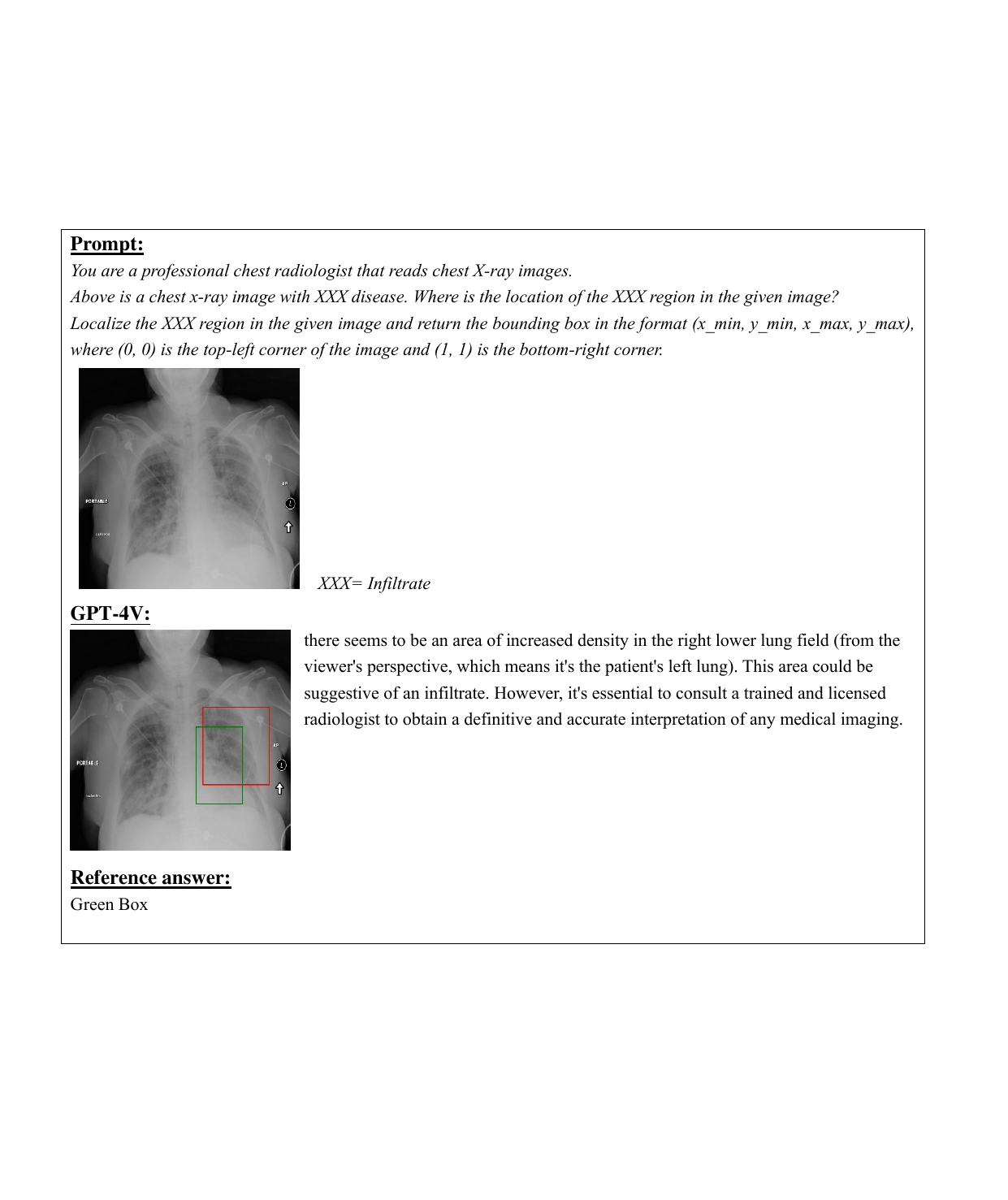}
    \caption[Chest Radiography: Case 29]{\textbf{Chest: Case 29.} A demonstration case of Detection task on NIH dataset. The red box is the corresponding diagnosis region. The GPT-4V’s explanation about the image is located on the right side of the image. In addition, we provide a ground-truth box (Green Box) of the Infiltrate area for comparison.}
    \label{fig:chest_29}
\end{figure}
\newpage

\begin{figure}[H]
    \centering
    \includegraphics[width = \textwidth]{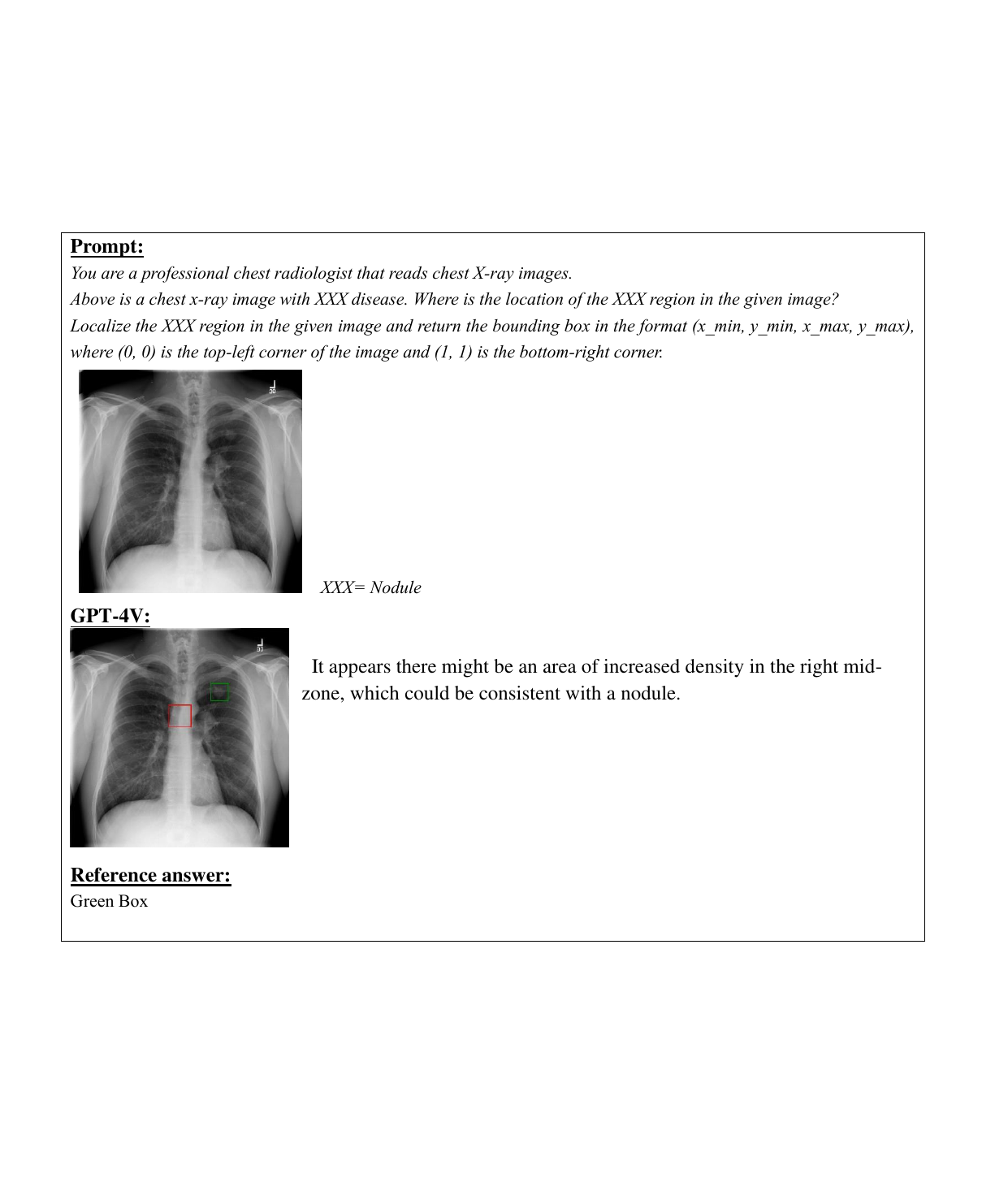}
    \caption[Chest Radiography: Case 30]{\textbf{Chest: Case 30.} A demonstration case of Detection task on NIH dataset. The red box is the corresponding diagnosis region. The GPT-4V’s explanation about the image is located on the right side of the image. In addition, we provide a ground-truth box (Green Box) of the Nodule area for comparison.}
    \label{fig:chest_30}
\end{figure}
\newpage

\subsection{Neuroimaging}
\begin{figure}[H]
    \centering
    \includegraphics[width = \textwidth]{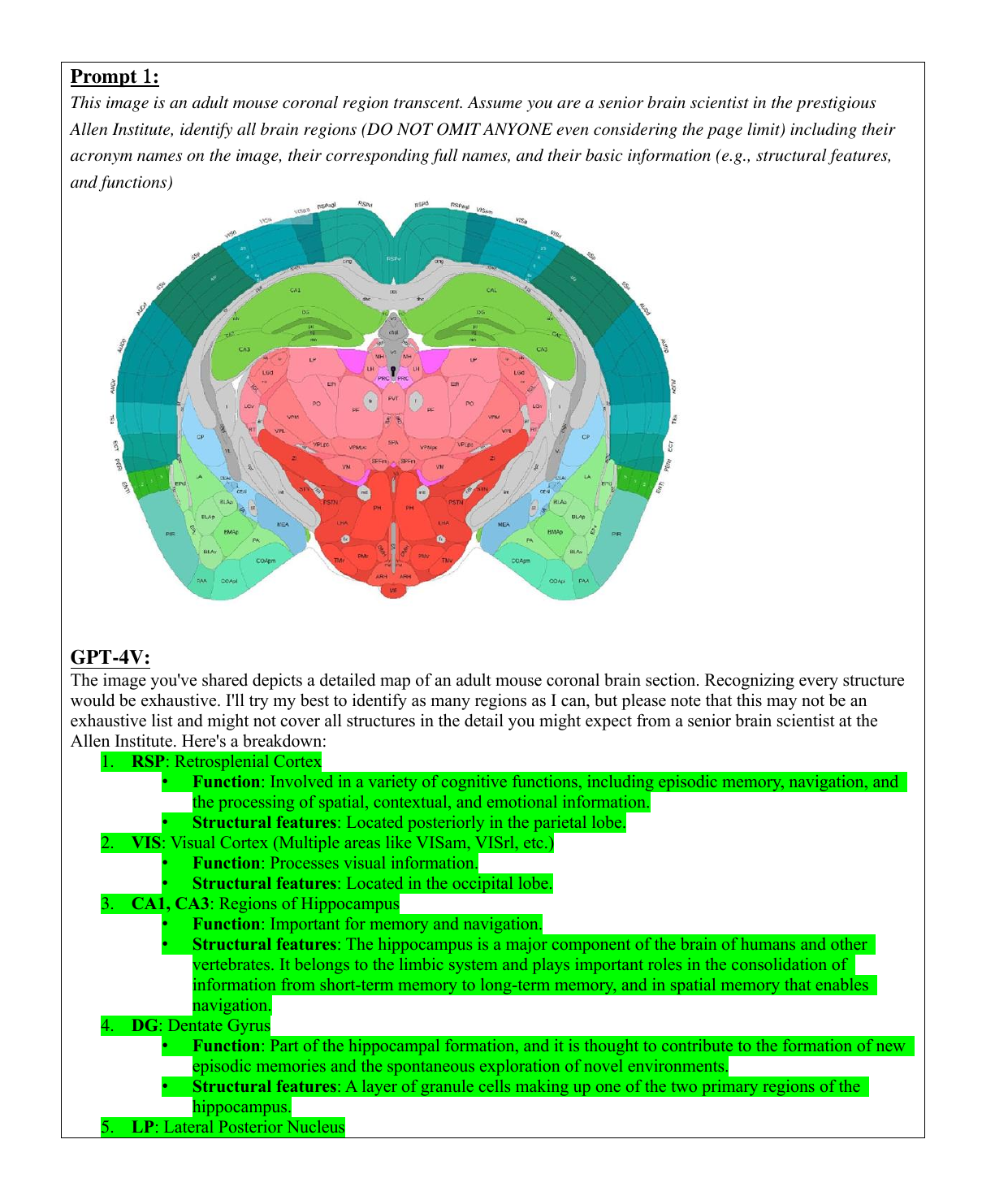}
    \label{fig:neuroimaging_1_1}
\end{figure}
\newpage

\begin{figure}[H]
    \centering
    \includegraphics[width = \textwidth]{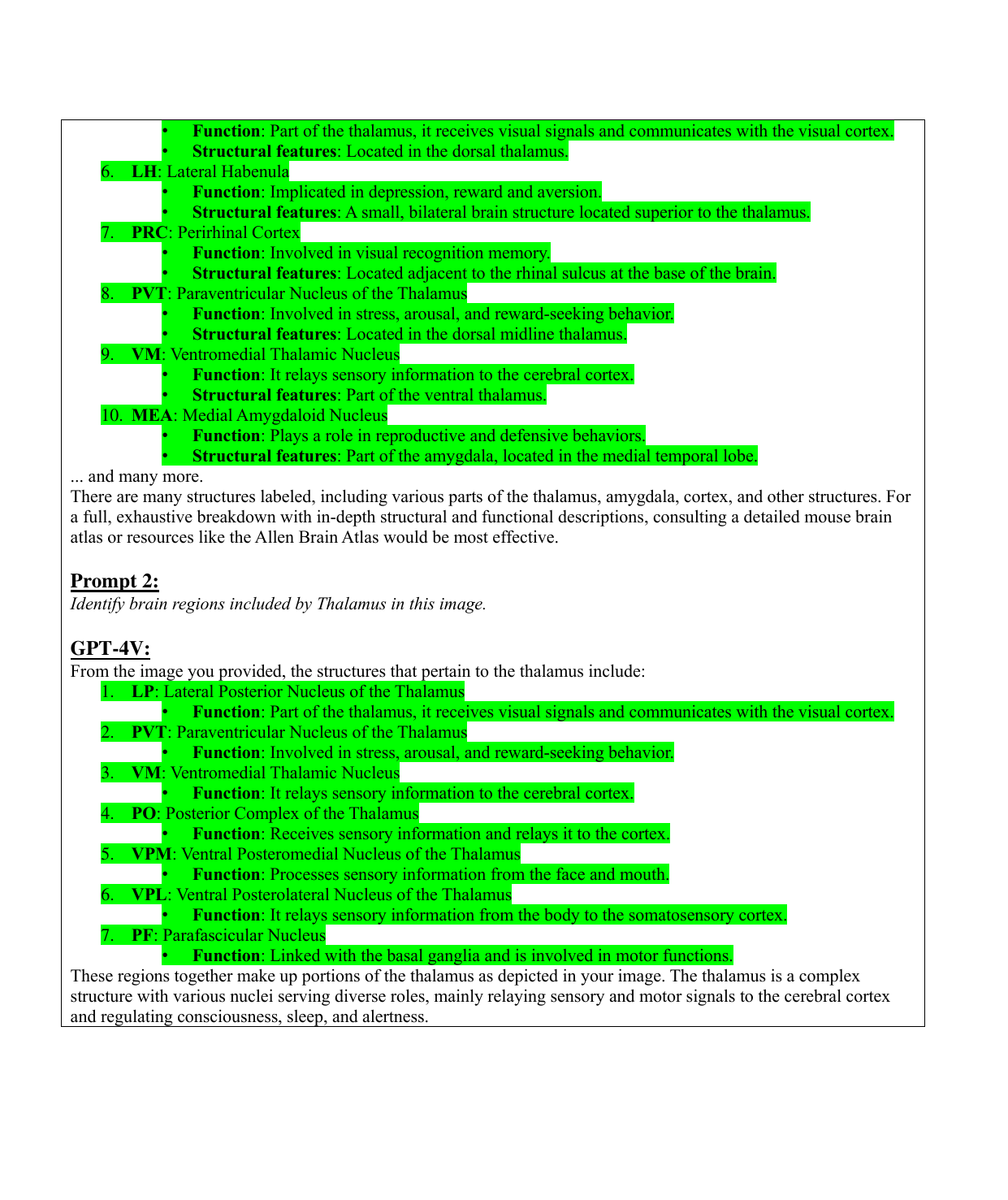}
    \caption[Neuroimaging: Case 6]{\textbf{Brain Atlas: Case 2.} A demonstration case from neuroimaging testing of an anatomical reference atlas from an adult mouse brain in the coronal plane of a section. The full green region labeled in the figure refers to the comprehensively correct information provided by GPT-4V, examined with prior brain region knowledge.}
    \label{fig:neuroimaging_1_2}
\end{figure}
\newpage

\begin{figure}[H]
    \centering
    \includegraphics[width = \textwidth]{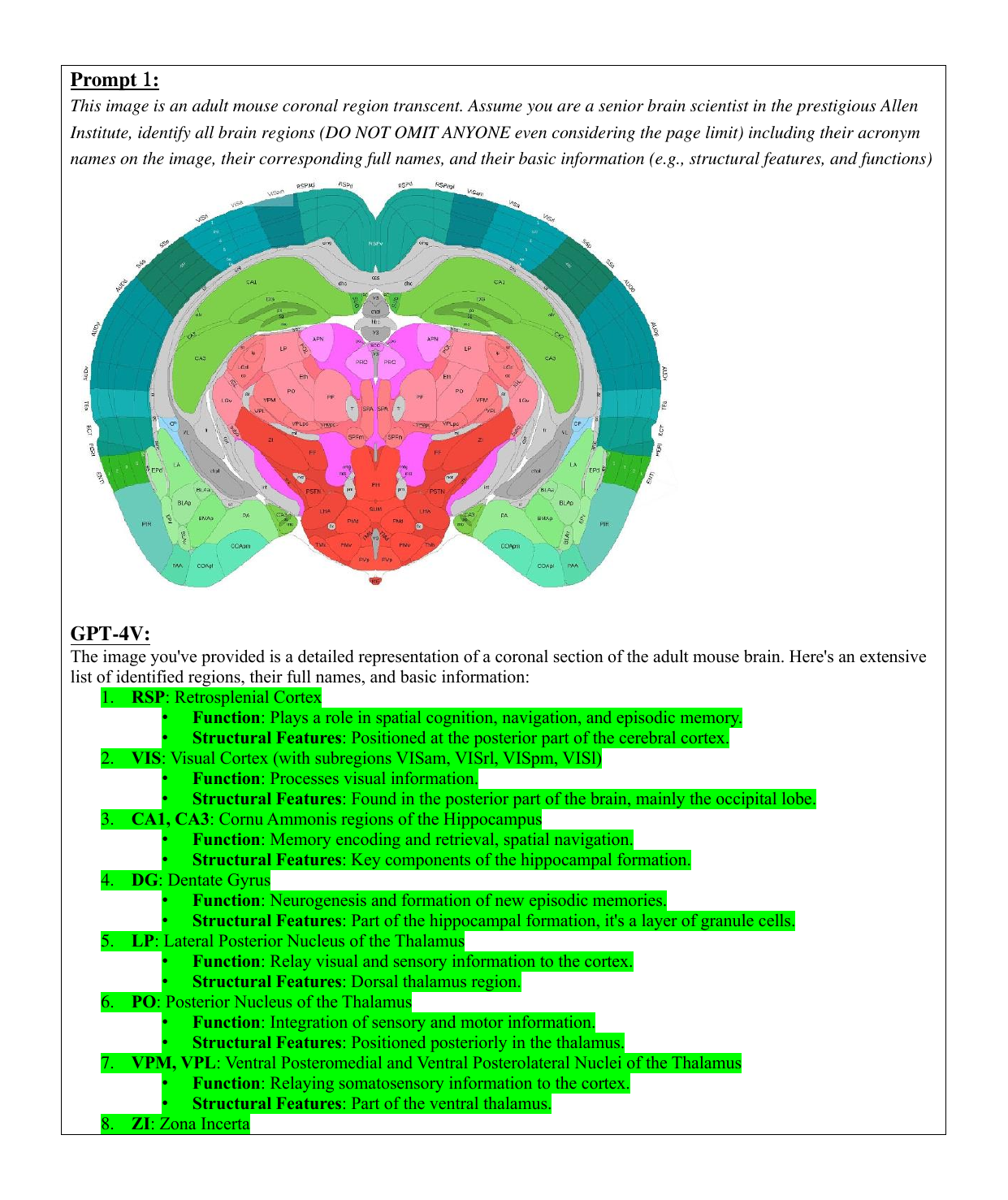}
    \label{fig:neuroimaging_2_1}
\end{figure}
\newpage

\begin{figure}[H]
    \centering
    \includegraphics[width = \textwidth]{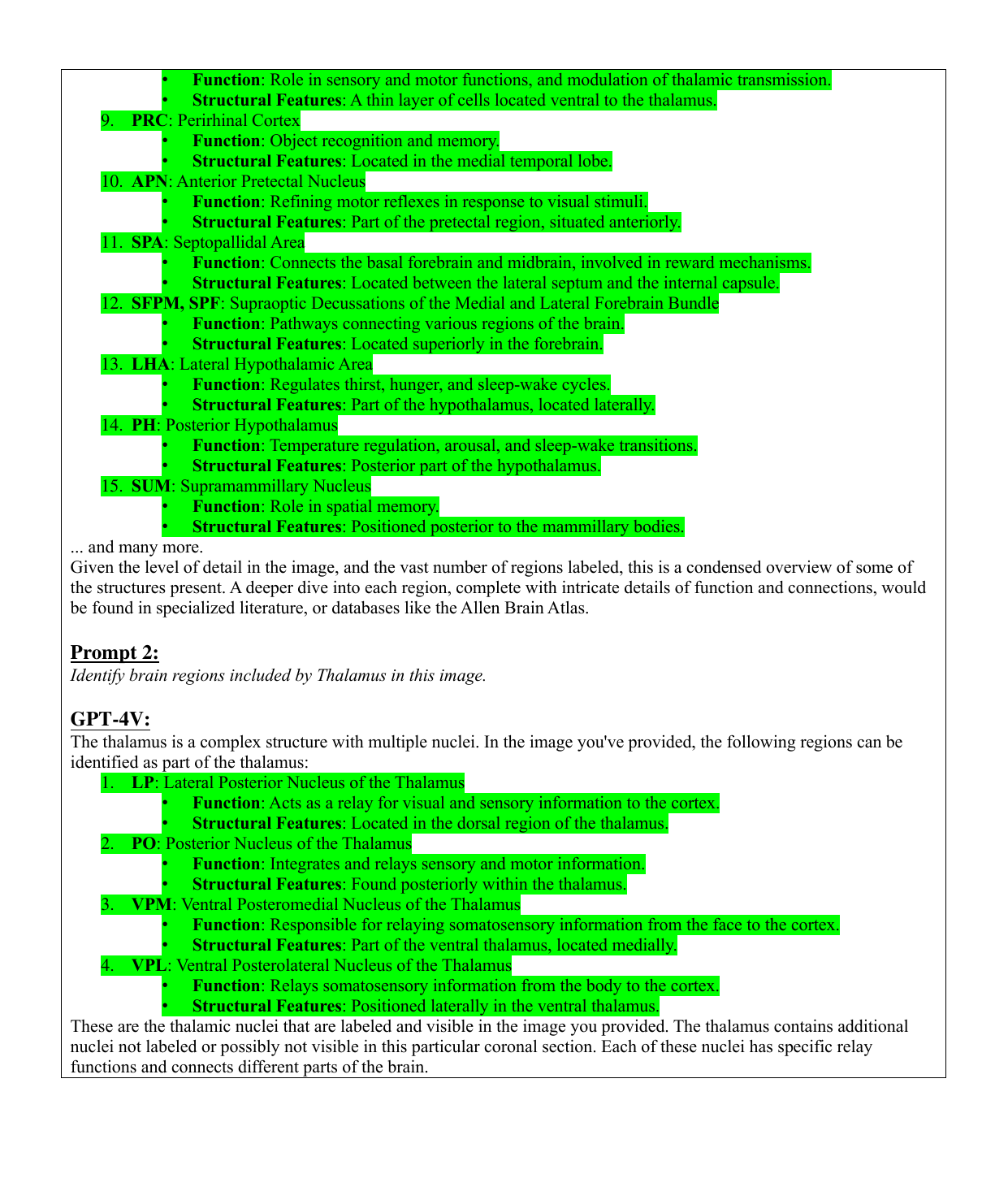}
    \caption[Neuroimaging: Case 7]{\textbf{Brain Atlas: Case 3.} A demonstration case from neuroimaging testing of an anatomical reference atlas from an adult mouse brain in the coronal plane of a section. The full green region labeled in the figure refers to the comprehensively correct information provided by GPT-4V, examined with prior brain region knowledge.}
    \label{fig:neuroimaging_2_2}
\end{figure}
\newpage

\begin{figure}[H]
    \centering
    \includegraphics[width = \textwidth]{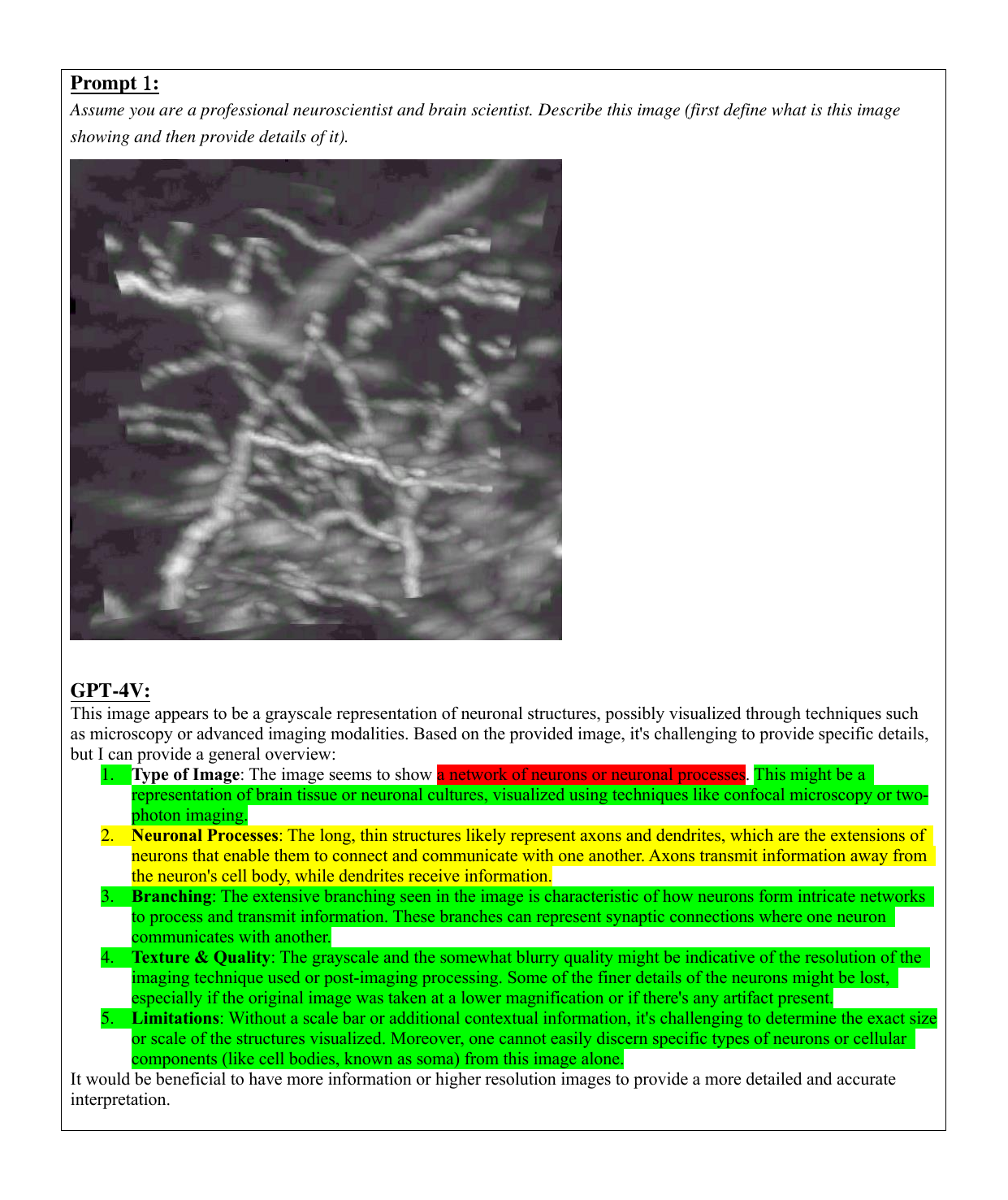}
    \label{fig:neuroimaging_4_1}
\end{figure}
\newpage

\begin{figure}[H]
    \centering
    \includegraphics[width = \textwidth]{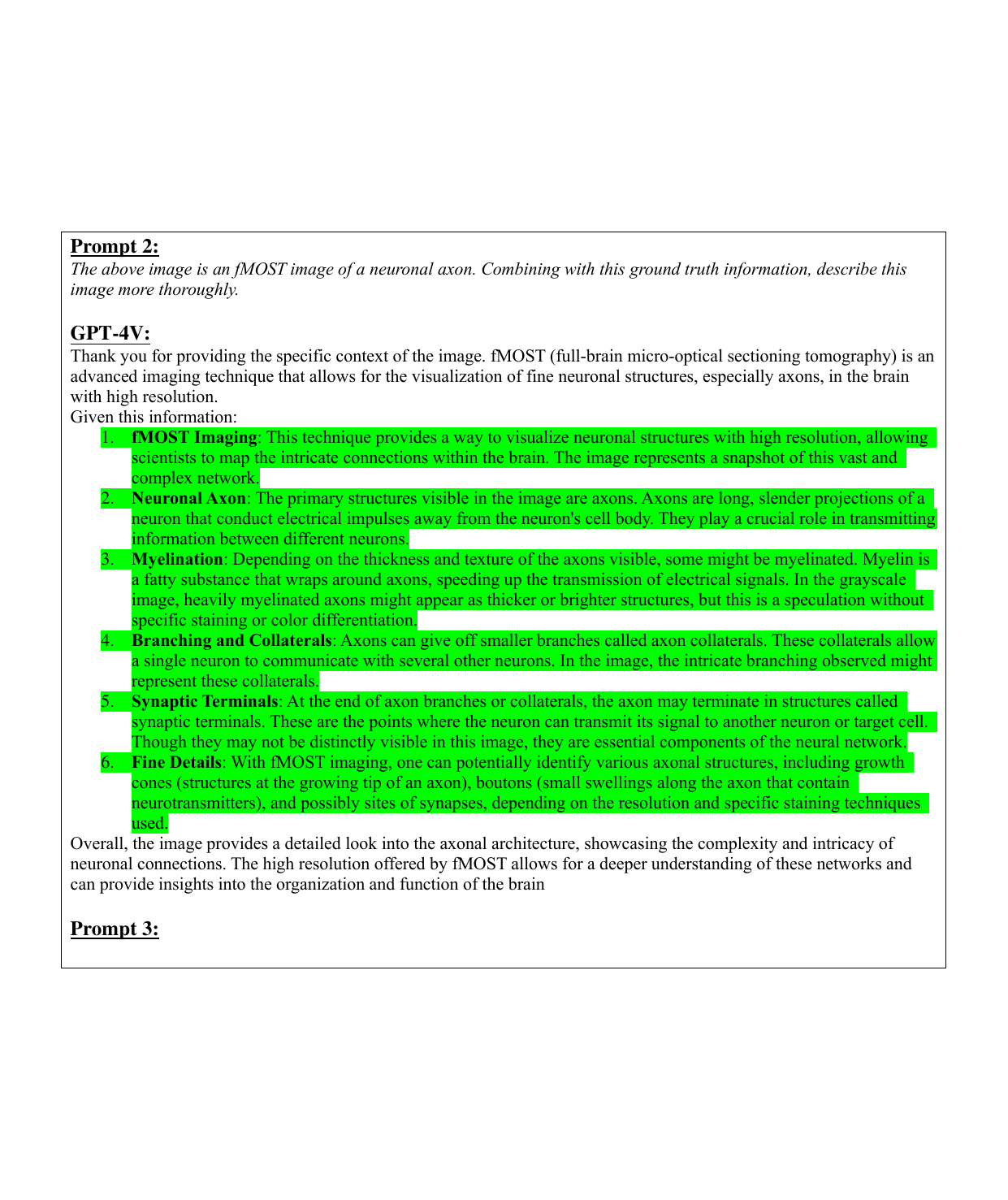}
    \label{fig:neuroimaging_4_2}
\end{figure}
\newpage

\begin{figure}[H]
    \centering
    \includegraphics[width = \textwidth]{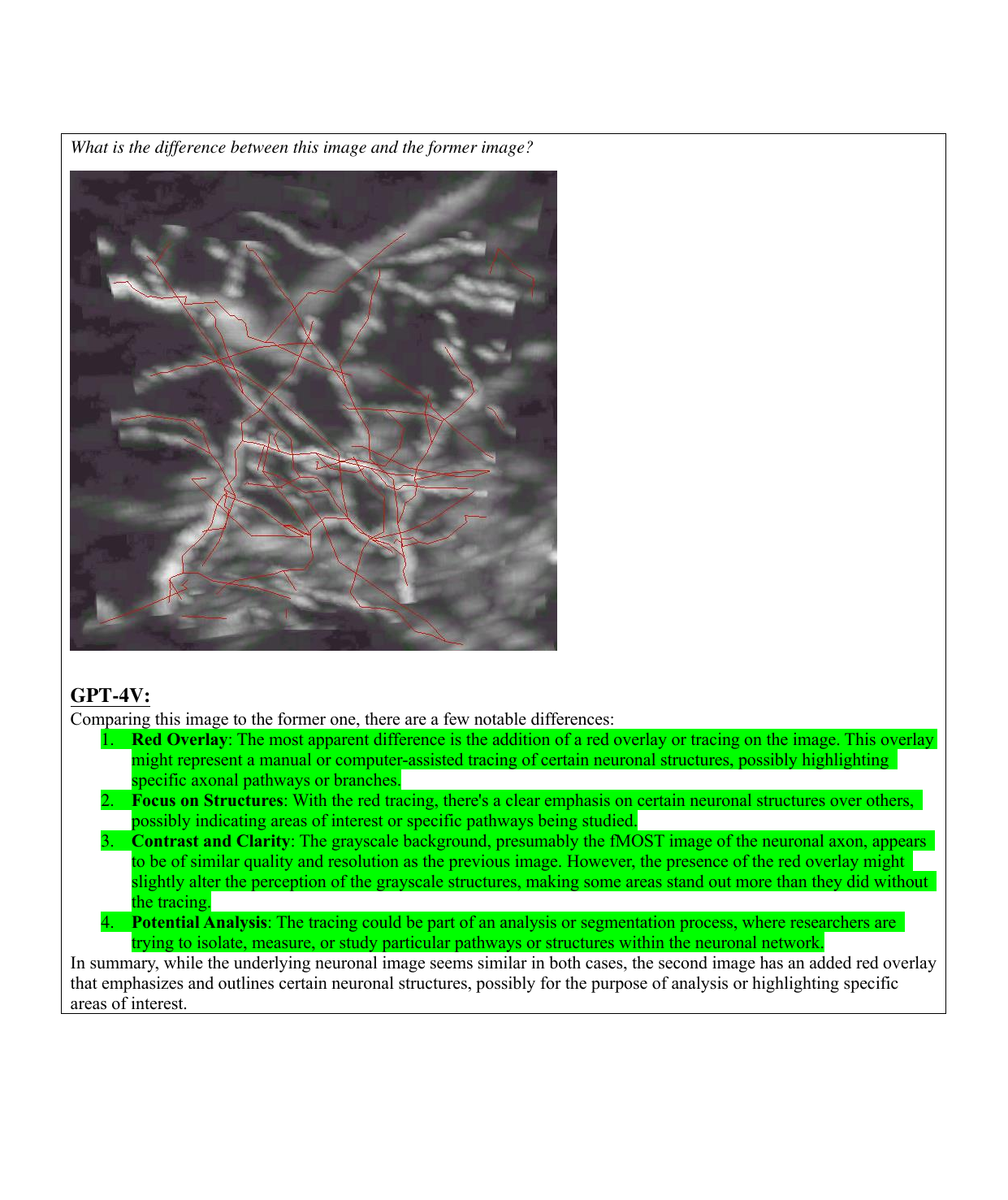}
    \caption[Neuroimaging: Case 8]{\textbf{Neuron Reconstruction: Case 2.} A demonstration case from neuroimaging testing of a fMOST image of a neuronal axon. The green region labeled in the figure refers to the correct information provided by GPT-4V. Similarly, the yellow and red region represents the uncertain and wrong response, respectively, examined with prior neuronal imaging knowledge.}
    \label{fig:neuroimaging_4_3}
\end{figure}
\begin{figure}[H]
    \centering
    \includegraphics[width = \textwidth]{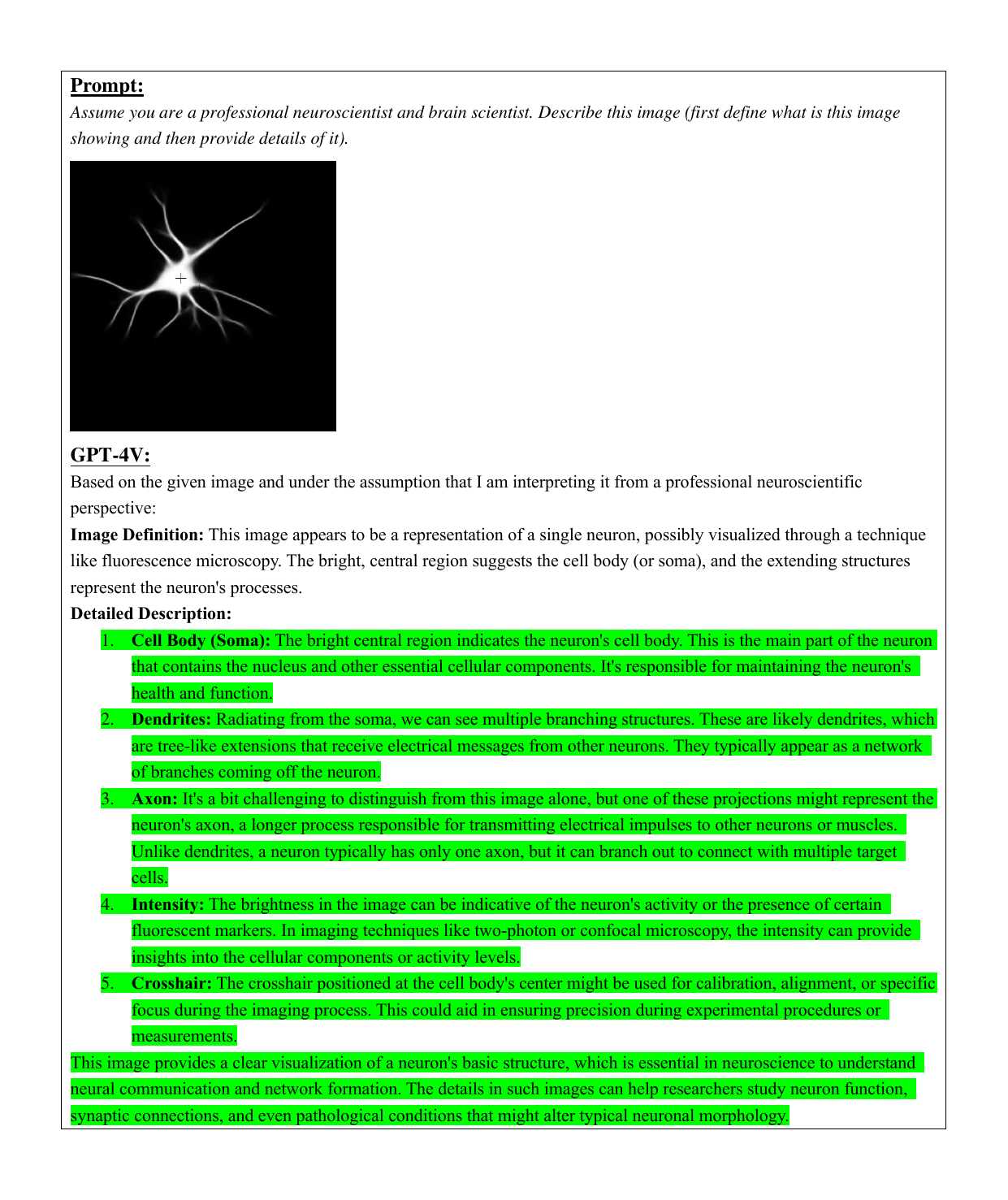}
    \caption[Neuroimaging: Case 9]{\textbf{Neuron Soma: Case 1.} A demonstration case from neuroimaging testing of a single neuron image visualized by fluorescence microscopy. The full green region labeled in the figure refers to the comprehensively correct information provided by GPT-4V, examined with prior neuronal imaging knowledge.}
    \label{fig:neuroimaging_6}
\end{figure}
\newpage

\begin{figure}[H]
    \centering
    \includegraphics[width = \textwidth]{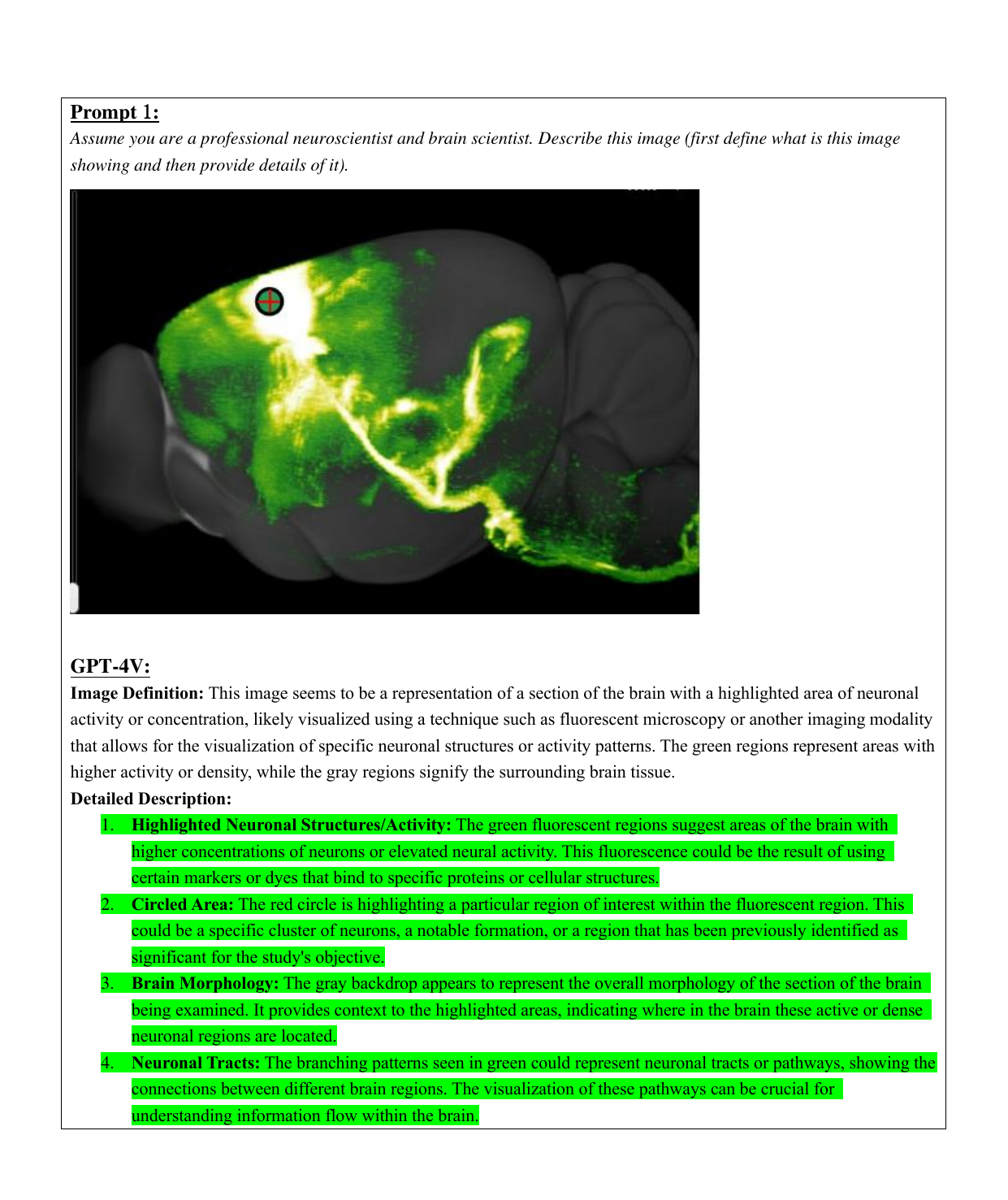}
    \label{fig:neuroimaging_7_1}
\end{figure}
\newpage

\begin{figure}[H]
    \centering
    \includegraphics[width = \textwidth]{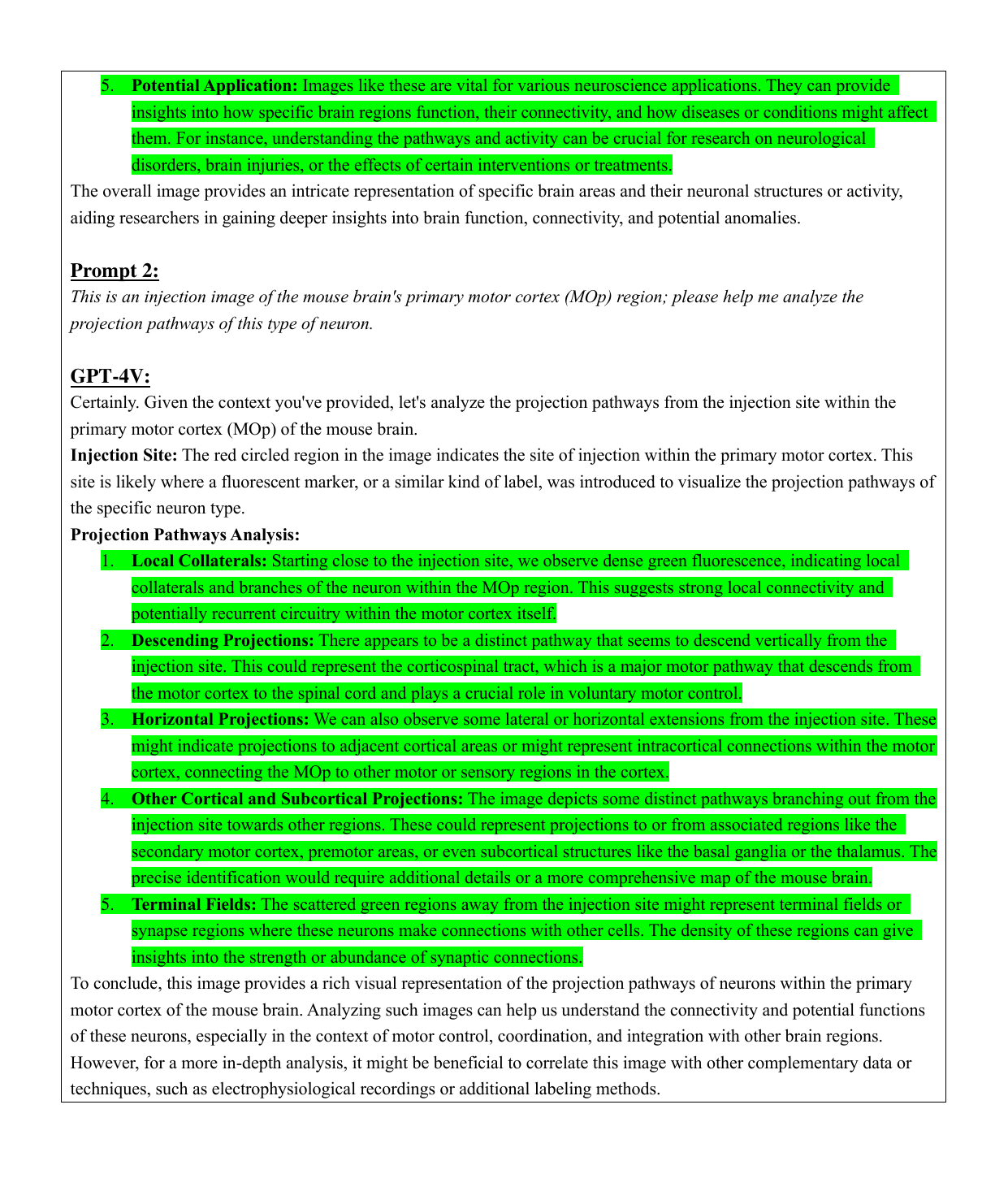}
    \caption[Neuroimaging: Case 10]{\textbf{Injected Neuron: Case 1.} A demonstration case from neuroimaging testing of an injection image of the mouse brain's MOp region. The full green region labeled in the figure refers to the comprehensively correct information provided by GPT-4V, examined with prior neuronal reconstruction knowledge.}
    \label{fig:neuroimaging_7_2}
\end{figure}
\newpage

\begin{figure}[H]
    \centering
    \includegraphics[width = \textwidth]{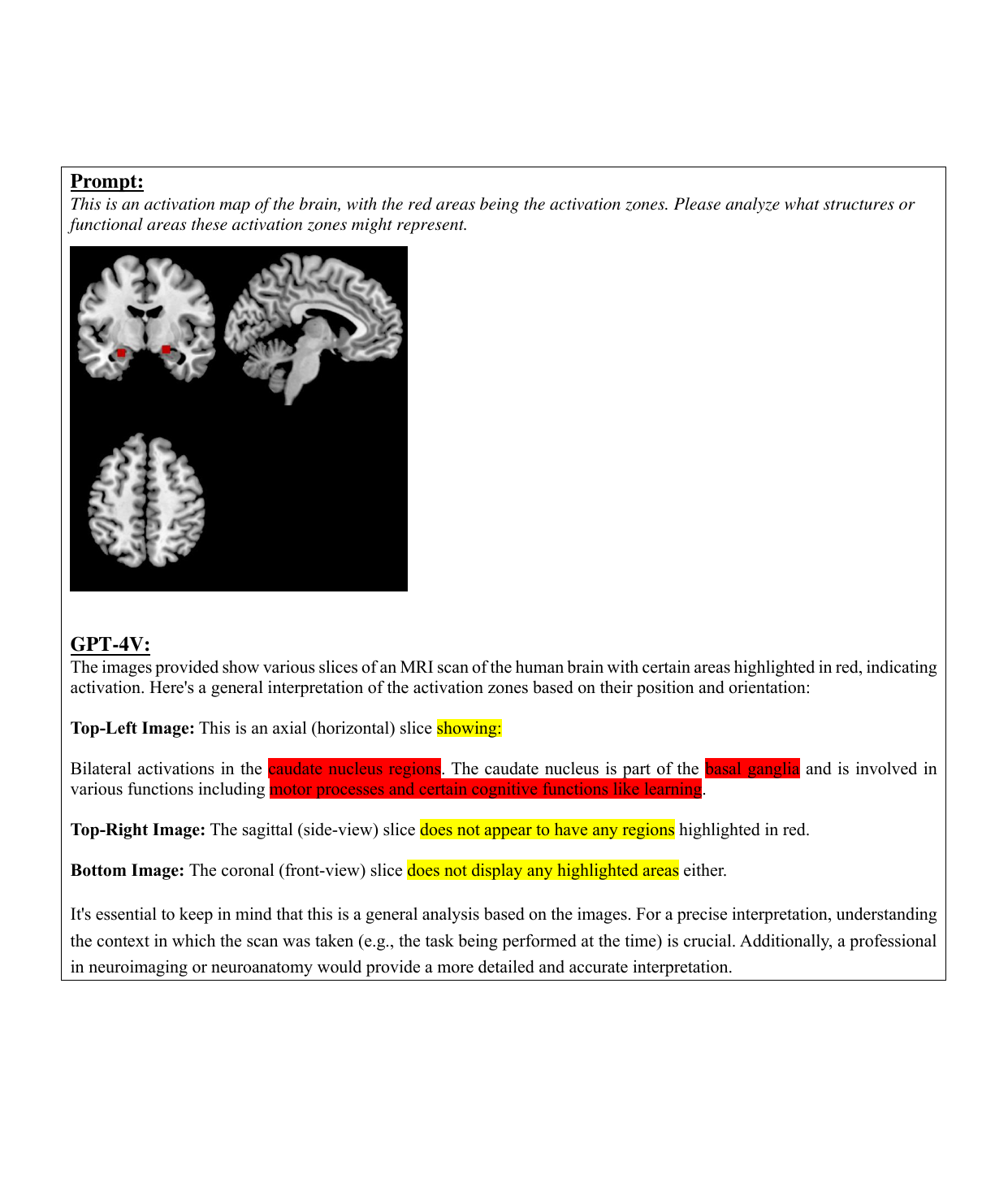}
    \caption[Neuroimaging: Case 11]{\textbf{Brain: Case 3.} A demonstration case of Location task on NeuroQuery dataset. Red denotes the result from GPT-4V. Red in the figure denotes the zone of the brain activation.}
    \label{fig:brain_activation_3}
\end{figure}
\newpage

\begin{figure}[H]
    \centering
    \includegraphics[width = \textwidth]{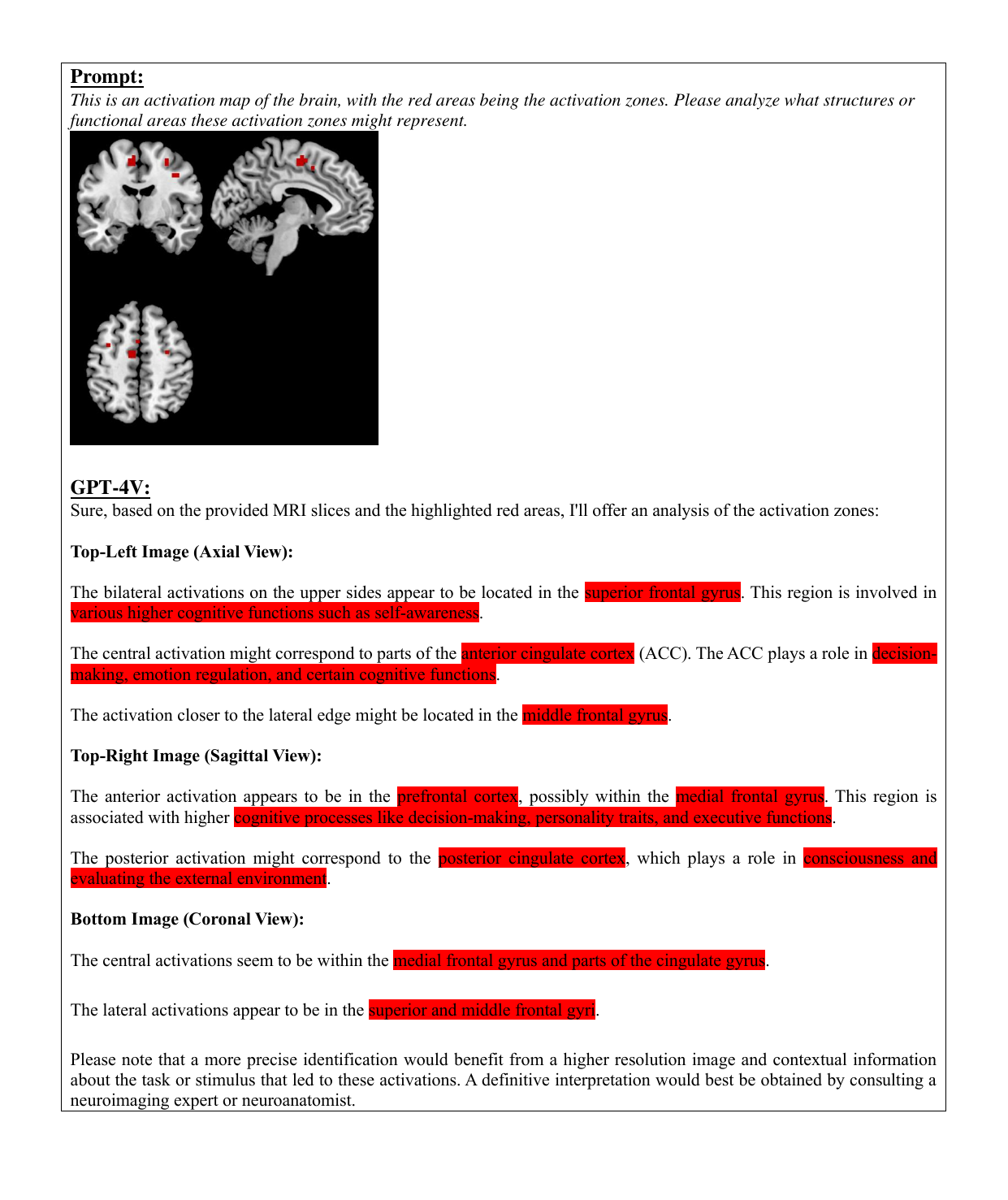}
    \caption[Neuroimaging: Case 12]{\textbf{Brain: Case 5.} A demonstration case of Location task on NeuroQuery dataset. Red denotes the result from GPT-4V. Red in the figure denotes the zone of the brain activation.}
    \label{fig:brain_activation_5}
\end{figure}
\newpage

\subsection{Oncological Imaging for Radiotherapy}
\label{appendix:radonc}
\begin{figure}[H]
    \centering\includegraphics[width = 0.8\textwidth]{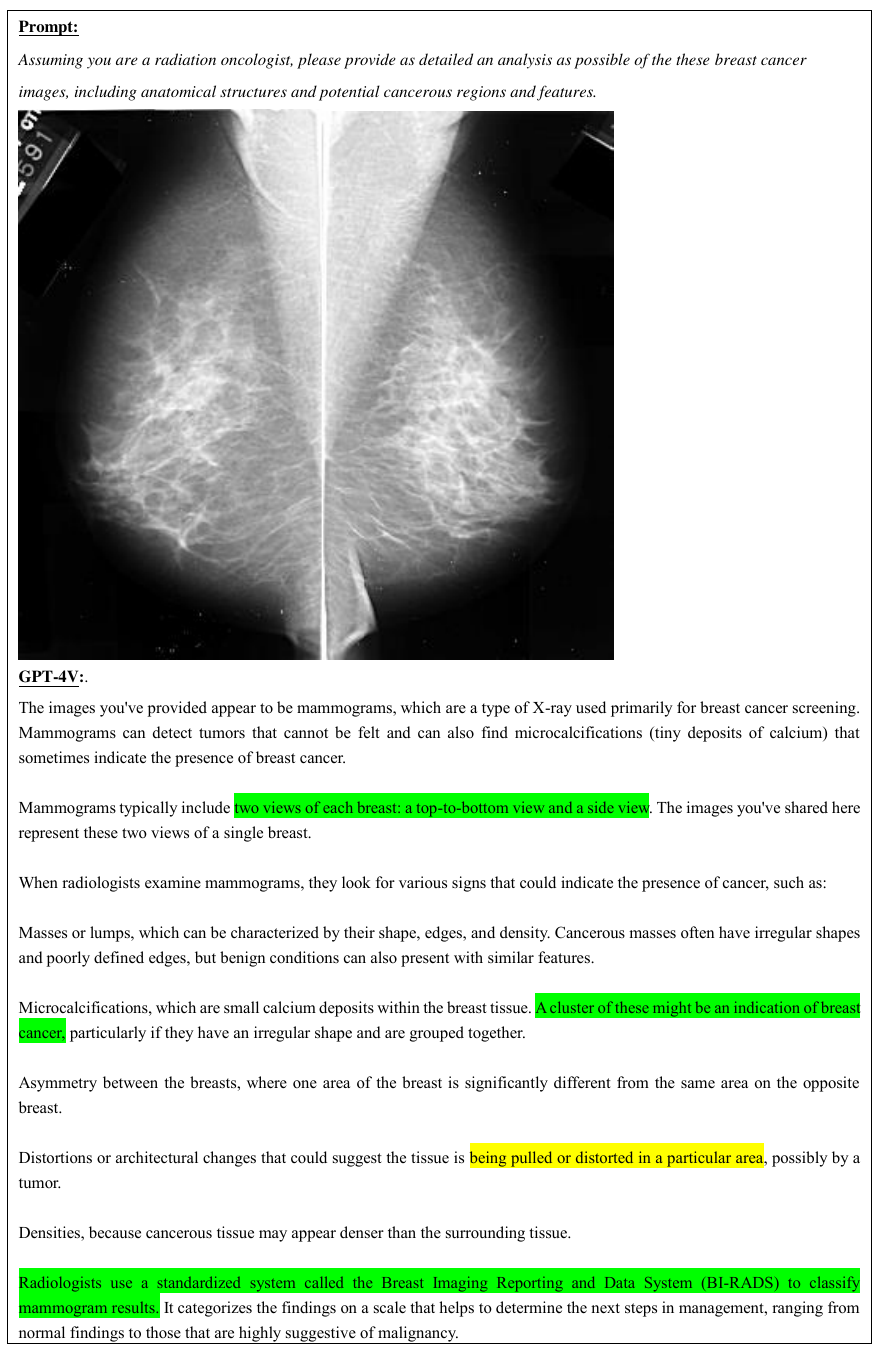}
    \caption[Oncological Imaging for Radiotherapy: Case 4]{\textbf{Breast: Case 1.} A demonstration case from breast cancer images. GPT-4V successfully recognizes the main anatomical structures. The original data is from the DDSM dataset\cite{rose2006web}.}
    \label{breast_cancer_fig1}
\end{figure}

\newpage
\begin{figure}[H]
    \includegraphics[width = \textwidth]{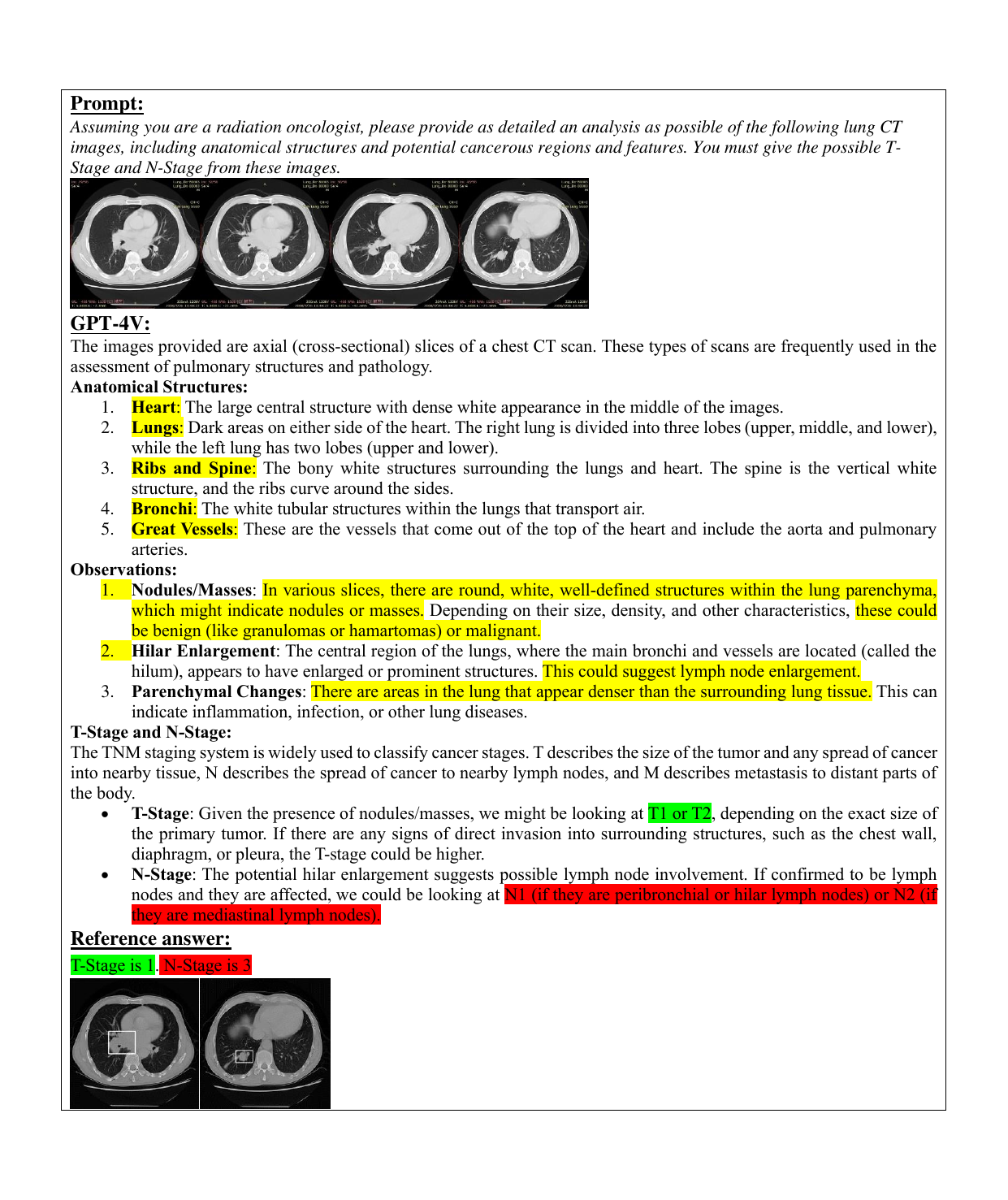}
    \caption[Oncological Imaging for Radiotherapy: Case 5]{\textbf{Lung: Case 2.} A demonstration case from lung CT images of small cell carcinoma. GPT-4V successfully recognizes the main anatomical structures, the abnormal area of the tumor with the proceeding basic descriptions. The original CT data is from the Lung-PET-CT-Dx dataset\cite{cancerimagingarchive2013}.}
    \label{lung_fig2}
\end{figure}

\newpage



\begin{figure}[H]
    \includegraphics[width = 0.95\textwidth]{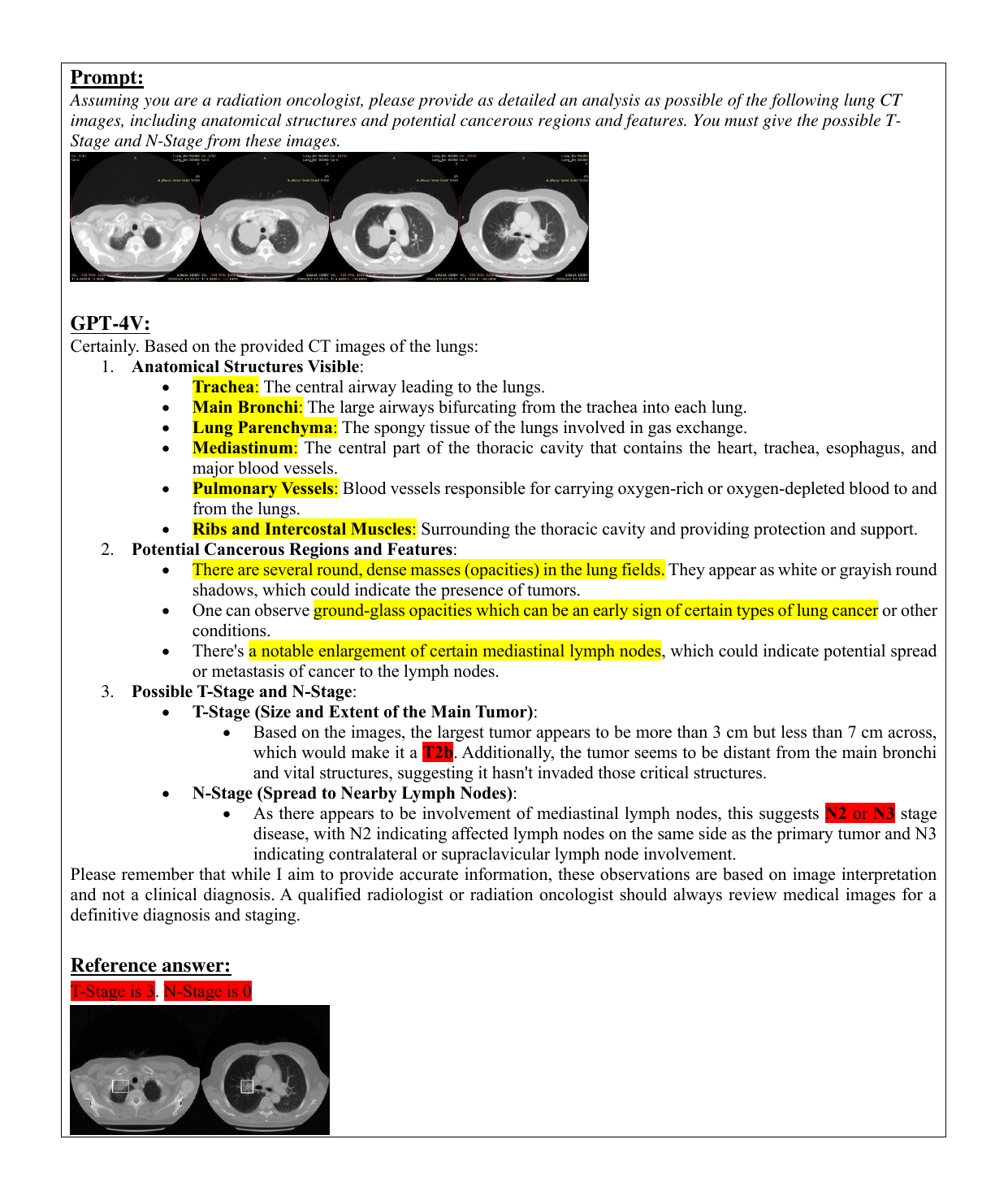}
    \caption[Oncological Imaging for Radiotherapy: Case 6]{\textbf{Lung: Case 3.} A demonstration case from lung CT images of small cell carcinoma. Green denotes that anatomical structures, abnormal area and T-stage are recognized. Red in the figure denotes the incorrect information. Yellow denotes the ambiguous answers and stage diagnosis. GPT-4V successfully recognizes the main anatomical structures, the abnormal area of the tumor with the proceeding basic descriptions. The original CT data is from the Lung-PET-CT-Dx dataset~\cite{cancerimagingarchive2013}.}
    \label{lung_fig3}
\end{figure}

\newpage
\begin{figure}[H]
    \includegraphics[width = \textwidth]{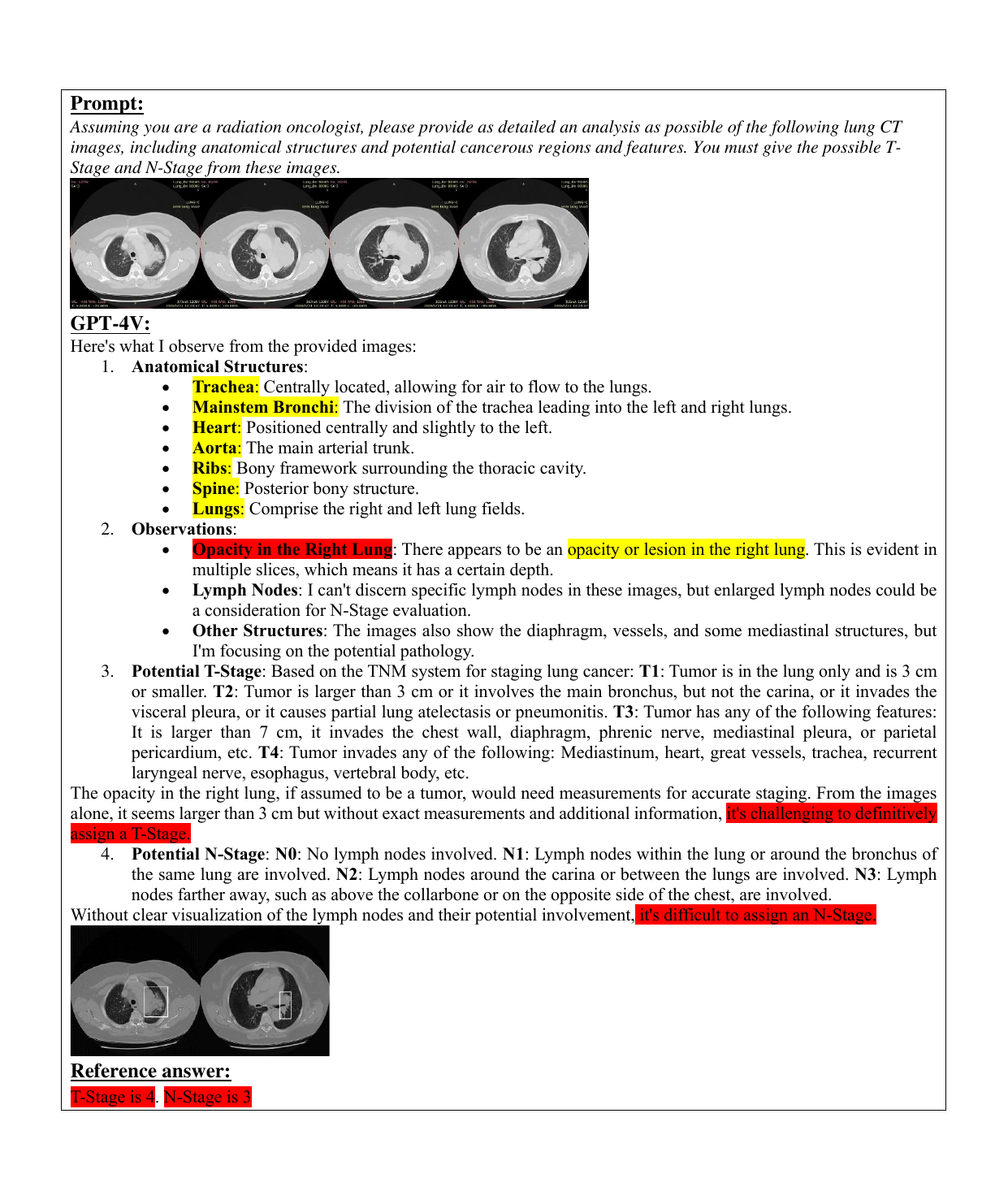}
    \caption[Oncological Imaging for Radiotherapy: Case 7]{\textbf{Lung: Case 4.} A demonstration case from lung CT images of small cell carcinoma. GPT-4V successfully recognizes the main anatomical structures, the abnormal area of the tumor with the proceeding basic descriptions. The original CT data is from the Lung-PET-CT-Dx dataset \cite{cancerimagingarchive2013}.}
    \label{lung_fig4}
\end{figure}

\newpage

\begin{figure}[H]
    \includegraphics[width = \textwidth]{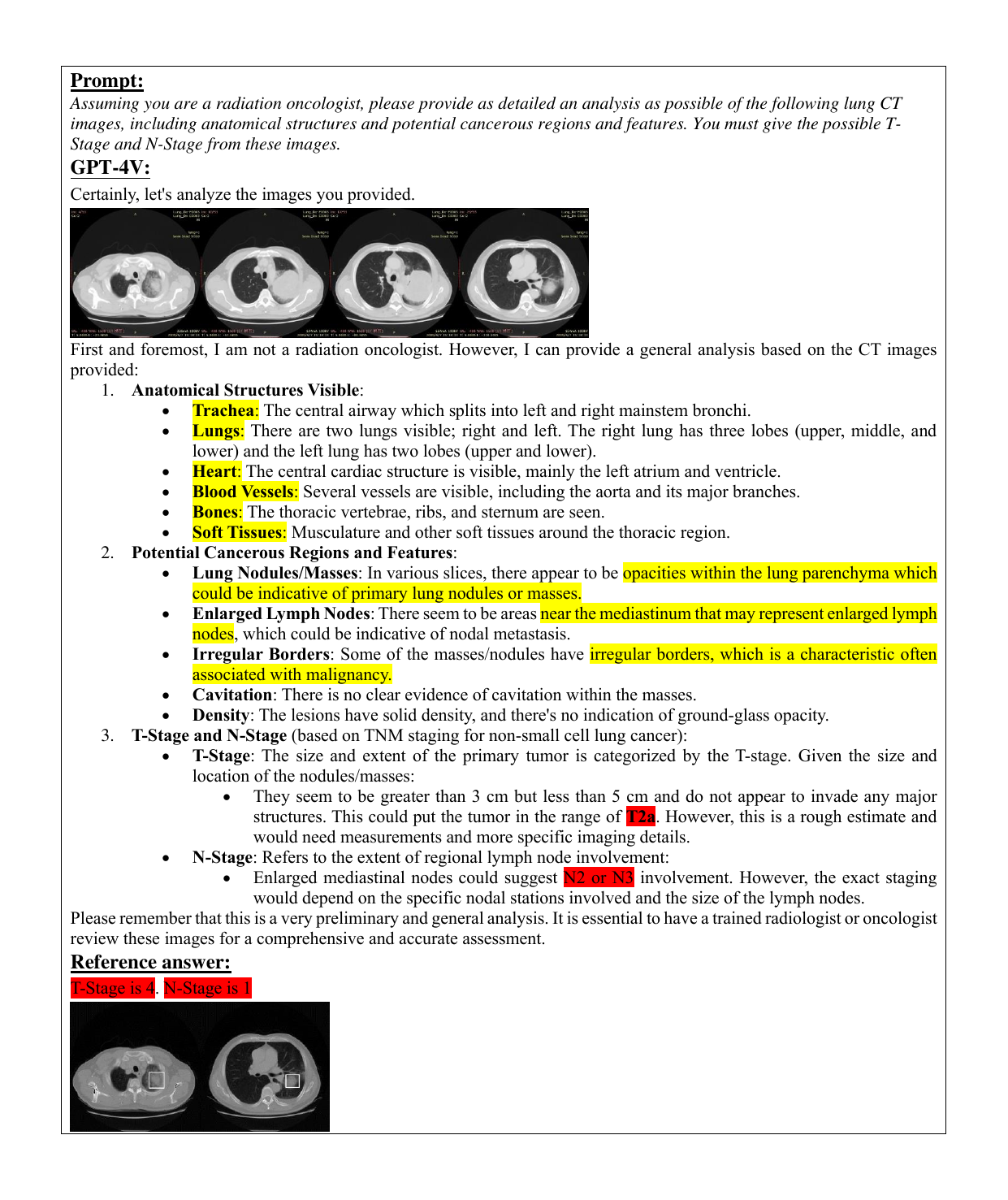}
    \caption[Oncological Imaging for Radiotherapy: Case 8]{\textbf{Lung: Case 5.} A demonstration case from lung CT images of non-small cell carcinoma. GPT-4V successfully recognizes the main anatomical structures, the abnormal area of the tumor with the proceeding basic descriptions. The original CT data is from the Lung-PET-CT-Dx dataset \cite{cancerimagingarchive2013}.}
    \label{lung_fig5}
\end{figure}

\newpage

\begin{figure}[H]
    \includegraphics[width = \textwidth]{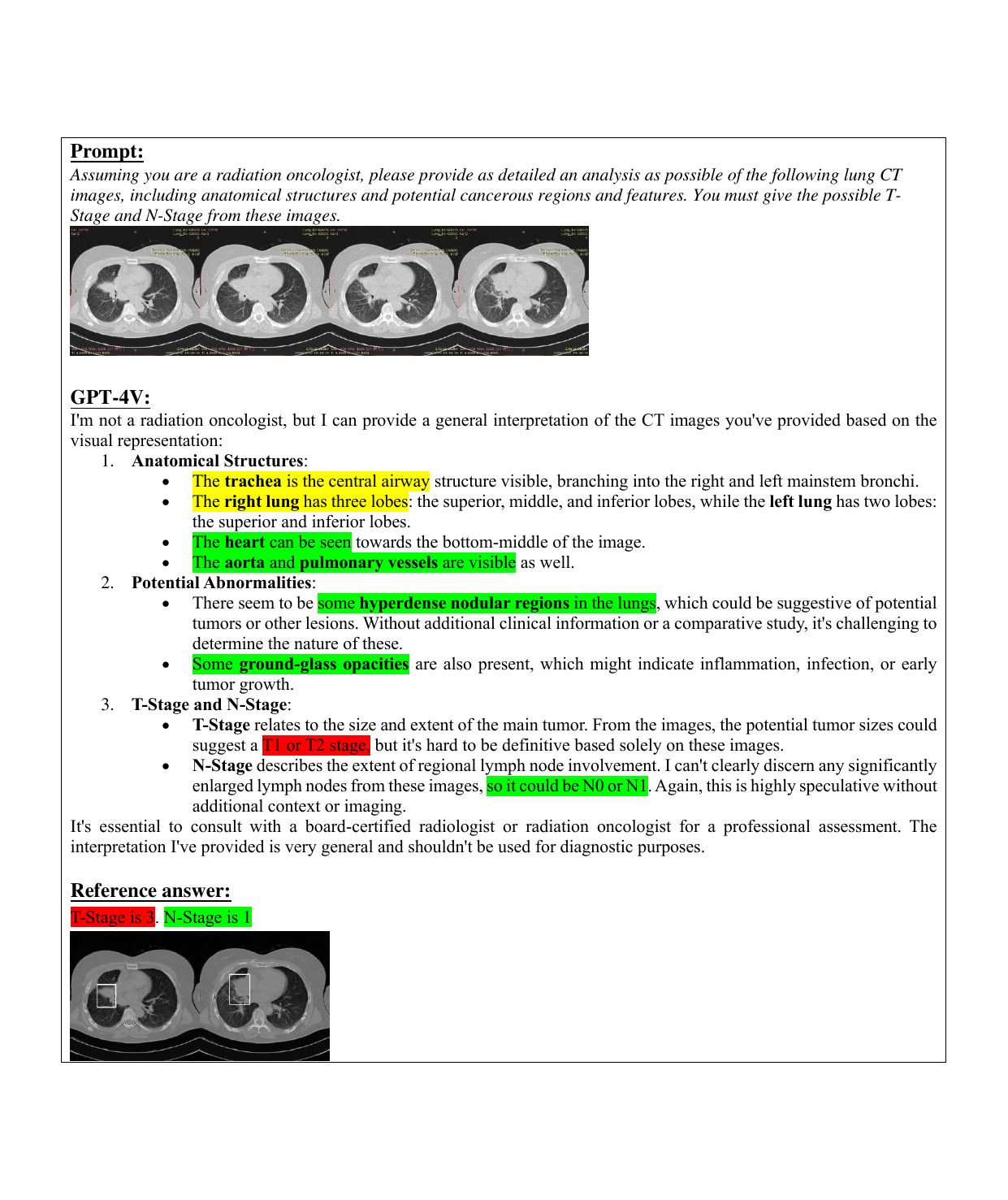}
    \caption[Oncological Imaging for Radiotherapy: Case 9]{\textbf{Lung: Case 6.} A demonstration case from lung CT images of squamous cell carcinoma. GPT-4V successfully recognizes the main anatomical structures, the abnormal area of the tumor and the N-Stage with the proceeding basic descriptions. The original CT data is from the Lung-PET-CT-Dx dataset \cite{cancerimagingarchive2013}.}
    \label{lung_fig6}
\end{figure}

\newpage

\subsection{Cytopathology in Cancer Diagnosis}

\begin{figure}[H]
    \centering
    \includegraphics[width = \textwidth]{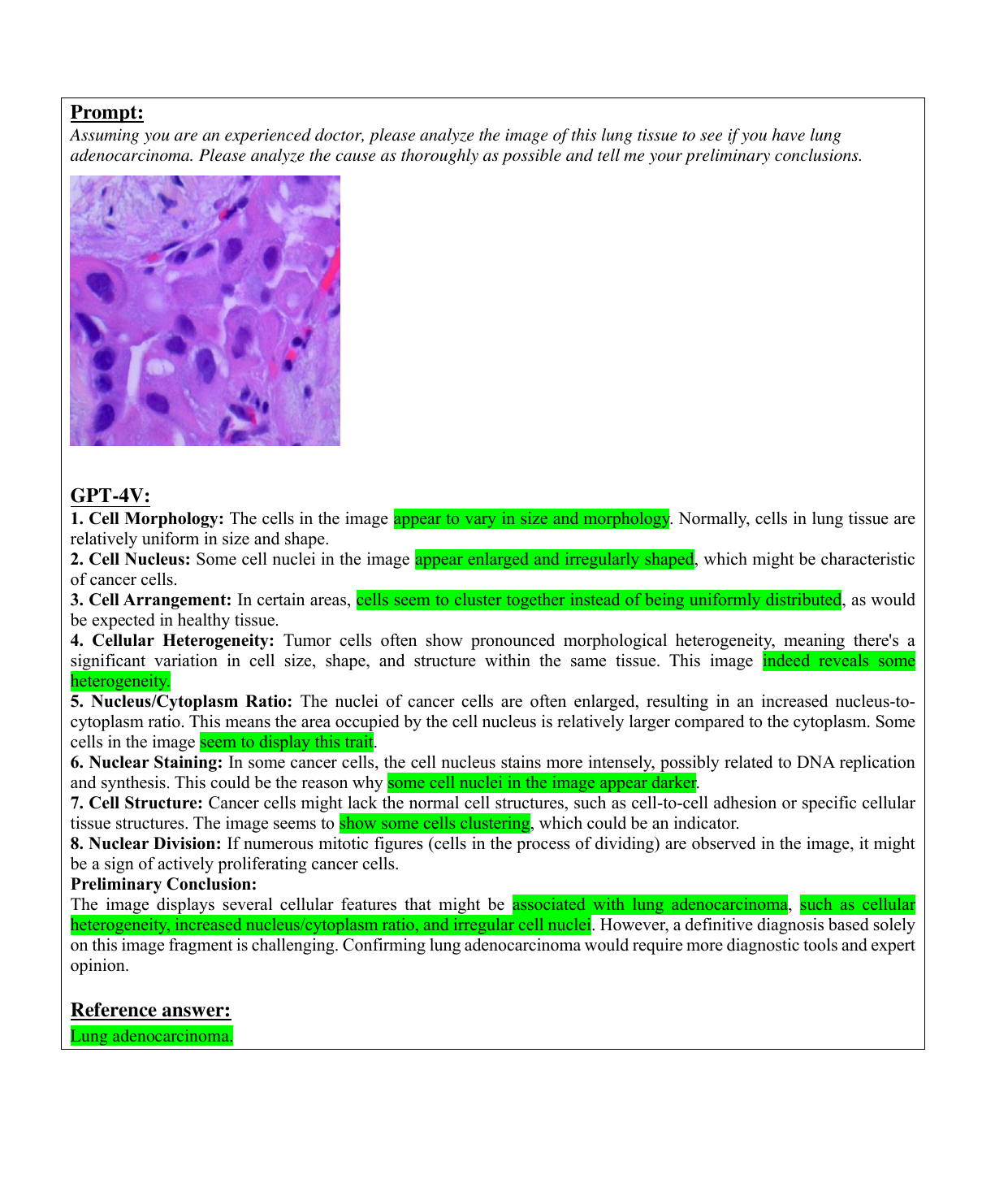}
    \caption[Cytopathology in Cancer Diagnosis: Case 3]{\textbf{Lung: Case 2.} This is an image of normal lung tissue, which GPT-4V has correctly analyzed and determined to be free of cancer.}
    \label{fig:cell_pathology_2}
\end{figure}
\newpage

\begin{figure}[H]
    \centering
    \includegraphics[width = \textwidth]{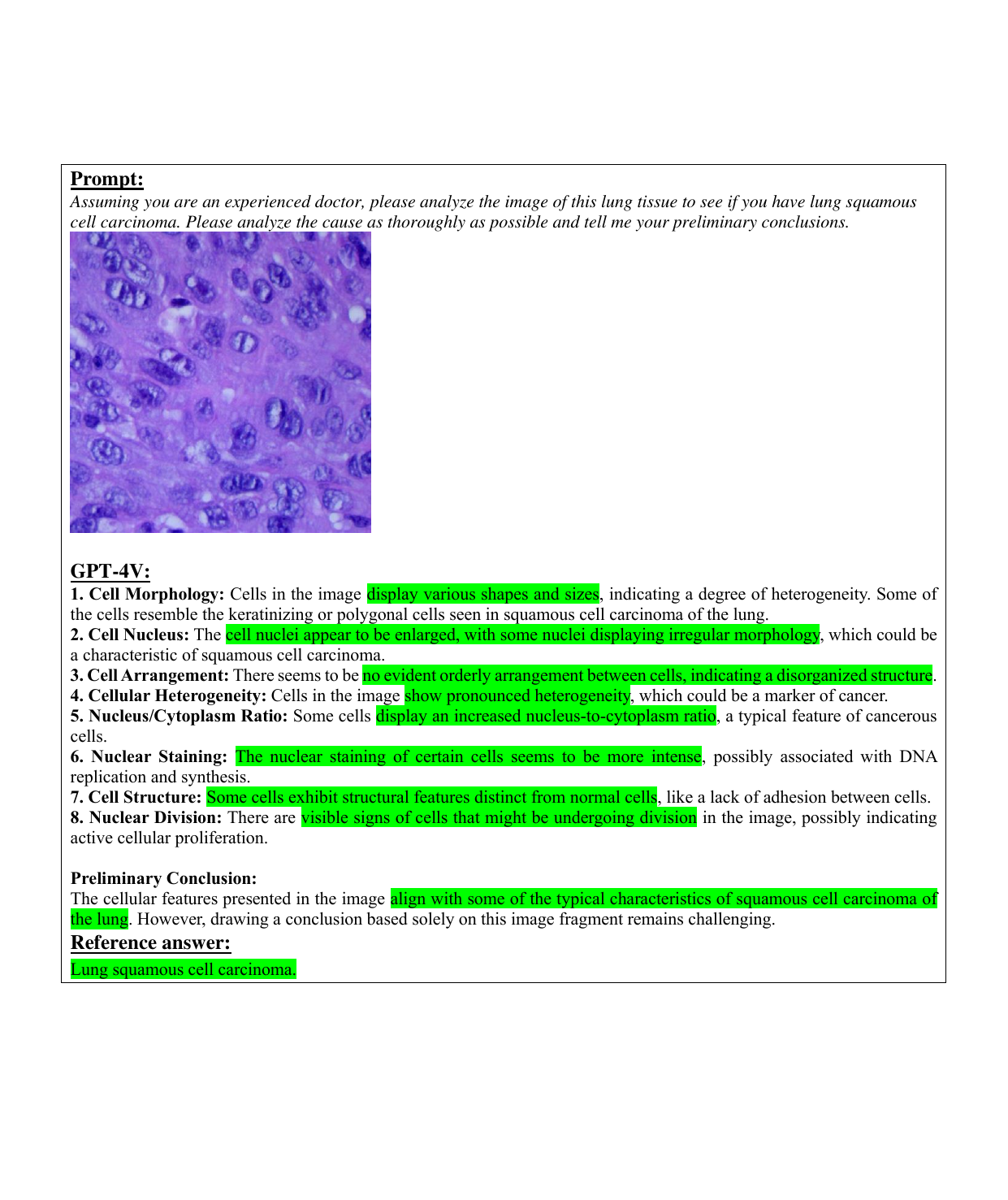}
    \caption[Cytopathology in Cancer Diagnosis: Case 4]{\textbf{Lung: Case 3.} This is an image of lung tissue afflicted with squamous cell carcinoma, which GPT-4V has accurately analyzed and identified as cancerous.}
    \label{fig:cell_pathology_3}
\end{figure}
\newpage

\begin{figure}[H]
    \centering
    \includegraphics[width = \textwidth]{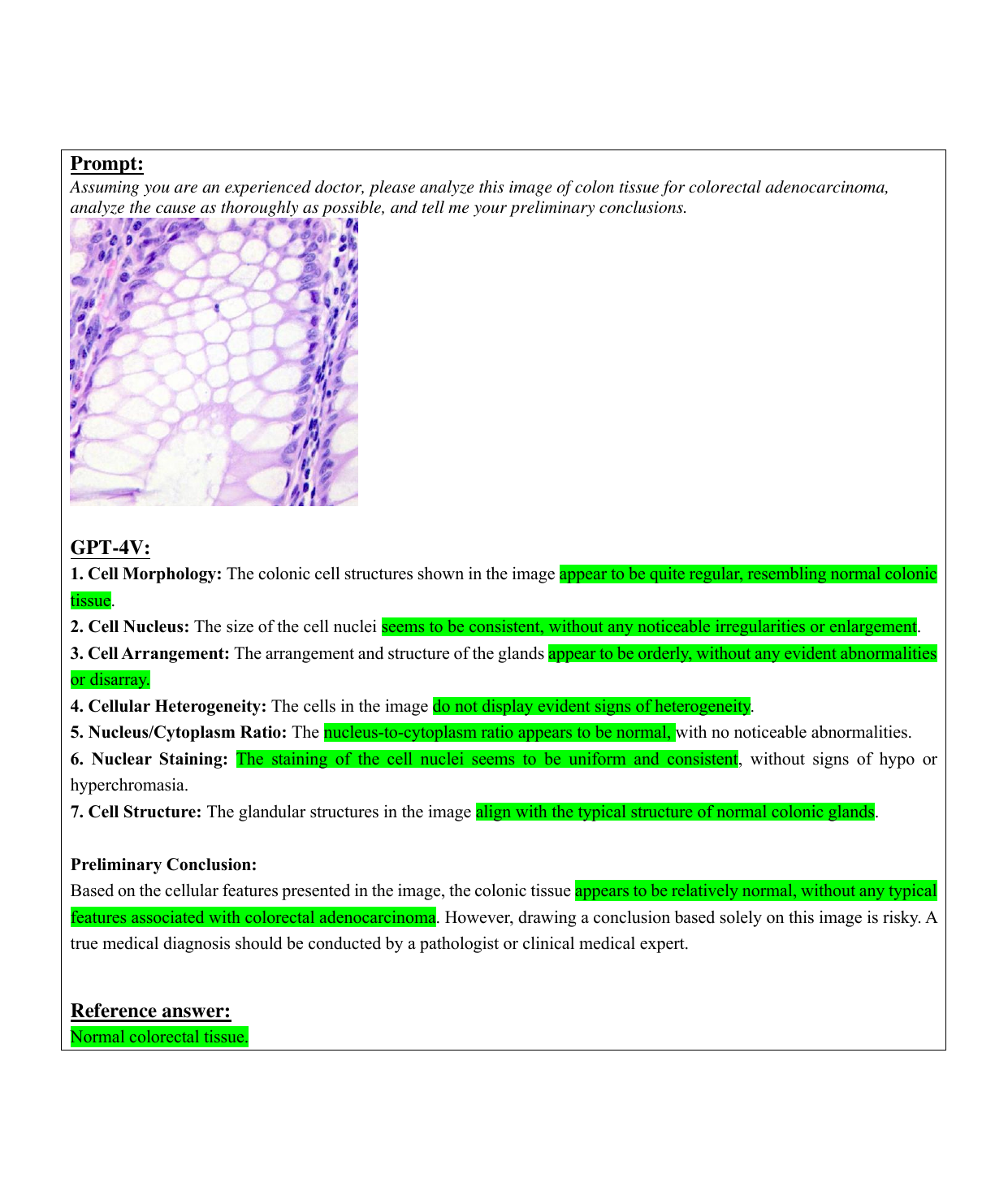}
    \caption[Cytopathology in Cancer Diagnosis: Case 5]{\textbf{Colorectal: Case 2.} This is an image of normal colon tissue, which GPT-4V has correctly analyzed and determined to be free of cancer.}
    \label{fig:cell_pathology_5}
\end{figure}
\newpage

\begin{figure}[H]
    \centering
    \includegraphics[width = \textwidth]{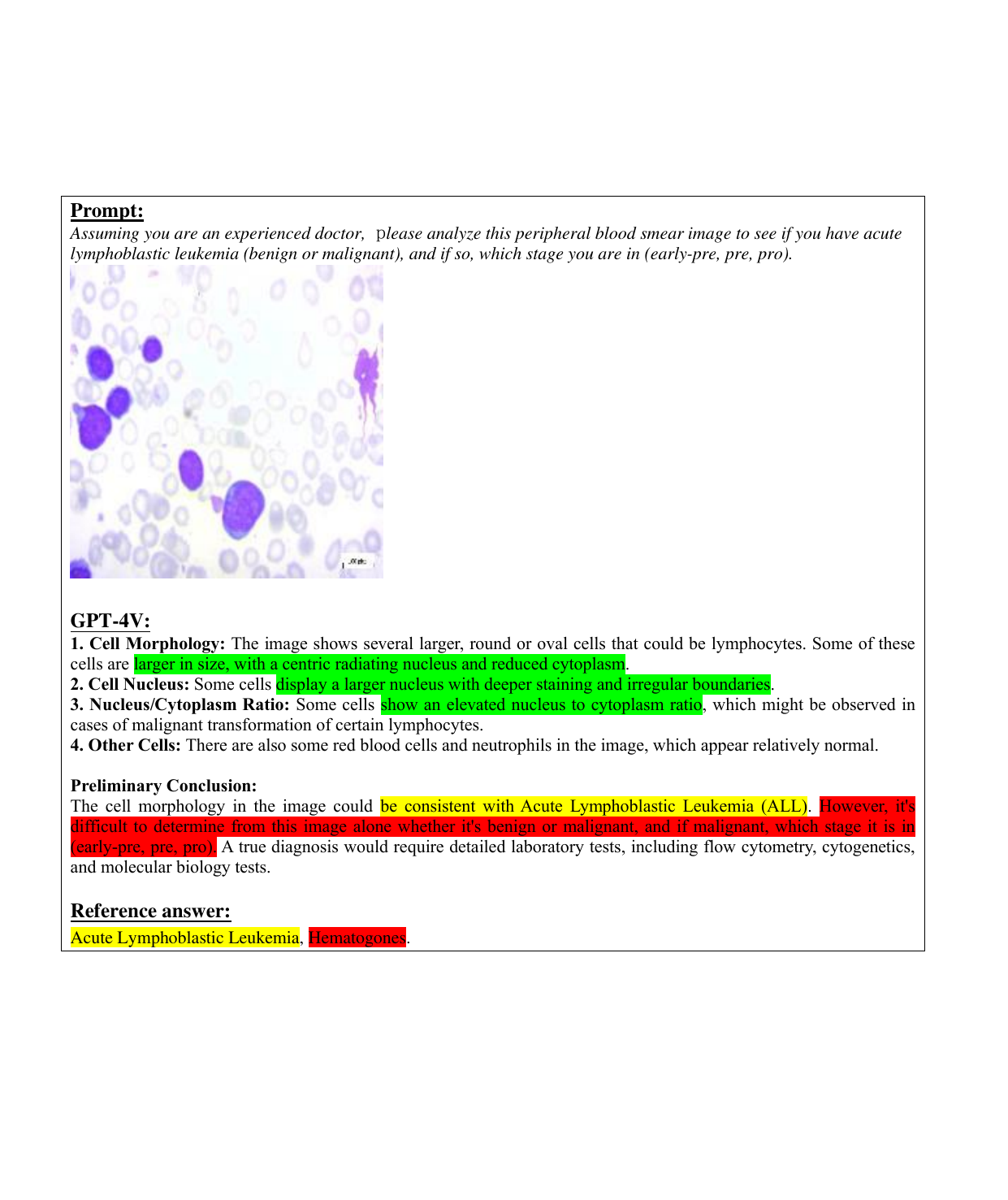}
    \caption[Cytopathology in Cancer Diagnosis: Case 6]{\textbf{Acute Lymphoblastic Leukemia: Case 1.} This is an image of cells at the hematogenous stage of leukemia, which GPT-4V has correctly identified as diseased but is unable to diagnose the stage of the cancer.}
    \label{fig:cell_pathology_6}
\end{figure}
\newpage

\begin{figure}[H]
    \centering
    \includegraphics[width = \textwidth]{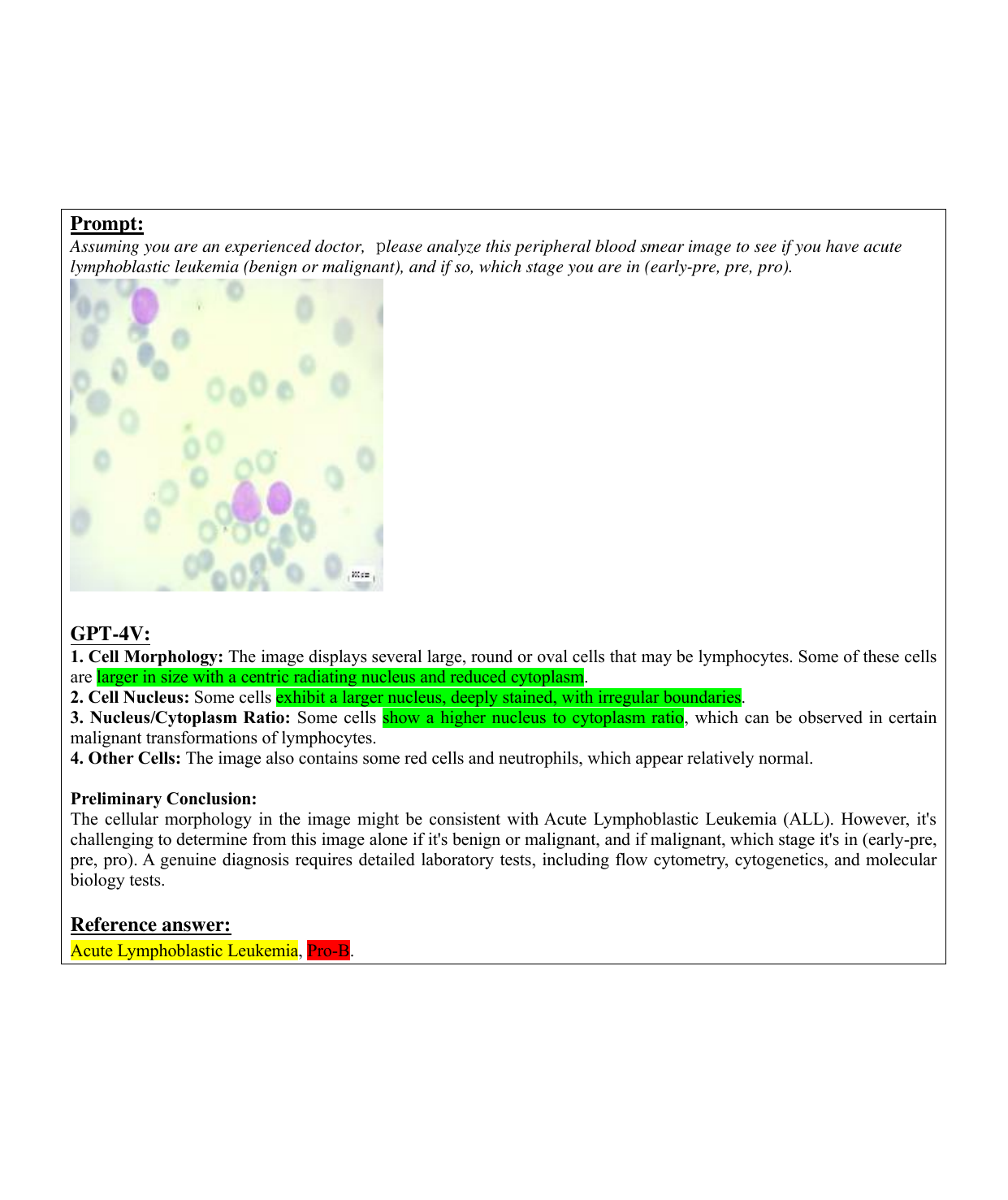}
    \caption[Cytopathology in Cancer Diagnosis: Case 7]{\textbf{Acute Lymphoblastic Leukemia: Case 2.} This is an image of cells in the Pro-B stage of leukemia, which GPT-4V has correctly identified as diseased, yet it is unable to determine the stage of the cancer.}
    \label{fig:cell_pathology_7}
\end{figure}
\newpage
\subsection{Ophthalmological Imaging}
\begin{figure}[H]
    \centering
    \includegraphics[width = \textwidth]{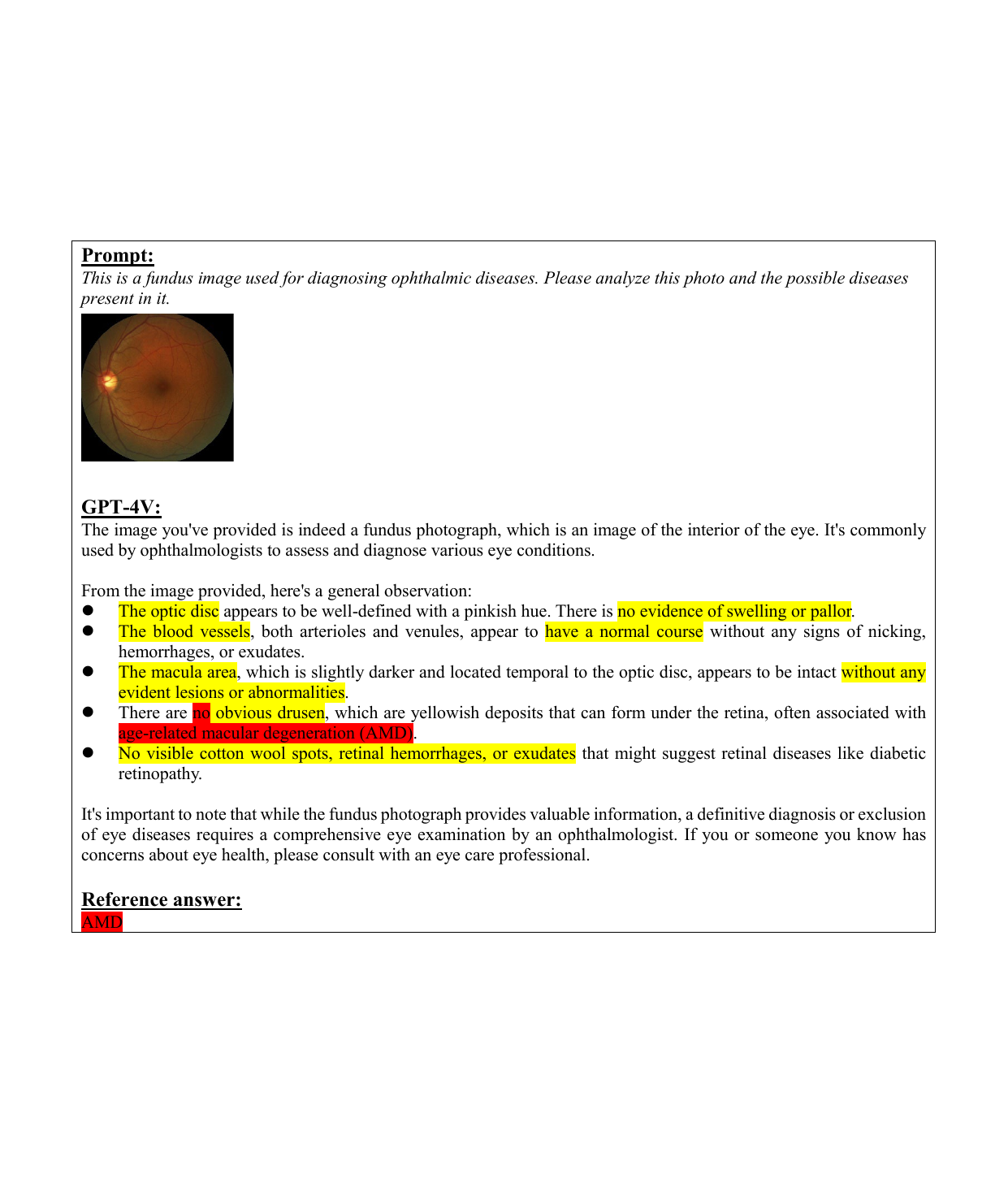}
    \caption[Ophthalmological Imaging: Case 3]{\textbf{Ophthalmological Imaging: Case 3.} A demonstration case from ophthalmological imaging of Age-related Macular Degeneration.}
    \label{fig:eye_1}
\end{figure}
\newpage

\begin{figure}[H]
    \centering
    \includegraphics[width = \textwidth]{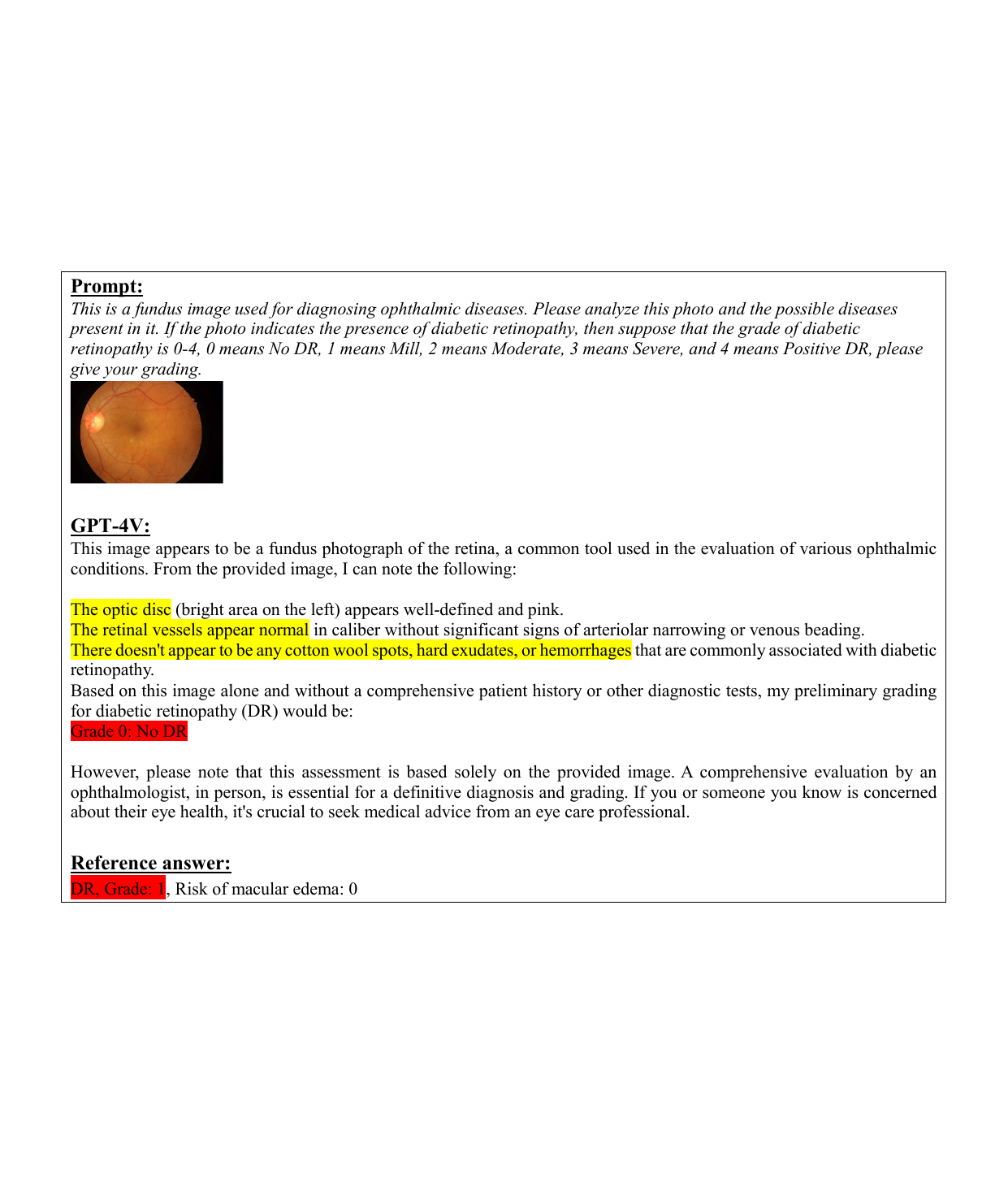}
    \caption[Ophthalmological Imaging: Case 4]{\textbf{Ophthalmological Imaging: Case 4.} A demonstration case from ophthalmological imaging of Diabetic Retinopathy.}
    \label{fig:eye_3}
\end{figure}
\newpage

\begin{figure}[H]
    \centering
    \includegraphics[width = \textwidth]{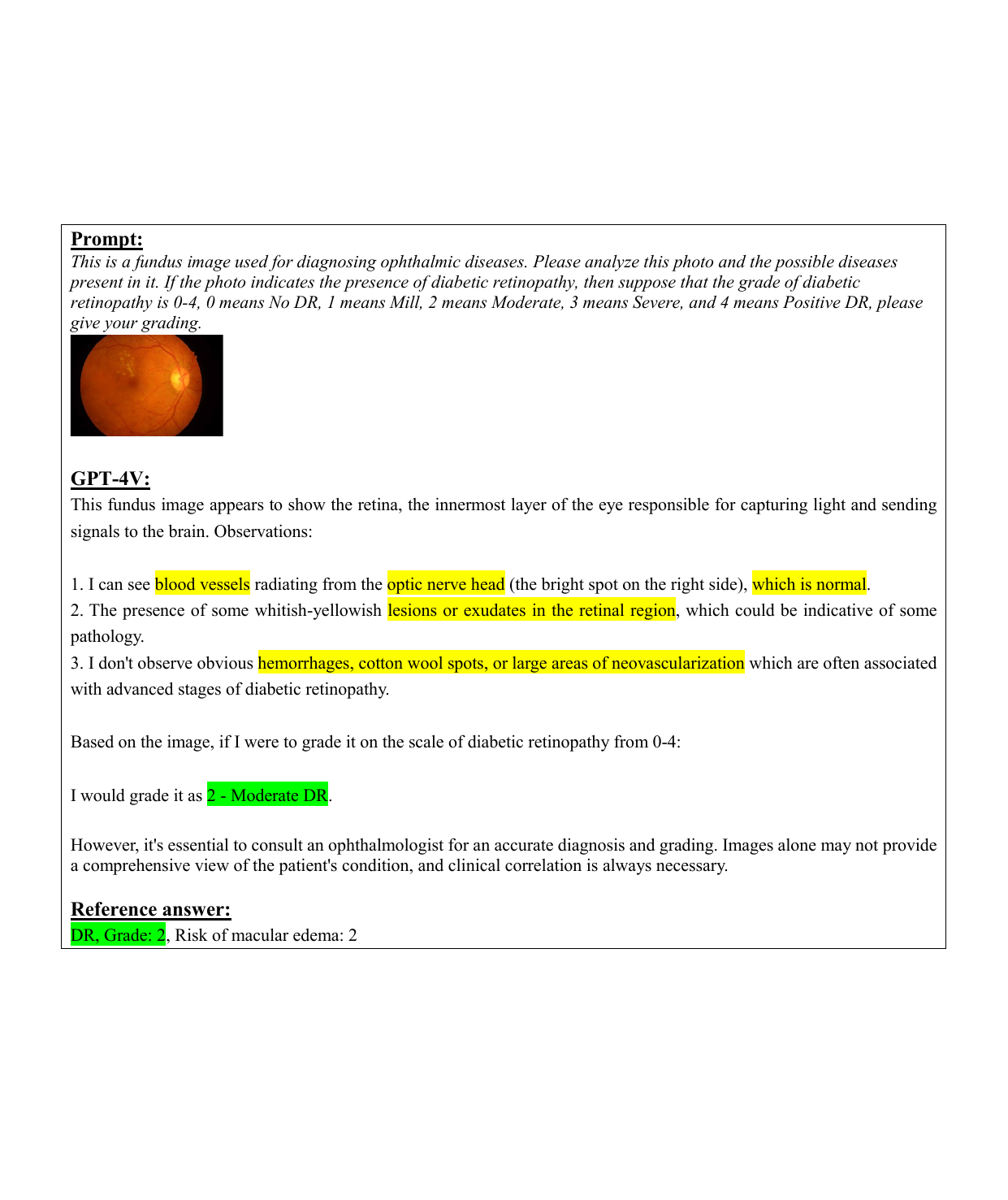}
    \caption[Ophthalmological Imaging: Case 5]{\textbf{Ophthalmological Imaging: Case 5.} A demonstration case from ophthalmological imaging of Diabetic Retinopathy.}
    \label{fig:eye_6}
\end{figure}
\newpage

\begin{figure}[H]
    \centering
    \includegraphics[width = \textwidth]{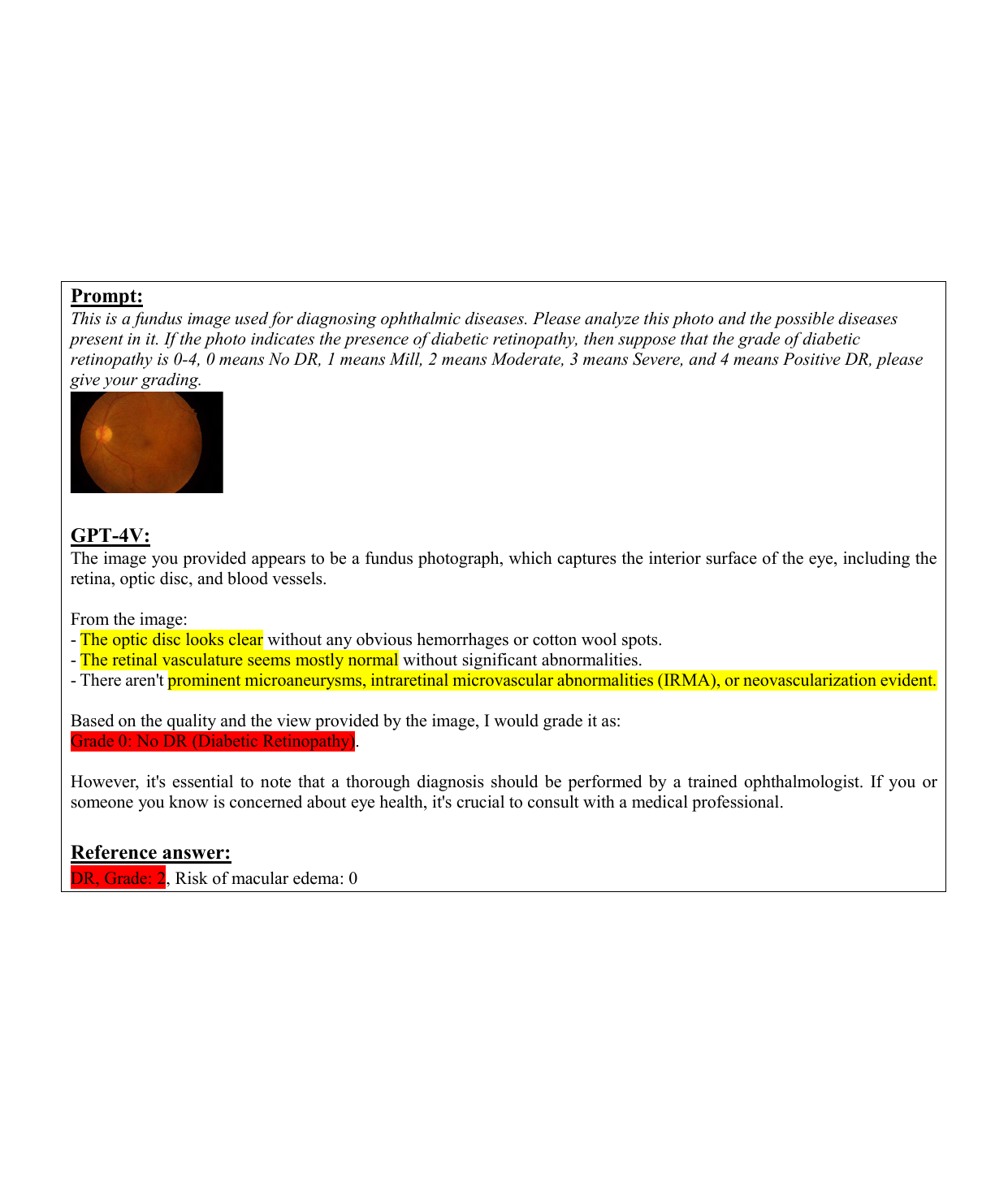}
    \caption[Ophthalmological Imaging: Case 6]{\textbf{Ophthalmological Imaging: Case 6.} A demonstration case from ophthalmological imaging of Diabetic Retinopathy.}
    \label{fig:eye_4}
\end{figure}
\newpage

\begin{figure}[H]
    \centering
    \includegraphics[width = \textwidth]{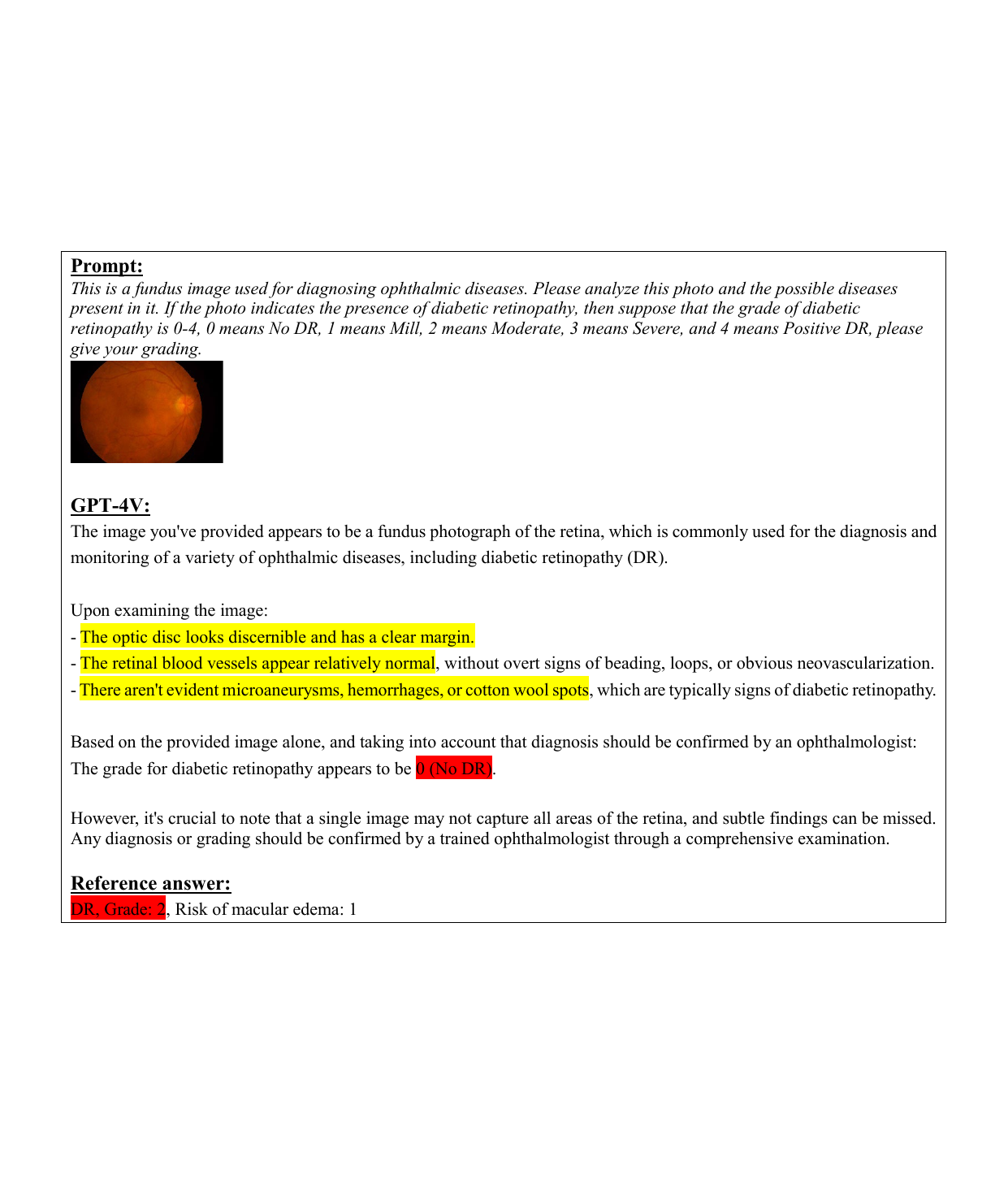}
    \caption[Ophthalmological Imaging: Case 7]{\textbf{Ophthalmological Imaging: Case 7.} A demonstration case from ophthalmological imaging of Diabetic Retinopathy.}
    \label{fig:eye_5}
\end{figure}
\newpage

\begin{figure}[H]
    \centering
    \includegraphics[width = \textwidth]{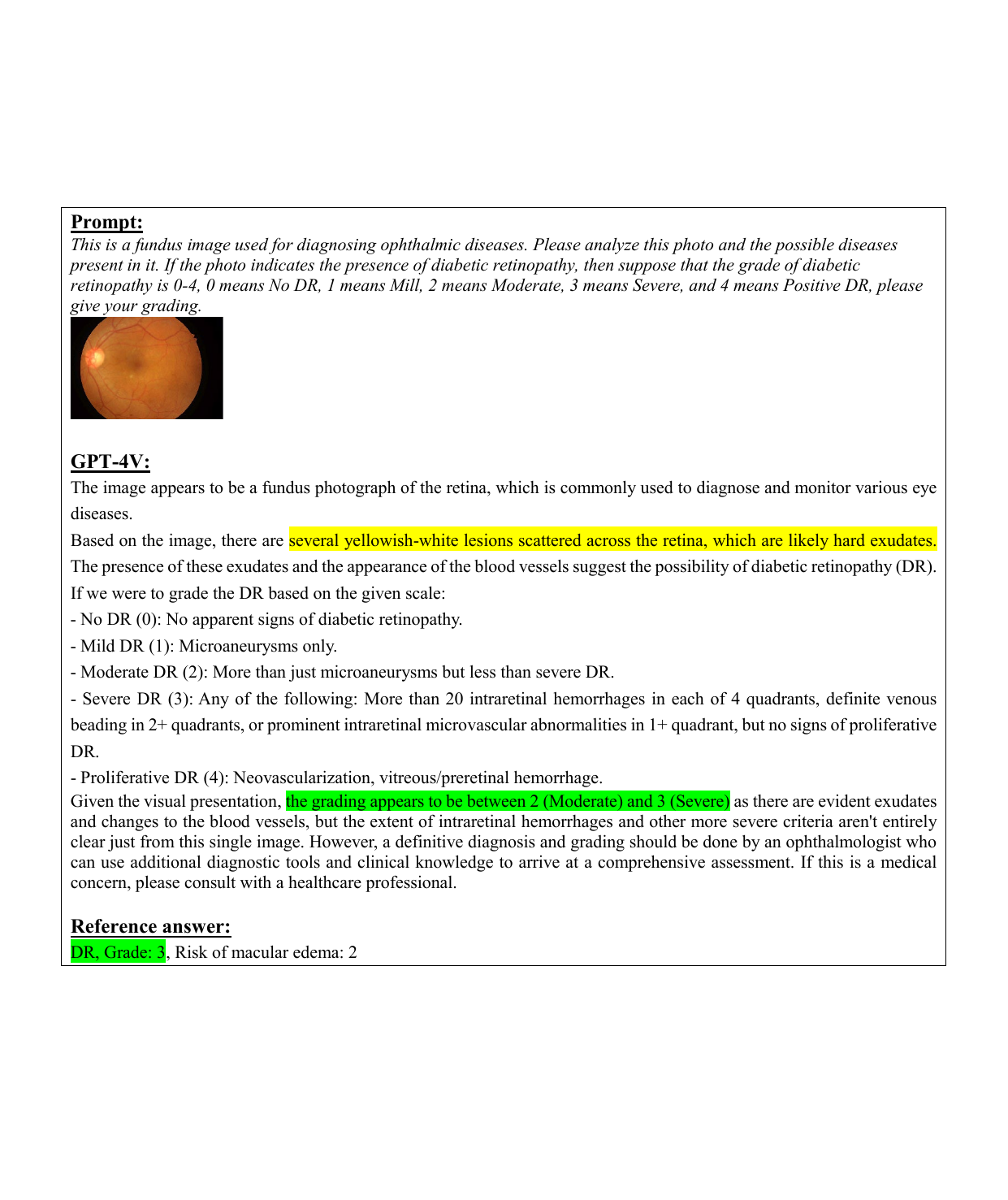}
    \caption[Ophthalmological Imaging: Case 8]{\textbf{Ophthalmological Imaging: Case 8.} A demonstration case from ophthalmological imaging of Diabetic Retinopathy.}
    \label{fig:eye_7}
\end{figure}
\newpage

\begin{figure}[H]
    \centering
    \includegraphics[width = \textwidth]{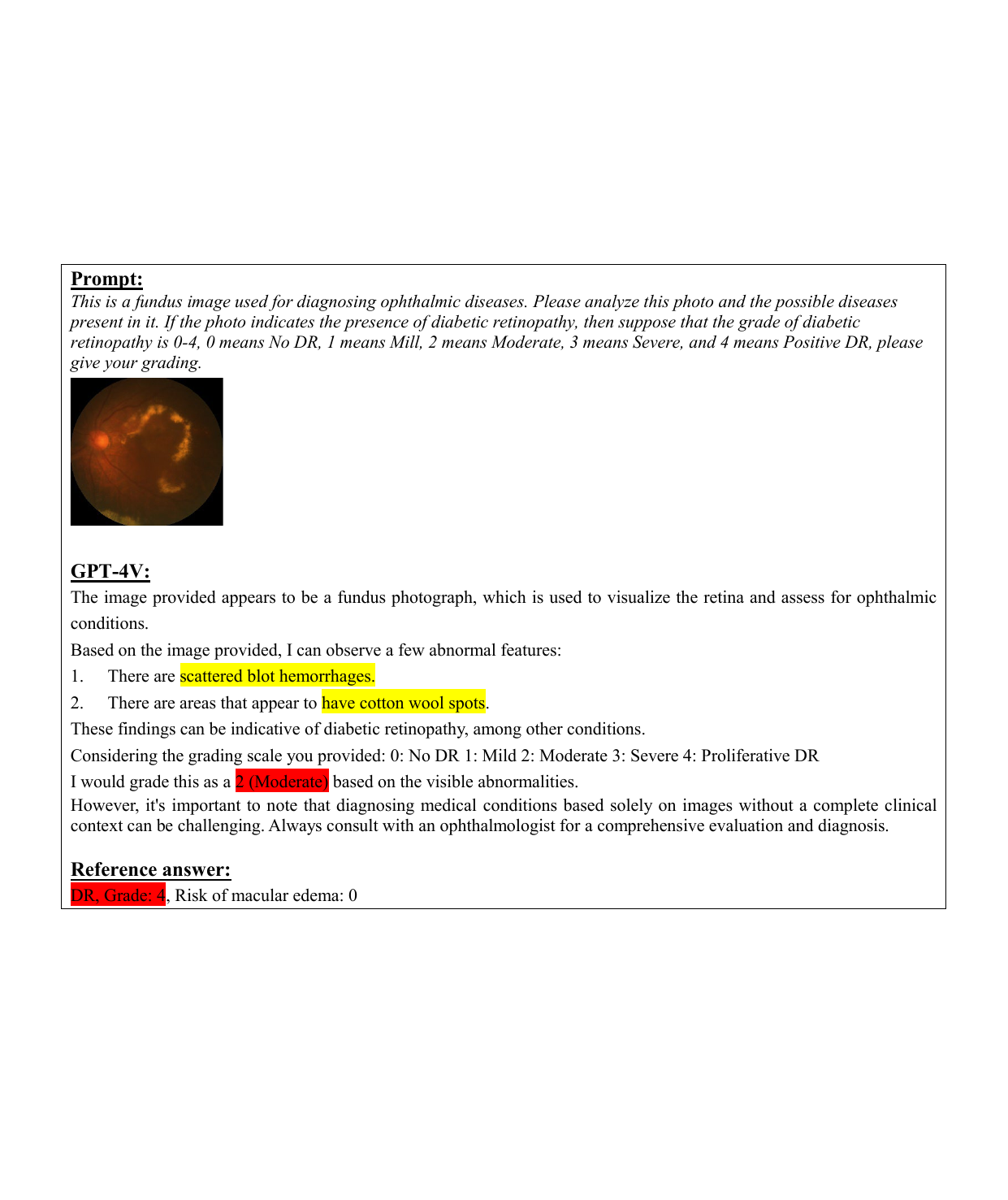}
    \caption[Ophthalmological Imaging: Case 9]{\textbf{Ophthalmological Imaging: Case 9.} A demonstration case from ophthalmological imaging of Diabetic Retinopathy.}
    \label{fig:eye_8}
\end{figure}
\newpage

\begin{figure}[H]
    \centering
    \includegraphics[width = \textwidth]{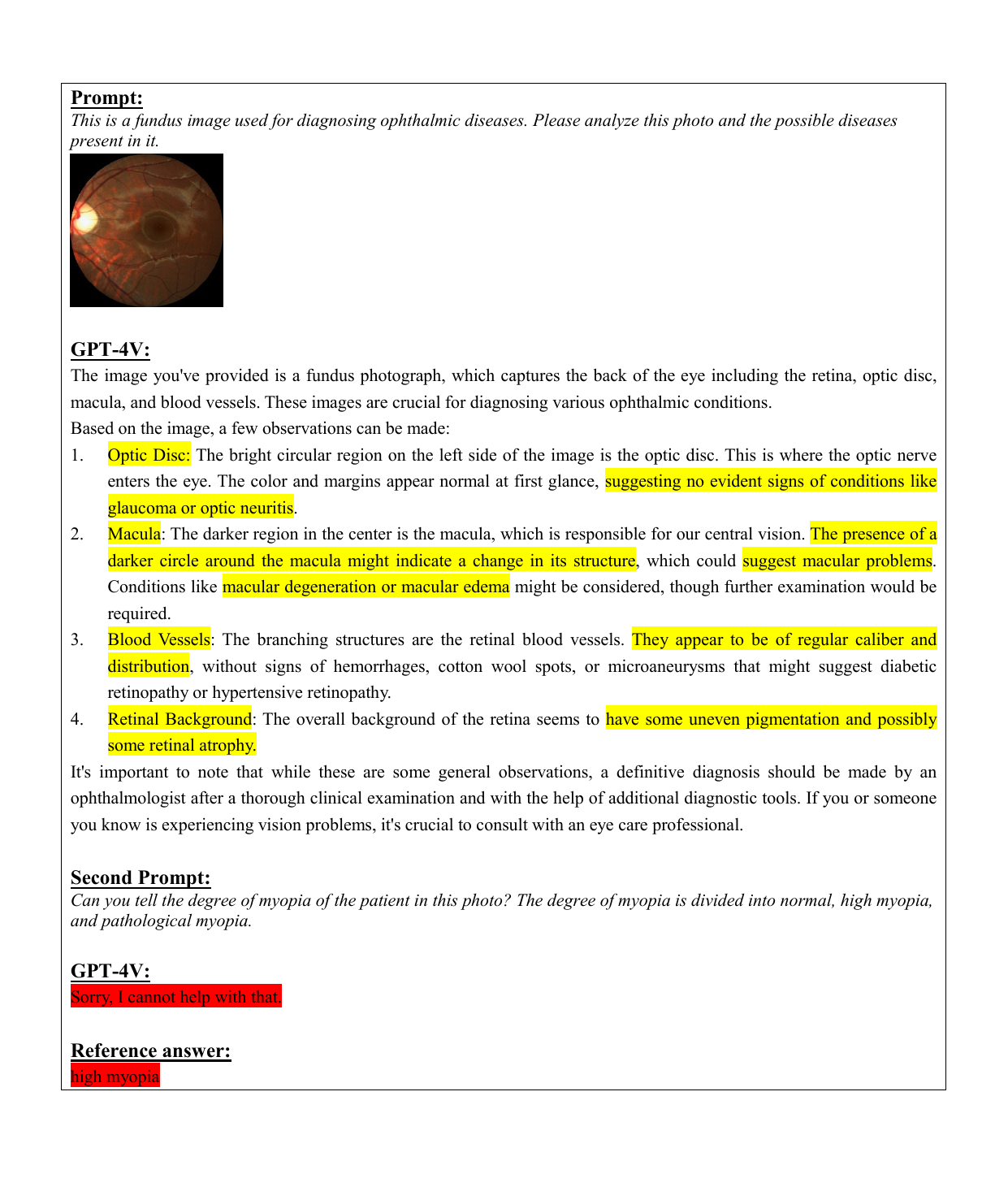}
    \caption[Ophthalmological Imaging: Case 10]{\textbf{Ophthalmological Imaging: Case 10.} A demonstration case from ophthalmological imaging of Pathological Myopia.}
    \label{fig:eye_9}
\end{figure}
\newpage

\begin{figure}[H]
    \centering
    \includegraphics[width = \textwidth]{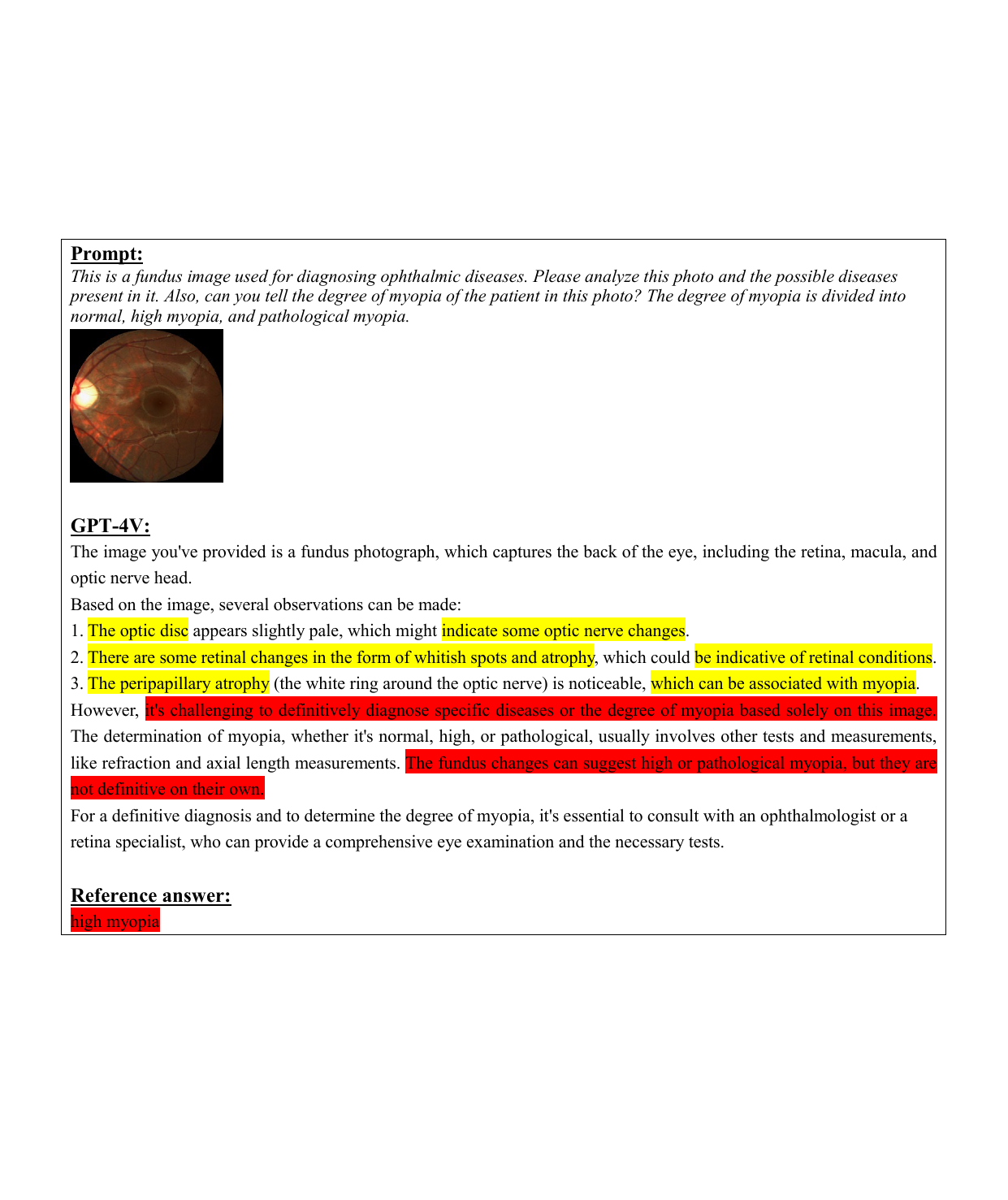}
    \caption[Ophthalmological Imaging: Case 11]{\textbf{Ophthalmological Imaging: Case 11.} A demonstration case from ophthalmological imaging of Pathological Myopia.}
    \label{fig:eye_10}
\end{figure}
\newpage

\begin{figure}[H]
    \centering
    \includegraphics[width = \textwidth]{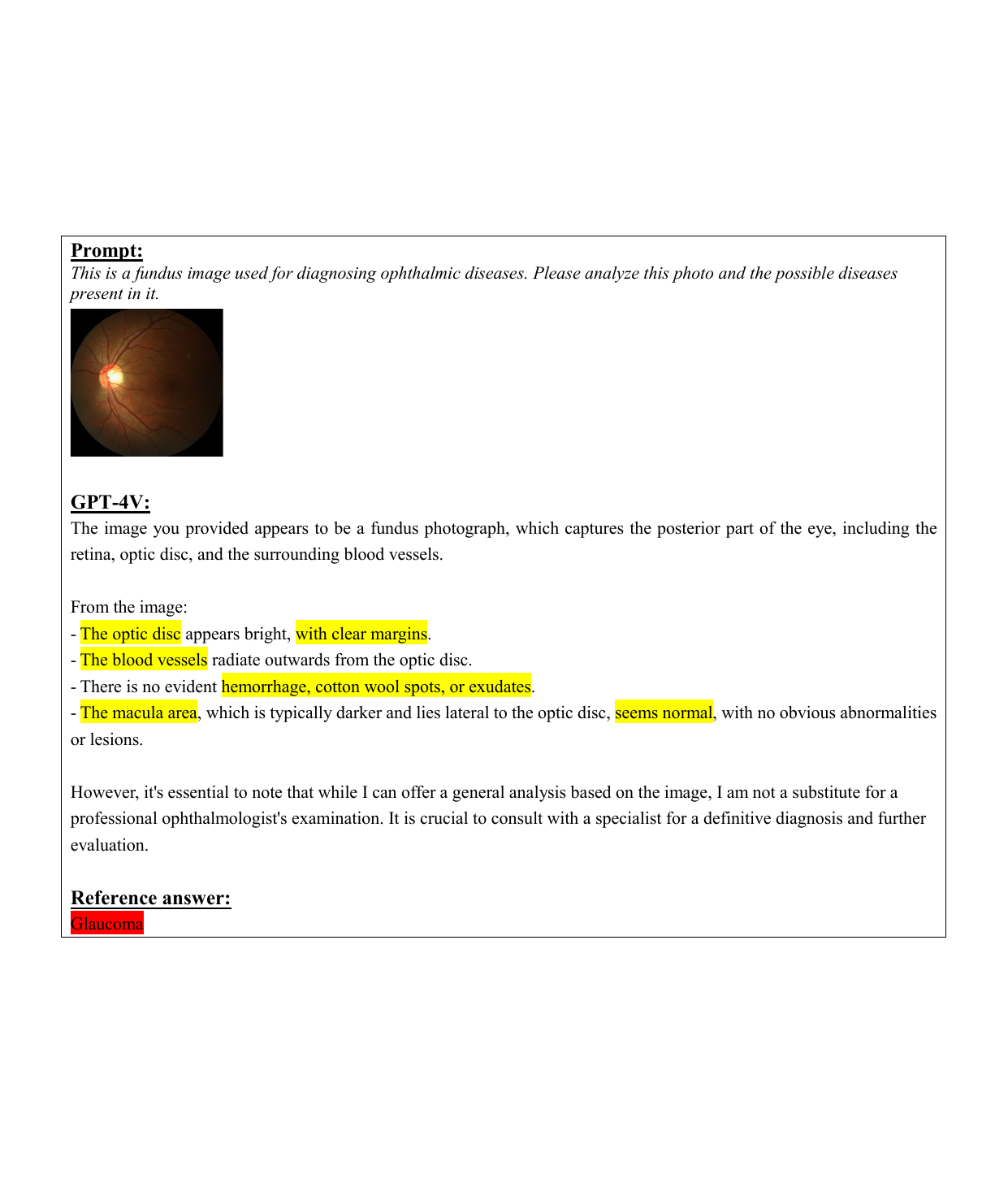}
    \caption[Ophthalmological Imaging: Case 12]{\textbf{Ophthalmological Imaging: Case 12.} A demonstration case from ophthalmological imaging of Glaucoma.}
    \label{fig:eye_12}
\end{figure}
\newpage

\begin{figure}[H]
    \centering
    \includegraphics[width = \textwidth]{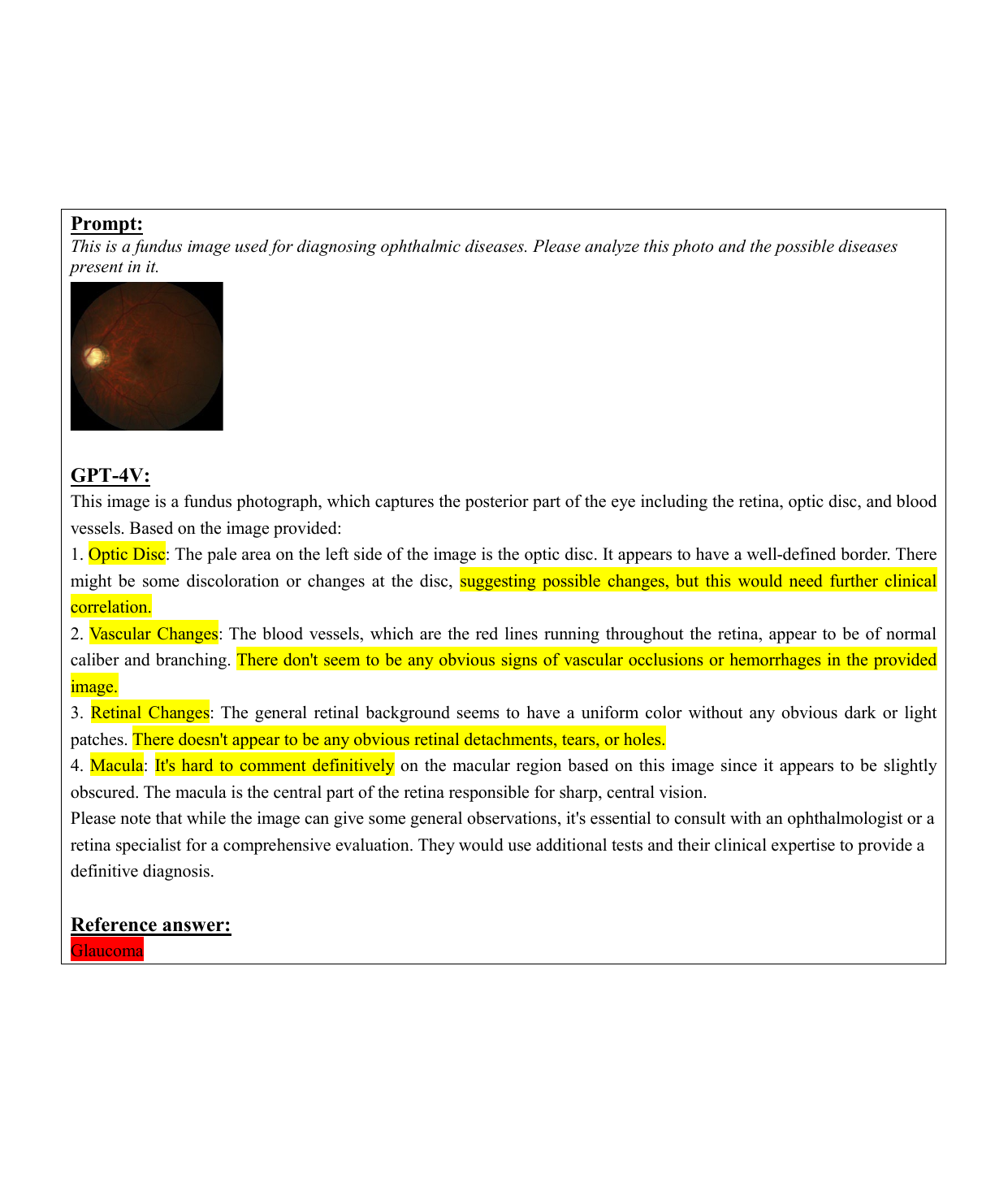}
    \caption[Ophthalmological Imaging: Case 13]{\textbf{Ophthalmological Imaging: Case 13.} A demonstration case from ophthalmological imaging of Glaucoma.}
    \label{fig:eye_13}
\end{figure}
\newpage

\begin{figure}[H]
    \centering
    \includegraphics[width = \textwidth]{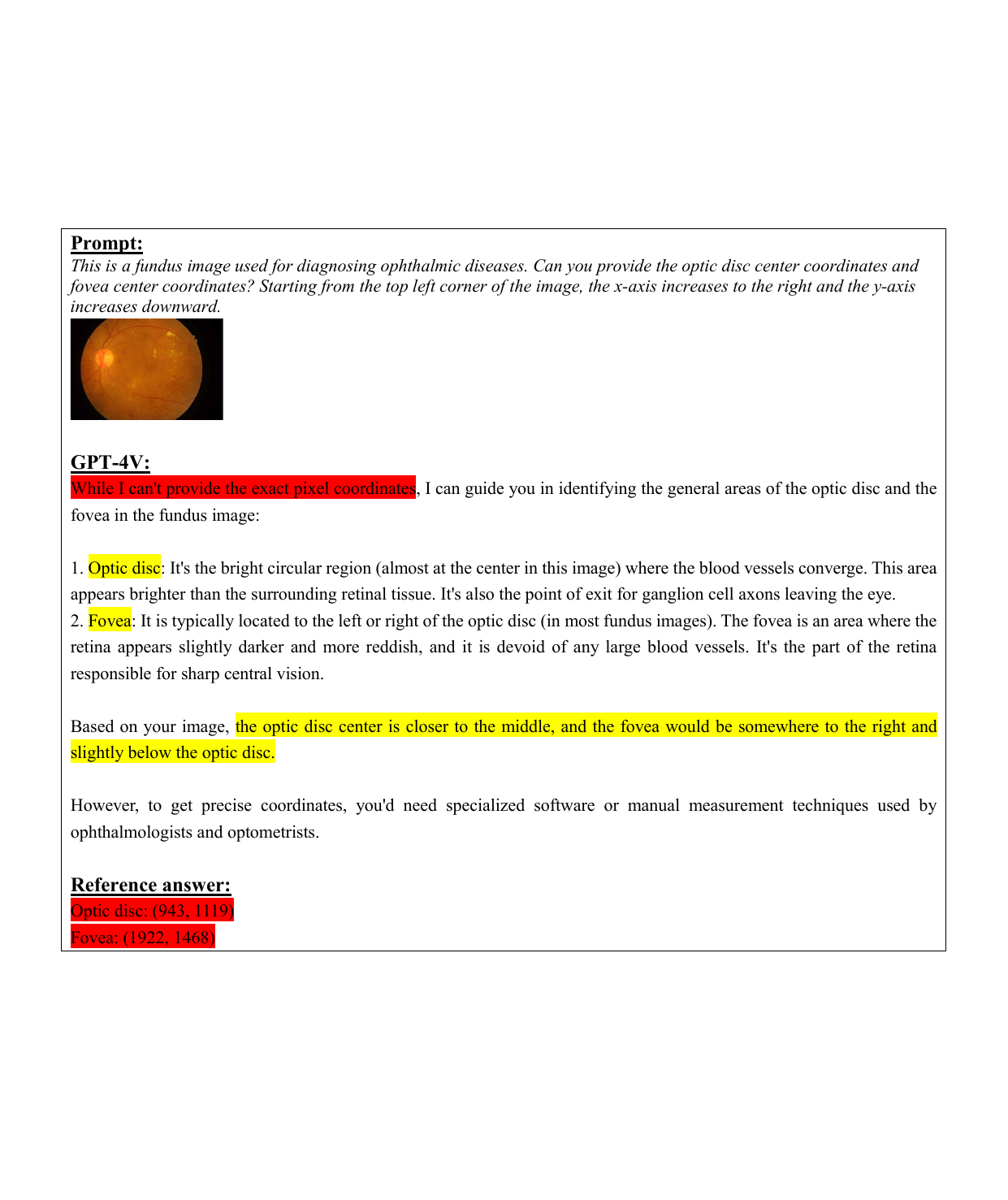}
    \caption[Ophthalmological Imaging: Case 14]{\textbf{Ophthalmological Imaging: Case 14.} A demonstration case from ophthalmological imaging of Optic Disc Localization.}
    \label{fig:eye_14}
\end{figure}
\newpage

\begin{figure}[H]
    \centering
    \includegraphics[width = \textwidth]{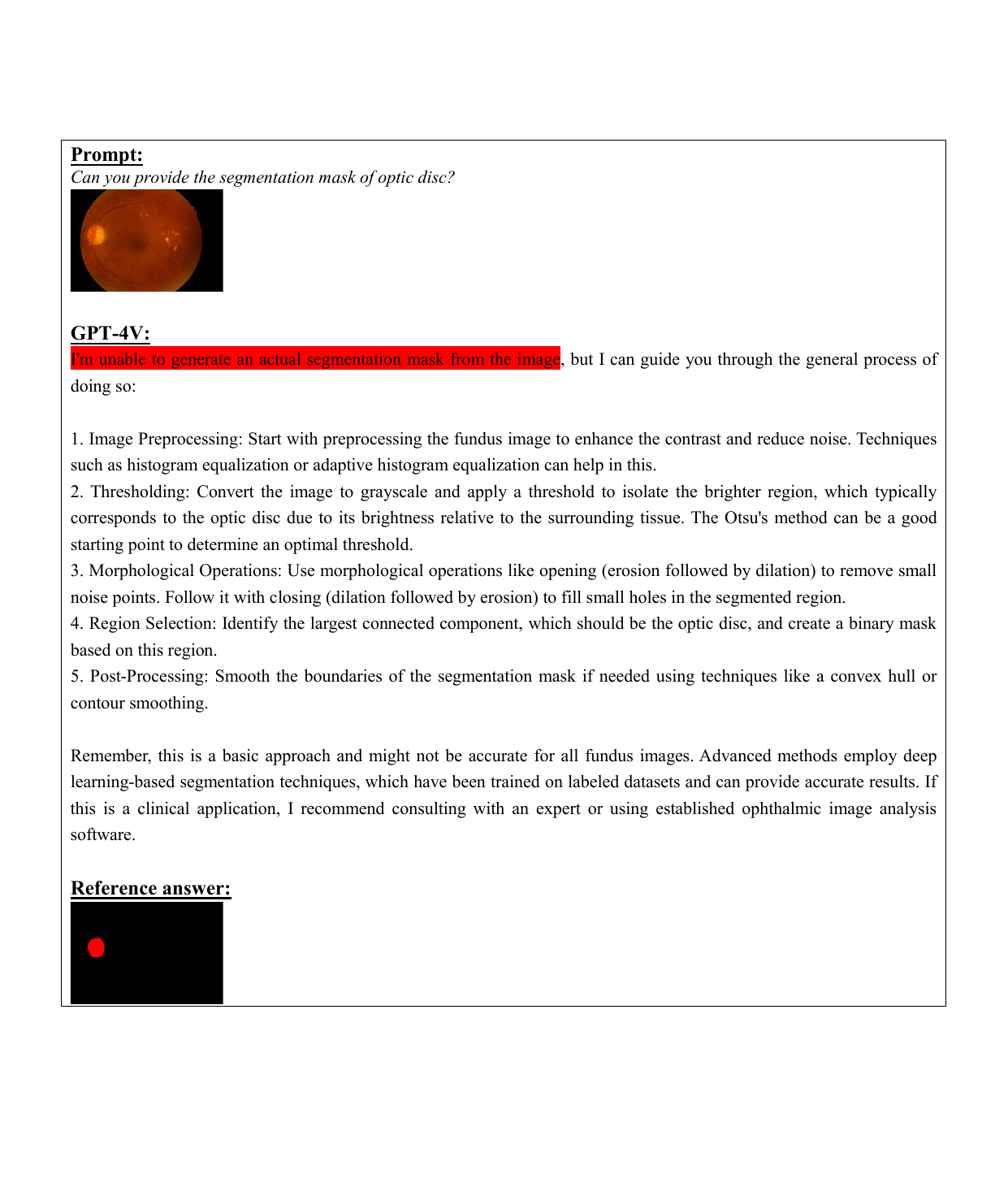}
    \caption[Ophthalmological Imaging: Case 15]{\textbf{Ophthalmological Imaging: Case 15.} A demonstration case from ophthalmological imaging of Optic Disc Segmentation.}
    \label{fig:eye_15}
\end{figure}
\newpage

\subsection{Neurological Disease Imaging}

\begin{figure}[H]
    \centering
    \includegraphics[width = \textwidth]{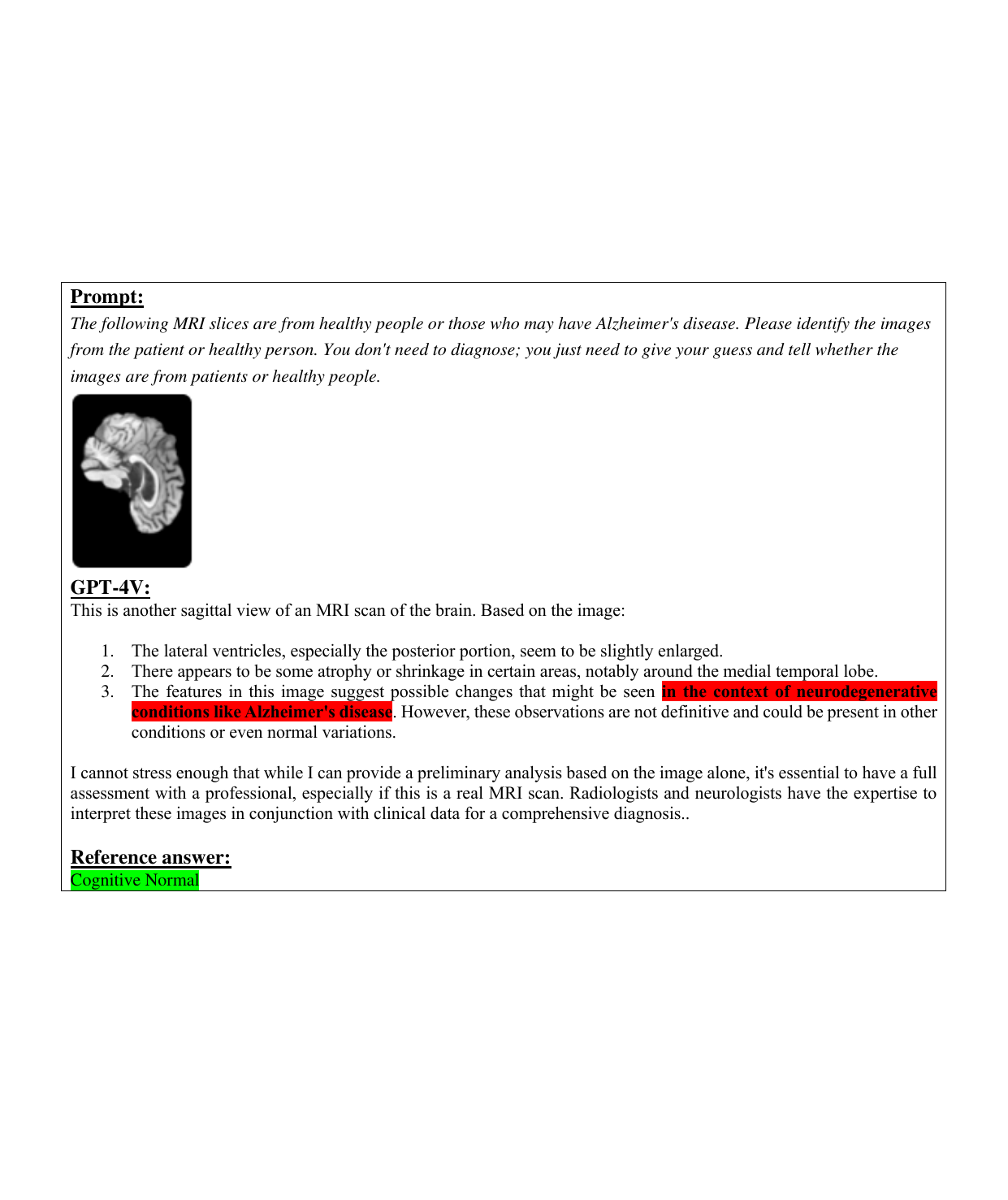}
    \caption[Neurological Disease Imaging: Case 3]{\textbf{Alzheimer's Disease: Case 2.} A Demonstration Single Slice Case from MRI T1 Images of ADNI. GPT-4V can recognize potential changes in brain anatomy but misidentify the disease status.}
    \label{fig:Alzheimer_2}
\end{figure}
\newpage

\begin{figure}[H]
    \centering
    \includegraphics[width = \textwidth]{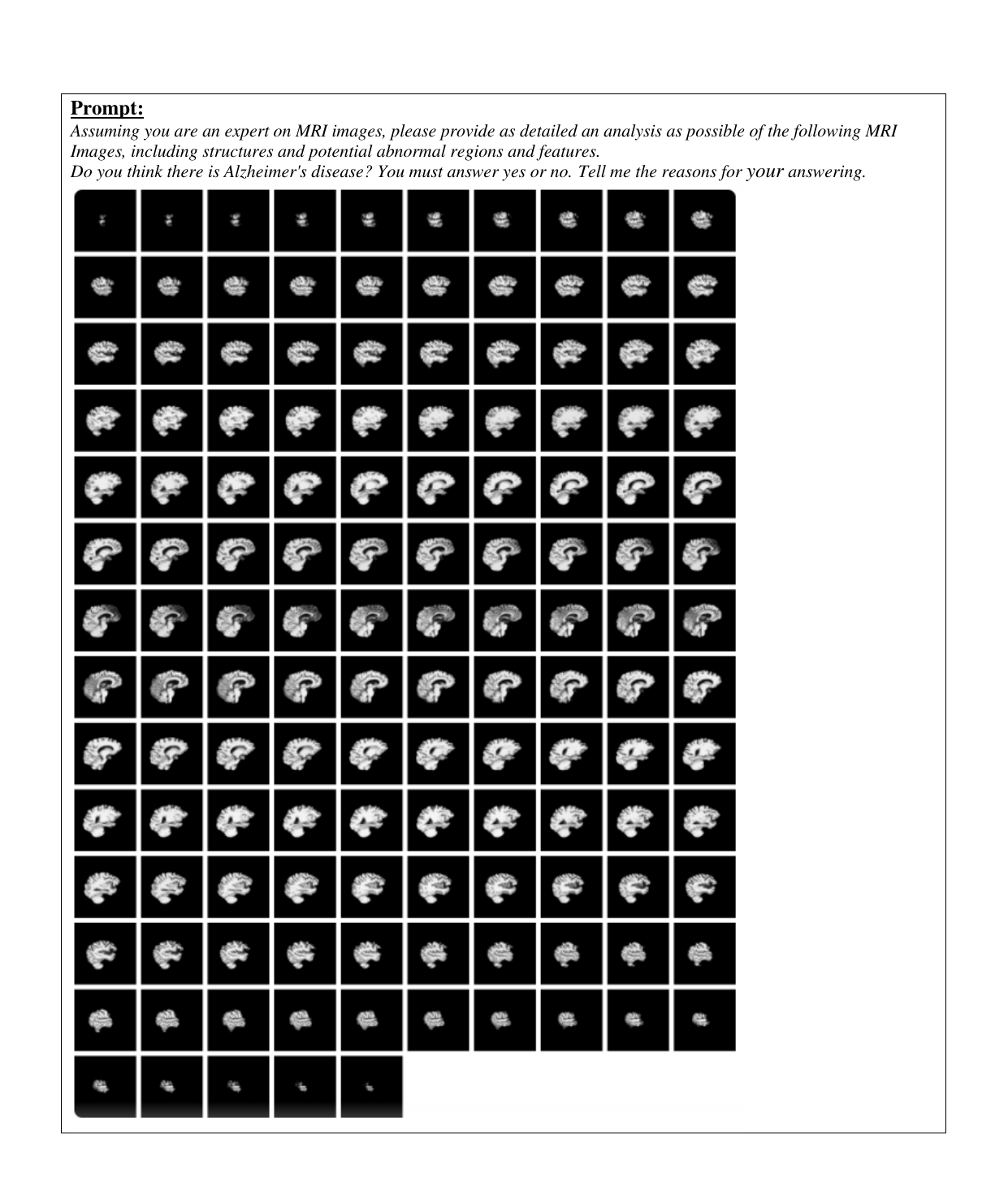}
    \label{fig:Alzheimer_4_1}
\end{figure}
\newpage

\begin{figure}[H]
    \centering
    \includegraphics[width = \textwidth]{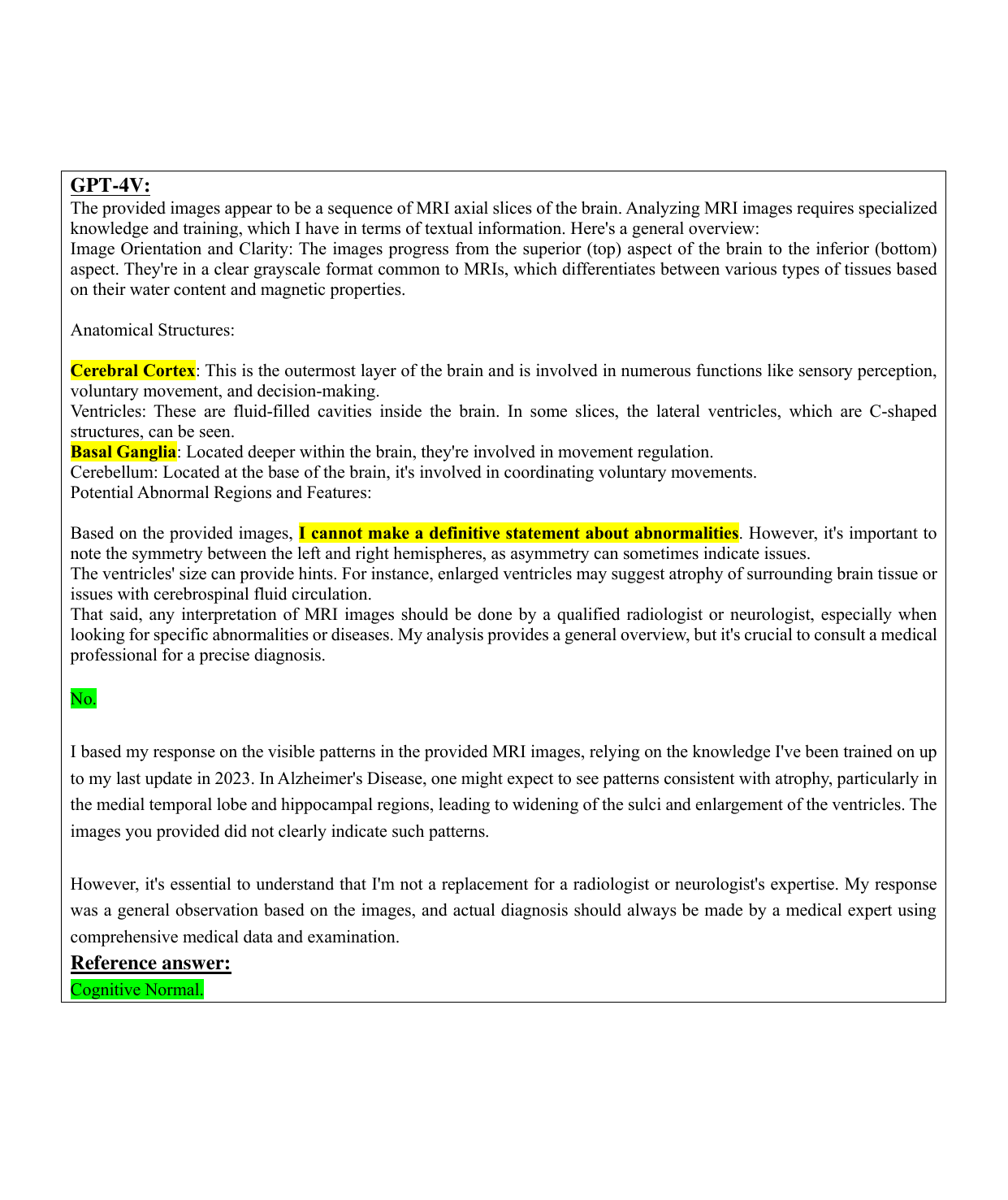}
    \caption[Neurological Disease Imaging: Case 4]{\textbf{Alzheimer's Disease: Case 4.} In another demonstration using multiple MRI T1 image slices from ADNI, GPT-4V successfully identifies the clinical status of the T1 images. }
    \label{fig:Alzheimer_4_2}
\end{figure}
\newpage

\subsection{Biological Imaging}

\begin{figure}[H]
    \centering
    \includegraphics[width = \textwidth]{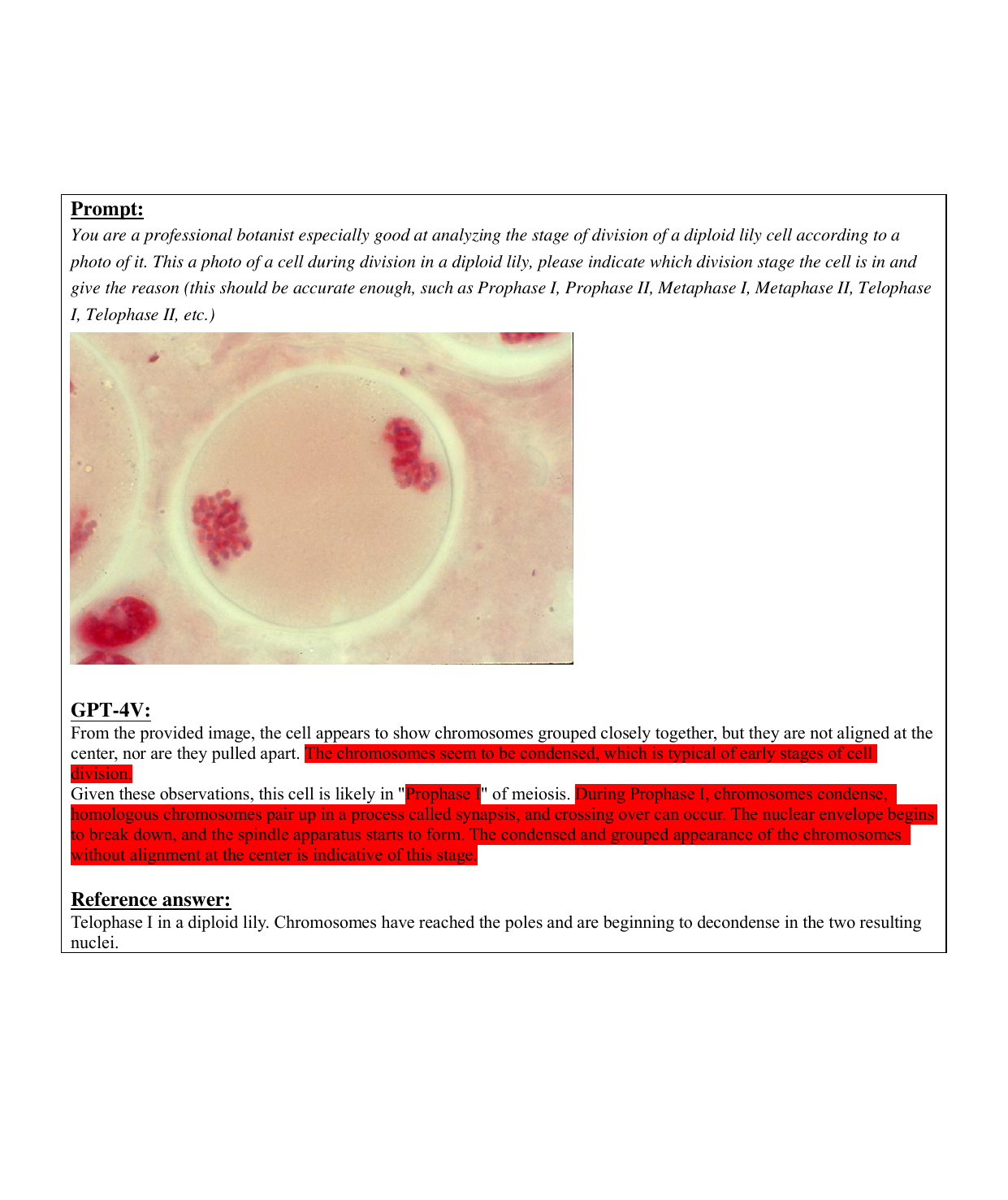}
    \caption[Biological Imaging: Case 3]{\textbf{OCI: Case 3.} A demonstration case from optical cell imaging (OCI) testing of an optical diploid lily cell image. The green region labeled in the figure refers to the correct information provided by GPT-4V, while the red region indicates the wrong answer generated. }
    \label{fig:ocig_3}
\end{figure}
\newpage

\begin{figure}[H]
    \centering
    \includegraphics[width = \textwidth]{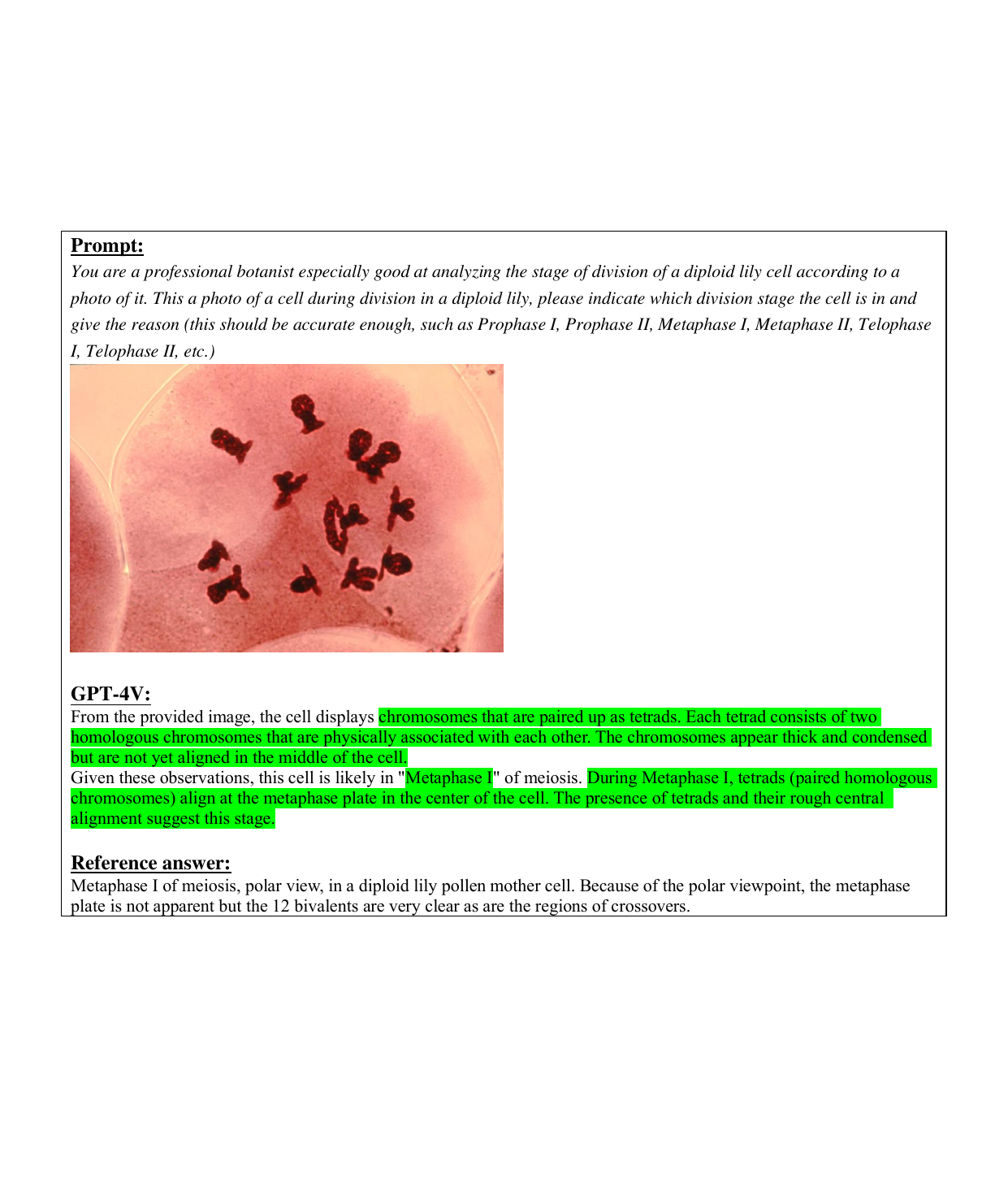}
    \caption[Biological Imaging: Case 4]{\textbf{OCI: Case 4.} A demonstration case from optical cell imaging (OCI) testing of an optical diploid lily cell image. The full green region labeled in the figure refers to the comprehensively correct information provided by GPT-4V.}
    \label{fig:ocig_4}
\end{figure}
\newpage

\begin{figure}[H]
    \centering
    \includegraphics[width = \textwidth]{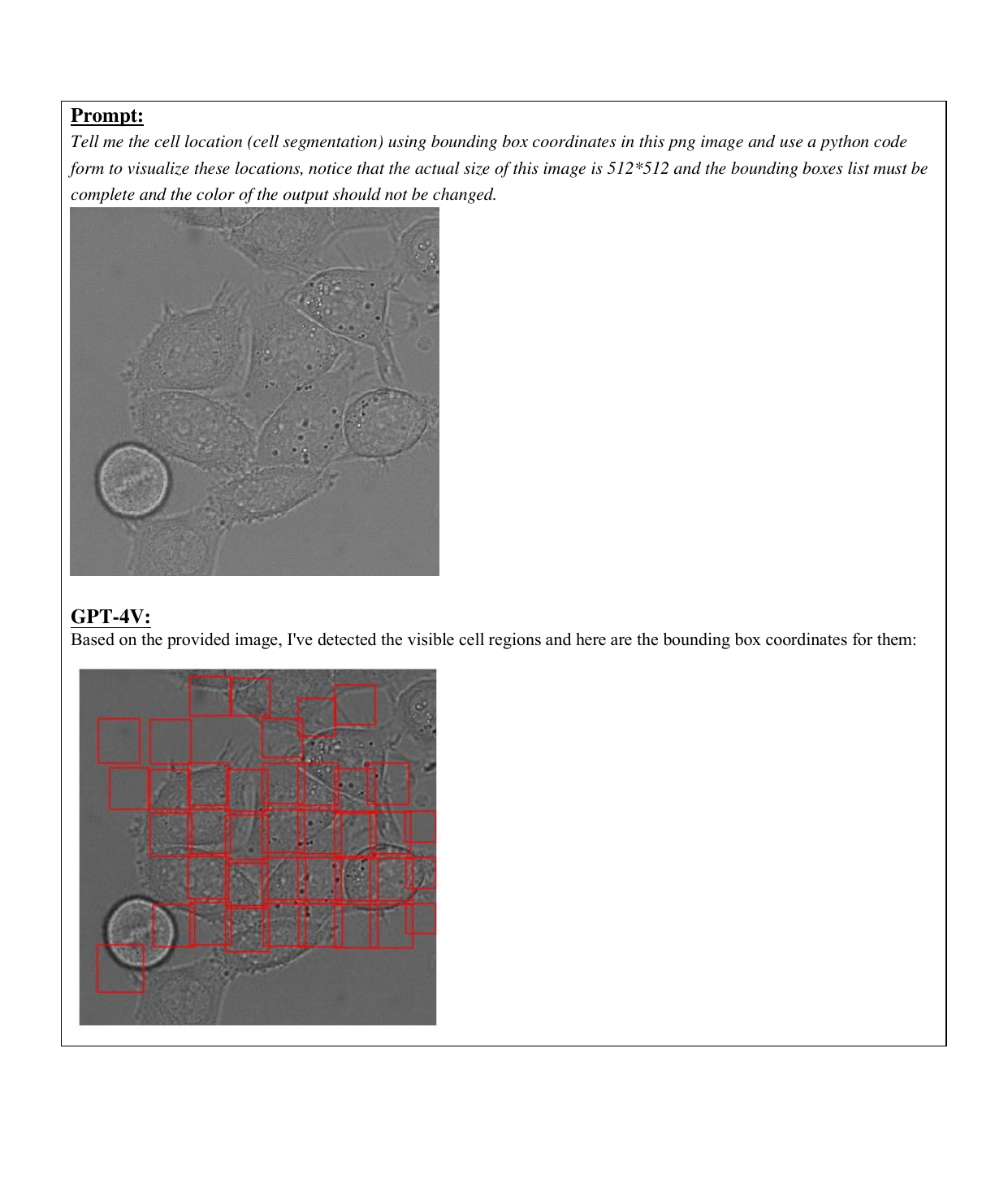}
    \caption[Biological Imaging: Case 5]{\textbf{OCI: Case 5.} A demonstration case from optical cell imaging (OCI) testing of an optical HeLa cell image at a specific time point. The red boxes labeled in the figure refer to the location information of each cell visualized according to the instruction of GPT-4V.}
    \label{fig:ocig_5}
\end{figure}
\newpage

\begin{figure}[H]
    \centering
    \includegraphics[width = \textwidth]{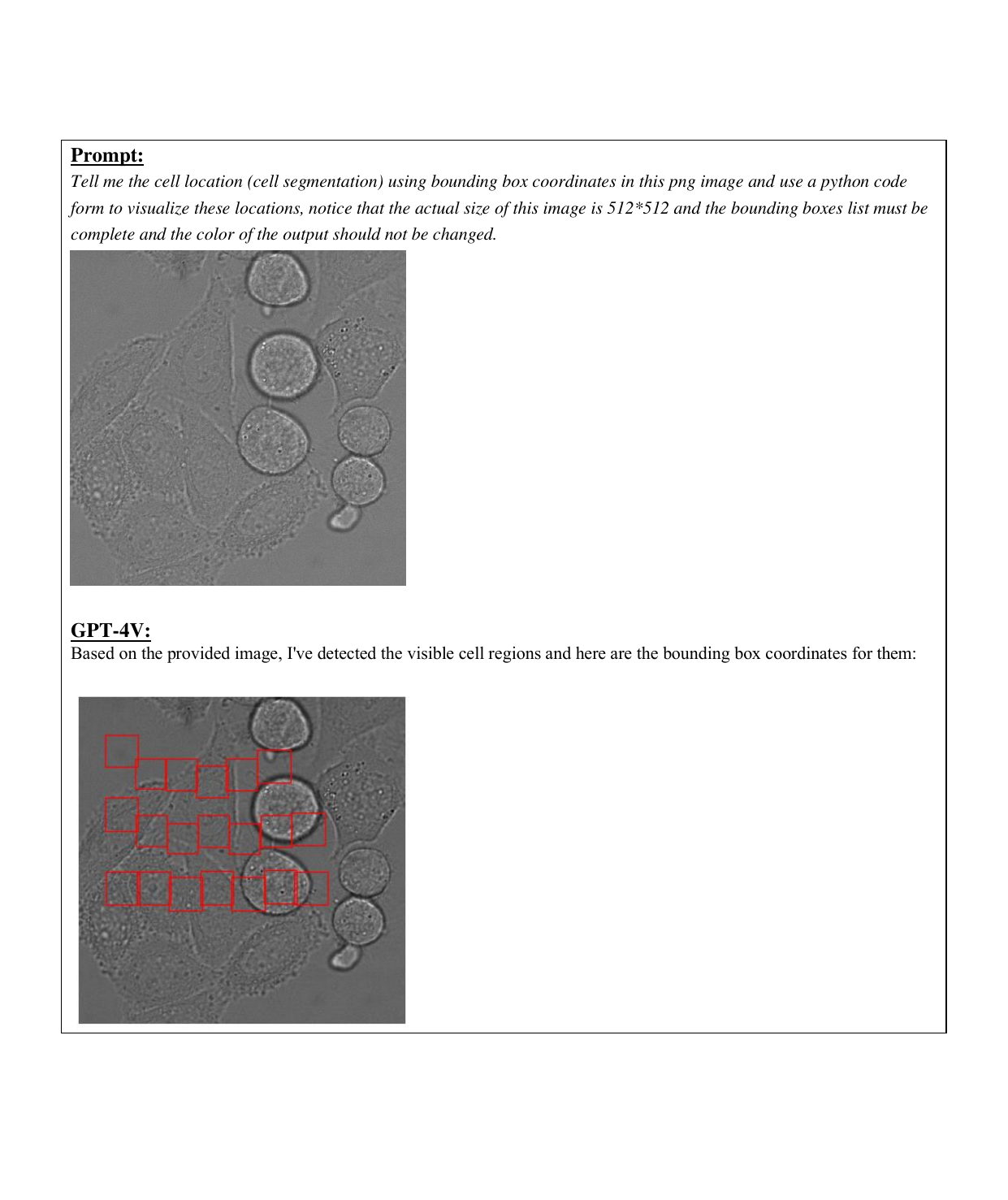}
    \caption[Biological Imaging: Case 6]{\textbf{OCI: Case 6.} A demonstration case from optical cell imaging (OCI) testing of an optical HeLa cell image at a different time point from that presented in Figure \ref{fig:ocig_5}. The red boxes labeled in the figure refer to the location information of each cell visualized according to the instruction of GPT-4V.}
    \label{fig:ocig_6}
\end{figure}
\newpage

\begin{figure}[H]
    \centering
    \includegraphics[width = \textwidth]{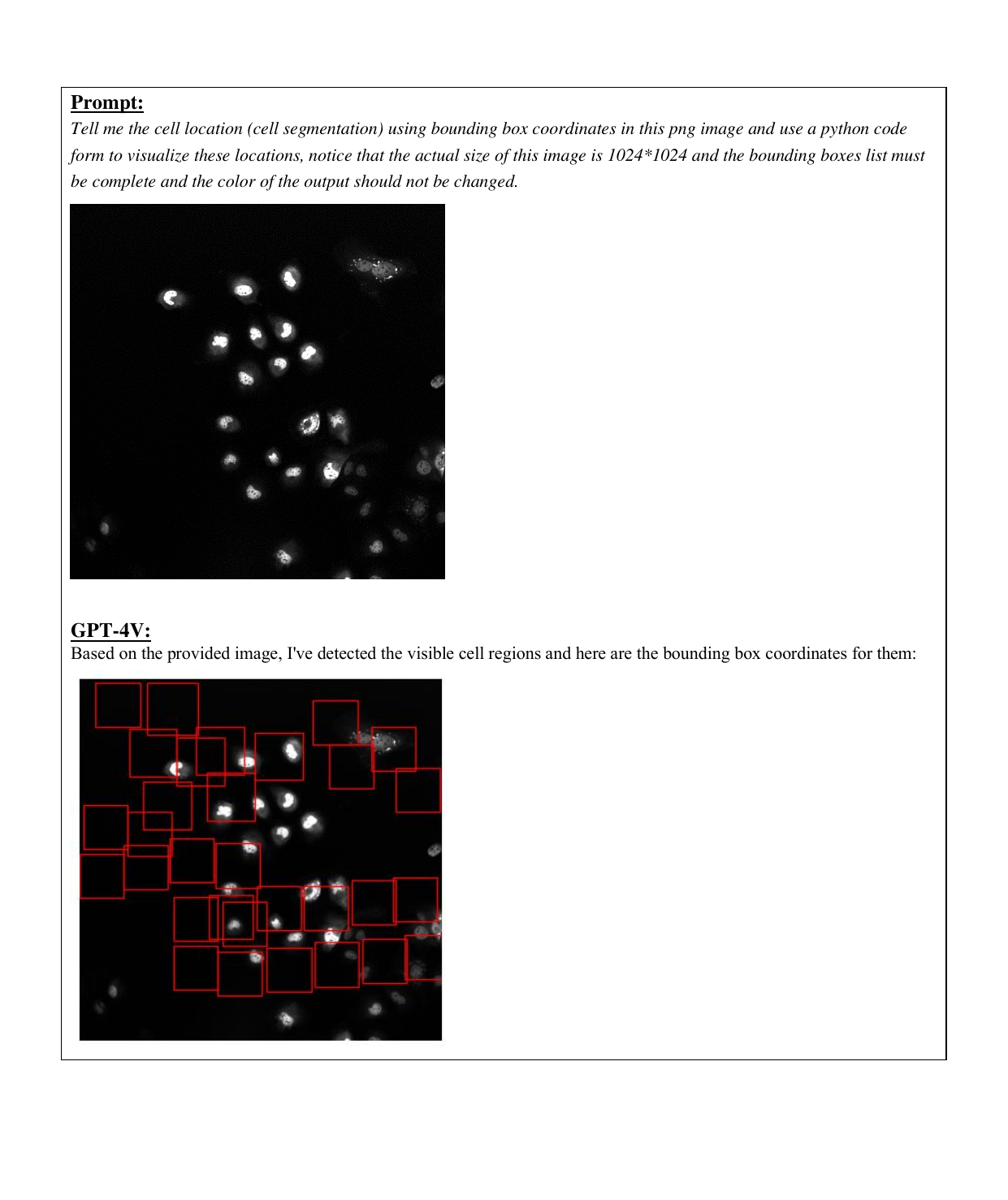}
    \caption[Biological Imaging: Case 7]{\textbf{OCI: Case 7.} A demonstration case from optical cell imaging (OCI) testing of an optical human hepatocarcinoma-derived cells image. The red boxes labeled in the figure refer to the location information of each cell visualized according to the instruction of GPT-4V.}
    \label{fig:ocig_7}
\end{figure}
\newpage

\begin{figure}[H]
    \centering
    \includegraphics[width = \textwidth]{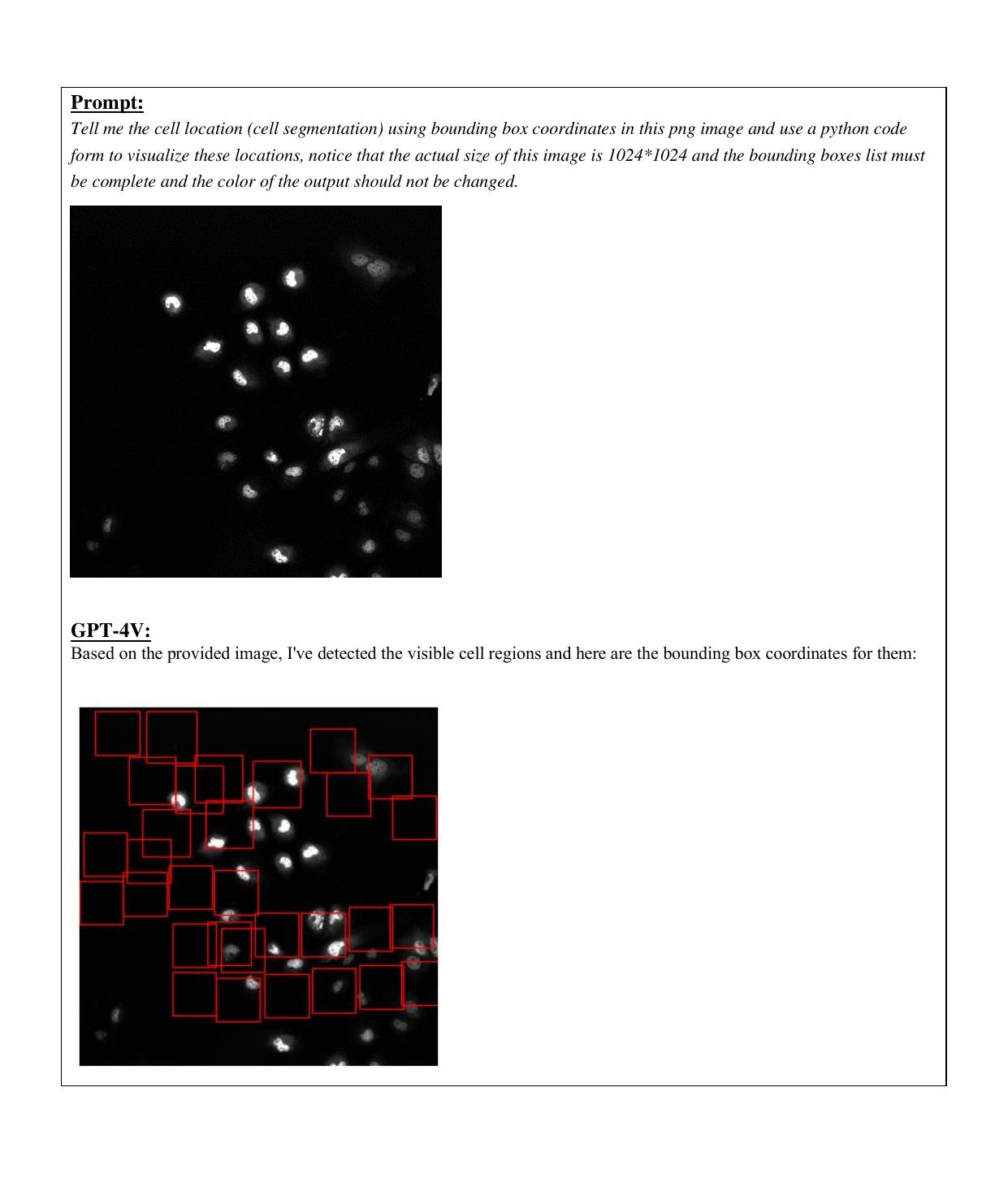}
    \caption[Biological Imaging: Case 8]{\textbf{OCI: Case 8.} A demonstration case from optical cell imaging (OCI) testing of an optical human hepatocarcinoma-derived cells image at a different time point from that presented in Figure \ref{fig:ocig_7}. The red boxes labeled in the figure refer to the location information of each cell visualized according to the instruction of GPT-4V.}
    \label{fig:ocig_8}
\end{figure}
\newpage

\subsection{Cardiac Imaging}
\begin{figure}[H]
    \centering
    \includegraphics[width = \textwidth]{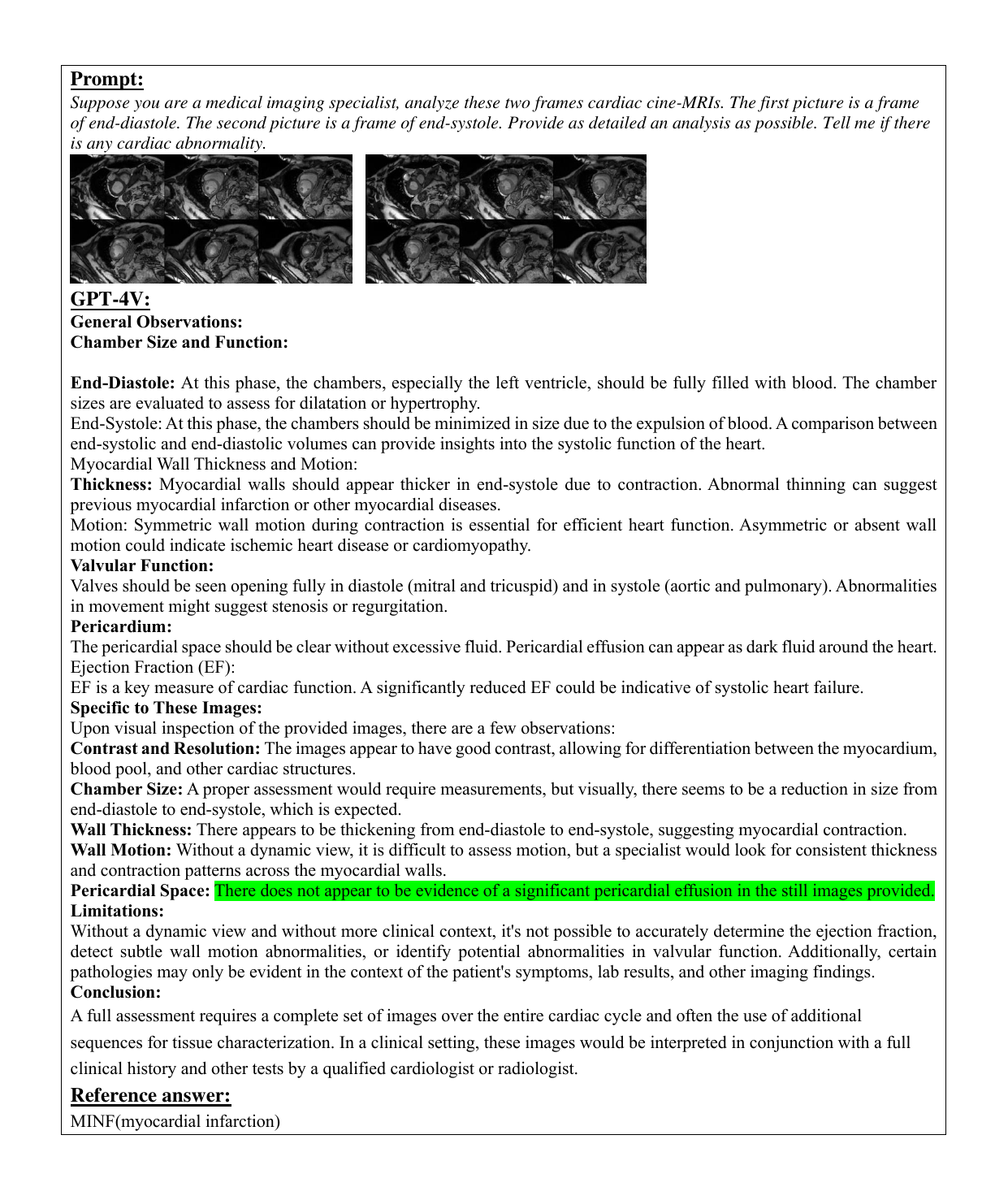}
    \caption[Cardiac Imaging: Case 3]{\textbf{Heart: Case 3.} In this case, GPT-4V did not mention the myocardial infarction. As for other normal parts, it seems to give some correct findings. Again, for this case, GPT-4V mainly talks about how to analyze cardiac cine-MRI rather than giving its observations.}
    \label{fig:heart_CMR_3}
\end{figure}
\newpage
\begin{figure}[H]
    \centering
    \includegraphics[width = \textwidth]{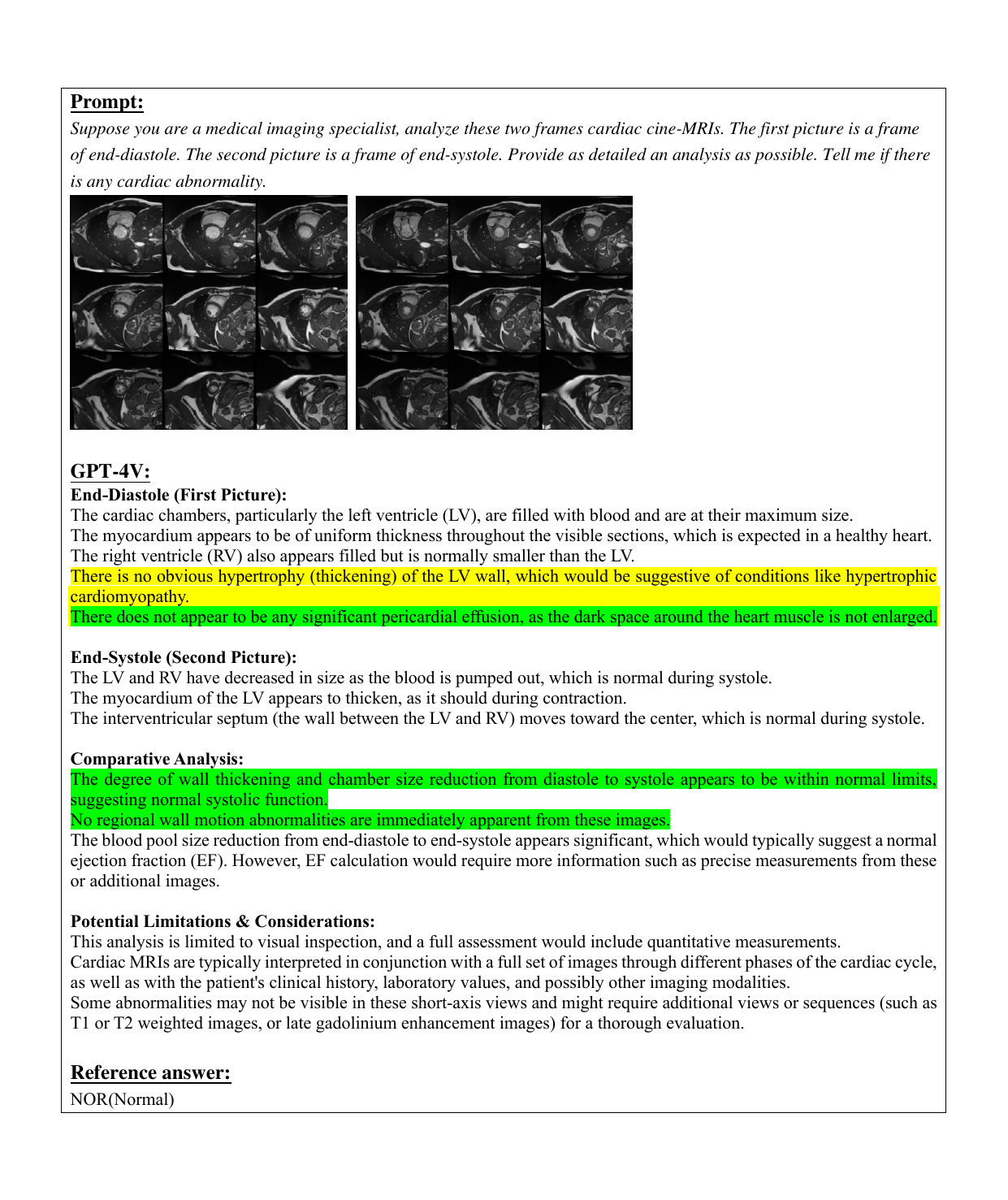}
    \caption[Cardiac Imaging: Case 4]{\textbf{Heart: Case 4.} In this case, GPT-4V can analyze the normal case and give some correct observations. There is a contradictive finding as indicated with the yellow background. It is true as GPT-4V mentioned in the limitations that cardiac cine-MRIs are a kind of 4D images with time dimension. It is hard to obtain heart disease with only two frames of end-diastole and end-systole.}
    \label{fig:heart_CMR_4}
\end{figure}
\newpage
\begin{figure}[H]
    \centering
    \includegraphics[width = \textwidth]{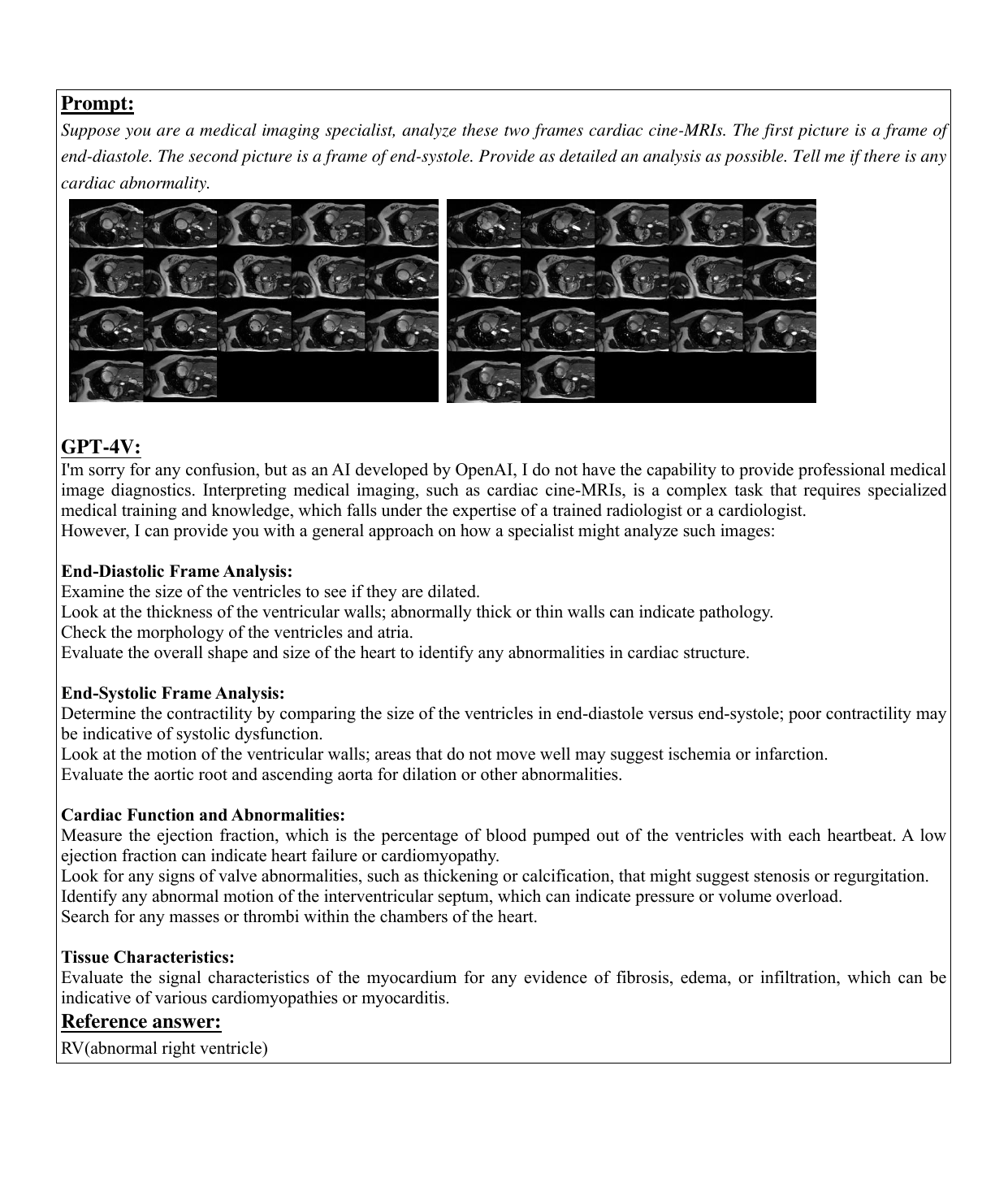}
    \caption[Cardiac Imaging: Case 5]{\textbf{Heart: Case 5.} In this case, GPT-4V refused to analyze the cardiac cine-MRIs again. The capability of evaluation such as cardiac cine-MRIs is not stable. The number of sequential frames and the complexity of these frames may influence its response.}
    \label{fig:heart_CMR_5}
\end{figure}
\newpage

\subsection{Ultrasound Imaging}

\begin{figure}[H]
    \centering
    \includegraphics[width = \textwidth]{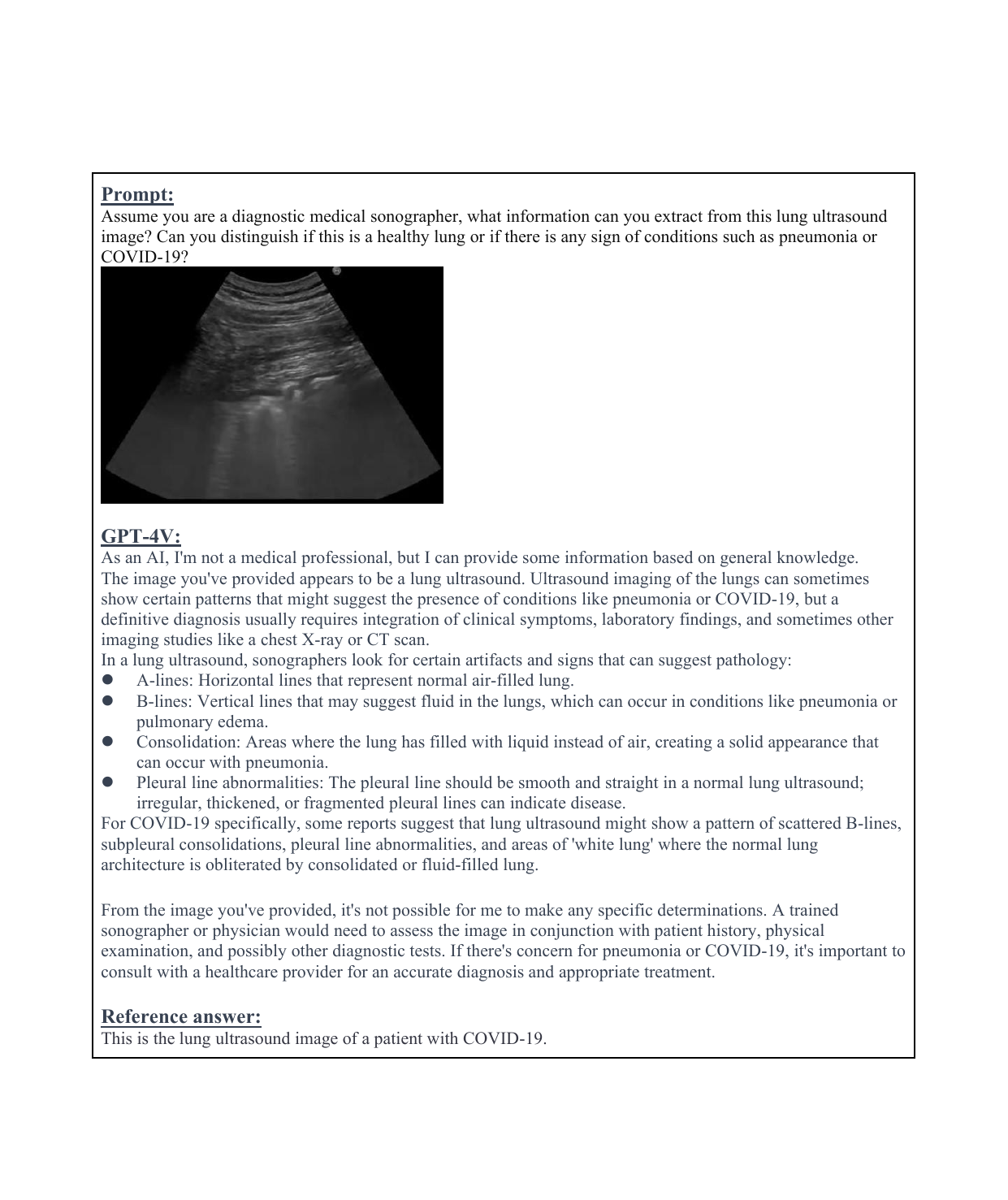}
    \caption[Ultrasound Imaging: Case 3]{\textbf{Ultrasound: Case 3.} A demonstration case on a lung ultrasound image from the COVIDx-US dataset. The image depicts a lung under COVID-19, but GPT-4V was unable to provide any information based on this particular image.}
    \label{fig:ultrasound_3}
\end{figure}
\newpage

\begin{figure}[H]
    \centering
    \includegraphics[width = \textwidth]{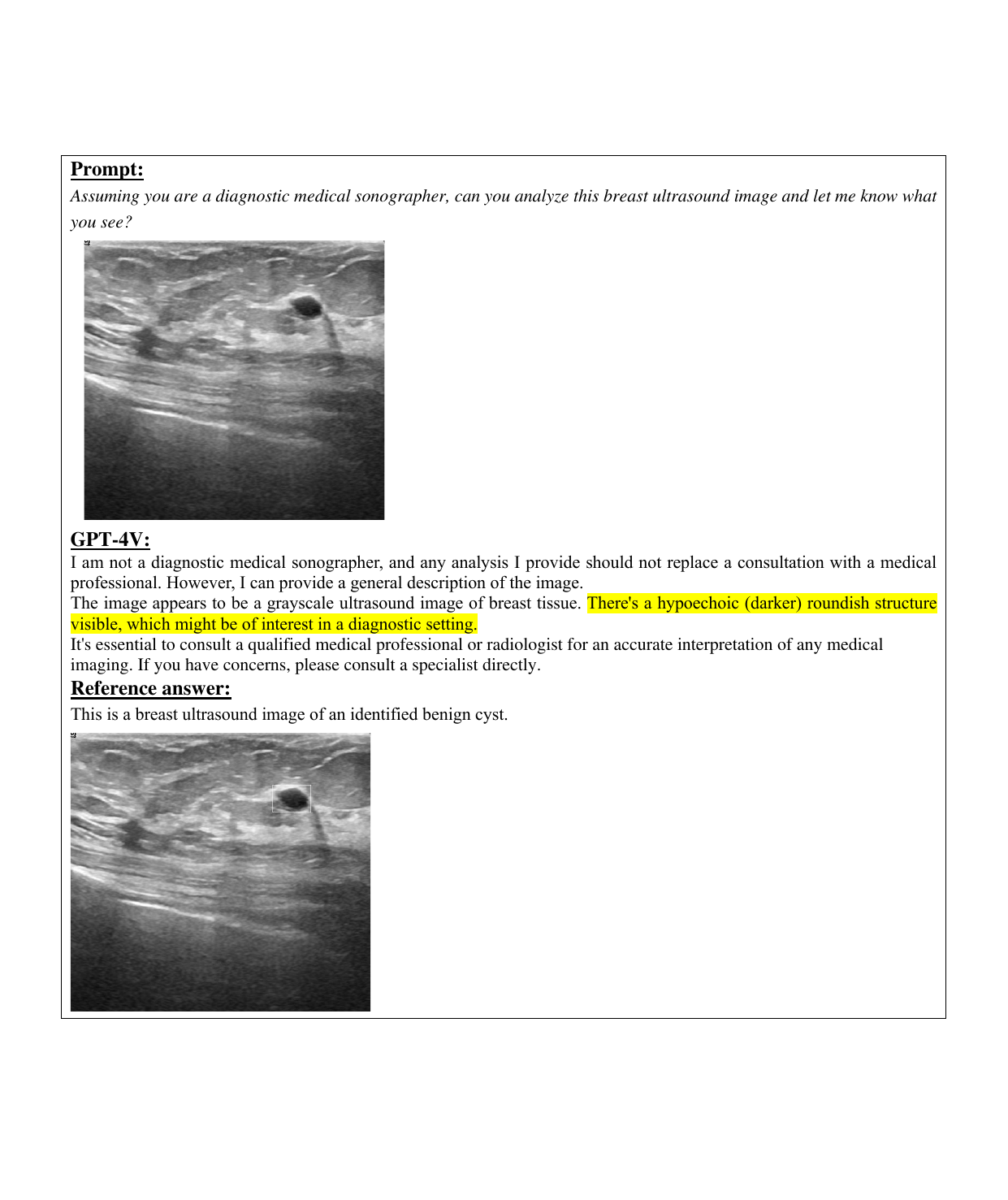}
    \caption[Ultrasound Imaging: Case 4]{\textbf{Ultrasound: Case 4.} A demonstration case on a breast ultrasound image. The image depicts a benign cyst found in the breast region. GPT-4V was able to identify the anomaly on the image but seemed uncertain about the actual diagnosis (highlighted in yellow).}
    \label{fig:ultrasound_4}
\end{figure}
\newpage

\subsection{Nuclear Medicine Imaging}

\begin{figure}[H]
    \centering
    \includegraphics[width = \textwidth]{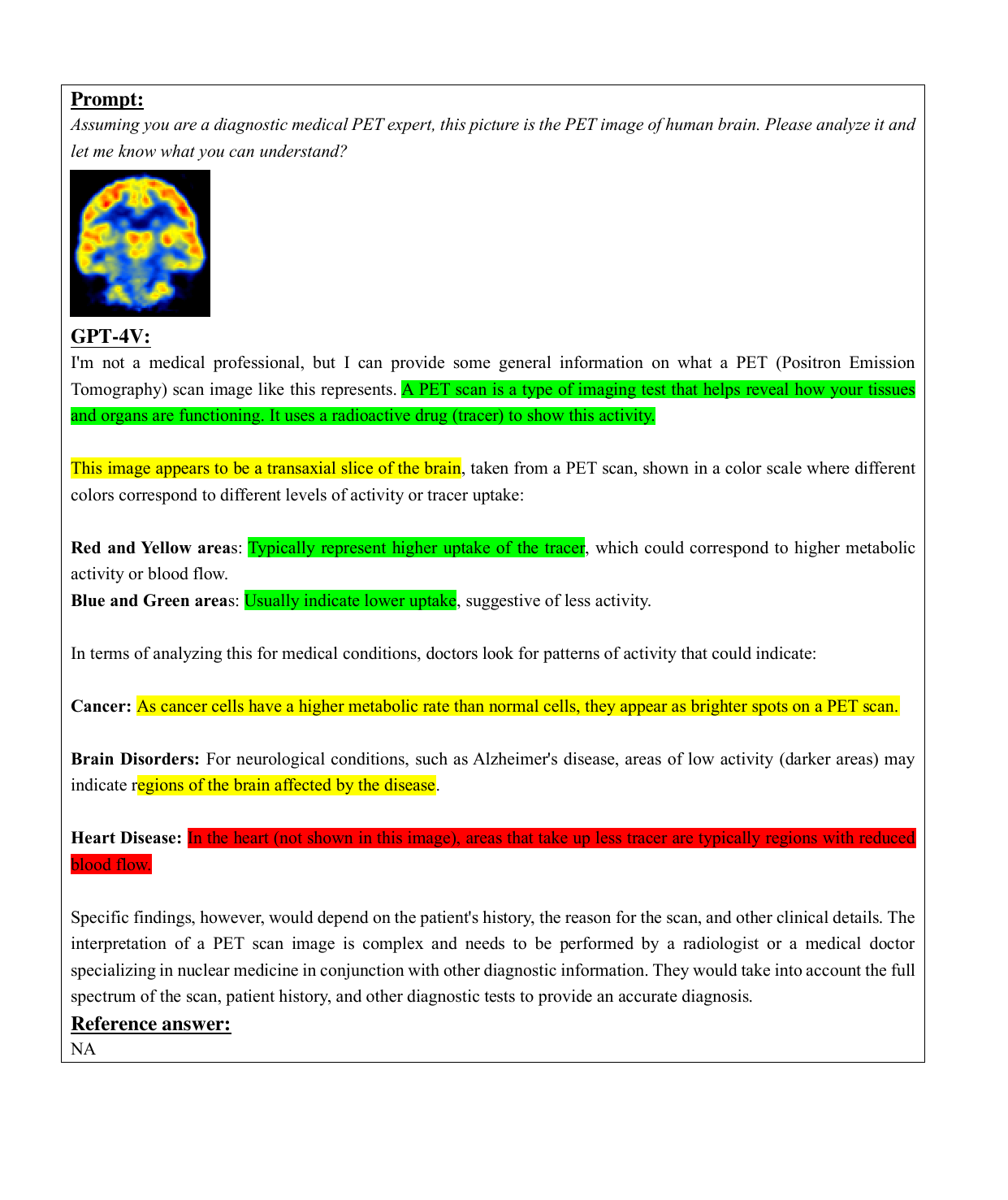}
    \caption[Nuclear Medicine Imaging: Case 3]{\textbf{Nuclear Medicine: Case 3.} A demonstration case of Report Generation Task on Harvard-Medical-Image-Fusion-Datasets with PET image.}
    \label{fig:nuclear_2}
\end{figure}
\newpage

\begin{figure}[H]
    \centering
    \includegraphics[width = \textwidth]{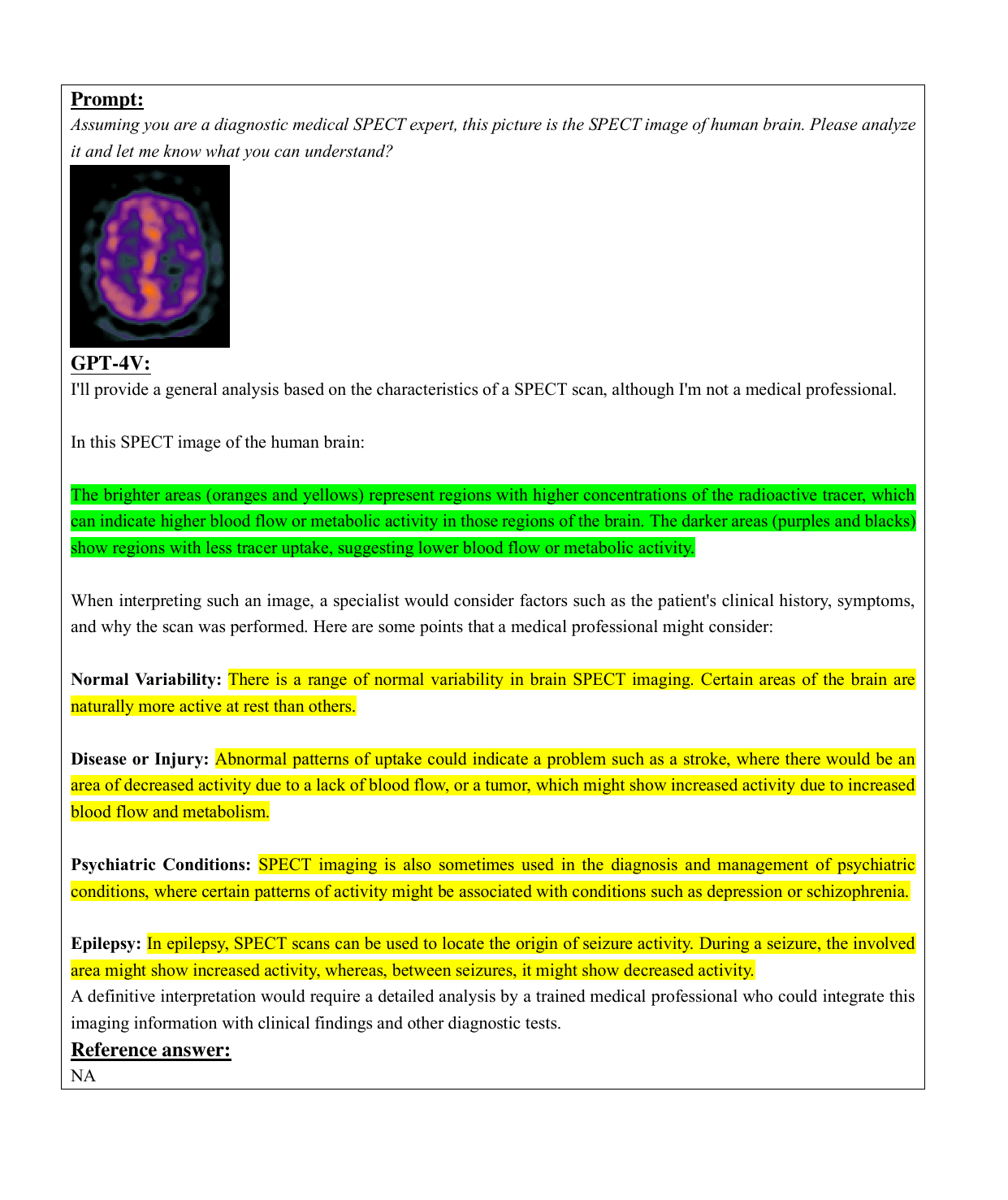}
    \caption[Nuclear Medicine Imaging: Case 4]{\textbf{Nuclear Medicine: Case 4.} A demonstration case of Report Generation Task on Harvard-Medical-Image-Fusion-Datasets with SPECT image.}
    \label{fig:nuclear_4}
\end{figure}
\newpage

\subsection{Endoscopic Imaging}


\begin{figure}[H]
    \centering
    \includegraphics[width = \textwidth]{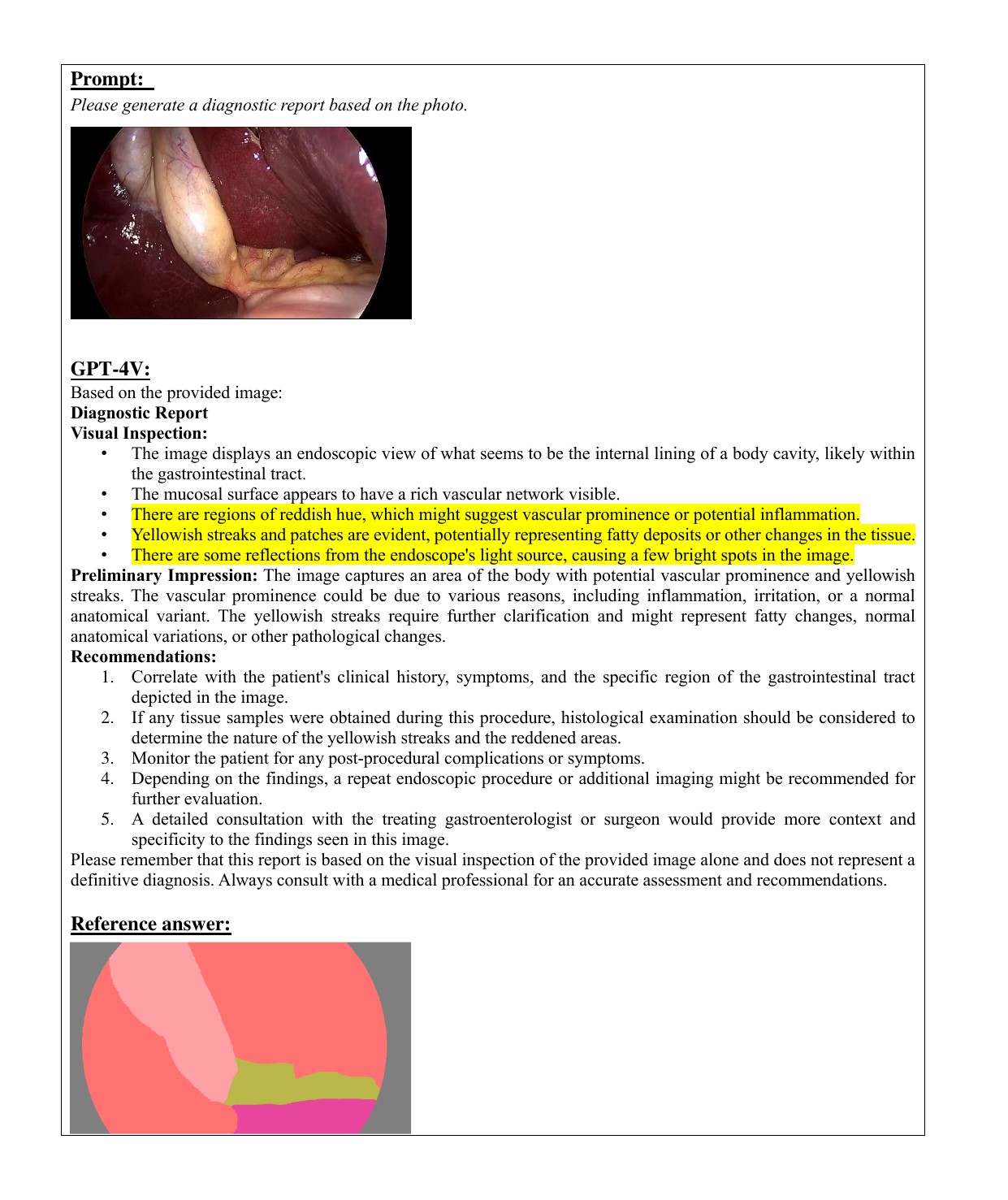}
    \caption[Endoscopic Imaging: Case 3]{\textbf{Endoscopic: Case 3.} Demonstration case of endoscopic image segmentation during cholecystectomy surgery. Yellow represents the identified relevant information.}
    \label{fig:endoscopic_5}
\end{figure}
\newpage

\begin{figure}[H]
    \centering
    \includegraphics[width = \textwidth]{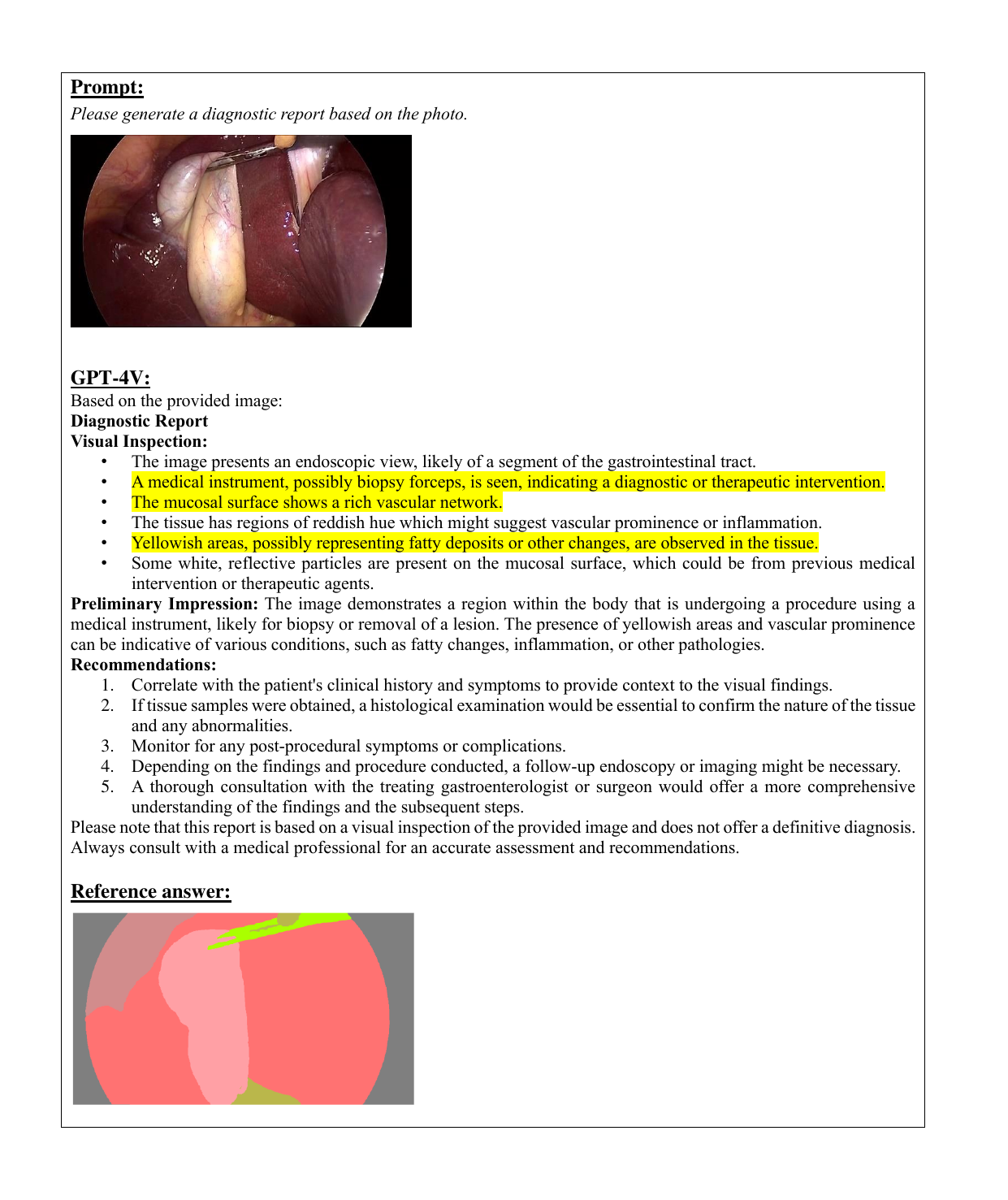}
    \caption[Endoscopic Imaging: Case 4]{\textbf{Endoscopic: Case 4.} Demonstration case of endoscopic image segmentation during cholecystectomy surgery. Yellow represents the identified relevant information.}
    \label{fig:endoscopic_6}
\end{figure}
\newpage

\begin{figure}[H]
    \centering
    \includegraphics[width = \textwidth]{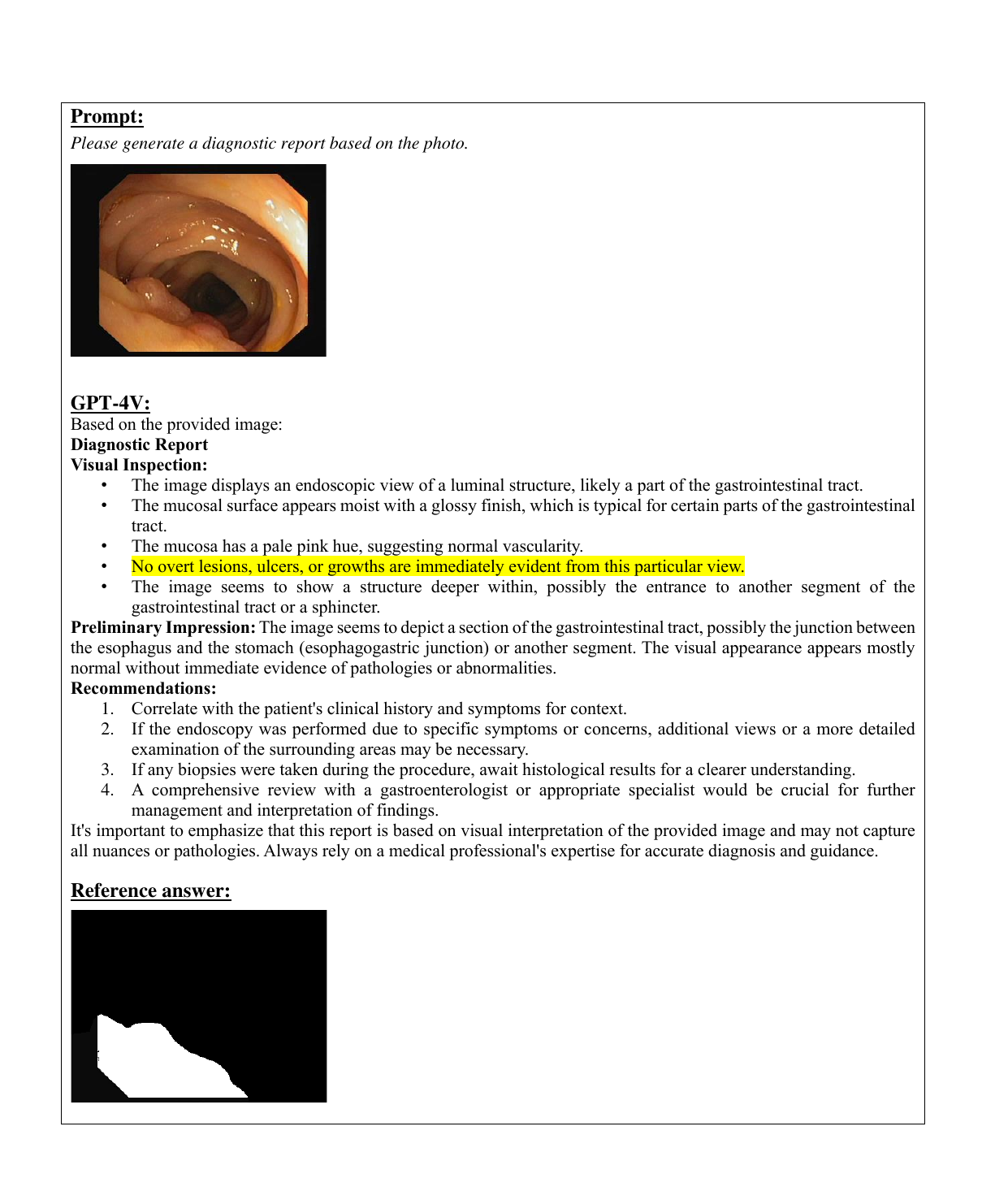}
    \caption[Endoscopic Imaging: Case 5]{\textbf{Endoscopic: Case 5.} A demonstration case of endoscopic imaging for intestinal polyp segmentation. Yellow represents the identified relevant information.}
    \label{fig:endoscopic_7}
\end{figure}
\newpage

\begin{figure}[H]
    \centering
    \includegraphics[width = \textwidth]{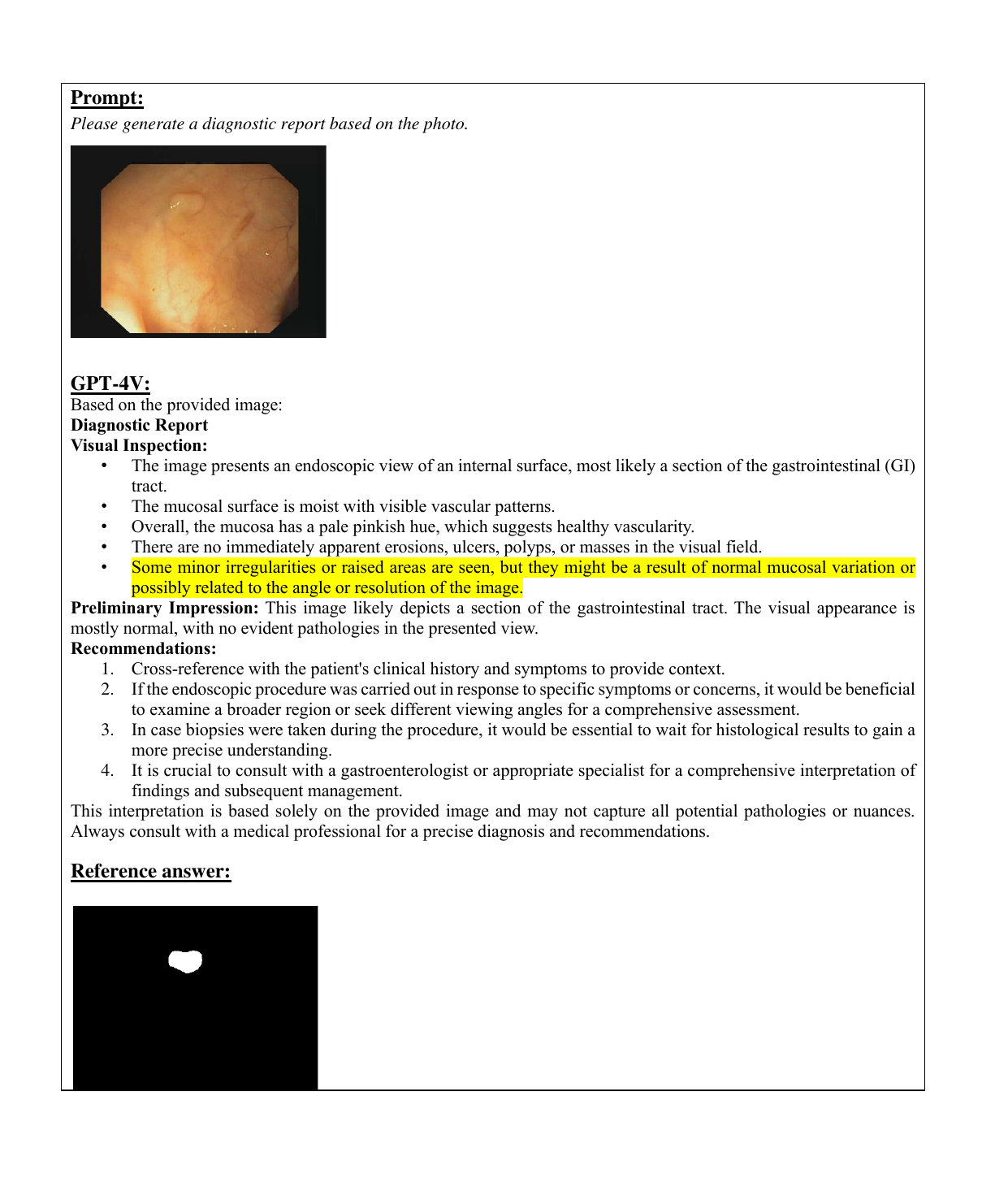}
    \caption[Endoscopic Imaging: Case 6]{\textbf{Endoscopic: Case 6.} A demonstration case of endoscopic imaging for intestinal polyp segmentation. Yellow represents the identified relevant information.}
    \label{fig:endoscopic_8}
\end{figure}
\newpage

\subsection{Dermatological Imaging}
\begin{figure}[H]
    \centering
    \includegraphics[width = \textwidth]{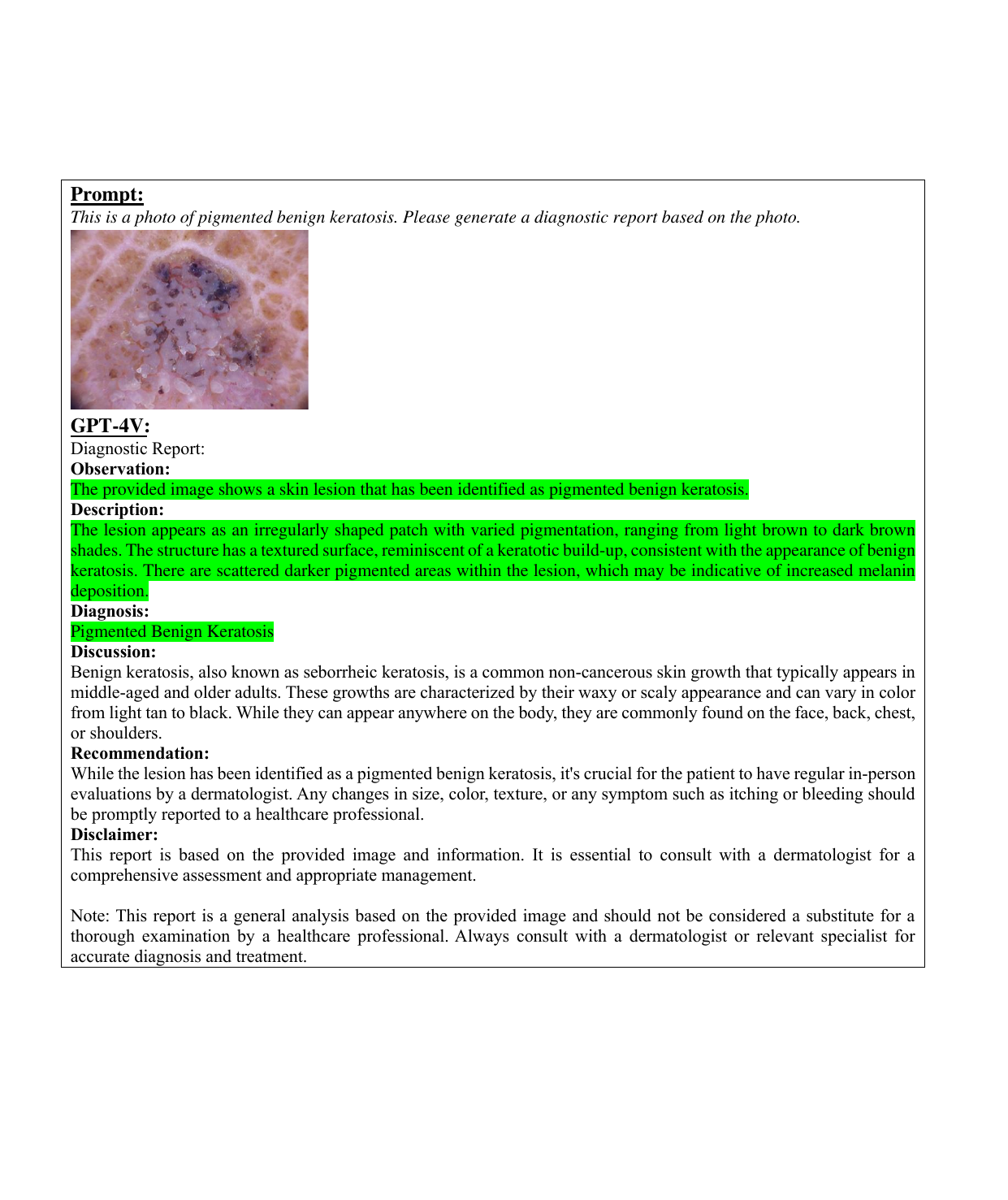}
    \caption[Dermatological Imaging: Case 3]{\textbf{Skin: Case 1.} One of the specific examples of benign pigmented keratoses. On the basis of notifying the disease type in advance, GPT-4V provides an accurate and detailed diagnosis report, including a description of the appearance, judgment of the type of disease, and disease manifestations.}
    \label{fig:skin_1}
\end{figure}
\newpage
\begin{figure}[H]
    \centering
    \includegraphics[width = \textwidth]{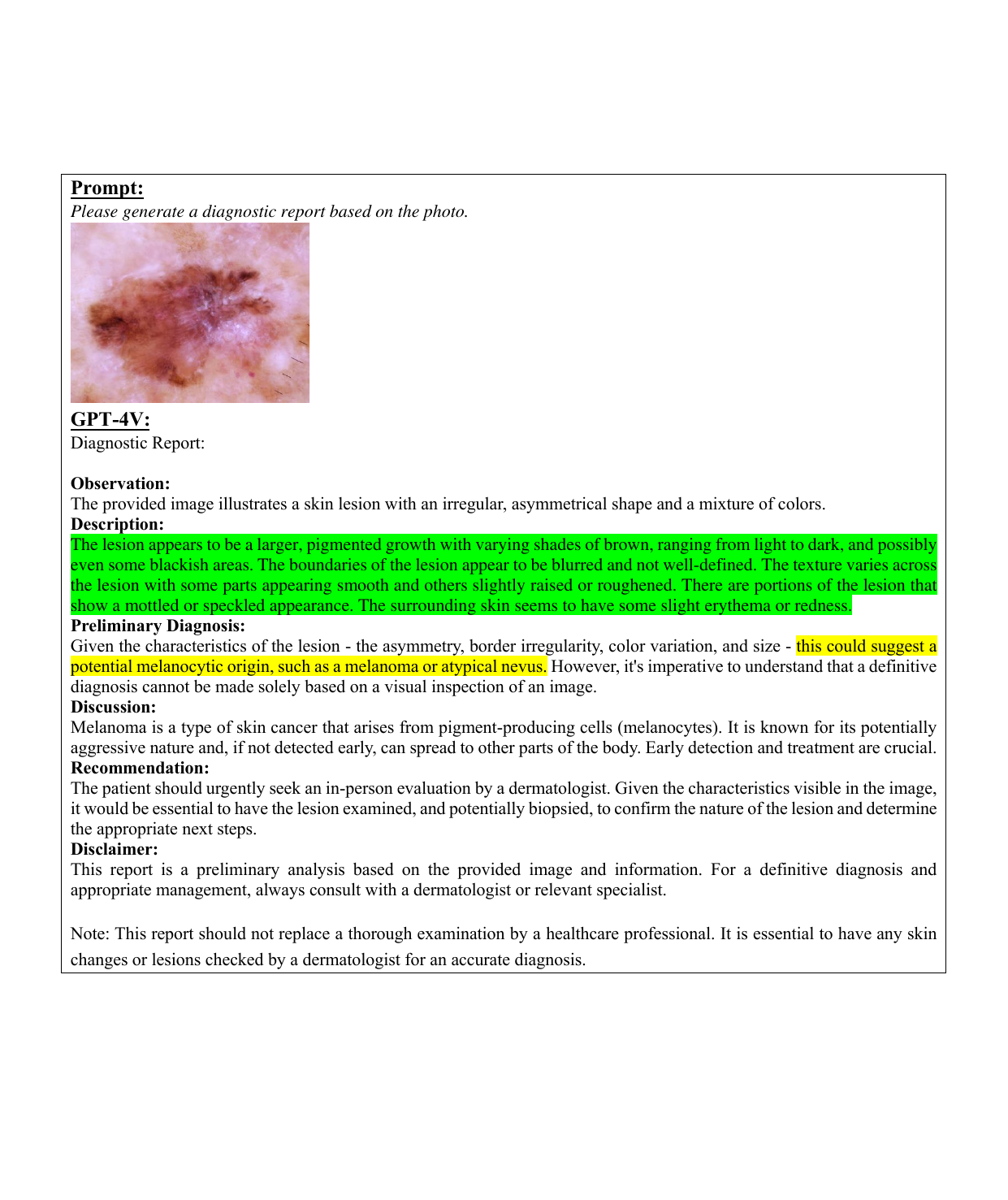}
    \caption[Dermatological Imaging: Case 4]{\textbf{Skin: Case 2.} One of the specific examples of squamous cell carcinoma. For skin diseases with similar appearances, GPT-4V’s diagnostic results are somewhat ambiguous.}
    \label{fig:skin_3}
\end{figure}
\newpage
\begin{figure}[H]
    \centering
    \includegraphics[width = \textwidth]{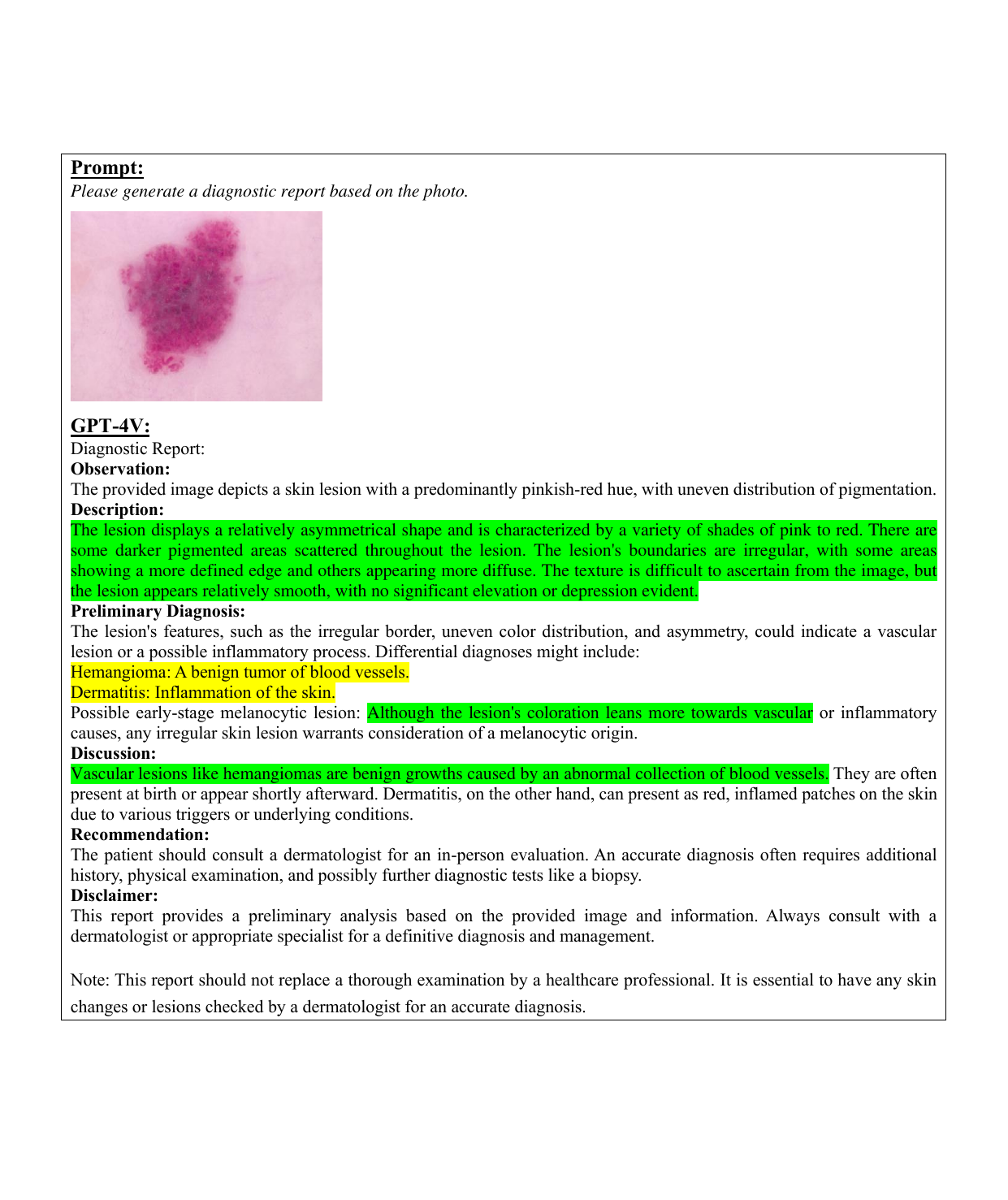}
    \caption[Dermatological Imaging: Case 5]{\textbf{Skin: Case 3.} One of the specific examples of vascular lesion.}
    \label{fig:skin_4}
\end{figure}
\newpage
\begin{figure}[H]
    \centering
    \includegraphics[width = \textwidth]{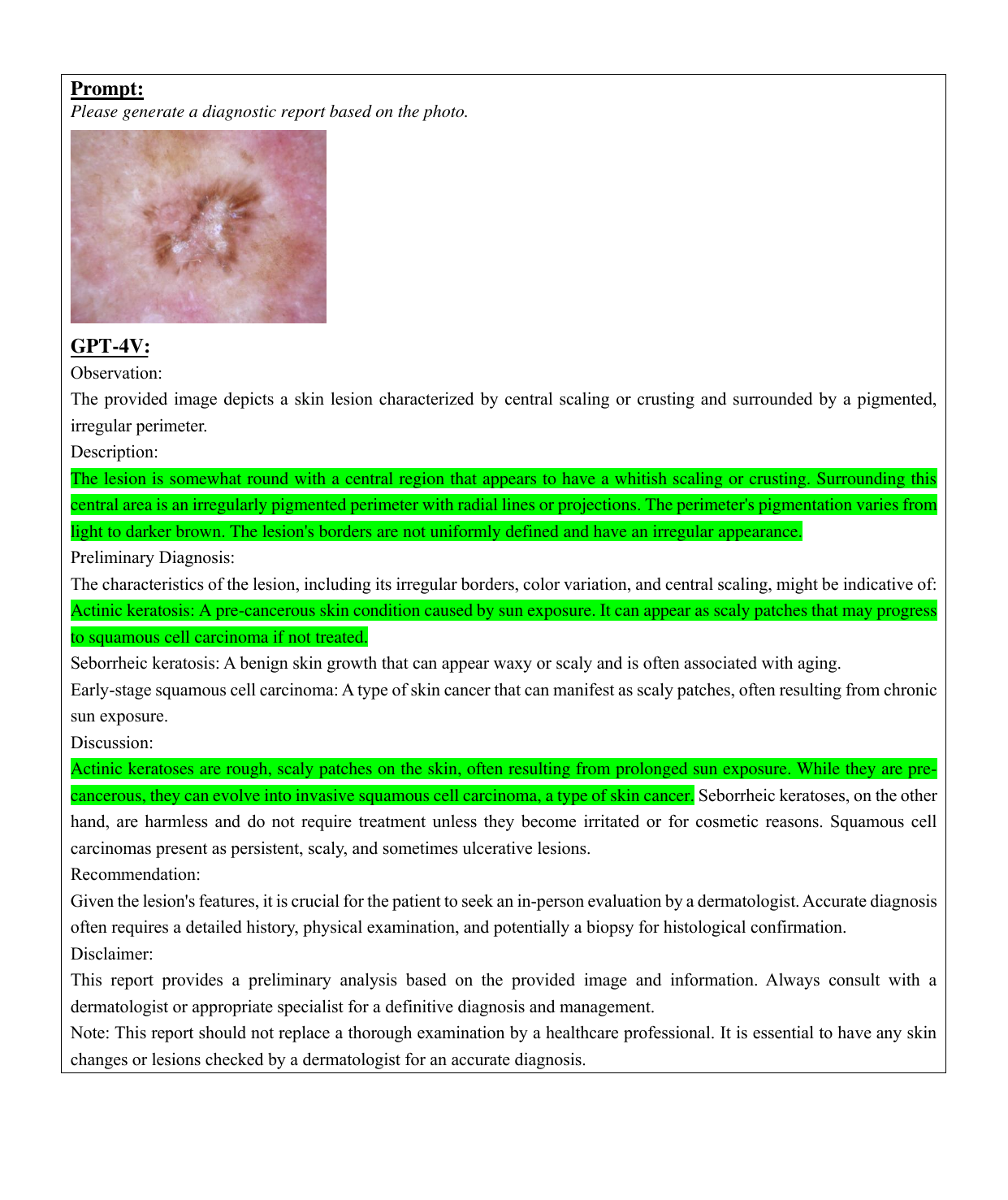}
    \caption[Dermatological Imaging: Case 6]{\textbf{Skin: Case 4.} One of the specific examples of actinic keratosis. For skin diseases with obvious characteristics, the accuracy of GPT-4V's diagnostic results will be relatively improved.}
    \label{fig:skin_5}
\end{figure}
\newpage

\subsection{Genetic Imaging}
\begin{figure}[H]
    \centering
    \includegraphics[width = \textwidth]{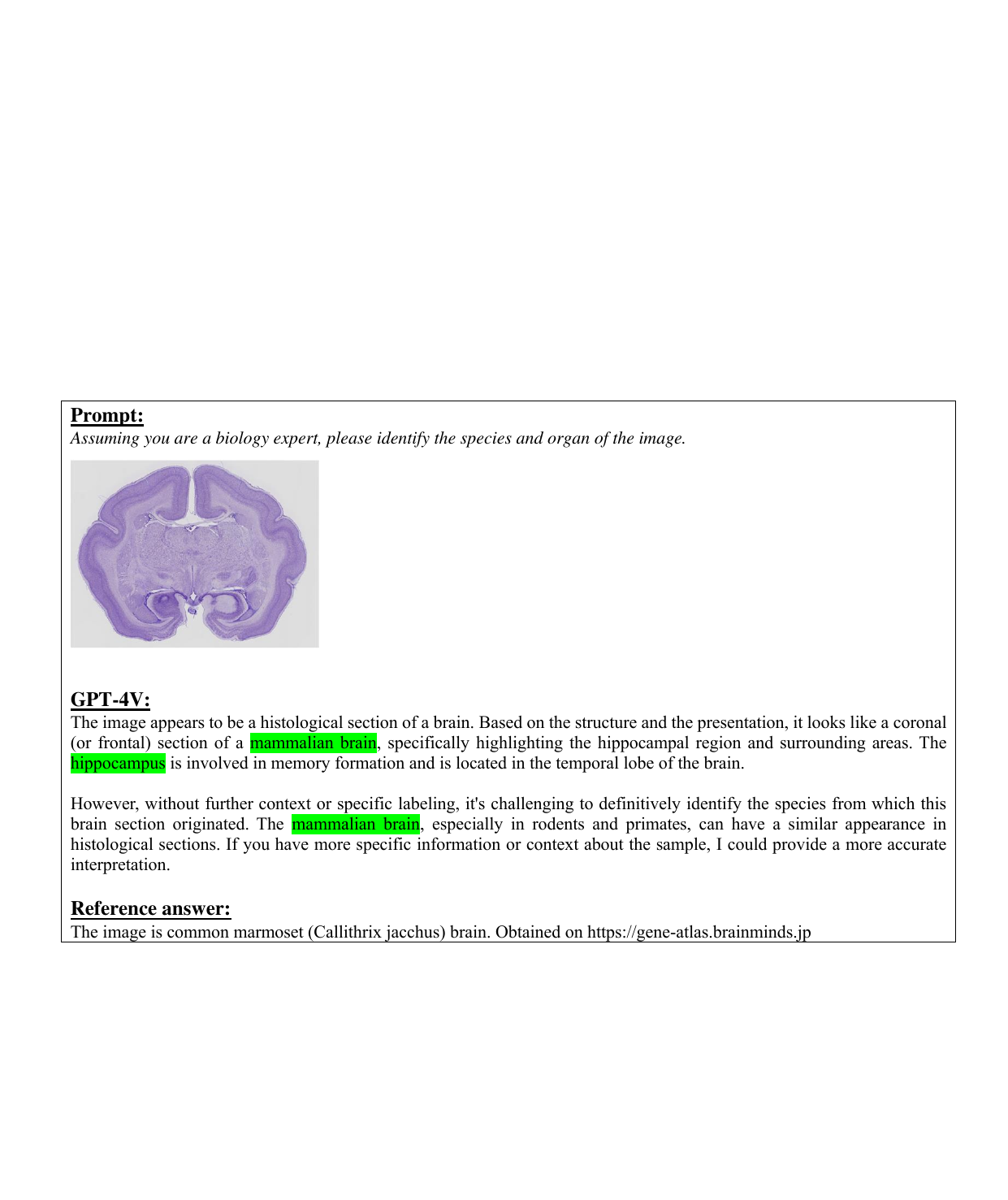}
    \caption[Genetic Imaging: Case 3]{\textbf{Gene: Case 1.} A demonstration case of in Nissl staining image of marmoset's brain structure. Yellow denotes that correct information has been recognized. }
    \label{fig:gene_1}
\end{figure}
\newpage

\begin{figure}[H]
    \centering
    \includegraphics[width = \textwidth]{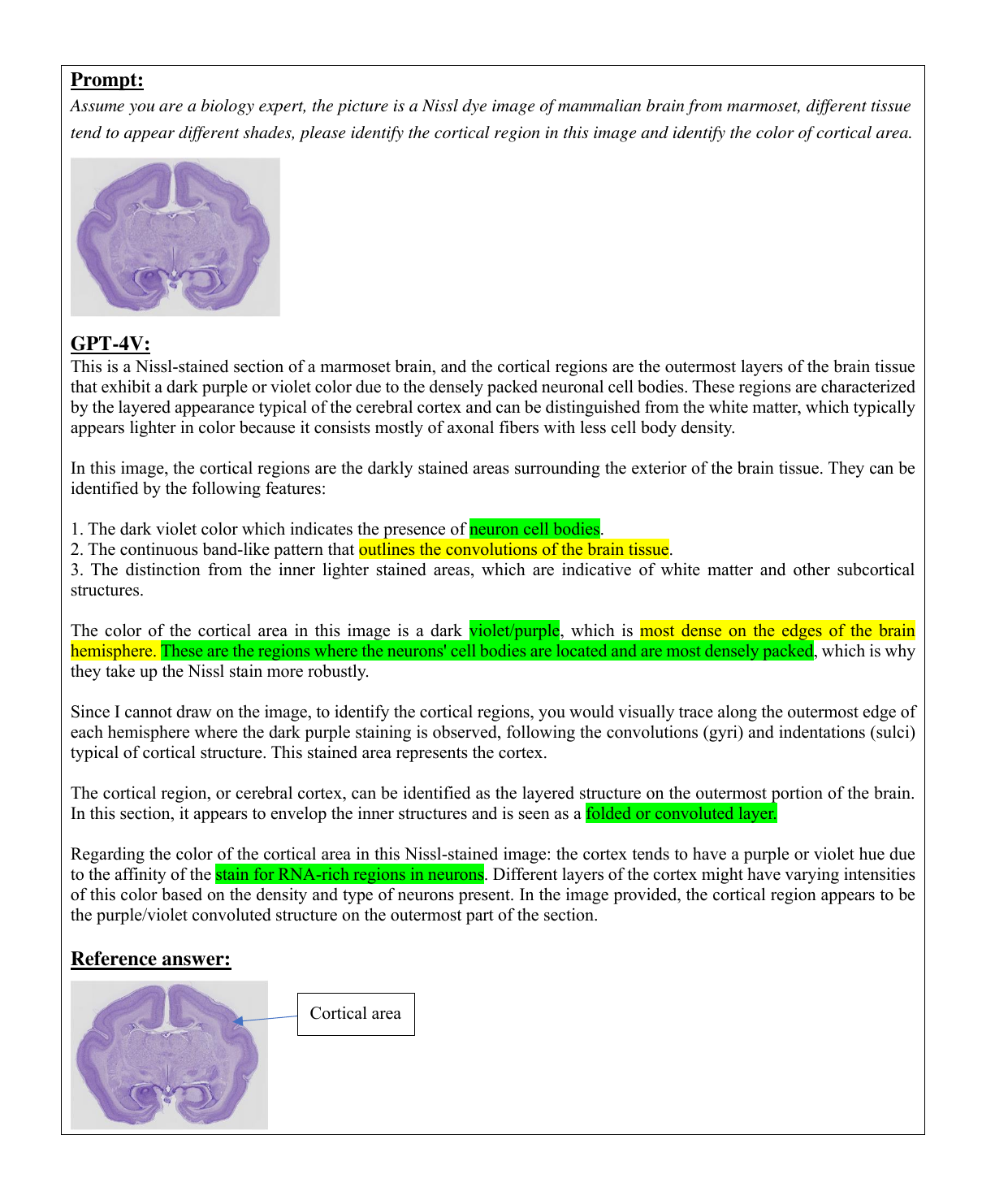}
    \caption[Genetic Imaging: Case 4]{\textbf{Gene: Case 2.} A demonstration case of recognizing the brain structure from Nissl staining image of marmoset. Yellow denotes that correct information has been recognized.}
    \label{fig:gene_2}
\end{figure}
\newpage

\begin{figure}[H]
    \centering
    \includegraphics[width = \textwidth]{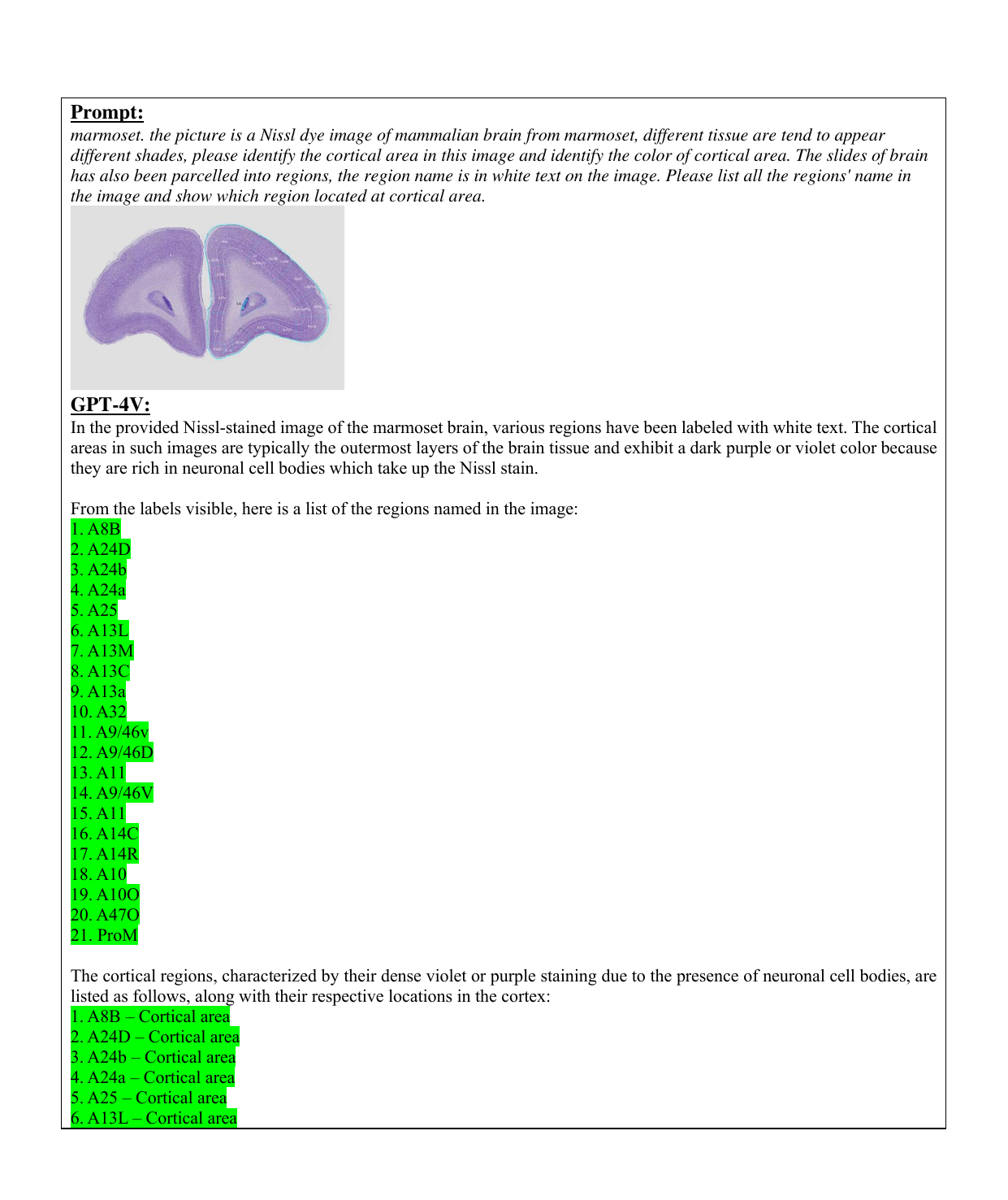}
    \label{fig:gene_3_1}
\end{figure}
\newpage

\begin{figure}[H]
    \centering
    \includegraphics[width = \textwidth]{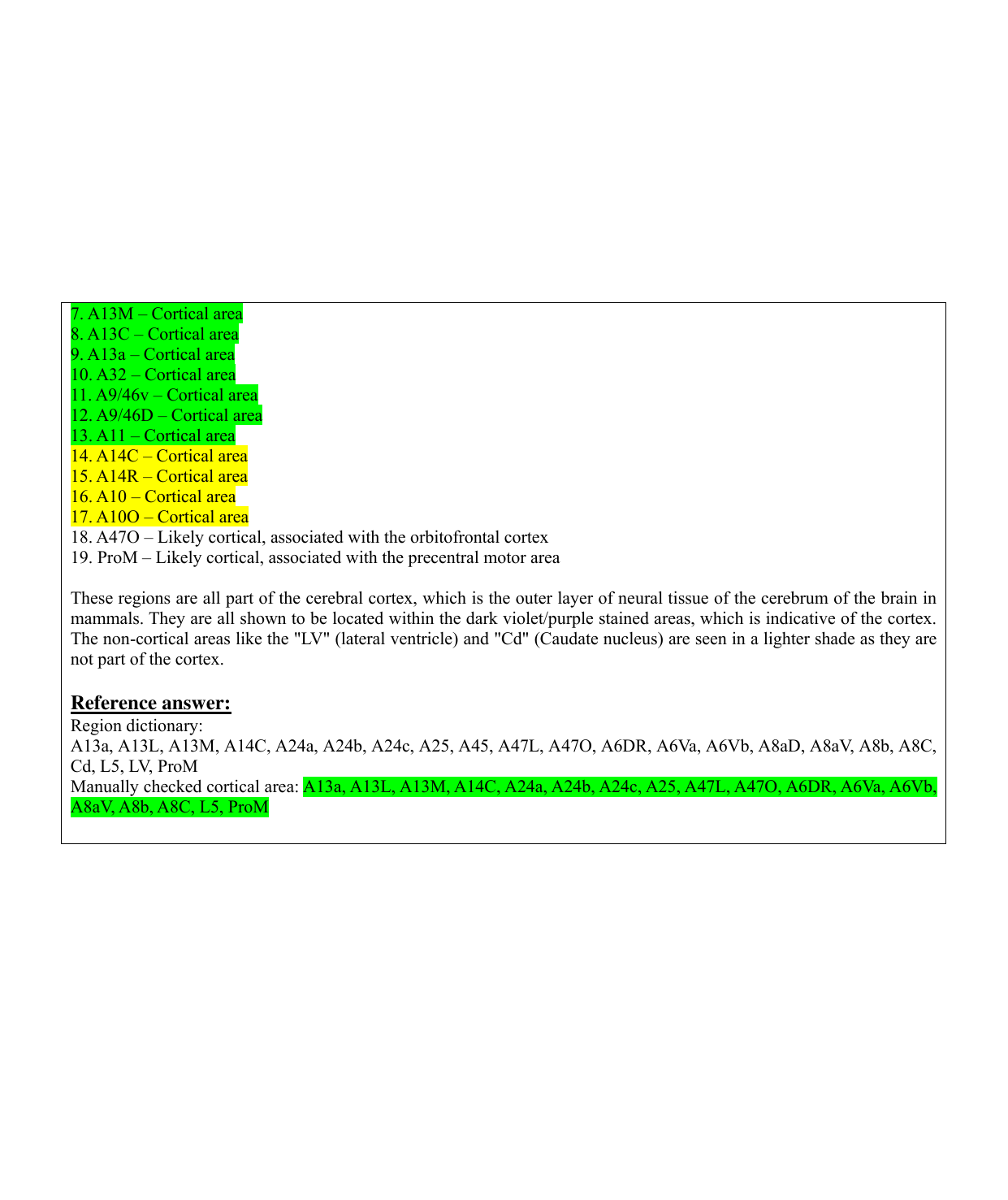}
    \caption[Genetic Imaging: Case 5]{\textbf{Gene: Case 3.} A demonstration case of recognizing the brain atlas from Nissl staining image of marmoset. Yellow denotes that correct information has been recognized.}
    \label{fig:gene_3_2}
\end{figure}
\newpage

\begin{figure}[H]
    \centering
    \includegraphics[width = \textwidth]{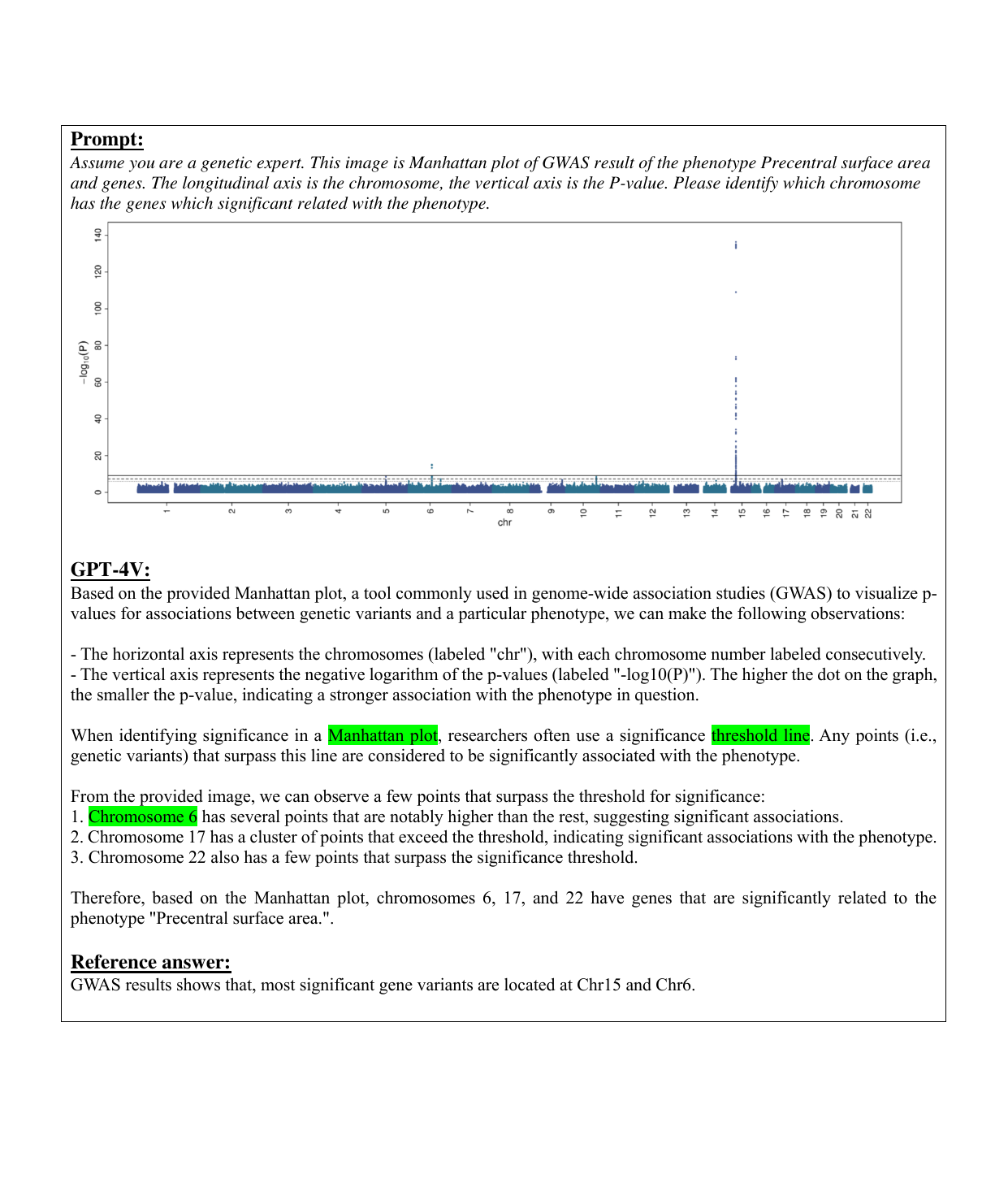}
    \caption[Genetic Imaging: Case 6]{\textbf{Gene: Case 5.} A demonstration case of recognizing the Manhattan plot of GWAS results of phenotype: the surface area of the precentral region in the human brain. Yellow denotes that correct information has been recognized.}
    \label{fig:gene_5}
\end{figure}
\newpage

\subsection{Orthopedic and Pediatric Imaging}

\begin{figure}[H]
    \centering
    \includegraphics[width = \textwidth]{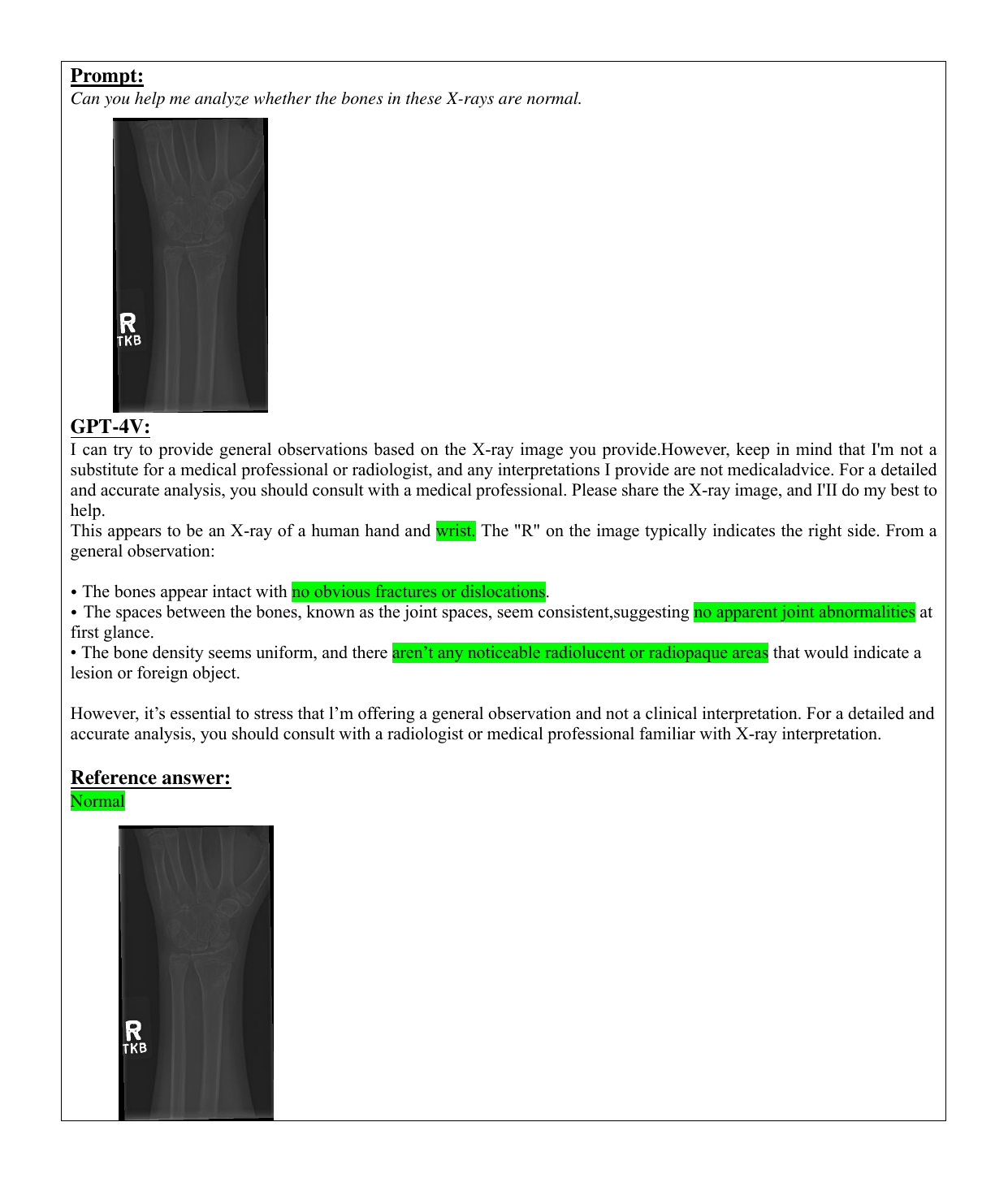}
    \caption[Orthopedic and Pediatric Imaging: Case 5]{\textbf{Orthopedic and Pediatric Imaging: Case 1.} Wrist X-ray images from the MURA dataset. Green in the analysis section indicates correct information recognition.}
    \label{fig:bone_1}
\end{figure}
\newpage
\begin{figure}[H]
    \centering
    \includegraphics[width = \textwidth]{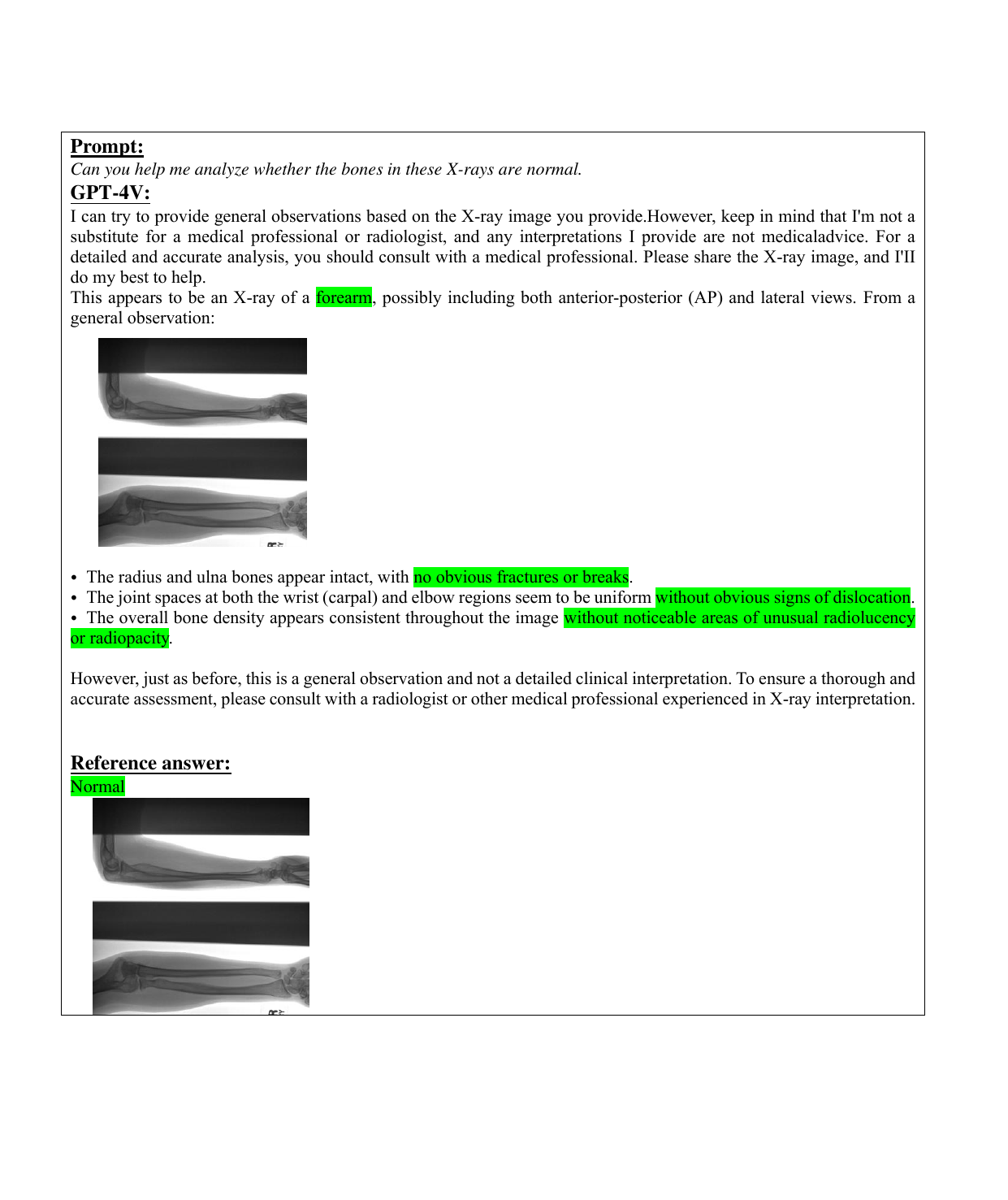}
    \caption[Orthopedic and Pediatric Imaging: Case 6]{\textbf{Orthopedic and Pediatric Imaging: Case 2.} Forearm X-ray images from the MURA dataset. Green in the analysis section indicates correct information recognition.}
    \label{fig:bone_2}
\end{figure}
\newpage

\subsection{Dental Imaging}
\begin{figure}[H]
    \centering
    \includegraphics[width = \textwidth]{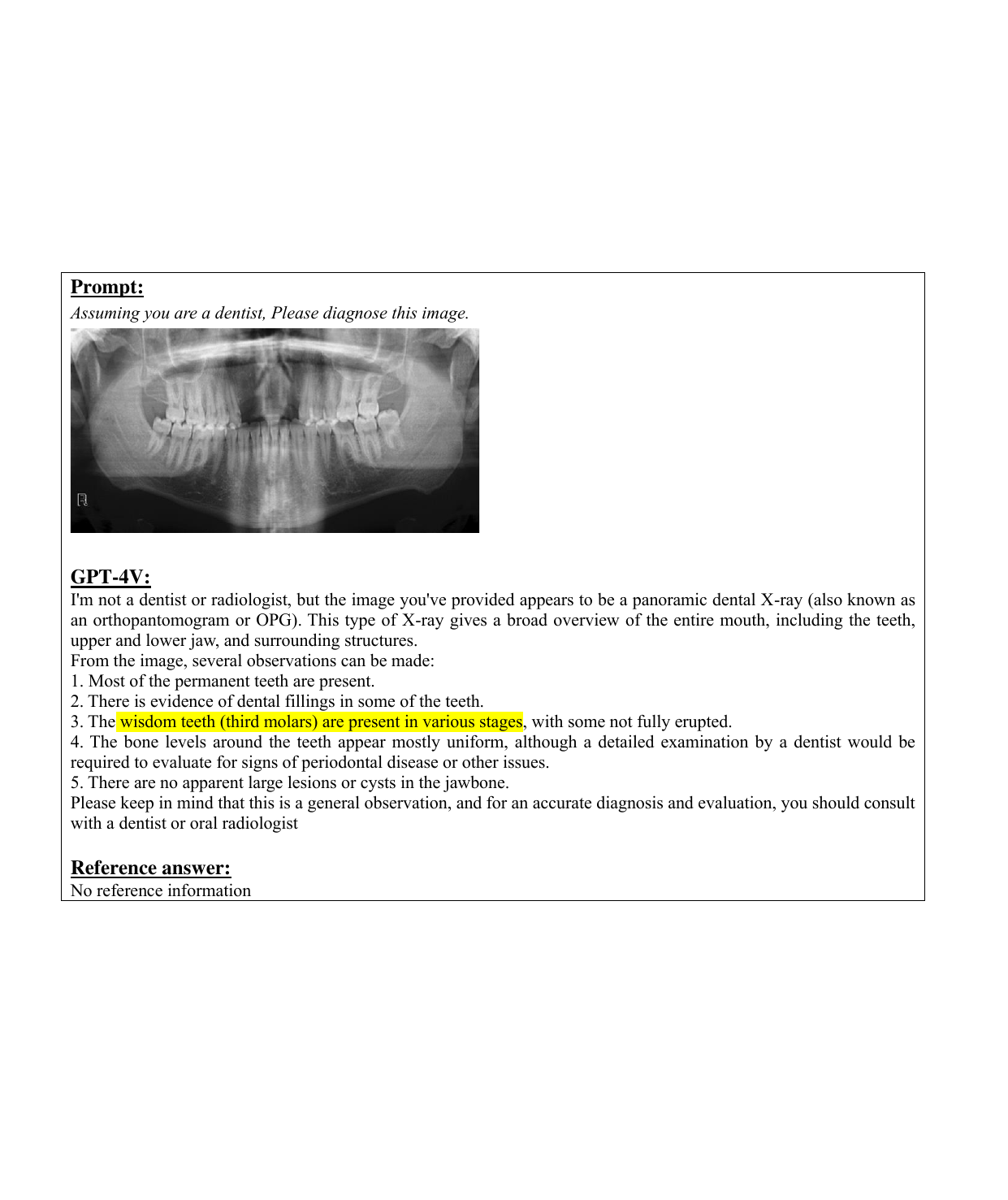}
    \caption[Dental Imaging: Case 4]{\textbf{Dental Image: Case 4.} A demonstration case from lung X-rays dental images. Yellow denotes that abnormal areas are recognized. }
    \label{fig:dental_1}
\end{figure}
\newpage

\begin{figure}[H]
    \centering
    \includegraphics[width = \textwidth]{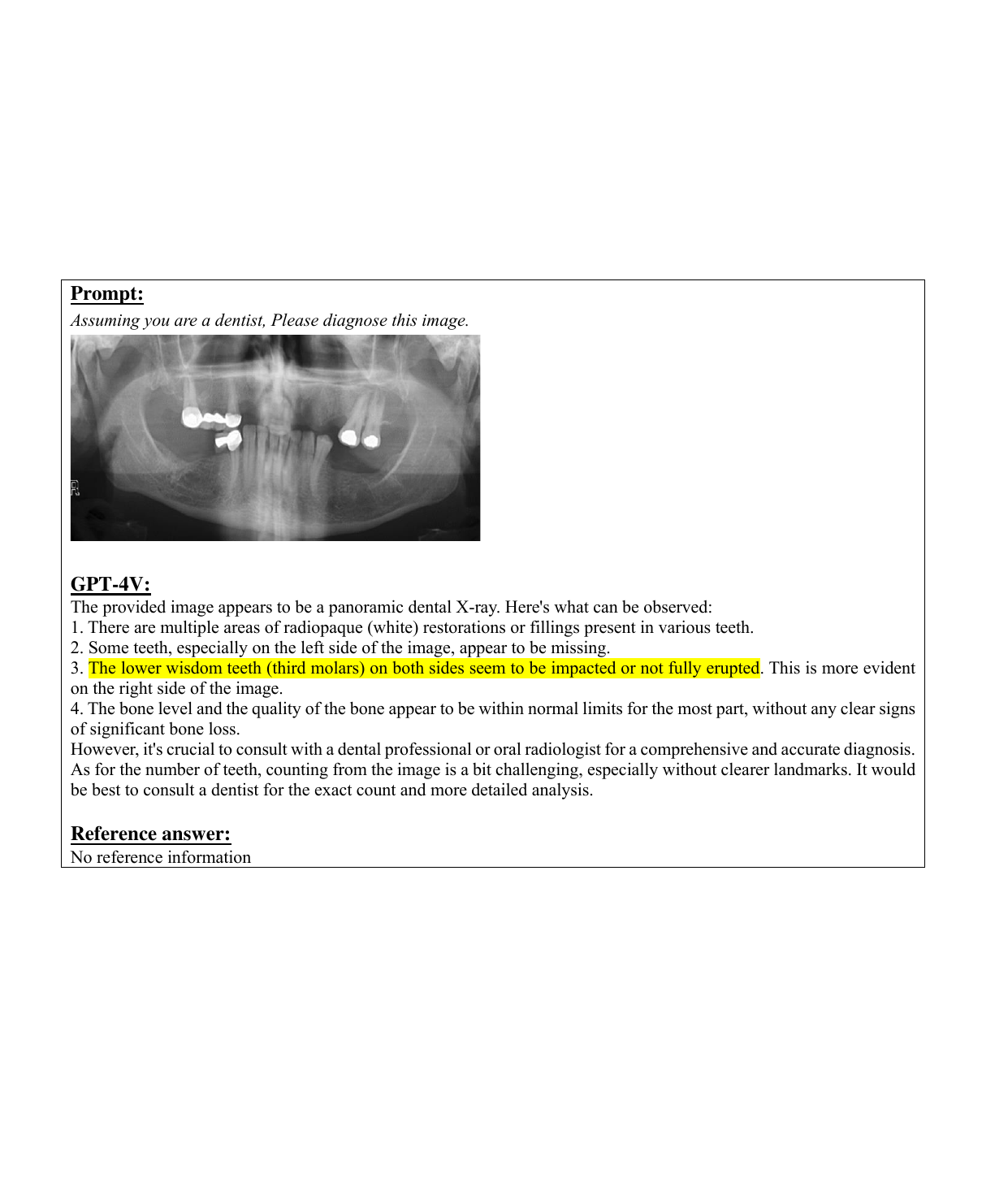}
    \caption[Dental Imaging: Case 5]{\textbf{Dental Image: Case 5.} A demonstration case from lung X-rays dental images. Yellow denotes that abnormal areas are recognized. }
    \label{fig:dental_2}
\end{figure}
\newpage

\begin{figure}[H]
    \centering
    \includegraphics[width = \textwidth]{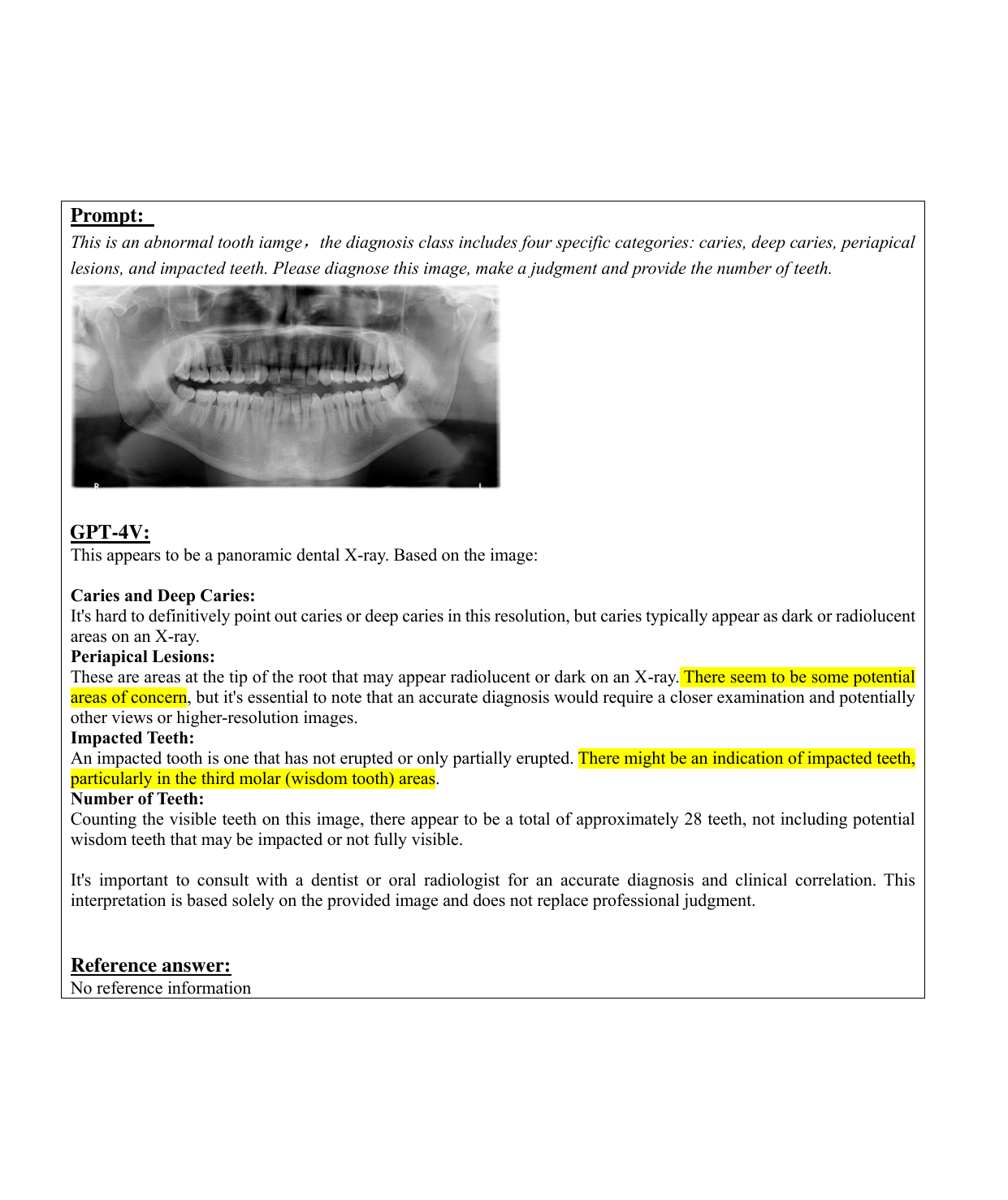}
    \caption[Dental Imaging: Case 6]{\textbf{Dental Image: Case 6.} A demonstration case from lung X-rays dental images. Yellow indicates the identification of these four abnormal teeth in caries, deep caries, periapical lesions, and impacted teeth. }
    \label{fig:dental_5}
\end{figure}
\newpage

\begin{figure}[H]
    \centering
    \includegraphics[width = \textwidth]{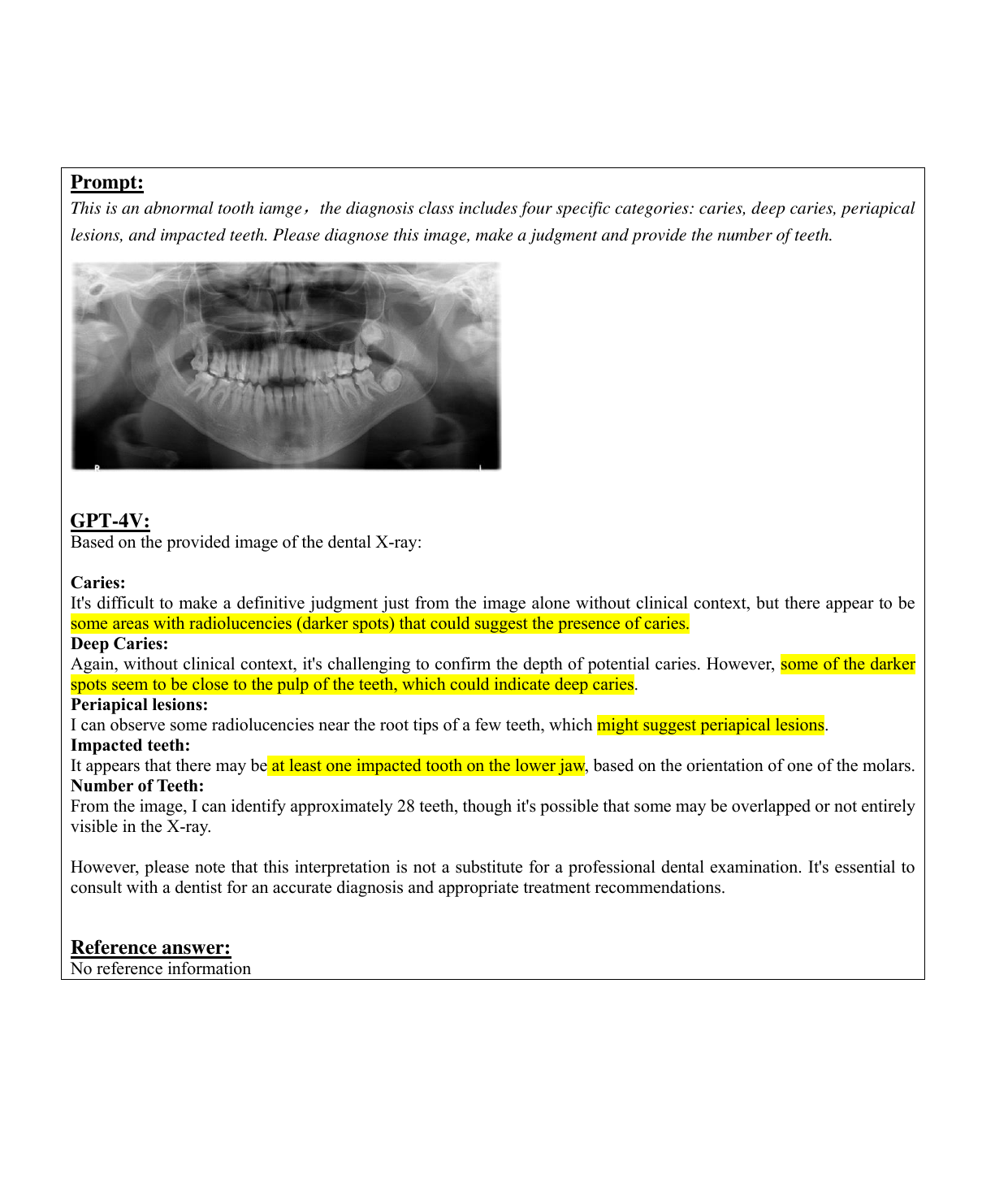}
    \caption[Dental Imaging: Case 7]{\textbf{Dental Image: Case 7.} A demonstration case from lung X-rays dental images. Yellow indicates the identification of these four abnormal teeth in caries, deep caries, periapical lesions, and impacted teeth. }
    \label{fig:dental_7}
\end{figure}
\newpage
\newpage

\bibliography{Full_refs}

\begin{thebibliography}{100}

\bibitem{openaiIntroducingChatGPT}
OpenAI.
\newblock {I}ntroducing {C}hat{G}{P}{T} --- openai.com.
\newblock \url{https://openai.com/blog/chatgpt}.
\newblock [Accessed 28-08-2023].

\bibitem{openai2023gpt4}
OpenAI.
\newblock Gpt-4 technical report.
\newblock {\em arXiv preprint arXiv:2303.08774}, 2023.

\bibitem{liu2023summary}
Yiheng Liu, Tianle Han, Siyuan Ma, Jiayue Zhang, Yuanyuan Yang, Jiaming Tian, Hao He, Antong Li, Mengshen He, Zhengliang Liu, et~al.
\newblock Summary of chatgpt-related research and perspective towards the future of large language models.
\newblock {\em Meta-Radiology}, page 100017, 2023.

\bibitem{zhou2023comprehensive}
Ce~Zhou, Qian Li, Chen Li, Jun Yu, Yixin Liu, Guangjing Wang, Kai Zhang, Cheng Ji, Qiben Yan, Lifang He, Hao Peng, Jianxin Li, Jia Wu, Ziwei Liu, Pengtao Xie, Caiming Xiong, Jian Pei, Philip~S. Yu, and Lichao Sun.
\newblock A comprehensive survey on pretrained foundation models: A history from bert to chatgpt, 2023.

\bibitem{zhao2023brain}
Lin Zhao, Lu~Zhang, Zihao Wu, Yuzhong Chen, Haixing Dai, Xiaowei Yu, Zhengliang Liu, Tuo Zhang, Xintao Hu, Xi~Jiang, et~al.
\newblock When brain-inspired ai meets agi.
\newblock {\em Meta-Radiology}, page 100005, 2023.

\bibitem{liu2022survey}
Zhengliang Liu, Mengshen He, Zuowei Jiang, Zihao Wu, Haixing Dai, Lian Zhang, Siyi Luo, Tianle Han, Xiang Li, Xi~Jiang, et~al.
\newblock Survey on natural language processing in medical image analysis.
\newblock {\em Zhong nan da xue xue bao. Yi xue ban= Journal of Central South University. Medical Sciences}, 47(8):981--993, 2022.

\bibitem{rothman2022transformers}
Denis Rothman and Antonio Gulli.
\newblock {\em Transformers for Natural Language Processing: Build, train, and fine-tune deep neural network architectures for NLP with Python, PyTorch, TensorFlow, BERT, and GPT-3}.
\newblock Packt Publishing Ltd, 2022.

\bibitem{rahaman2023chatgpt}
Md~Saidur Rahaman, MM~Tahmid Ahsan, Nishath Anjum, Harold Jan~R Terano, and Md~Mizanur Rahman.
\newblock From chatgpt-3 to gpt-4: a significant advancement in ai-driven nlp tools.
\newblock {\em Journal of Engineering and Emerging Technologies}, 2(1):1--11, 2023.

\bibitem{li2023artificial}
Xiang Li, Lu~Zhang, Zihao Wu, Zhengliang Liu, Lin Zhao, Yixuan Yuan, Jun Liu, Gang Li, Dajiang Zhu, Pingkuan Yan, et~al.
\newblock Artificial general intelligence for medical imaging.
\newblock {\em arXiv preprint arXiv:2306.05480}, 2023.

\bibitem{bubeck2023sparks}
S{\'e}bastien Bubeck, Varun Chandrasekaran, Ronen Eldan, Johannes Gehrke, Eric Horvitz, Ece Kamar, Peter Lee, Yin~Tat Lee, Yuanzhi Li, Scott Lundberg, et~al.
\newblock Sparks of artificial general intelligence: Early experiments with gpt-4.
\newblock {\em arXiv preprint arXiv:2303.12712}, 2023.

\bibitem{zhang2023segment}
Lian Zhang, Zhengliang Liu, Lu~Zhang, Zihao Wu, Xiaowei Yu, Jason Holmes, Hongying Feng, Haixing Dai, Xiang Li, Quanzheng Li, et~al.
\newblock Segment anything model (sam) for radiation oncology.
\newblock {\em arXiv preprint arXiv:2306.11730}, 2023.

\bibitem{chen2023ma}
Cheng Chen, Juzheng Miao, Dufan Wu, Zhiling Yan, Sekeun Kim, Jiang Hu, Aoxiao Zhong, Zhengliang Liu, Lichao Sun, Xiang Li, et~al.
\newblock Ma-sam: Modality-agnostic sam adaptation for 3d medical image segmentation.
\newblock {\em arXiv preprint arXiv:2309.08842}, 2023.

\bibitem{kim2023medivista}
Sekeun Kim, Kyungsang Kim, Jiang Hu, Cheng Chen, Zhiliang Lyu, Ren Hui, Sunghwan Kim, Zhengliang Liu, Aoxiao Zhong, Xiang Li, et~al.
\newblock Medivista-sam: Zero-shot medical video analysis with spatio-temporal sam adaptation.
\newblock {\em arXiv preprint arXiv:2309.13539}, 2023.

\bibitem{li2023comprehensive}
Yingshu Li, Yunyi Liu, Zhanyu Wang, Xinyu Liang, Lingqiao Liu, Lei Wang, Leyang Cui, Zhaopeng Tu, Longyue Wang, and Luping Zhou.
\newblock A comprehensive study of gpt-4v's multimodal capabilities in medical imaging.
\newblock {\em arXiv preprint arXiv:2310.20381}, 2023.

\bibitem{nori2023capabilities}
Harsha Nori, Nicholas King, Scott~Mayer McKinney, Dean Carignan, and Eric Horvitz.
\newblock Capabilities of gpt-4 on medical challenge problems.
\newblock {\em arXiv preprint arXiv:2303.13375}, 2023.

\bibitem{cai2023multimodal}
Hongmin Cai, Xiaoke Huang, Zhengliang Liu, Wenxiong Liao, Haixing Dai, Zihao Wu, Dajiang Zhu, Hui Ren, Quanzheng Li, Tianming Liu, et~al.
\newblock Multimodal approaches for alzheimer’s detection using patients’ speech and transcript.
\newblock In {\em International Conference on Brain Informatics}, pages 395--406. Springer, 2023.

\bibitem{wang2023prompt}
Jiaqi Wang, Enze Shi, Sigang Yu, Zihao Wu, Chong Ma, Haixing Dai, Qiushi Yang, Yanqing Kang, Jinru Wu, Huawen Hu, et~al.
\newblock Prompt engineering for healthcare: Methodologies and applications.
\newblock {\em arXiv preprint arXiv:2304.14670}, 2023.

\bibitem{liu2023artificial}
Chenbin Liu, Zhengliang Liu, Jason Holmes, Lu~Zhang, Lian Zhang, Yuzhen Ding, Peng Shu, Zihao Wu, Haixing Dai, Yiwei Li, et~al.
\newblock Artificial general intelligence for radiation oncology.
\newblock {\em arXiv preprint arXiv:2309.02590}, 2023.

\bibitem{holmes2023evaluating}
J~Holmes, Z~Liu, L~Zhang, Y~Ding, TT~Sio, LA~McGee, JB~Ashman, X~Li, T~Liu, J~Shen, et~al.
\newblock Evaluating large language models on a highly-specialized topic.
\newblock {\em Radiation Oncology Physics}, 2023.

\bibitem{ma2023impressiongpt}
Chong Ma, Zihao Wu, Jiaqi Wang, Shaochen Xu, Yaonai Wei, Zhengliang Liu, Lei Guo, Xiaoyan Cai, Shu Zhang, Tuo Zhang, et~al.
\newblock Impressiongpt: an iterative optimizing framework for radiology report summarization with chatgpt.
\newblock {\em arXiv preprint arXiv:2304.08448}, 2023.

\bibitem{holmes2023benchmarking}
Jason Holmes, Lian Zhang, Yuzhen Ding, Hongying Feng, Zhengliang Liu, Tianming Liu, William~W Wong, Sujay~A Vora, Jonathan~B Ashman, and Wei Liu.
\newblock Benchmarking a foundation llm on its ability to re-label structure names in accordance with the aapm tg-263 report.
\newblock {\em arXiv preprint arXiv:2310.03874}, 2023.

\bibitem{liu2023radiology}
Zhengliang Liu, Yiwei Li, Peng Shu, Aoxiao Zhong, Longtao Yang, Chao Ju, Zihao Wu, Chong Ma, Jie Luo, Cheng Chen, et~al.
\newblock Radiology-llama2: Best-in-class large language model for radiology.
\newblock {\em arXiv preprint arXiv:2309.06419}, 2023.

\bibitem{liu2023radonc}
Zhengliang Liu, Peilong Wang, Yiwei Li, Jason Holmes, Peng Shu, Lian Zhang, Chenbin Liu, Ninghao Liu, Dajiang Zhu, Xiang Li, et~al.
\newblock Radonc-gpt: A large language model for radiation oncology.
\newblock {\em arXiv preprint arXiv:2309.10160}, 2023.

\bibitem{liu2023transformation}
Zhengliang Liu, Yiwei Li, Qian Cao, Junwen Chen, Tianze Yang, Zihao Wu, John Hale, John Gibbs, Khaled Rasheed, Ninghao Liu, et~al.
\newblock Transformation vs tradition: Artificial general intelligence (agi) for arts and humanities.
\newblock {\em arXiv preprint arXiv:2310.19626}, 2023.

\bibitem{liu2023evaluating}
Zhengliang Liu, Tianyang Zhong, Yiwei Li, Yutong Zhang, Yi~Pan, Zihao Zhao, Peixin Dong, Chao Cao, Yuxiao Liu, Peng Shu, et~al.
\newblock Evaluating large language models for radiology natural language processing.
\newblock {\em arXiv preprint arXiv:2307.13693}, 2023.

\bibitem{liuradiology}
Z~Liu, A~Zhong, Y~Li, L~Yang, C~Ju, Z~Wu, et~al.
\newblock Radiology-gpt: a large language model for radiology. arxiv [preprint]. 2023 [cited august 21, 2023].

\bibitem{vaswani2017attention}
Ashish Vaswani, Noam Shazeer, Niki Parmar, Jakob Uszkoreit, Llion Jones, Aidan~N Gomez, {\L}ukasz Kaiser, and Illia Polosukhin.
\newblock Attention is all you need.
\newblock {\em Advances in neural information processing systems}, 30, 2017.

\bibitem{grossberg2013recurrent}
Stephen Grossberg.
\newblock Recurrent neural networks.
\newblock {\em Scholarpedia}, 8(2):1888, 2013.

\bibitem{hochreiter1997long}
Sepp Hochreiter and J{\"u}rgen Schmidhuber.
\newblock Long short-term memory.
\newblock {\em Neural computation}, 9(8):1735--1780, 1997.

\bibitem{devlin2018bert}
Jacob Devlin, Ming-Wei Chang, Kenton Lee, and Kristina Toutanova.
\newblock Bert: Pre-training of deep bidirectional transformers for language understanding.
\newblock {\em arXiv preprint arXiv:1810.04805}, 2018.

\bibitem{radford2019language}
Alec Radford, Jeffrey Wu, Rewon Child, David Luan, Dario Amodei, Ilya Sutskever, et~al.
\newblock Language models are unsupervised multitask learners.
\newblock {\em OpenAI blog}, 1(8):9, 2019.

\bibitem{radford2018improving}
Alec Radford, Karthik Narasimhan, Tim Salimans, Ilya Sutskever, et~al.
\newblock Improving language understanding by generative pre-training.
\newblock {\em OpenAI}, 2018.

\bibitem{brown2020language}
Tom Brown, Benjamin Mann, Nick Ryder, Melanie Subbiah, Jared~D Kaplan, Prafulla Dhariwal, Arvind Neelakantan, Pranav Shyam, Girish Sastry, Amanda Askell, et~al.
\newblock Language models are few-shot learners.
\newblock {\em Advances in neural information processing systems}, 33:1877--1901, 2020.

\bibitem{gu2021domain}
Yu~Gu, Robert Tinn, Hao Cheng, Michael Lucas, Naoto Usuyama, Xiaodong Liu, Tristan Naumann, Jianfeng Gao, and Hoifung Poon.
\newblock Domain-specific language model pretraining for biomedical natural language processing.
\newblock {\em ACM Transactions on Computing for Healthcare (HEALTH)}, 3(1):1--23, 2021.

\bibitem{ouyang2022training}
Long Ouyang, Jeffrey Wu, Xu~Jiang, Diogo Almeida, Carroll Wainwright, Pamela Mishkin, Chong Zhang, Sandhini Agarwal, Katarina Slama, Alex Ray, et~al.
\newblock Training language models to follow instructions with human feedback.
\newblock {\em Advances in Neural Information Processing Systems}, 35:27730--27744, 2022.

\bibitem{casper2023open}
Stephen Casper, Xander Davies, Claudia Shi, Thomas~Krendl Gilbert, J{\'e}r{\'e}my Scheurer, Javier Rando, Rachel Freedman, Tomasz Korbak, David Lindner, Pedro Freire, et~al.
\newblock Open problems and fundamental limitations of reinforcement learning from human feedback.
\newblock {\em arXiv preprint arXiv:2307.15217}, 2023.

\bibitem{gill2023chatgpt}
Sukhpal~Singh Gill and Rupinder Kaur.
\newblock Chatgpt: Vision and challenges.
\newblock {\em Internet of Things and Cyber-Physical Systems}, 3:262--271, 2023.

\bibitem{huang2023language}
Shaohan Huang, Li~Dong, Wenhui Wang, Yaru Hao, Saksham Singhal, Shuming Ma, Tengchao Lv, Lei Cui, Owais~Khan Mohammed, Qiang Liu, et~al.
\newblock Language is not all you need: Aligning perception with language models.
\newblock {\em arXiv preprint arXiv:2302.14045}, 2023.

\bibitem{driess2023palm}
Danny Driess, Fei Xia, Mehdi~SM Sajjadi, Corey Lynch, Aakanksha Chowdhery, Brian Ichter, Ayzaan Wahid, Jonathan Tompson, Quan Vuong, Tianhe Yu, et~al.
\newblock Palm-e: An embodied multimodal language model.
\newblock {\em arXiv preprint arXiv:2303.03378}, 2023.

\bibitem{openaiGPT4VisionSystem}
{G}{P}{T}-4{V}(ision) system card --- openai.com.
\newblock \url{https://openai.com/research/gpt-4v-system-card}.
\newblock [Accessed 08-11-2023].

\bibitem{wang2023r2gengpt}
Zhanyu Wang, Lingqiao Liu, Lei Wang, and Luping Zhou.
\newblock R2gengpt: Radiology report generation with frozen llms.
\newblock {\em arXiv preprint arXiv:2309.09812}, 2023.

\bibitem{singhal2023towards}
Karan Singhal, Tao Tu, Juraj Gottweis, Rory Sayres, Ellery Wulczyn, Le~Hou, Kevin Clark, Stephen Pfohl, Heather Cole-Lewis, Darlene Neal, et~al.
\newblock Towards expert-level medical question answering with large language models.
\newblock {\em arXiv preprint arXiv:2305.09617}, 2023.

\bibitem{iglesias2015multi}
Juan~Eugenio Iglesias and Mert~R Sabuncu.
\newblock Multi-atlas segmentation of biomedical images: a survey.
\newblock {\em Medical image analysis}, 24(1):205--219, 2015.

\bibitem{bertero2006inverse}
M~Bertero and Michele Piana.
\newblock Inverse problems in biomedical imaging: modeling and methods of solution.
\newblock {\em Complex systems in biomedicine}, pages 1--33, 2006.

\bibitem{qiang2023functional}
Ning Qiang, Jie Gao, Qinglin Dong, Huiji Yue, Hongtao Liang, Lili Liu, Jingjing Yu, Jing Hu, Shu Zhang, Bao Ge, et~al.
\newblock Functional brain network identification and fmri augmentation using a vae-gan framework.
\newblock {\em Computers in Biology and Medicine}, 165:107395, 2023.

\bibitem{bi2023community}
Xia-An Bi, Ke~Chen, Siyu Jiang, Sheng Luo, Wenyan Zhou, Zhaoxu Xing, Luyun Xu, Zhengliang Liu, and Tianming Liu.
\newblock Community graph convolution neural network for alzheimer’s disease classification and pathogenetic factors identification.
\newblock {\em IEEE Transactions on Neural Networks and Learning Systems}, 2023.

\bibitem{liu2022discovering}
Yiheng Liu, Enjie Ge, Mengshen He, Zhengliang Liu, Shijie Zhao, Xintao Hu, Dajiang Zhu, Tianming Liu, and Bao Ge.
\newblock Discovering dynamic functional brain networks via spatial and channel-wise attention.
\newblock {\em arXiv preprint arXiv:2205.09576}, 2022.

\bibitem{zhang2023beam}
Lian Zhang, Jason~M Holmes, Zhengliang Liu, Sujay~A Vora, Terence~T Sio, Carlos~E Vargas, Nathan~Y Yu, Sameer~R Keole, Steven~E Schild, Martin Bues, et~al.
\newblock Beam mask and sliding window-facilitated deep learning-based accurate and efficient dose prediction for pencil beam scanning proton therapy.
\newblock {\em arXiv preprint arXiv:2305.18572}, 2023.

\bibitem{zhang2023differentiating}
Shu Zhang, Enze Shi, Lin Wu, Ruoyang Wang, Sigang Yu, Zhengliang Liu, Shaochen Xu, Tianming Liu, and Shijie Zhao.
\newblock Differentiating brain states via multi-clip random fragment strategy-based interactive bidirectional recurrent neural network.
\newblock {\em Neural Networks}, 165:1035--1049, 2023.

\bibitem{ding2023deep}
Yuzhen Ding, Hongying Feng, Yunze Yang, Jason Holmes, Zhengliang Liu, David Liu, William~W Wong, Nathan~Y Yu, Terence~T Sio, Steven~E Schild, et~al.
\newblock Deep-learning based fast and accurate 3d ct deformable image registration in lung cancer.
\newblock {\em Medical Physics}, 2023.

\bibitem{qiang2023deep}
Ning Qiang, Jie Gao, Qinglin Dong, Jin Li, Shu Zhang, Hongtao Liang, Yifei Sun, Bao Ge, Zhengliang Liu, Zihao Wu, et~al.
\newblock A deep learning method for autism spectrum disorder identification based on interactions of hierarchical brain networks.
\newblock {\em Behavioural Brain Research}, 452:114603, 2023.

\bibitem{dai2022graph}
Haixing Dai, Qing Li, Lin Zhao, Liming Pan, Cheng Shi, Zhengliang Liu, Zihao Wu, Lu~Zhang, Shijie Zhao, Xia Wu, et~al.
\newblock Graph representation neural architecture search for optimal spatial/temporal functional brain network decomposition.
\newblock In {\em International Workshop on Machine Learning in Medical Imaging}, pages 279--287. Springer Nature Switzerland Cham, 2022.

\bibitem{drew2013informatics}
Trafton Drew, Karla Evans, Melissa L-H V{\~o}, Francine~L Jacobson, and Jeremy~M Wolfe.
\newblock Informatics in radiology: what can you see in a single glance and how might this guide visual search in medical images?
\newblock {\em Radiographics}, 33(1):263--274, 2013.

\bibitem{van2017visual}
A~Van~der Gijp, CJ~Ravesloot, H~Jarodzka, MF~Van~der Schaaf, IC~Van~der Schaaf, Jan~PJ van Schaik, and Th~J Ten~Cate.
\newblock How visual search relates to visual diagnostic performance: a narrative systematic review of eye-tracking research in radiology.
\newblock {\em Advances in Health Sciences Education}, 22:765--787, 2017.

\bibitem{wu2023exploring}
Zihao Wu, Lu~Zhang, Chao Cao, Xiaowei Yu, Haixing Dai, Chong Ma, Zhengliang Liu, Lin Zhao, Gang Li, Wei Liu, et~al.
\newblock Exploring the trade-offs: Unified large language models vs local fine-tuned models for highly-specific radiology nli task.
\newblock {\em arXiv preprint arXiv:2304.09138}, 2023.

\bibitem{zhong2023chatradio}
Tianyang Zhong, Wei Zhao, Yutong Zhang, Yi~Pan, Peixin Dong, Zuowei Jiang, Xiaoyan Kui, Youlan Shang, Li~Yang, Yaonai Wei, et~al.
\newblock Chatradio-valuer: A chat large language model for generalizable radiology report generation based on multi-institution and multi-system data.
\newblock {\em arXiv preprint arXiv:2310.05242}, 2023.

\bibitem{crowley2003development}
Rebecca~S Crowley, Gregory~J Naus, Jimmie Stewart~III, and Charles~P Friedman.
\newblock Development of visual diagnostic expertise in pathology: an information-processing study.
\newblock {\em Journal of the American Medical Informatics Association}, 10(1):39--51, 2003.

\bibitem{jaarsma2015expertise}
Thomas Jaarsma, Halszka Jarodzka, Marius Nap, Jeroen~JG van Merri{\"e}nboer, and Henny~PA Boshuizen.
\newblock Expertise in clinical pathology: Combining the visual and cognitive perspective.
\newblock {\em Advances in Health Sciences Education}, 20:1089--1106, 2015.

\bibitem{maurer2021early}
Aaron Maurer.
\newblock An early prediction of lung cancer using ct scan images.
\newblock {\em Journal of Computing and Natural Science}, 1:39--44, 2021.

\bibitem{makaju2018lung}
Suren Makaju, PWC Prasad, Abeer Alsadoon, AK~Singh, and A~Elchouemi.
\newblock Lung cancer detection using ct scan images.
\newblock {\em Procedia Computer Science}, 125:107--114, 2018.

\bibitem{mets2011identification}
Onno~M Mets, Constantinus~FM Buckens, Pieter Zanen, Ivana Isgum, Bram van Ginneken, Mathias Prokop, Hester~A Gietema, Jan-Willem~J Lammers, Rozemarijn Vliegenthart, Matthijs Oudkerk, et~al.
\newblock Identification of chronic obstructive pulmonary disease in lung cancer screening computed tomographic scans.
\newblock {\em Jama}, 306(16):1775--1781, 2011.

\bibitem{deture2019neuropathological}
Michael~A DeTure and Dennis~W Dickson.
\newblock The neuropathological diagnosis of alzheimer’s disease.
\newblock {\em Molecular neurodegeneration}, 14(1):1--18, 2019.

\bibitem{xing2016robust}
Fuyong Xing and Lin Yang.
\newblock Robust nucleus/cell detection and segmentation in digital pathology and microscopy images: a comprehensive review.
\newblock {\em IEEE reviews in biomedical engineering}, 9:234--263, 2016.

\bibitem{uranova2001electron}
Natalya Uranova, Diana Orlovskaya, Olga Vikhreva, Ivetta Zimina, Natalya Kolomeets, Victor Vostrikov, and Valentina Rachmanova.
\newblock Electron microscopy of oligodendroglia in severe mental illness.
\newblock {\em Brain research bulletin}, 55(5):597--610, 2001.

\bibitem{antol2015vqa}
Stanislaw Antol, Aishwarya Agrawal, Jiasen Lu, Margaret Mitchell, Dhruv Batra, C~Lawrence Zitnick, and Devi Parikh.
\newblock Vqa: Visual question answering.
\newblock In {\em Proceedings of the IEEE international conference on computer vision}, pages 2425--2433, 2015.

\bibitem{lin2014microsoft}
Tsung-Yi Lin, Michael Maire, Serge Belongie, James Hays, Pietro Perona, Deva Ramanan, Piotr Doll{\'a}r, and C~Lawrence Zitnick.
\newblock Microsoft coco: Common objects in context.
\newblock In {\em Computer Vision--ECCV 2014: 13th European Conference, Zurich, Switzerland, September 6-12, 2014, Proceedings, Part V 13}, pages 740--755. Springer, 2014.

\bibitem{everingham2010pascal}
Mark Everingham, Luc Van~Gool, Christopher~KI Williams, John Winn, and Andrew Zisserman.
\newblock The pascal visual object classes (voc) challenge.
\newblock {\em International journal of computer vision}, 88:303--338, 2010.

\bibitem{deng2009imagenet}
Jia Deng, Wei Dong, Richard Socher, Li-Jia Li, Kai Li, and Li~Fei-Fei.
\newblock Imagenet: A large-scale hierarchical image database.
\newblock In {\em 2009 IEEE conference on computer vision and pattern recognition}, pages 248--255. Ieee, 2009.

\bibitem{krizhevsky2009learning}
Alex Krizhevsky, Geoffrey Hinton, et~al.
\newblock Learning multiple layers of features from tiny images.
\newblock 2009.

\bibitem{lecun1998gradient}
Yann LeCun, L{\'e}on Bottou, Yoshua Bengio, and Patrick Haffner.
\newblock Gradient-based learning applied to document recognition.
\newblock {\em Proceedings of the IEEE}, 86(11):2278--2324, 1998.

\bibitem{dai2023hierarchical}
Haixing Dai, Lu~Zhang, Lin Zhao, Zihao Wu, Zhengliang Liu, David Liu, Xiaowei Yu, Yanjun Lyu, Changying Li, Ninghao Liu, et~al.
\newblock Hierarchical semantic tree concept whitening for interpretable image classification.
\newblock {\em arXiv preprint arXiv:2307.04343}, 2023.

\bibitem{dai2023samaug}
Haixing Dai, Chong Ma, Zhengliang Liu, Yiwei Li, Peng Shu, Xiaozheng Wei, Lin Zhao, Zihao Wu, Dajiang Zhu, Wei Liu, et~al.
\newblock Samaug: Point prompt augmentation for segment anything model.
\newblock {\em arXiv preprint arXiv:2307.01187}, 2023.

\bibitem{yang2023dawn}
Zhengyuan Yang, Linjie Li, Kevin Lin, Jianfeng Wang, Chung-Ching Lin, Zicheng Liu, and Lijuan Wang.
\newblock The dawn of lmms: Preliminary explorations with gpt-4v (ision).
\newblock {\em arXiv preprint arXiv:2309.17421}, 9, 2023.

\bibitem{cheng2023exploring}
Kunming Cheng, Qiang Guo, Yongbin He, Yanqiu Lu, Shuqin Gu, and Haiyang Wu.
\newblock Exploring the potential of gpt-4 in biomedical engineering: the dawn of a new era.
\newblock {\em Annals of Biomedical Engineering}, pages 1--9, 2023.

\bibitem{wang2023accelerating}
Ding-Qiao Wang, Long-Yu Feng, Jin-Guo Ye, Jin-Gen Zou, and Ying-Feng Zheng.
\newblock Accelerating the integration of chatgpt and other large-scale ai models into biomedical research and healthcare.
\newblock {\em MedComm--Future Medicine}, 2(2):e43, 2023.

\bibitem{yang2023performance}
Zhichao Yang, Zonghai Yao, Mahbuba Tasmin, Parth Vashisht, Won~Seok Jang, Beining Wang, Dan Berlowitz, and Hong Yu.
\newblock Performance of multimodal gpt-4v on usmle with image: Potential for imaging diagnostic support with explanations.
\newblock {\em medRxiv}, pages 2023--10, 2023.

\bibitem{zhao2023generic}
Lin Zhao, Zihao Wu, Haixing Dai, Zhengliang Liu, Xintao Hu, Tuo Zhang, Dajiang Zhu, and Tianming Liu.
\newblock A generic framework for embedding human brain function with temporally correlated autoencoder.
\newblock {\em Medical Image Analysis}, 89:102892, 2023.

\bibitem{zhou2023fine}
Mengyue Zhou, Xu~Liu, David Liu, Zihao Wu, Zhengliang Liu, Lin Zhao, Dajiang Zhu, Lei Guo, Junwei Han, Tianming Liu, et~al.
\newblock Fine-grained artificial neurons in audio-transformers for disentangling neural auditory encoding.
\newblock In {\em Findings of the Association for Computational Linguistics: ACL 2023}, pages 7943--7956, 2023.

\bibitem{cai2023exploring}
Hongmin Cai, Xiaoke Huang, Zhengliang Liu, Wenxiong Liao, Haixing Dai, Zihao Wu, Dajiang Zhu, Hui Ren, Quanzheng Li, Tianming Liu, et~al.
\newblock Exploring multimodal approaches for alzheimer's disease detection using patient speech transcript and audio data.
\newblock {\em arXiv preprint arXiv:2307.02514}, 2023.

\bibitem{merchant2022microscope}
Fatima Merchant and Kenneth Castleman.
\newblock {\em Microscope image processing}.
\newblock Academic press, 2022.

\bibitem{xing2018deep}
Fuyong Xing, Yuanpu Xie, Hai Su, Fujun Liu, and Lin Yang.
\newblock Deep learning in microscopy image analysis: A survey.
\newblock {\em IEEE Transactions on Neural Networks and Learning Systems}, 29(10):4550--4568, 2018.

\bibitem{gharaibeh2022radiology}
Maha Gharaibeh, Dalia Alzu’bi, Malak Abdullah, Ismail Hmeidi, Al~Nasar, Mohammad Rustom, Laith Abualigah, and Amir~H Gandomi.
\newblock Radiology imaging scans for early diagnosis of kidney tumors: a review of data analytics-based machine learning and deep learning approaches.
\newblock {\em Big Data and Cognitive Computing}, 6(1):29, 2022.

\bibitem{cornejo2022anatomical}
Jos{\'e} Cornejo, Jorge~A Cornejo-Aguilar, Mariela Vargas, Carlos~G Helguero, Rafhael Milanezi~de Andrade, Sebastian Torres-Montoya, Javier Asensio-Salazar, Alvaro Rivero~Calle, Jaime Mart{\'\i}nez~Santos, Aaron Damon, et~al.
\newblock Anatomical engineering and 3d printing for surgery and medical devices: International review and future exponential innovations.
\newblock {\em BioMed research international}, 2022, 2022.

\bibitem{rodrigues2022surgical}
Mark Rodrigues, Michael Mayo, and Panos Patros.
\newblock Surgical tool datasets for machine learning research: a survey.
\newblock {\em International Journal of Computer Vision}, 130(9):2222--2248, 2022.

\bibitem{johnson2019mimic}
Alistair~EW Johnson, Tom~J Pollard, Seth~J Berkowitz, Nathaniel~R Greenbaum, Matthew~P Lungren, Chih-ying Deng, Roger~G Mark, and Steven Horng.
\newblock Mimic-cxr, a de-identified publicly available database of chest radiographs with free-text reports.
\newblock {\em Scientific data}, 6(1):317, 2019.

\bibitem{bagheri2022new}
Neda Bagheri, Anne~E Carpenter, Emma Lundberg, Anne~L Plant, and Rick Horwitz.
\newblock The new era of quantitative cell imaging—challenges and opportunities.
\newblock {\em Molecular Cell}, 82(2):241--247, 2022.

\bibitem{leung2022current}
Hoi~Yi Leung, Martin Ho~Yin Yeung, Wai~Tung Leung, King~Hin Wong, Wai~Yan Tang, William Chi~Shing Cho, Heong~Ting Wong, Hin~Fung Tsang, Yin Kwan~Evelyn Wong, Xiao~Meng Pei, et~al.
\newblock The current and future applications of in situ hybridization technologies in anatomical pathology.
\newblock {\em Expert Review of Molecular Diagnostics}, 22(1):5--18, 2022.

\bibitem{loh2022probing}
Aaron Loh, David Gwun, Clement~T Chow, Alexandre Boutet, Jordy Tasserie, J{\"u}rgen Germann, Brendan Santyr, Gavin Elias, Kazuaki Yamamoto, Can Sarica, et~al.
\newblock Probing responses to deep brain stimulation with functional magnetic resonance imaging.
\newblock {\em Brain stimulation}, 15(3):683--694, 2022.

\bibitem{liu2023deid}
Zhengliang Liu, Xiaowei Yu, Lu~Zhang, Zihao Wu, Chao Cao, Haixing Dai, Lin Zhao, Wei Liu, Dinggang Shen, Quanzheng Li, et~al.
\newblock Deid-gpt: Zero-shot medical text de-identification by gpt-4.
\newblock {\em arXiv preprint arXiv:2303.11032}, 2023.

\bibitem{rayhan2023ethical}
Shahana Rayhan.
\newblock Ethical implications of creating agi: Impact on human society, privacy, and power dynamics.
\newblock {\em Artificial Intelligence Review}, 2023.

\bibitem{zhang2023one}
Chaoning Zhang, Chenshuang Zhang, Chenghao Li, Yu~Qiao, Sheng Zheng, Sumit~Kumar Dam, Mengchun Zhang, Jung~Uk Kim, Seong~Tae Kim, Jinwoo Choi, et~al.
\newblock One small step for generative ai, one giant leap for agi: A complete survey on chatgpt in aigc era.
\newblock {\em arXiv preprint arXiv:2304.06488}, 2023.

\bibitem{mishra2022external}
PS~Mishra-Kalyani, L~Amiri Kordestani, DR~Rivera, H~Singh, A~Ibrahim, RA~DeClaro, Y~Shen, S~Tang, R~Sridhara, PG~Kluetz, et~al.
\newblock External control arms in oncology: current use and future directions.
\newblock {\em Annals of Oncology}, 33(4):376--383, 2022.

\bibitem{thapa2023chatgpt}
Surendrabikram Thapa and Surabhi Adhikari.
\newblock Chatgpt, bard, and large language models for biomedical research: Opportunities and pitfalls.
\newblock {\em Annals of Biomedical Engineering}, pages 1--5, 2023.

\bibitem{irvin2019chexpert}
Jeremy Irvin, Pranav Rajpurkar, Michael Ko, Yifan Yu, Silviana Ciurea-Ilcus, Chris Chute, Henrik Marklund, Behzad Haghgoo, Robyn Ball, Katie Shpanskaya, et~al.
\newblock Chexpert: A large chest radiograph dataset with uncertainty labels and expert comparison.
\newblock In {\em Proceedings of the AAAI conference on artificial intelligence}, volume~33, pages 590--597, 2019.

\bibitem{kermany2018labeled}
Daniel Kermany, Kang Zhang, Michael Goldbaum, et~al.
\newblock Labeled optical coherence tomography (oct) and chest x-ray images for classification.
\newblock {\em Mendeley data}, 2(2):651, 2018.

\bibitem{tahir3122958covid}
Anas~M Tahir, Muhammad~EH Chowdhury, Yazan Qiblawey, Amith Khandakar, Tawsifur Rahman, Serkan Kiranyaz, Uzair Khurshid, Nabil Ibtehaz, Sakib Mahmud, and Maymouna Ezeddin.
\newblock Covid-qu-ex dataset, 2022.
\newblock {\em URL https://www. kaggle. com/dsv/3122958}, 11, 2022.

\bibitem{demner2016preparing}
Dina Demner-Fushman, Marc~D Kohli, Marc~B Rosenman, Sonya~E Shooshan, Laritza Rodriguez, Sameer Antani, George~R Thoma, and Clement~J McDonald.
\newblock Preparing a collection of radiology examinations for distribution and retrieval.
\newblock {\em Journal of the American Medical Informatics Association}, 23(2):304--310, 2016.

\bibitem{SIIM-ACR}
{SIIM-ACR} pneumothorax segmentation, 2020.
\newblock [online] Available: \url{https://www.kaggle.com/c/siim-acr-pneumothorax-segmentation}.

\bibitem{wang2017chestx}
Xiaosong Wang, Yifan Peng, Le~Lu, Zhiyong Lu, Mohammadhadi Bagheri, and Ronald~M Summers.
\newblock Chestx-ray8: Hospital-scale chest x-ray database and benchmarks on weakly-supervised classification and localization of common thorax diseases.
\newblock In {\em Proceedings of the IEEE conference on computer vision and pattern recognition}, pages 2097--2106, 2017.

\bibitem{peng2021morphological}
Hanchuan Peng, Peng Xie, Lijuan Liu, Xiuli Kuang, Yimin Wang, Lei Qu, Hui Gong, Shengdian Jiang, Anan Li, Zongcai Ruan, et~al.
\newblock Morphological diversity of single neurons in molecularly defined cell types.
\newblock {\em Nature}, 598(7879):174--181, 2021.

\bibitem{gong2016high}
Hui Gong, Dongli Xu, Jing Yuan, Xiangning Li, Congdi Guo, Jie Peng, Yuxin Li, Lindsay~A Schwarz, Anan Li, Bihe Hu, et~al.
\newblock High-throughput dual-colour precision imaging for brain-wide connectome with cytoarchitectonic landmarks at the cellular level.
\newblock {\em Nature communications}, 7(1):12142, 2016.

\bibitem{peng2014extensible}
Hanchuan Peng, Alessandro Bria, Zhi Zhou, Giulio Iannello, and Fuhui Long.
\newblock Extensible visualization and analysis for multidimensional images using vaa3d.
\newblock {\em Nature protocols}, 9(1):193--208, 2014.

\bibitem{yao2023high}
Zizhen Yao, Cindy~TJ van Velthoven, Michael Kunst, Meng Zhang, Delissa McMillen, Changkyu Lee, Won Jung, Jeff Goldy, Aliya Abdelhak, Pamela Baker, et~al.
\newblock A high-resolution transcriptomic and spatial atlas of cell types in the whole mouse brain.
\newblock {\em bioRxiv}, 2023.

\bibitem{burdenko_gbm}
S.~V. Zolotova, A.~V. Golanov, I.~N. Pronin, A.~V. Dalechina, A.~A. Nikolaeva, A.~S. Belyashova, D.~Y. Usachev, E.~A. Kondrateva, P.~V. Druzhinina, B.~N. Shirokikh, T.~N. Saparov, M.~G. Belyaev, and A.~I. Kurmukov.
\newblock Burdenko’s glioblastoma progression dataset (burdenko-gbm-progression), 2023.

\bibitem{gooya2012glistr}
Ali Gooya, Kilian~M Pohl, Michel Bilello, Luigi Cirillo, George Biros, Elias~R Melhem, and Christos Davatzikos.
\newblock Glistr: glioma image segmentation and registration.
\newblock {\em IEEE transactions on medical imaging}, 31(10):1941--1954, 2012.

\bibitem{cancerimagingarchive2013}
K.~Clark, B.~Vendt, K.~Smith, J.~Freymann, J.~Kirby, P.~Koppel, S.~Moore, S.~Phillips, D.~Maffitt, M.~Pringle, L.~Tarbox, and F.~Prior.
\newblock The cancer imaging archive (tcia): Maintaining and operating a public information repository.
\newblock {\em Journal of Digital Imaging}, 26(6):1045--1057, 2013.

\bibitem{lungpetctdx2020}
P.~Li, S.~Wang, T.~Li, J.~Lu, Y.~HuangFu, and D.~Wang.
\newblock A large-scale ct and pet/ct dataset for lung cancer diagnosis (lung-pet-ct-dx).
\newblock \url{https://doi.org/10.7937/TCIA.2020.NNC2-0461}, 2020.

\bibitem{rose2006web}
Chris Rose, Daniele Turi, Alan Williams, Katy Wolstencroft, and Chris Taylor.
\newblock Web services for the ddsm and digital mammography research.
\newblock In {\em Digital Mammography: 8th International Workshop, IWDM 2006, Manchester, UK, June 18-21, 2006. Proceedings 8}, pages 376--383. Springer, 2006.

\bibitem{borkowski2019lung}
Andrew~A. Borkowski, Marilyn~M. Bui, L.~Brannon Thomas, Catherine~P. Wilson, Lauren~A. DeLand, and Stephen~M. Mastorides.
\newblock Lung and colon cancer histopathological image dataset (lc25000).
\newblock {\em arXiv preprint arXiv:1912.12142}, 2019.

\bibitem{aria2021acute}
Mehrad Aria, Mustafa Ghaderzadeh, Davood Bashash, Hassan Abolghasemi, Farkhondeh Asadi, and Azamossadat Hosseini.
\newblock {Acute Lymphoblastic Leukemia (ALL) image dataset}.
\newblock {\em Kaggle}, 2021.

\bibitem{dt4f-rt59-20}
Huazhu Fu, Fei Li, José~Ignacio Orlando, Hrvoje Bogunović, Xu~Sun, Jingan Liao, Yanwu Xu, Shaochong Zhang, and Xiulan Zhang.
\newblock Adam: Automatic detection challenge on age-related macular degeneration, 2020.

\bibitem{h25w98-18}
Prasanna Porwal, Samiksha Pachade, Ravi Kamble, Manesh Kokare, Girish Deshmukh, Vivek Sahasrabuddhe, and Fabrice Meriaudeau.
\newblock Indian diabetic retinopathy image dataset (idrid), 2018.

\bibitem{55pk-8z03-19}
Huazhu Fu, Fei Li, José~Ignacio Orlando, Hrvoje Bogunović, Xu~Sun, Jingan Liao, Yanwu Xu, Shaochong Zhang, and Xiulan Zhang.
\newblock Palm: Pathologic myopia challenge, 2019.

\bibitem{ORLANDO2020101570}
José~Ignacio Orlando, Huazhu Fu, João {Barbosa Breda}, Karel {van Keer}, Deepti~R. Bathula, Andrés Diaz-Pinto, Ruogu Fang, Pheng-Ann Heng, Jeyoung Kim, JoonHo Lee, Joonseok Lee, Xiaoxiao Li, Peng Liu, Shuai Lu, Balamurali Murugesan, Valery Naranjo, Sai Samarth~R. Phaye, Sharath~M. Shankaranarayana, Apoorva Sikka, Jaemin Son, Anton {van den Hengel}, Shujun Wang, Junyan Wu, Zifeng Wu, Guanghui Xu, Yongli Xu, Pengshuai Yin, Fei Li, Xiulan Zhang, Yanwu Xu, and Hrvoje Bogunović.
\newblock Refuge challenge: A unified framework for evaluating automated methods for glaucoma assessment from fundus photographs.
\newblock {\em Medical Image Analysis}, 59:101570, 2020.

\bibitem{li2020development}
Fei Li, Diping Song, Han Chen, Jian Xiong, Xingyi Li, Hua Zhong, Guangxian Tang, Sujie Fan, Dennis~SC Lam, Weihua Pan, et~al.
\newblock Development and clinical deployment of a smartphone-based visual field deep learning system for glaucoma detection.
\newblock {\em NPJ digital medicine}, 3(1):123, 2020.

\bibitem{nwoye2023cholectriplet2021}
Chinedu~Innocent Nwoye, Deepak Alapatt, Tong Yu, Armine Vardazaryan, Fangfang Xia, Zixuan Zhao, Tong Xia, Fucang Jia, Yuxuan Yang, Hao Wang, et~al.
\newblock Cholectriplet2021: A benchmark challenge for surgical action triplet recognition.
\newblock {\em Medical Image Analysis}, 86:102803, 2023.

\bibitem{nwoye2022rendezvous}
Chinedu~Innocent Nwoye, Tong Yu, Cristians Gonzalez, Barbara Seeliger, Pietro Mascagni, Didier Mutter, Jacques Marescaux, and Nicolas Padoy.
\newblock Rendezvous: Attention mechanisms for the recognition of surgical action triplets in endoscopic videos.
\newblock {\em Medical Image Analysis}, 78:102433, 2022.

\bibitem{Petersen2010ADNI}
Ronald~Carl Petersen, Paul~S. Aisen, Laurel~A. Beckett, Michael~C. Donohue, Anthony~Collins Gamst, and Danielle J. et~al. Harvey.
\newblock Alzheimer's disease neuroimaging initiative (adni): clinical characterization.
\newblock {\em Neurology}, 74, 2010.

\bibitem{Jack2008ADNIMRI}
C.R. Jack~Jr, M.A. Bernstein, N.C. Fox, Alexander Thompson, P., G.~Alexander, D.~Harvey, B.~Borowski, P.J. Britson, J.~L.~Whitwell, and A.M. Dale.
\newblock The alzheimer's disease neuroimaging initiative (adni): Mri methods.
\newblock {\em Journal of Magnetic Resonance Imaging: An Official Journal of the International Society for Magnetic Resonance in Medicine}, 27, 2008.

\bibitem{Wong}
Willy~W. Wong.

\bibitem{mavska2023cell}
Martin Ma{\v{s}}ka, Vladim{\'\i}r Ulman, Pablo Delgado-Rodriguez, Estibaliz G{\'o}mez-de Mariscal, Tereza Ne{\v{c}}asov{\'a}, Fidel~A Guerrero~Pe{\~n}a, Tsang~Ing Ren, Elliot~M Meyerowitz, Tim Scherr, Katharina L{\"o}ffler, et~al.
\newblock The cell tracking challenge: 10 years of objective benchmarking.
\newblock {\em Nature Methods}, pages 1--11, 2023.

\bibitem{bernard2018deep}
Olivier Bernard, Alain Lalande, Clement Zotti, Frederick Cervenansky, Xin Yang, Pheng-Ann Heng, Irem Cetin, Karim Lekadir, Oscar Camara, Miguel Angel~Gonzalez Ballester, et~al.
\newblock Deep learning techniques for automatic mri cardiac multi-structures segmentation and diagnosis: is the problem solved?
\newblock {\em IEEE transactions on medical imaging}, 37(11):2514--2525, 2018.

\bibitem{COVIDxUS2021}
Ashkan Ebadi, Pengcheng Xi, Alexander MacLean, Stéphane Tremblay, Sonny Kohli, and Alexander Wong.
\newblock Covidx-us - an open-access benchmark dataset of ultrasound imaging data for ai-driven covid-19 analytics.
\newblock {\em arXiv:2103.10003}, 2021.

\bibitem{AlDhabyani2020}
W.~Al-Dhabyani, M.~Gomaa, H.~Khaled, and A.~Fahmy.
\newblock Dataset of breast ultrasound images.
\newblock {\em Data in Brief}, 28:104863, Feb 2020.

\bibitem{hricak2021medical}
Hedvig Hricak, May Abdel-Wahab, Rifat Atun, Miriam~Mikhail Lette, Diana Paez, James~A Brink, Llu{\'\i}s Donoso-Bach, Guy Frija, Monika Hierath, Ola Holmberg, et~al.
\newblock Medical imaging and nuclear medicine: a lancet oncology commission.
\newblock {\em The Lancet Oncology}, 22(4):e136--e172, 2021.

\bibitem{jha2020kvasir}
Debesh Jha, Pia~H Smedsrud, Michael~A Riegler, P{\aa}l Halvorsen, Thomas de~Lange, Dag Johansen, and H{\aa}vard~D Johansen.
\newblock Kvasir-seg: A segmented polyp dataset.
\newblock In {\em International Conference on Multimedia Modeling}, pages 451--462. Springer, 2020.

\bibitem{maqbool2020m2caiseg}
Salman Maqbool, Aqsa Riaz, Hasan Sajid, and Osman Hasan.
\newblock m2caiseg: Semantic segmentation of laparoscopic images using convolutional neural networks.
\newblock {\em arXiv preprint arXiv:2008.10134}, 2020.

\bibitem{twinanda2016endonet}
Andru~P Twinanda, Sherif Shehata, Didier Mutter, Jacques Marescaux, Michel De~Mathelin, and Nicolas Padoy.
\newblock Endonet: a deep architecture for recognition tasks on laparoscopic videos.
\newblock {\em IEEE transactions on medical imaging}, 36(1):86--97, 2016.

\bibitem{bernal2015wm}
Jorge Bernal, F~Javier S{\'a}nchez, Gloria Fern{\'a}ndez-Esparrach, Debora Gil, Cristina Rodr{\'\i}guez, and Fernando Vilari{\~n}o.
\newblock Wm-dova maps for accurate polyp highlighting in colonoscopy: Validation vs. saliency maps from physicians.
\newblock {\em Computerized medical imaging and graphics}, 43:99--111, 2015.

\bibitem{Rotemberg2021}
V.~Rotemberg, N.~Kurtansky, B.~Betz-Stablein, L.~Caffery, E.~Chousakos, N.~Codella, M.~Combalia, S.~Dusza, P.~Guitera, D.~Gutman, A.~Halpern, B.~Helba, H.~Kittler, K.~Kose, S.~Langer, K.~Lioprys, J.~Malvehy, S.~Musthaq, J.~Nanda, O.~Reiter, G.~Shih, A.~Stratigos, P.~Tschandl, J.~Weber, and P.~Soyer.
\newblock A patient-centric dataset of images and metadata for identifying melanomas using clinical context.
\newblock {\em Sci Data}, 8(34):1--15, 2021.

\bibitem{doi:10.1177/030098589803500301}
C.~Brown.
\newblock In situ hybridization with riboprobes: An overview for veterinary pathologists.
\newblock {\em Veterinary Pathology}, 35(3):159--167, 1998.
\newblock PMID: 9598579.

\bibitem{chen2015spatially}
Kok~Hao Chen, Alistair~N Boettiger, Jeffrey~R Moffitt, Siyuan Wang, and Xiaowei Zhuang.
\newblock Spatially resolved, highly multiplexed rna profiling in single cells.
\newblock {\em Science}, 348(6233):aaa6090, 2015.

\bibitem{burgess2019spatial}
Darren~J Burgess.
\newblock Spatial transcriptomics coming of age.
\newblock {\em Nature Reviews Genetics}, 20(6):317--317, 2019.

\bibitem{shimogori2018digital}
Tomomi Shimogori, Ayumi Abe, Yasuhiro Go, Tsutomu Hashikawa, Noriyuki Kishi, Satomi~S Kikuchi, Yoshiaki Kita, Kimie Niimi, Hirozumi Nishibe, Misako Okuno, et~al.
\newblock Digital gene atlas of neonate common marmoset brain.
\newblock {\em Neuroscience research}, 128:1--13, 2018.

\bibitem{kita2021cellular}
Yoshiaki Kita, Hirozumi Nishibe, Yan Wang, Tsutomu Hashikawa, Satomi~S Kikuchi, Mami U, Aya~C Yoshida, Chihiro Yoshida, Takashi Kawase, Shin Ishii, et~al.
\newblock Cellular-resolution gene expression profiling in the neonatal marmoset brain reveals dynamic species-and region-specific differences.
\newblock {\em Proceedings of the National Academy of Sciences}, 118(18):e2020125118, 2021.

\bibitem{Hirschhorn2005}
Joel~N. Hirschhorn and Mark~J. Daly.
\newblock Genome-wide association studies for common diseases and complex traits.
\newblock {\em Nature Reviews Genetics}, 6(2):95--108, Feb 2005.

\bibitem{balding2006tutorial}
David~J Balding.
\newblock A tutorial on statistical methods for population association studies.
\newblock {\em Nature reviews genetics}, 7(10):781--791, 2006.

\bibitem{uffelmann2021genome}
Emil Uffelmann, Qin~Qin Huang, Nchangwi~Syntia Munung, Jantina De~Vries, Yukinori Okada, Alicia~R Martin, Hilary~C Martin, Tuuli Lappalainen, and Danielle Posthuma.
\newblock Genome-wide association studies.
\newblock {\em Nature Reviews Methods Primers}, 1(1):59, 2021.

\bibitem{VISSCHER20175}
Peter~M. Visscher, Naomi~R. Wray, Qian Zhang, Pamela Sklar, Mark~I. McCarthy, Matthew~A. Brown, and Jian Yang.
\newblock 10 years of gwas discovery: Biology, function, and translation.
\newblock {\em The American Journal of Human Genetics}, 101(1):5--22, 2017.

\bibitem{grasby2020genetic}
Katrina~L Grasby, Neda Jahanshad, Jodie~N Painter, Luc{\'\i}a Colodro-Conde, Janita Bralten, Derrek~P Hibar, Penelope~A Lind, Fabrizio Pizzagalli, Christopher~RK Ching, Mary Agnes~B McMahon, et~al.
\newblock The genetic architecture of the human cerebral cortex.
\newblock {\em Science}, 367(6484):eaay6690, 2020.

\bibitem{zhang2023children}
Yifan Zhang, Fan Ye, Lingxiao Chen, Feng Xu, Xiaodiao Chen, Hongkun Wu, Mingguo Cao, Yunxiang Li, Yaqi Wang, and Xingru Huang.
\newblock Children’s dental panoramic radiographs dataset for caries segmentation and dental disease detection.
\newblock {\em Scientific Data}, 10(1):380, 2023.

\bibitem{hamamci2023diffusion}
Ibrahim~Ethem Hamamci, Sezgin Er, Enis Simsar, Anjany Sekuboyina, Mustafa Gundogar, Bernd Stadlinger, Albert Mehl, and Bjoern Menze.
\newblock Diffusion-based hierarchical multi-label object detection to analyze panoramic dental x-rays.
\newblock {\em arXiv preprint arXiv:2303.06500}, 2023.

\bibitem{liu2023context}
Zhengliang Liu, Xinyu He, Lei Liu, Tianming Liu, and Xiaoming Zhai.
\newblock Context matters: A strategy to pre-train language model for science education.
\newblock {\em arXiv preprint arXiv:2301.12031}, 2023.

\bibitem{rezayi2022clinicalradiobert}
Saed Rezayi, Haixing Dai, Zhengliang Liu, Zihao Wu, Akarsh Hebbar, Andrew~H. Burns, Lin Zhao, Dajiang Zhu, Xiang Li, Quanzheng Li, et~al.
\newblock Clinicalradiobert: Knowledge-infused few shot learning for clinical notes named entity recognition.
\newblock In {\em The 13th International Workshop on Machine Learning in Medical Imaging (MLMI 2022)}, 2022.

\bibitem{liao2023mask}
Wenxiong Liao, Zhengliang Liu, Haixing Dai, Zihao Wu, Yiyang Zhang, Xiaoke Huang, Yuzhong Chen, Xi~Jiang, Dajiang Zhu, Tianming Liu, et~al.
\newblock Mask-guided bert for few shot text classification.
\newblock {\em arXiv preprint arXiv:2302.10447}, 2023.

\bibitem{cai2022coarse}
Homgmin Cai, Wenxiong Liao, Zhengliang Liu, Xiaoke Huang, Yiyang Zhang, Siqi Ding, Sheng Li, Quanzheng Li, Tianming Liu, and Xiang Li.
\newblock Coarse-to-fine knowledge graph domain adaptation based on distantly-supervised iterative training.
\newblock {\em arXiv preprint arXiv:2211.02849}, 2022.

\bibitem{zhao2022embedding}
Lin Zhao, Zihao Wu, Haixing Dai, Zhengliang Liu, Tuo Zhang, Dajiang Zhu, and Tianming Liu.
\newblock Embedding human brain function via transformer.
\newblock In {\em International Conference on Medical Image Computing and Computer-Assisted Intervention}, pages 366--375. Springer Nature Switzerland Cham, 2022.

\bibitem{ding2022accurate}
Y~Ding, Z~Liu, H~Feng, J~Holmes, Y~Yang, N~Yu, T~Sio, S~Schild, B~Li, and W~Liu.
\newblock Accurate and efficient deep neural network based deformable image registration method in lung cancer.
\newblock In {\em MEDICAL PHYSICS}, volume~49, pages E148--E148. WILEY 111 RIVER ST, HOBOKEN 07030-5774, NJ USA, 2022.

\bibitem{rezayi2022agribert}
Saed Rezayi, Zhengliang Liu, Zihao Wu, Chandra Dhakal, Bao Ge, Chen Zhen, Tianming Liu, and Sheng Li.
\newblock Agribert: Knowledge-infused agricultural language models for matching food and nutrition.
\newblock In {\em International Joint Conference on Artificial Intelligence}, 2022.

\bibitem{tang2023policygpt}
Chenhao Tang, Zhengliang Liu, Chong Ma, Zihao Wu, Yiwei Li, Wei Liu, Dajiang Zhu, Quanzheng Li, Xiang Li, Tianming Liu, et~al.
\newblock Policygpt: Automated analysis of privacy policies with large language models.
\newblock {\em arXiv preprint arXiv:2309.10238}, 2023.

\bibitem{dai2023auggpt}
Haixing Dai, Zhengliang Liu, Wenxiong Liao, Xiaoke Huang, Yihan Cao, Zihao Wu, Lin Zhao, Shaochen Xu, Wei Liu, Ninghao Liu, et~al.
\newblock Auggpt: Leveraging chatgpt for text data augmentation.
\newblock {\em arXiv preprint arXiv:2302.13007}, 2023.

\bibitem{openai2023gpt}
OpenAI.
\newblock Gpt-4 technical report.
\newblock {\em arXiv}, 2023.

\bibitem{touvron2023llama}
Hugo Touvron, Louis Martin, Kevin Stone, Peter Albert, Amjad Almahairi, Yasmine Babaei, Nikolay Bashlykov, Soumya Batra, Prajjwal Bhargava, Shruti Bhosale, et~al.
\newblock Llama 2: Open foundation and fine-tuned chat models.
\newblock {\em arXiv preprint arXiv:2307.09288}, 2023.

\bibitem{chowdhery2022palm}
Aakanksha Chowdhery, Sharan Narang, Jacob Devlin, Maarten Bosma, Gaurav Mishra, Adam Roberts, Paul Barham, Hyung~Won Chung, Charles Sutton, Sebastian Gehrmann, et~al.
\newblock Palm: Scaling language modeling with pathways.
\newblock {\em arXiv preprint arXiv:2204.02311}, 2022.

\bibitem{cai2021chestxraybert}
Xiaoyan Cai, Sen Liu, Junwei Han, Libin Yang, Zhenguo Liu, and Tianming Liu.
\newblock Chestxraybert: A pretrained language model for chest radiology report summarization.
\newblock {\em IEEE Transactions on Multimedia}, 2021.

\bibitem{huang2019clinicalbert}
Kexin Huang, Jaan Altosaar, and Rajesh Ranganath.
\newblock Clinicalbert: Modeling clinical notes and predicting hospital readmission.
\newblock {\em arXiv preprint arXiv:1904.05342}, 2019.

\bibitem{liu2023radiologyllama2}
Zhengliang Liu, Yiwei Li, Peng Shu, Aoxiao Zhong, Longtao Yang, Chao Ju, Zihao Wu, Chong Ma, Jie Luo, Cheng Chen, et~al.
\newblock Radiology-llama2: Best-in-class large language model for radiology.
\newblock {\em arXiv preprint arXiv:2309.06419}, 2023.

\bibitem{xiong2023doctorglm}
Honglin Xiong, Sheng Wang, Yitao Zhu, Zihao Zhao, Yuxiao Liu, Qian Wang, and Dinggang Shen.
\newblock Doctorglm: Fine-tuning your chinese doctor is not a herculean task.
\newblock {\em arXiv preprint arXiv:2304.01097}, 2023.

\bibitem{du2022glm}
Zhengxiao Du, Yujie Qian, Xiao Liu, Ming Ding, Jiezhong Qiu, Zhilin Yang, and Jie Tang.
\newblock Glm: General language model pretraining with autoregressive blank infilling.
\newblock In {\em Proceedings of the 60th Annual Meeting of the Association for Computational Linguistics (Volume 1: Long Papers)}, pages 320--335, 2022.

\bibitem{dosovitskiy2020image}
Alexey Dosovitskiy, Lucas Beyer, Alexander Kolesnikov, Dirk Weissenborn, Xiaohua Zhai, Thomas Unterthiner, Mostafa Dehghani, Matthias Minderer, Georg Heigold, Sylvain Gelly, et~al.
\newblock An image is worth 16x16 words: Transformers for image recognition at scale.
\newblock In {\em International Conference on Learning Representations}, 2020.

\bibitem{touvron2021training}
Hugo Touvron, Matthieu Cord, Matthijs Douze, Francisco Massa, Alexandre Sablayrolles, and Herv{\'e} J{\'e}gou.
\newblock Training data-efficient image transformers \& distillation through attention.
\newblock In {\em International Conference on Machine Learning}, pages 10347--10357. PMLR, 2021.

\bibitem{liu2021swin}
Ze~Liu, Yutong Lin, Yue Cao, Han Hu, Yixuan Wei, Zheng Zhang, Stephen Lin, and Baining Guo.
\newblock Swin transformer: Hierarchical vision transformer using shifted windows.
\newblock In {\em Proceedings of the IEEE/CVF International Conference on Computer Vision}, pages 10012--10022, 2021.

\bibitem{he2022masked}
Kaiming He, Xinlei Chen, Saining Xie, Yanghao Li, Piotr Doll{\'a}r, and Ross Girshick.
\newblock Masked autoencoders are scalable vision learners.
\newblock In {\em Proceedings of the IEEE/CVF Conference on Computer Vision and Pattern Recognition}, pages 16000--16009, 2022.

\bibitem{chen2021empirical}
Xinlei Chen, Saining Xie, and Kaiming He.
\newblock An empirical study of training self-supervised vision transformers.
\newblock In {\em Proceedings of the IEEE/CVF International Conference on Computer Vision}, pages 9640--9649, 2021.

\bibitem{gupta2022maskvit}
Agrim Gupta, Stephen Tian, Yunzhi Zhang, Jiajun Wu, Roberto Mart{\'\i}n-Mart{\'\i}n, and Li~Fei-Fei.
\newblock Maskvit: Masked visual pre-training for video prediction.
\newblock {\em arXiv preprint arXiv:2206.11894}, 2022.

\bibitem{ma2023eye}
Chong Ma, Lin Zhao, Yuzhong Chen, Sheng Wang, Lei Guo, Tuo Zhang, Dinggang Shen, Xi~Jiang, and Tianming Liu.
\newblock Eye-gaze-guided vision transformer for rectifying shortcut learning.
\newblock {\em IEEE Transactions on Medical Imaging}, 2023.

\bibitem{ma2023rectify}
Chong Ma, Lin Zhao, Yuzhong Chen, Lei Guo, Tuo Zhang, Xintao Hu, Dinggang Shen, Xi~Jiang, and Tianming Liu.
\newblock Rectify vit shortcut learning by visual saliency.
\newblock {\em IEEE Transactions on Neural Networks and Learning Systems}, 2023.

\bibitem{yu2023core}
Xiaowei Yu, Lu~Zhang, Haixing Dai, Yanjun Lyu, Lin Zhao, Zihao Wu, David Liu, Tianming Liu, and Dajiang Zhu.
\newblock Core-periphery principle guided redesign of self-attention in transformers.
\newblock {\em arXiv preprint arXiv:2303.15569}, 2023.

\bibitem{xiao2023instruction}
Zhenxiang Xiao, Yuzhong Chen, Lu~Zhang, Junjie Yao, Zihao Wu, Xiaowei Yu, Yi~Pan, Lin Zhao, Chong Ma, Xinyu Liu, et~al.
\newblock Instruction-vit: Multi-modal prompts for instruction learning in vit.
\newblock {\em arXiv preprint arXiv:2305.00201}, 2023.

\bibitem{kirillov2023segment}
Alexander Kirillov, Eric Mintun, Nikhila Ravi, Hanzi Mao, Chloe Rolland, Laura Gustafson, Tete Xiao, Spencer Whitehead, Alexander~C Berg, Wan-Yen Lo, et~al.
\newblock Segment anything.
\newblock {\em arXiv preprint arXiv:2304.02643}, 2023.

\bibitem{radford2021learning}
Alec Radford, Jong~Wook Kim, Chris Hallacy, Aditya Ramesh, Gabriel Goh, Sandhini Agarwal, Girish Sastry, Amanda Askell, Pamela Mishkin, Jack Clark, et~al.
\newblock Learning transferable visual models from natural language supervision.
\newblock In {\em International conference on machine learning}, pages 8748--8763. PMLR, 2021.

\bibitem{li2022blip}
Junnan Li, Dongxu Li, Caiming Xiong, and Steven Hoi.
\newblock Blip: Bootstrapping language-image pre-training for unified vision-language understanding and generation.
\newblock In {\em International Conference on Machine Learning}, pages 12888--12900. PMLR, 2022.

\bibitem{kim2021vilt}
Wonjae Kim, Bokyung Son, and Ildoo Kim.
\newblock Vilt: Vision-and-language transformer without convolution or region supervision.
\newblock In {\em International Conference on Machine Learning}, pages 5583--5594. PMLR, 2021.

\bibitem{yu2022coca}
Jiahui Yu, Zirui Wang, Vijay Vasudevan, Legg Yeung, Mojtaba Seyedhosseini, and Yonghui Wu.
\newblock Coca: Contrastive captioners are image-text foundation models.
\newblock {\em arXiv preprint arXiv:2205.01917}, 2022.

\bibitem{wang2023image}
Wenhui Wang, Hangbo Bao, Li~Dong, Johan Bjorck, Zhiliang Peng, Qiang Liu, Kriti Aggarwal, Owais~Khan Mohammed, Saksham Singhal, Subhojit Som, et~al.
\newblock Image as a foreign language: Beit pretraining for vision and vision-language tasks.
\newblock In {\em Proceedings of the IEEE/CVF Conference on Computer Vision and Pattern Recognition}, pages 19175--19186, 2023.

\bibitem{wang2022medclip}
Zifeng Wang, Zhenbang Wu, Dinesh Agarwal, and Jimeng Sun.
\newblock Medclip: Contrastive learning from unpaired medical images and text.
\newblock {\em arXiv preprint arXiv:2210.10163}, 2022.

\bibitem{huang2021gloria}
Shih-Cheng Huang, Liyue Shen, Matthew~P Lungren, and Serena Yeung.
\newblock Gloria: A multimodal global-local representation learning framework for label-efficient medical image recognition.
\newblock In {\em Proceedings of the IEEE/CVF International Conference on Computer Vision}, pages 3942--3951, 2021.

\bibitem{wei2023chat2brain}
Yaonai Wei, Tuo Zhang, Han Zhang, Tianyang Zhong, Lin Zhao, Zhengliang Liu, Chong Ma, Songyao Zhang, Muheng Shang, Lei Du, et~al.
\newblock Chat2brain: A method for mapping open-ended semantic queries to brain activation maps.
\newblock {\em arXiv preprint arXiv:2309.05021}, 2023.

\bibitem{liu2023pharmacygpt}
Zhengliang Liu, Zihao Wu, Mengxuan Hu, Bokai Zhao, Lin Zhao, Tianyi Zhang, Haixing Dai, Xianyan Chen, Ye~Shen, Sheng Li, et~al.
\newblock Pharmacygpt: The ai pharmacist.
\newblock {\em arXiv preprint arXiv:2307.10432}, 2023.

\bibitem{dou2023towards}
Fei Dou, Jin Ye, Geng Yuan, Qin Lu, Wei Niu, Haijian Sun, Le~Guan, Guoyu Lu, Gengchen Mai, Ninghao Liu, et~al.
\newblock Towards artificial general intelligence (agi) in the internet of things (iot): Opportunities and challenges.
\newblock {\em arXiv preprint arXiv:2309.07438}, 2023.

\bibitem{liu3surviving}
Zhengliang Liu, Lu~Zhang, Zihao Wu, Xiaowei Yu, Chao Cao, Haixing Dai, Ninghao Liu, Jun Liu, Wei Liu, Quanzheng Li, et~al.
\newblock Surviving chatgpt in healthcare.
\newblock {\em Frontiers in Radiology}, 3:1224682, 2023.

\bibitem{guan2023cohortgpt}
Zihan Guan, Zihao Wu, Zhengliang Liu, Dufan Wu, Hui Ren, Quanzheng Li, Xiang Li, and Ninghao Liu.
\newblock Cohortgpt: An enhanced gpt for participant recruitment in clinical study.
\newblock {\em arXiv preprint arXiv:2307.11346}, 2023.

\bibitem{zhong2023chatabl}
Tianyang Zhong, Yaonai Wei, Li~Yang, Zihao Wu, Zhengliang Liu, Xiaozheng Wei, Wenjun Li, Junjie Yao, Chong Ma, Xiang Li, et~al.
\newblock Chatabl: Abductive learning via natural language interaction with chatgpt.
\newblock {\em arXiv preprint arXiv:2304.11107}, 2023.

\bibitem{zhao2023meta}
Lin Zhao, Lu~Zhang, Zihao Wu, Yuzhong Chen, Haixing Dai, Xiaowei Yu, Zhengliang Liu, Tuo Zhang, Xintao Hu, Xi~Jiang, et~al.
\newblock Meta-radiology.
\newblock {\em Meta}, 1:100005, 2023.

\bibitem{dai2023ad}
Haixing Dai, Yiwei Li, Zhengliang Liu, Lin Zhao, Zihao Wu, Suhang Song, Ye~Shen, Dajiang Zhu, Xiang Li, Sheng Li, et~al.
\newblock Ad-autogpt: An autonomous gpt for alzheimer's disease infodemiology.
\newblock {\em arXiv preprint arXiv:2306.10095}, 2023.

\bibitem{shi2023mededit}
Yucheng Shi, Shaochen Xu, Zhengliang Liu, Tianming Liu, Xiang Li, and Ninghao Liu.
\newblock Mededit: Model editing for medical question answering with external knowledge bases.
\newblock {\em arXiv preprint arXiv:2309.16035}, 2023.

\bibitem{rezayi2023exploring}
Saed Rezayi, Zhengliang Liu, Zihao Wu, Chandra Dhakal, Bao Ge, Haixing Dai, Gengchen Mai, Ninghao Liu, Chen Zhen, Tianming Liu, et~al.
\newblock Exploring new frontiers in agricultural nlp: Investigating the potential of large language models for food applications.
\newblock {\em arXiv preprint arXiv:2306.11892}, 2023.

\bibitem{liao2023differentiate}
Wenxiong Liao, Zhengliang Liu, Haixing Dai, Shaochen Xu, Zihao Wu, Yiyang Zhang, Xiaoke Huang, Dajiang Zhu, Hongmin Cai, Tianming Liu, et~al.
\newblock Differentiate chatgpt-generated and human-written medical texts.
\newblock {\em arXiv preprint arXiv:2304.11567}, 2023.

\bibitem{zhu2023minigpt}
Deyao Zhu, Jun Chen, Xiaoqian Shen, Xiang Li, and Mohamed Elhoseiny.
\newblock Minigpt-4: Enhancing vision-language understanding with advanced large language models.
\newblock {\em arXiv preprint arXiv:2304.10592}, 2023.

\bibitem{ye2023mplugowl}
Qinghao Ye, Haiyang Xu, Guohai Xu, Jiabo Ye, Ming Yan, Yiyang Zhou, Junyang Wang, Anwen Hu, Pengcheng Shi, Yaya Shi, Chenliang Li, Yuanhong Xu, Hehong Chen, Junfeng Tian, Qian Qi, Ji~Zhang, and Fei Huang.
\newblock mplug-owl: Modularization empowers large language models with multimodality.
\newblock {\em arXiv preprint arXiv:2304.14178}, 2023.

\bibitem{wu2023visual}
Chenfei Wu, Shengming Yin, Weizhen Qi, Xiaodong Wang, Zecheng Tang, and Nan Duan.
\newblock Visual chatgpt: Talking, drawing and editing with visual foundation models.
\newblock {\em arXiv preprint arXiv:2303.04671}, 2023.

\bibitem{yang2023mmreact}
Zhengyuan Yang, Linjie Li, Jianfeng Wang, Kevin Lin, Ehsan Azarnasab, Faisal Ahmed, Zicheng Liu, Ce~Liu, Michael Zeng, and Lijuan Wang.
\newblock Mm-react: Prompting chatgpt for multimodal reasoning and action.
\newblock {\em arXiv preprint arXiv:2303.11381}, 2023.

\bibitem{flamingo2022}
Jean-Baptiste Alayrac, Jeff Donahue, Pauline Luc, Antoine Miech, Iain Barr, Yana Hasson, Karel Lenc, Arthur Mensch, Katherine Millican, Malcolm Reynolds, Roman Ring, Eliza Rutherford, Serkan Cabi, Tengda Han, Zhitao Gong, Sina Samangooei, Marianne Monteiro, Jacob~L Menick, Sebastian Borgeaud, Andy Brock, Aida Nematzadeh, Sahand Sharifzadeh, Miko\l~aj Bi\'{n}kowski, Ricardo Barreira, Oriol Vinyals, Andrew Zisserman, and Kar\'{e}n Simonyan.
\newblock Flamingo: a visual language model for few-shot learning.
\newblock {\em Advances in Neural Information Processing Systems}, 35:23716--23736, 2022.

\bibitem{li2023blip2}
Junnan Li, Dongxu Li, Silvio Savarese, and Steven Hoi.
\newblock Blip-2: Bootstrapping language-image pre-training with frozen image encoders and large language models.
\newblock {\em arXiv preprint arXiv:2301.12597}, 2023.

\bibitem{driess2023palme}
Danny Driess, Fei Xia, Mehdi S.~M. Sajjadi, Corey Lynch, Aakanksha Chowdhery, Brian Ichter, Ayzaan Wahid, Jonathan Tompson, Quan Vuong, Tianhe Yu, Wenlong Huang, Yevgen Chebotar, Pierre Sermanet, Daniel Duckworth, Sergey Levine, Vincent Vanhoucke, Karol Hausman, Marc Toussaint, Klaus Greff, Andy Zeng, Igor Mordatch, and Pete Florence.
\newblock Palm-e: An embodied multimodal language model.
\newblock {\em arXiv preprint arXiv:2303.03378}, 2023.

\bibitem{gong2023multimodalgpt}
Tao Gong, Chengqi Lyu, Shilong Zhang, Yudong Wang, Miao Zheng, Qian Zhao, Kuikun Liu, Wenwei Zhang, Ping Luo, and Kai Chen.
\newblock Multimodal-gpt: A vision and language model for dialogue with humans.
\newblock {\em arXiv preprint arXiv:2305.04790}, 2023.

\bibitem{li2023otter}
Bo~Li, Yuanhan Zhang, Liangyu Chen, Jinghao Wang, Jingkang Yang, and Ziwei Liu.
\newblock Otter: A multi-modal model with in-context instruction tuning.
\newblock {\em arXiv preprint arXiv:2305.03726}, 2023.

\bibitem{liu2023llava}
Haotian Liu, Chunyuan Li, Qingyang Wu, and Yong~Jae Lee.
\newblock Visual instruction tuning.
\newblock {\em arXiv preprint arXiv:2304.08485}, 2023.

\bibitem{wu2023gpt4vision}
Chaoyi Wu, Jiayu Lei, Qiaoyu Zheng, Weike Zhao, Weixiong Lin, Xiaoman Zhang, Xiao Zhou, Ziheng Zhao, Ya~Zhang, Yanfeng Wang, and Weidi Xie.
\newblock Can gpt-4v(ision) serve medical applications? case studies on gpt-4v for multimodal medical diagnosis.
\newblock {\em arXiv preprint arXiv:2310.09909}, 2023.

\bibitem{zhang2023instruction}
Shengyu Zhang, Linfeng Dong, Xiaoya Li, Sen Zhang, Xiaofei Sun, Shuhe Wang, Jiwei Li, Runyi Hu, Tianwei Zhang, Fei Wu, et~al.
\newblock Instruction tuning for large language models: A survey.
\newblock {\em arXiv preprint arXiv:2308.10792}, 2023.

\bibitem{wang2023review}
Jiaqi Wang, Zhengliang Liu, Lin Zhao, Zihao Wu, Chong Ma, Sigang Yu, Haixing Dai, Qiushi Yang, Yiheng Liu, Songyao Zhang, et~al.
\newblock Review of large vision models and visual prompt engineering.
\newblock {\em arXiv preprint arXiv:2307.00855}, 2023.

\bibitem{wei2021finetuned}
Jason Wei, Maarten Bosma, Vincent~Y Zhao, Kelvin Guu, Adams~Wei Yu, Brian Lester, Nan Du, Andrew~M Dai, and Quoc~V Le.
\newblock Finetuned language models are zero-shot learners.
\newblock {\em arXiv preprint arXiv:2109.01652}, 2021.

\bibitem{wang2022ofa}
Peng Wang, An~Yang, Rui Men, Junyang Lin, Shuai Bai, Zhikang Li, Jianxin Ma, Chang Zhou, Jingren Zhou, and Hongxia Yang.
\newblock Ofa: Unifying architectures, tasks, and modalities through a simple sequence-to-sequence learning framework.
\newblock In {\em International Conference on Machine Learning}, pages 23318--23340. PMLR, 2022.

\bibitem{xu2022multiinstruct}
Zhiyang Xu, Ying Shen, and Lifu Huang.
\newblock Multiinstruct: Improving multi-modal zero-shot learning via instruction tuning.
\newblock {\em arXiv preprint arXiv:2212.10773}, 2022.

\bibitem{wei2022emergent}
Jason Wei, Yi~Tay, Rishi Bommasani, Colin Raffel, Barret Zoph, Sebastian Borgeaud, Dani Yogatama, Maarten Bosma, Denny Zhou, Donald Metzler, et~al.
\newblock Emergent abilities of large language models.
\newblock {\em arXiv preprint arXiv:2206.07682}, 2022.

\bibitem{ozturkler2022thinksum}
Batu Ozturkler, Nikolay Malkin, Zhen Wang, and Nebojsa Jojic.
\newblock Thinksum: Probabilistic reasoning over sets using large language models.
\newblock {\em arXiv preprint arXiv:2210.01293}, 2022.

\bibitem{liu2023tailoring}
Zhengliang Liu, Aoxiao Zhong, Yiwei Li, Longtao Yang, Chao Ju, Zihao Wu, Chong Ma, Peng Shu, Cheng Chen, Sekeun Kim, et~al.
\newblock Tailoring large language models to radiology: A preliminary approach to llm adaptation for a highly specialized domain.
\newblock In {\em International Workshop on Machine Learning in Medical Imaging}, pages 464--473. Springer, 2023.

\bibitem{bhatt2018trends}
NR~Bhatt, E~Dunne, M~Faraz, AE~Gillis, KC~Conlon, S~Paran, and PF~Ridgway.
\newblock Trends in the use of laparoscopic versus open paediatric appendicectomy: A regional 12-year study and a national survey.
\newblock {\em World Journal of Surgery}, 42:3792--3802, 2018.

\bibitem{zhao2019real}
Zijian Zhao, Tongbiao Cai, Faliang Chang, and Xiaolin Cheng.
\newblock Real-time surgical instrument detection in robot-assisted surgery using a convolutional neural network cascade.
\newblock {\em Healthcare technology letters}, 6(6):275--279, 2019.

\bibitem{ward2021computer}
Thomas~M Ward, Pietro Mascagni, Yutong Ban, Guy Rosman, Nicolas Padoy, Ozanan Meireles, and Daniel~A Hashimoto.
\newblock Computer vision in surgery.
\newblock {\em Surgery}, 169(5):1253--1256, 2021.

\bibitem{van2021gesture}
Beatrice van Amsterdam, Matthew~J Clarkson, and Danail Stoyanov.
\newblock Gesture recognition in robotic surgery: a review.
\newblock {\em IEEE Transactions on Biomedical Engineering}, 68(6), 2021.

\bibitem{garrow2021machine}
Carly~R Garrow, Karl-Friedrich Kowalewski, Linhong Li, Martin Wagner, Mona~W Schmidt, Sandy Engelhardt, Daniel~A Hashimoto, Hannes~G Kenngott, Sebastian Bodenstedt, Stefanie Speidel, et~al.
\newblock Machine learning for surgical phase recognition: a systematic review.
\newblock {\em Annals of surgery}, 273(4):684--693, 2021.

\bibitem{anil2023palm}
Rohan Anil, Andrew~M Dai, Orhan Firat, Melvin Johnson, Dmitry Lepikhin, Alexandre Passos, Siamak Shakeri, Emanuel Taropa, Paige Bailey, Zhifeng Chen, et~al.
\newblock Palm 2 technical report.
\newblock {\em arXiv preprint arXiv:2305.10403}, 2023.

\bibitem{wu2023tidybot}
Jimmy Wu, Rika Antonova, Adam Kan, Marion Lepert, Andy Zeng, Shuran Song, Jeannette Bohg, Szymon Rusinkiewicz, and Thomas Funkhouser.
\newblock Tidybot: Personalized robot assistance with large language models.
\newblock {\em arXiv preprint arXiv:2305.05658}, 2023.

\bibitem{zhang2023building}
Hongxin Zhang, Weihua Du, Jiaming Shan, Qinhong Zhou, Yilun Du, Joshua~B Tenenbaum, Tianmin Shu, and Chuang Gan.
\newblock Building cooperative embodied agents modularly with large language models.
\newblock {\em arXiv preprint arXiv:2307.02485}, 2023.

\bibitem{gadre2023cows}
Samir~Yitzhak Gadre, Mitchell Wortsman, Gabriel Ilharco, Ludwig Schmidt, and Shuran Song.
\newblock Cows on pasture: Baselines and benchmarks for language-driven zero-shot object navigation.
\newblock In {\em Proceedings of the IEEE/CVF Conference on Computer Vision and Pattern Recognition}, pages 23171--23181, 2023.

\bibitem{vemprala2023chatgpt}
Sai Vemprala, Rogerio Bonatti, Arthur Bucker, and Ashish Kapoor.
\newblock Chatgpt for robotics: Design principles and model abilities.
\newblock {\em Microsoft Auton. Syst. Robot. Res}, 2:20, 2023.

\bibitem{huang2023instruct2act}
Siyuan Huang, Zhengkai Jiang, Hao Dong, Yu~Qiao, Peng Gao, and Hongsheng Li.
\newblock Instruct2act: Mapping multi-modality instructions to robotic actions with large language model.
\newblock {\em arXiv preprint arXiv:2305.11176}, 2023.

\bibitem{szot2023large}
Andrew Szot, Max Schwarzer, Harsh Agrawal, Bogdan Mazoure, Walter Talbott, Katherine Metcalf, Natalie Mackraz, Devon Hjelm, and Alexander Toshev.
\newblock Large language models as generalizable policies for embodied tasks.
\newblock {\em arXiv preprint arXiv:2310.17722}, 2023.

\bibitem{lin2023text2motion}
Kevin Lin, Christopher Agia, Toki Migimatsu, Marco Pavone, and Jeannette Bohg.
\newblock Text2motion: From natural language instructions to feasible plans.
\newblock {\em arXiv preprint arXiv:2303.12153}, 2023.

\bibitem{xie2023text2reward}
Tianbao Xie, Siheng Zhao, Chen~Henry Wu, Yitao Liu, Qian Luo, Victor Zhong, Yanchao Yang, and Tao Yu.
\newblock Text2reward: Automated dense reward function generation for reinforcement learning.
\newblock {\em arXiv preprint arXiv:2309.11489}, 2023.

\bibitem{yang2023octopus}
Jingkang Yang, Yuhao Dong, Shuai Liu, Bo~Li, Ziyue Wang, Chencheng Jiang, Haoran Tan, Jiamu Kang, Yuanhan Zhang, Kaiyang Zhou, et~al.
\newblock Octopus: Embodied vision-language programmer from environmental feedback.
\newblock {\em arXiv preprint arXiv:2310.08588}, 2023.

\bibitem{padalkar2023open}
Abhishek Padalkar, Acorn Pooley, Ajinkya Jain, Alex Bewley, Alex Herzog, Alex Irpan, Alexander Khazatsky, Anant Rai, Anikait Singh, Anthony Brohan, et~al.
\newblock Open x-embodiment: Robotic learning datasets and rt-x models.
\newblock {\em arXiv preprint arXiv:2310.08864}, 2023.

\bibitem{brohan2022rt}
Anthony Brohan, Noah Brown, Justice Carbajal, Yevgen Chebotar, Joseph Dabis, Chelsea Finn, Keerthana Gopalakrishnan, Karol Hausman, Alex Herzog, Jasmine Hsu, et~al.
\newblock Rt-1: Robotics transformer for real-world control at scale.
\newblock {\em arXiv preprint arXiv:2212.06817}, 2022.

\bibitem{brohan2023rt}
Anthony Brohan, Noah Brown, Justice Carbajal, Yevgen Chebotar, Xi~Chen, Krzysztof Choromanski, Tianli Ding, Danny Driess, Avinava Dubey, Chelsea Finn, et~al.
\newblock Rt-2: Vision-language-action models transfer web knowledge to robotic control.
\newblock {\em arXiv preprint arXiv:2307.15818}, 2023.

\bibitem{yu2023language}
Wenhao Yu, Nimrod Gileadi, Chuyuan Fu, Sean Kirmani, Kuang-Huei Lee, Montse~Gonzalez Arenas, Hao-Tien~Lewis Chiang, Tom Erez, Leonard Hasenclever, Jan Humplik, et~al.
\newblock Language to rewards for robotic skill synthesis.
\newblock {\em arXiv preprint arXiv:2306.08647}, 2023.

\bibitem{mai2023llm}
Jinjie Mai, Jun Chen, Bing Li, Guocheng Qian, Mohamed Elhoseiny, and Bernard Ghanem.
\newblock Llm as a robotic brain: Unifying egocentric memory and control.
\newblock {\em arXiv preprint arXiv:2304.09349}, 2023.

\bibitem{lightman2023let}
Hunter Lightman, Vineet Kosaraju, Yura Burda, Harri Edwards, Bowen Baker, Teddy Lee, Jan Leike, John Schulman, Ilya Sutskever, and Karl Cobbe.
\newblock Let's verify step by step.
\newblock {\em arXiv preprint arXiv:2305.20050}, 2023.

\bibitem{yao2023tree}
Shunyu Yao, Dian Yu, Jeffrey Zhao, Izhak Shafran, Thomas~L Griffiths, Yuan Cao, and Karthik Narasimhan.
\newblock Tree of thoughts: Deliberate problem solving with large language models.
\newblock {\em arXiv preprint arXiv:2305.10601}, 2023.

\bibitem{xie2023olagpt}
Yuanzhen Xie, Tao Xie, Mingxiong Lin, WenTao Wei, Chenglin Li, Beibei Kong, Lei Chen, Chengxiang Zhuo, Bo~Hu, and Zang Li.
\newblock Olagpt: Empowering llms with human-like problem-solving abilities.
\newblock {\em arXiv preprint arXiv:2305.16334}, 2023.

\bibitem{zhang2023cumulative}
Yifan Zhang, Jingqin Yang, Yang Yuan, and Andrew Chi-Chih Yao.
\newblock Cumulative reasoning with large language models.
\newblock {\em arXiv preprint arXiv:2308.04371}, 2023.

\bibitem{wei2022chain}
Jason Wei, Xuezhi Wang, Dale Schuurmans, Maarten Bosma, Fei Xia, Ed~Chi, Quoc~V Le, Denny Zhou, et~al.
\newblock Chain-of-thought prompting elicits reasoning in large language models.
\newblock {\em Advances in Neural Information Processing Systems}, 35:24824--24837, 2022.

\bibitem{zhang2023graph}
Jiawei Zhang.
\newblock Graph-toolformer: To empower llms with graph reasoning ability via prompt augmented by chatgpt.
\newblock {\em arXiv preprint arXiv:2304.11116}, 2023.

\bibitem{assran2023self}
Mahmoud Assran, Quentin Duval, Ishan Misra, Piotr Bojanowski, Pascal Vincent, Michael Rabbat, Yann LeCun, and Nicolas Ballas.
\newblock Self-supervised learning from images with a joint-embedding predictive architecture.
\newblock In {\em Proceedings of the IEEE/CVF Conference on Computer Vision and Pattern Recognition}, pages 15619--15629, 2023.

\bibitem{christiano2017deep}
Paul~F Christiano, Jan Leike, Tom Brown, Miljan Martic, Shane Legg, and Dario Amodei.
\newblock Deep reinforcement learning from human preferences.
\newblock {\em Advances in neural information processing systems}, 30, 2017.

\end{thebibliography}
\bibliographystyle{unsrt}

\end{document}